\documentclass[11pt]{article}
\usepackage[margin=2cm]{geometry}
\pdfoutput=1

\usepackage[colorlinks=true,citecolor=darkred,urlcolor=darkred, pdfborder={0 0 0}]{hyperref}
\hypersetup{linkcolor=blue}
\usepackage{amssymb}
\usepackage{amsmath}
\usepackage{amsfonts}
\usepackage{cite}
\usepackage{bbold}
\usepackage{bm} 
\usepackage{xfrac} 
\usepackage{color}
\usepackage{colordvi}
\usepackage{mathtools,braket,blkarray}
\usepackage{multirow}
\usepackage{float}
\usepackage{comment}
\usepackage{slashed}
\usepackage{enumitem}  
\usepackage{cancel}
\usepackage{graphicx}
\usepackage{booktabs}
\usepackage{longtable}
\usepackage[ruled,linesnumbered]{algorithm2e}
\usepackage{siunitx}
\usepackage{amsthm}

\usepackage{xcolor}
\usepackage{multirow}
\usepackage{braket}
\usepackage{comment}
\usepackage{mathtools,extarrows}
\usepackage{setspace}

\allowdisplaybreaks
\newcommand {\ignore}[1]{}

\definecolor{darkred}{rgb}{0.6,0,0}
\definecolor{dgreen}{rgb}{0,0.5,0}

\newcommand {\black} {\color{black}}

\def\vev#1{\left\langle #1\right\rangle}

\def\e6{$\mathrm{E(6)}$ }
\def\10{$\mathrm{SO(10)}$ }
\def\21{$\mathrm{SU(2)_L \otimes U(1)_Y}$ }
\def\31{$\mathrm{SU(3)_c \otimes U(1)_Q}$ }
\def\SM{$\mathrm{SU(3)_c \otimes SU(2)_L \otimes U(1)_Y}$ }
\newcommand{\sm}{{Standard Model }}

\def\3211{$\mathrm{SU(3) \otimes SU(2)_L \otimes U(1)_R \otimes U(1)_{B-L}}$ }
\def\321{$\mathrm{SU(3) \otimes SU(2) \otimes U(1)}$ }
\def\422{$\mathrm{SU(4) \otimes SU(2) \otimes SU(2)_R}$ }

\def \znbb {$0\nu\beta\beta$ }
\def \zn4b {$0\nu4\beta$ }
\def\lfv{lepton flavour violation }
\def\lnv{lepton number violation }

\makeatletter
\@addtoreset{equation}{section}
\makeatother

\begin{document}

\title{\Large\bf The symmetry approach to quark and lepton masses and mixing }

\author{Gui-Jun~Ding$^{a}$
\thanks{E-mail: {\tt dinggj@ustc.edu.cn}},\
Jos\'e W.F. Valle$^{b}$
\thanks{E-mail: {\tt valle@ific.uv.es}} \\*[20pt]
\centerline{
\begin{minipage}{\linewidth}
\begin{center}
$^a${\it \small
Department of Modern Physics, University of Science and Technology of China,\\
Hefei, Anhui 230026, China}\\[2mm]
$^b${\it \small
AHEP Group, Instituto de F\'{\i}sica Corpuscular -
CSIC/Universitat de Val{\`e}ncia,
Parque Cient{\'i}fico\\ C/Catedratico Jos\'e Beltr\'an, 2, ~~E-46980 Paterna (Val\`{e}ncia) - SPAIN}\\
\end{center}
\end{minipage}}}

\date{}

\maketitle

\begin{abstract}
The Standard Model lacks an organizing principle to describe quark and lepton ``flavours''. Neutrino oscillation experiments show that leptons mix very differently from quarks, adding a major challenge to the flavour puzzle. We briefly sketch the seesaw and the dark-matter-mediated ``scotogenic'' neutrino mass generation approaches.
We discuss the limitations of popular neutrino mixing patterns and examine the possibility that they arise from symmetry, giving a bottom-up approach to residual flavour and CP symmetries. We show how such family and/or CP symmetries can yield novel, viable and predictive mixing patterns. Model-independent ways to predict lepton mixing and neutrino mass sum rules are reviewed. We also discuss UV-complete flavour theories in four and more space-time dimensions. As benchmark examples we present an $A_4$ scotogenic construction with trimaximal mixing pattern TM2 and another with $S_4$ flavour symmetry and generalized CP symmetry. Higher-dimensional flavour completions are also briefly discussed, such as 5-D warped flavordynamics with a $T^\prime$ symmetry yielding a TM1
mixing pattern, detectable neutrinoless double beta decay rates and a very good global fit of flavour observables. We also mention 6-D orbifolds as a way to fix the structure of the 4-D family symmetry. We give a scotogenic benchmark orbifold model predicting the ``golden'' quark-lepton mass relation, stringent neutrino oscillation parameter regions, and an excellent global flavour fit, including quark observables. Finally, we discuss promising recent progress in tackling the flavor issue through the use of modular symmetries.
\end{abstract}

{\it Keywords}:{ Fermion mixing, CP violation, generalized CP, flavor and modular symmetry, \\ \indent ~~~~~~~~~~~~~~ orbifolds, warped-flavordynamics}

\newpage


\tableofcontents

\section{Introduction}                                                     
\label{sec:introduction}                                                   

Elucidating the spontaneous mechanism of symmetry breaking~\cite{Yang:1954ek} within the Standard Model (SM)~\cite{Glashow:1961tr,Weinberg:1967tq,Salam:1968rm,Glashow:1970gm}, i.e. the existence of a physical Higgs boson~\cite{Guralnik:1964eu,Englert:1964et,Higgs:1966ev}, has so far been the main accomplished goal of the successful LHC programme. This was achieved, at least partially, with the discovery of a scalar particle with properties closely resembling those of the SM Higgs boson, by the ATLAS~\cite{ATLAS:2012yve} and CMS~\cite{CMS:2012qbp} experiments at CERN. More is expected from future studies, for example, at the upcoming FCC facility~\cite{FCC:2018bvk,FCC:2018byv} and other complementary lepton accelerators~\cite{CLICdp:2018cto,Bambade:2019fyw,ILC:2013jhg,Linssen:2012hp,CEPCStudyGroup:2018ghi}.

Another major milestone in elementary particle physics has been the discovery of neutrino oscillations in solar and atmospheric neutrino experiments~\cite{McDonald:2016ixn,Kajita:2016cak}.
These were followed by reactor- and accelerator-based studies, e.g.~\cite{KamLAND:2002uet,DayaBay:2012fng,RENO:2012mkc,T2K:2011ypd} which, altogether, imply nonzero neutrino masses, as well as large mixing angles in the lepton sector~\cite{deSalas:2020pgw,10.5281/zenodo.4726908}~\footnote{There is a fairly good agreement with the other determinations, by the Bari group~\cite{Capozzi:2021fjo} and NuFit~\cite{Esteban:2020cvm}.}. This comes as a surprise, when compared with the pattern seen in the Cabibbo-Kobayashi-Maskawa (CKM) matrix describing quark mixing and CP violation~\cite{Workman:2022ynf}. It also means that, even after the discovery of the Higgs boson, the architecture of particle physics is still quite far from ``complete''.

The origin of neutrino masses is one of the deepest secrets of modern particle physics~\cite{Weinberg:1979sa}. The most general neutrino mass generation template is given by the \SM gauge theory framework characterizing the SM~\cite{Schechter:1980gr,Cheng:1980qt,Schechter:1981cv}. In this review we will be mainly concerned with bottom-up neutrino mass generation approaches, for which this is the most appropriate choice. The seesaw has been widely discussed in terms of left-right theories and SO(10)~\cite{Minkowski:1977sc,Gell-Mann:1979vob,Mohapatra:1979ia,Magg:1980ut,Lazarides:1980nt,Mohapatra:1980yp}.
Nonetheless we stress that, besides the high-scale seesaw, one can also have mechanisms generating neutrino masses at low scales~\cite{Mohapatra:1986bd,Gonzalez-Garcia:1988okv,Akhmedov:1995vm,Akhmedov:1995ip,Malinsky:2005bi,Boucenna:2014zba,Cai:2017jrq}.

Following the gauge principle that underlies the SM construction, neutrinos should also get mass through spontaneous symmetry breaking (SSB). Small neutrino masses would be understood dynamically through new vacuum expectation values (VEVs). A specially interesting case is that of spontaneous violation of lepton number~\cite{Chikashige:1980ui,Schechter:1981cv}. Within the SM picture three of the fundamental interactions of nature (electromagnetic, weak and strong) all have a gauge description. It would be very appealing if these could also have a common origin at very high energies~\cite{Georgi:1974sy,Georgi:1974yf,Fritzsch:1974nn,Georgi:1979df,Gell-Mann:1979vob,Dimopoulos:1981zb}, though no hard evidence for this beautiful idea has ever been found. Despite their potential in providing an all-encompassing unified description including also gravity~\cite{Green:1987sp,Polchinski:1998rr,book:971009}, superstring theories have so-far also failed to provide a phenomenologically convincing roadmap.

Another major drawback of the SM construction is that it fails to explain family replication, fermion mass hierarchies and mixing pattern. The discovery of oscillations has only exacerbated this fact. The disparity observed in the pattern of quark and lepton mixing parameters appears to us unlikely to be the result of pure chance. As a result, we will not examine the possibility of neutrino mixing anarchy~\cite{Hall:1999sn,deGouvea:2012ac}. Although viable, we find this hypothesis theoretically unsatisfactory. Instead, our main common thread in this review will be the symmetry approach to the ``flavour problem''.

Here we will be mainly concerned with explaining the detailed pattern of the weak interactions of quarks and leptons within family-symmetry extended theories based on the \SM gauge group, with special emphasis on the case of leptons. Early attempts to understand the lepton mixing pattern starting from the quark sector have now become obsolete, since the discovery of neutrino oscillations. A fully successful flavour theory should explain not only the observed large mixing angles in the lepton sector but also the CKM mixing pattern governing quark mixing and CP violation. Likewise, it should also account for the pattern of quark and lepton masses. In this review we will illustrate how this may, at least partially, be achieved either within 4-dimensional renormalizable gauge field theories~\cite{Ma:2001dn,Babu:2002dz,Hirsch:2003dr,Ma:2006sk,Hirsch:2008mg,Hirsch:2008rp,Hirsch:2009mx,Hirsch:2010ru,Ding:2011gt,King:2011zj,Shimizu:2011xg,Boucenna:2011tj,deMedeirosVarzielas:2012apl,Morisi:2013eca,King:2013hj,Morisi:2013qna,Hirsch:2005mc,Hirsch:2007kh} or in the context of theories with extra spacetime dimensions~\cite{Altarelli:2005yp,Altarelli:2008bg,Csaki:2008qq,Chen:2009gy,Burrows:2009pi,Kadosh:2010rm,Kadosh:2011id,Kadosh:2013nra,Chen:2015jta,Chen:2020udk,Ding:2013eca,CarcamoHernandez:2015iao,delAguila:2010vg,Hagedorn:2011un,Hagedorn:2011pw}. To set up notation for the following chapters here we start with some preliminaries on the gauge-theoretic description of quark and lepton mixing, followed by a very brief critique of the SM drawbacks.

We should stress that there are already several excellent reviews on the discrete flavor symmetry approach to address the SM flavor puzzle~\cite{Ishimori:2010au,Altarelli:2010gt,Morisi:2012fg,King:2013eh,King:2014nza,King:2017guk,Feruglio:2019ybq,Xing:2020ijf,Chauhan:2023faf}. Apart from providing an update to these, the present review will focus on how the residual flavor and CP symmetries can constrain the fermion mixing angles and CP violation phases independently of the details of a specific implementation. We discuss in detail both theory and phenomenological predictions. In addition, we discuss several other topics and predictive benchmark flavor model examples, both in four dimensions and extra dimensions. The latter include warped 5D flavordynamics as well as orbifold-based scenarios, which are reviewed here for the first time.
We discuss extensively the predictions of a broad class of symmetry-based theories of flavor. Finally, we also give a brief discussion of theories based on modular symmetries and how these may help with the vacuum alignment problem.

\subsection{Quark masses and mixing}
\label{sec:quark-masses-mixing}

We recall that in a gauge field theory like the \sm all the fermion masses arise from spontaneous symmetry breaking. The most general quark Yukawa interactions with the Higgs doublet $H$ allowed by symmetry are given by~\cite{Peskin:1995ev,Burgess:2006hbd,Donoghue:1992dd},
\begin{equation}
\mathcal{L}^q_{\text{Yuk}}=-(y_{D})_{ij}\overline{D^{i}_R}H^{\dagger}Q^{j}_L-
(y_{U})_{ij}\overline{U^{i}_R}\widetilde{H}^{\dagger}Q^{j}_L+\text{h.c.}\,,
\end{equation}
where $\widetilde{H}=i\tau_2H^{*}$ and $Q^{i}_L=\left(U^i_L, D^i_L\right)^T$, and $\tau_2=\begin{pmatrix}
0  ~& -i\\
i  ~& 0
\end{pmatrix}$ is a Pauli matrix. The Yukawa matrices $y_D$ and $y_U$ are arbitrary complex matrices in flavour space. Upon electroweak symmetry breaking, the Higgs vacuum expectation value $\vev{ H}=(0, v/\sqrt{2})^{T}$ with $v\simeq 246$GeV gives us the quark mass terms
\begin{equation}
\label{eq:quark_mass_ci_dirac}\mathcal{L}^{q}_{\text{mass}}=-\overline{U}_{R}m_{U}U_{L}-\overline{D}_{R}m_{D}D_{L} +\text{h.c.}\,,
\end{equation}
where $U_L \equiv (u_L, c_L, t_L)^T$, $D_L \equiv (d_L, s_L, b_L)^T$, $U_R \equiv (u_R, c_R, t_R)^T$ and $D_R \equiv (d_R, s_R, b_R)^T$ denote the three generations of left and right-handed up- and down-type quark fields, respectively. The mass matrices are determined by the Yukawa couplings and the Higgs VEV as follows
\begin{equation}
m_U=\frac{v}{\sqrt{2}}y_U, ~~~m_D=\frac{v}{\sqrt{2}}y_D\,.
\end{equation}
Gauge invariance does not constrain the flavour structure of the Yukawa couplings $y_D$ and $y_U$ and therefore $m_U$ and $m_D$ are arbitrary complex matrices.
These matrices can be brought into diagonal form by separate unitary transformations on the left and right fermions, i.e.
\begin{equation}
W_{u}^{\dagger}m_{U}V_{u}=\text{diag}(m_{u},m_{c},m_{t}),\quad W_{d}^{\dagger}m_{D}V_{d}=\text{diag}(m_{d},m_{s},m_{b})\,,
\end{equation}
which implies
\begin{align}
\label{eq:mudmu_diag}&V^{\dagger}_{u}m^{\dagger}_{U}m_{U}V_{u}=\textrm{diag}\left(m^2_{u}, m^2_{c}, m^2_{t}\right)\,,\\
\label{eq:mddmd_diag}&V^{\dagger}_{d}m^{\dagger}_{D}m_{D}V_{d}=\textrm{diag}\left(m^2_{d}, m^2_{s}, m^2_{b}\right)\,.
\end{align}
By performing such unitary transformations $U_{L}\to V_{u}U_{L}$, $D_{L}\to V_{d}D_{L}$, $U_{R}\to W_{u}U_{R}$ and $D_{R}\to  W_{d}D_{R}$ one can go to the physical mass eigenstates. The resulting flavour-changing quark charged current weak interaction reads
\begin{equation}
\mathcal{L}^{q}_{CC}=\frac{g}{\sqrt{2}}\overline{U}_{L}\gamma^{\mu}V_{CKM}D_{L}W_{\mu}^{+}+\text{h.c.}\,,
\end{equation}
where $V_{CKM}\equiv V^{\dagger}_{u}V_{d}$ is the CKM mixing matrix~\cite{Cabibbo:1963yz,Kobayashi:1973fv}. In what follows we adopt the standard form for the CKM matrix describing quark mixing, i.e.
\begin{equation}  \label{eq:CKM}
V_{CKM}=\left(\begin{array}{ccc}
c^q_{12} c^q_{13} & s^q_{12} c^q_{13}  & s^q_{13} e^{-i\delta^q}
\\
-s^q_{12} c^q_{23}- c^q_{12} s^q_{13} s^q_{23} e^{ i \delta^q} & c^q_{12} c^q_{23} - s^q_{12} s^q_{13} s^q_{23} e^{ i \delta^q } & c^q_{13} s^q_{23} \\
s^q_{12} s^q_{23} - c^q_{12} s^q_{13} c^q_{23} e^{ i\delta^q } & - c^q_{12} s^q_{23}  - s^q_{12} s^q_{13} c^q_{23} e^{i\delta^q } & c^q_{13} c^q_{23}
\end{array}
\right)\,,
\end{equation}
with $c^{q}_{ij}\equiv\cos\theta^{q}_{ij}$, $s^{q}_{ij}\equiv\sin\theta^{q}_{ij}$.
This parameterization is the one adopted by the Review of Particle Physics of the Particle Data Group (PDG)~\cite{Workman:2022ynf}, and supplements the original proposal in Ref.~\cite{Schechter:1980gr} by specifying a convenient factor ordering. Notice that only one physical phase remains after the allowed quark phase redefinitions. The CKM-phase-parameter $\delta^q$ is expressed in a neat and rephasing-invariant way as~\cite{Rodejohann:2011vc},
\begin{equation}
  \delta^{q}=\phi_{13}^q-\phi_{12}^q-\phi_{23}^q~,
  \label{eq:dellq}
\end{equation}
in terms of the fundamental $\phi_{ij}^q$ phases. These are the relative phases between up- and down-type diagonalization matrices. From Ref.~\cite{Workman:2022ynf} we extract the following allowed ranges~\cite{UTfit:2006vpt,UTfit} for the CKM parameters,
\begin{equation}\scriptsize
V_{CKM}=\begin{pmatrix}
0.97431\pm0.00012   &  0.22514\pm0.00055   &   (0.00365\pm0.00010)e^{i(-66.8\pm2.0)^{\circ}} \\
(-0.22500\pm0.00054)e^{i(0.0351\pm0.0010)^{\circ}}   &  (0.97344\pm0.00012)e^{i(-0.001880\pm0.000052)^{\circ}}   &  0.04241\pm 0.00065\\
(0.00869\pm0.00014)e^{i(-22.23\pm0.63)^{\circ}}  &  (-0.04124\pm 0.00056)e^{i(1.056\pm0.032)^{\circ}}  &  0.999112\pm0.000024
\end{pmatrix}\,.
\end{equation}

\subsection{Lepton masses and mixing}
\label{sec:lepton-masses-mixing}

It is a characteristic feature of the Standard Model that spontaneous gauge symmetry breaking leaves neutrinos as massless fermions, since right-handed neutrinos are absent. However, the discovery of neutrino oscillations~\cite{McDonald:2016ixn,Kajita:2016cak} implies non-zero neutrino masses and neutrino mixing. One can introduce three right-handed neutrino fields $\nu^i_{R}$ for $i=1, 2, 3$ as full \SM singlets. In analogy with the quark sector the neutrino and charged lepton masses are described by the Yukawa interactions,
\begin{equation}
\label{eq:mdir}
\mathcal{L}^l_{\text{Yuk}}=-(y_{l})_{ij}\overline{l^{i}_R}H^{\dagger}L^{j}_L-
(y_{\nu})_{ij}\overline{\nu^{i}_R}\widetilde{H}^{\dagger}L^{j}_L+\text{h.c.}\,,
\end{equation}
where $L^{i}_L=\left(\nu^i_L, l^i_L\right)^T$ are the lepton doublet fields. In this case neutrinos would be Dirac particles. This requires the \textit{ad hoc} imposition of lepton number symmetry to forbid the right-handed Majorana mass term $M_{ij}\overline{\nu^{i}_R}\;(\nu^j_R)^c$ allowed by SM gauge invariance. Moreover, the Yukawa coupling $y_{\nu}$ should be of order $10^{-11}$ in order to accommodate the neutrino masses below the eV scale.

Under the assumption of lepton number conservation one gets, after electroweak symmetry breaking, the following charged lepton and neutrino mass terms,
\begin{equation}
\label{eq:charge-lepton-neutrino-dirac-mass}\mathcal{L}^{l}_{\text{mass}}=-\bar{l}_{R}m_{l}l_{L}-\overline{\nu}_{R}m_{\nu}\nu_{L}+\text{h.c.}\,.
\end{equation}
However, without \textit{a priori} assumptions, gauge theories suggest that neutrinos are Majorana particles~\cite{Schechter:1980gr}~\footnote{Note that the most general description of neutrinos is in terms of two-component spinors, Dirac fermions being just a particular case of Majorana~\cite{Valle:2015pba}. Although the two-component approach is universal and often more insightful, in this review we will adopt the more familiar four-component formalism used in most textbooks, for example~\cite{Fukugita:2003en,Giunti:2007ry,Xing:2011zza}, in which Majorana and Dirac neutrinos appear as separate ``cases''.}. However, this important issue of the nature of neutrinos must be settled by experiment. A prime example is to search for neutrinoless double beta decay~\cite{Jones:2021cga,Cirigliano:2022oqy,Dolinski:2019nrj,Vergados:2012xy}. A positive detection would imply, by the black-box theorem, the Majorana nature of at least one of the neutrinos~\cite{Schechter:1981bd}.

Majorana neutrino masses can be effectively described by the non-renormalizable Weinberg operator $(y_{\nu})_{ij}\left(\overline{(L^{i}_L)^c}\;i\tau_2H\right)\left(H^Ti\tau_2 L^j_{L}\right)/(2\Lambda)$~\cite{Weinberg:1979sa}. As a result the charged lepton and neutrino mass terms take of the following form
\begin{equation}
\label{eq:charge-lepton-neutrino-maj-mass}\mathcal{L}^{l}_{\text{mass}}=-\bar{l}_{R}m_{l}l_{L}-\frac{1}{2}\overline{\nu^c_{L}}m_{\nu}\nu_{L}+\text{h.c.}\,,
\end{equation}
where $m_{\nu}=y_{\nu}v^2/(2\Lambda)$. The smallness of neutrino mass may be ascribed to the large new physics scale $\Lambda$. There are also attractive low-scale realizations of the seesaw mechanism, as we comment below. Both charged lepton and neutrino mass matrices are diagonalized through unitary transformations as follows
\begin{equation}
\label{eq:mch_diag}U^{\dagger}_{l}m^{\dagger}_{l}m_{l}U_{l}=\textrm{diag}\left(m^2_{e}, m^2_{\mu}, m^2_{\tau}\right)\,,
\end{equation}
and
\begin{eqnarray}
\nonumber&&U^{T}_{\nu}m_{\nu}U_{\nu}=\textrm{diag}\left(m_{1}, m_{2}, m_{3}\right),                 ~~~~\quad \text{for Majorana neutrinos}\,,\\
\label{eq:mnu_diag}&&U^{\dagger}_{\nu}m^{\dagger}_{\nu}m_{\nu}U_{\nu}=\textrm{diag}\left(m^2_{1}, m^2_{2}, m^2_{3}\right),\quad \text{for Dirac neutrinos}\,.
\end{eqnarray}
where the light neutrino masses $m_{1,2,3}$ are real and non-negative. The nonzero mass squared differences measured in neutrino oscillation experiments imply a non-degenerate mass spectrum $m_1\neq m_2\neq m_3$. As a consequence, we can express the lepton mass matrices in terms of $U_{l}$, $U_{\nu}$ and the mass eigenvalues as
\begin{align}
\label{eq:mldag_ml_general}& m^{\dagger}_{l}m_{l}=U_{l}\,\textrm{diag}\left(m^2_{e}, m^2_{\mu}, m^2_{\tau}\right)U^{\dagger}_{l}\,,\\
\label{eq:mnu_general}&m_{\nu}=U^{*}_{\nu}\textrm{diag}\left(m_{1}, m_{2}, m_{3}\right)U^{\dagger}_{\nu},                ~~~~\quad \text{for Majorana neutrinos}\,,\\
\label{eq:mnudag_mnu_general}&m^{\dagger}_{\nu}m_{\nu}=U_{\nu}\textrm{diag}\left(m^2_{1}, m^2_{2}, m^2_{3}\right)U^{\dagger}_{\nu},\quad \text{for Dirac neutrinos}\,.
\end{align}
Transforming to the charged lepton and neutrino mass eigenstates, one obtains the leptonic charged-current weak interaction as~\cite{Schechter:1980gr}
\begin{equation}
\mathcal{L}^{l}_{CC}=\frac{g}{\sqrt{2}}\bar{l}_{L}\gamma^{\mu}U^{\dagger}_{l}U_{\nu}\nu_{L}W_{\mu}^{-}+h.c.\,,
\end{equation}
where the combination $U^{\dagger}_{l}U_{\nu}$ defines the lepton mixing matrix, i.e.
\begin{equation}
U_{}=U^{\dagger}_{l}U_{\nu}\,.
\end{equation}
Like the CKM matrix describing quark mixing, the lepton mixing matrix $U$ arises from the mismatch between the diagonalizations of charged leptons and neutrinos mass matrices. Assuming unitarity, the matrix $U$ is characterized by three angles and three physical CP phases~\cite{Schechter:1980gr},
\begin{equation} \small
\label{eq:lepton-mixing}
U = \left(
\begin{array}{ccc}
c^{\ell}_{12} c^{\ell}_{13} & s^{\ell}_{12} c^{\ell}_{13} e^{ - i \phi_{12} } & s^{\ell}_{13} e^{ -i \phi_{13} } \\
-s^{\ell}_{12} c^{\ell}_{23} e^{ i \phi_{12} } - c^{\ell}_{12} s^{\ell}_{13} s^{\ell}_{23} e^{ -i (\phi_{23}-\phi_{13} ) } & c^{\ell}_{12} c^{\ell}_{23} - s^{\ell}_{12} s^{\ell}_{13} s^{\ell}_{23} e^{ -i (\phi_{23}+\phi_{12} - \phi_{13} ) } & c^{\ell}_{13} s^{\ell}_{23} e^{- i \phi_{23} } \\
s^{\ell}_{12} s^{\ell}_{23} e^{ i ( \phi_{23} + \phi_{12} ) } - c^{\ell}_{12} s^{\ell}_{13} c^{\ell}_{23} e^{i\phi_{13} } & - c^{\ell}_{12} s^{\ell}_{23} e^{ i \phi_{23} } - s^{\ell}_{12} s^{\ell}_{13} c^{\ell}_{23} e^{-i(\phi_{12}-\phi_{13})} & c^{\ell}_{13} c^{\ell}_{23}
\end{array}
\right)\,,
\end{equation}
with the abbreviation $c^{\ell}_{ij}\equiv \cos\theta^{\ell}_{ij}$ and $s^{\ell}_{ij}\equiv \sin\theta^{\ell}_{ij}$. The above universal symmetrical presentation for the matrix $U$ in Eq.~\eqref{eq:lepton-mixing} is very convenient both for the description of quark as well as lepton mixing~\cite{Schechter:1980gr}. Notice that there is an important difference between leptons and quarks concerning CP violation, namely, all three CP phases in Eq.~(\ref{eq:lepton-mixing}) are physical parameters. One of them is the lepton analogue of the CKM phase in Eq.~(\ref{eq:dellq}), also called \textit{Dirac phase}.
\begin{equation}
\delta^{\ell}=\phi_{13} -\phi_{12}-\phi_{23}~.
\label{eq:dell}
\end{equation}
The others are the so-called \textit{Majorana phases}, which can not be eliminated by field redefinitions~\cite{Schechter:1980gr} if neutrinos are Majorana particles. Only the Dirac phase affects conventional neutrino oscillation probabilities, while the Majorana phases can only affect lepton number violating processes~\cite{Schechter:1980gk,Bilenky:1980cx}. Note that in what follows we will also use the notation $\delta_{CP}$ for the lepton Dirac CP phase $\delta^{\ell}$.
It will be an experimental challenge to obtain robust information on their magnitudes.

Current neutrino oscillation data restrict the elements of the lepton mixing matrix. One finds, at $3\sigma$, the following ranges~\cite{deSalas:2020pgw,10.5281/zenodo.4726908}
\begin{equation} \tiny
U^{\rm NO}=\left(\begin{array}{ccc} 0.7838 \to 0.8442 & 0.5133 \to 0.6004 & (-0.1568 \to 0.1489) + i (-0.1182 \to 0.1520) \\ (-0.4831 \to -0.2394) + i (-0.0749 \to 0.0963) & (0.4635 \to 0.6749) + i (-0.0521 \to 0.0668) & 0.6499 \to 0.7719 \\ (0.3068 \to 0.5391) + i (-0.0643 \to 0.0933) & (-0.6897 \to -0.4821) + i (-0.0446 \to 0.0644) & 0.6161 \to 0.7434 \\ \end{array}\right)
\label{eq:U-NO}
\end{equation}
for the case of normal neutrino mass-ordering. On the other hand, for inverted-ordered neutrino masses one has~\cite{deSalas:2020pgw,10.5281/zenodo.4726908},
\begin{equation} \tiny
U^{\rm IO}=\left(\begin{array}{ccc} 0.7835 \to 0.8440 & 0.5133 \to 0.6005 & (-0.1423 \to 0.1490) + i (0.0191 \to 0.1553) \\ (-0.4806 \to -0.2682) + i (0.0114 \to 0.0990) & (0.4546 \to 0.6395) + i (0.0074 \to 0.0695) & 0.6493 \to 0.7711 \\ (0.3102 \to 0.5133) + i (0.0094 \to 0.0947) & (-0.6956 \to -0.5248) + i (0.0057 \to 0.0654) & 0.6171 \to 0.7436 \\ \end{array}\right)
\label{eq:U-IO}
\end{equation}
Current determinations of the mass-ordering and the atmospheric octant are not yet fully robust.
On the other hand we still struggle with a very poor determination of the magnitude of the Dirac CP phase (see below).

\subsection{Neutrino oscillation recap}
\label{sec:global-osc}

\begin{figure}[h]
\centering
\includegraphics[width=1.0\textwidth]{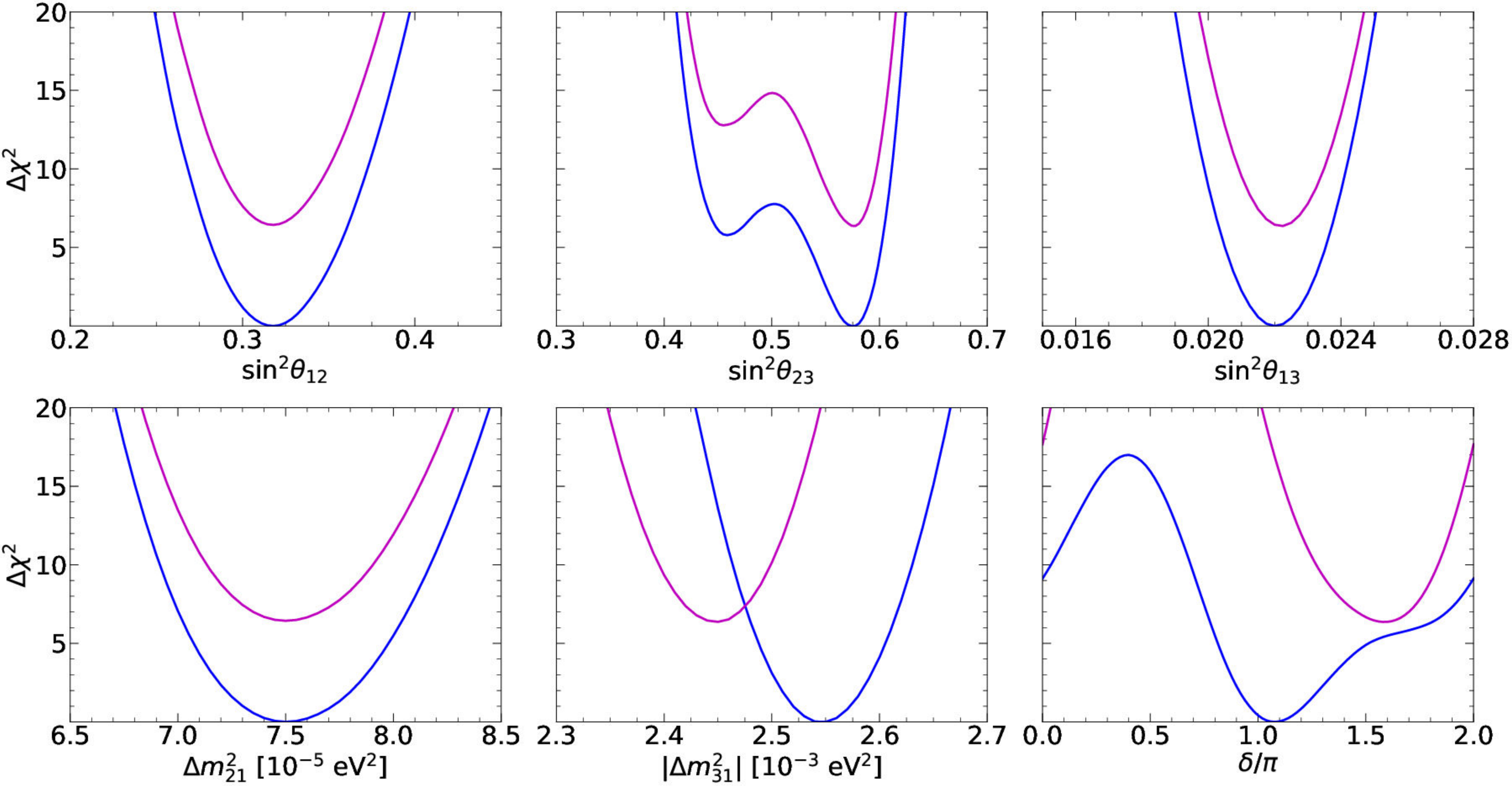}
\caption{Current summary of neutrino oscillation parameters, where $\delta\equiv\delta_{CP}$. From~\cite{deSalas:2020pgw,10.5281/zenodo.4726908}.}
\label{fig:osc}
\end{figure}

We now give a ``drone view'' of the current status of neutrino oscillation parameters. The basic discovery made in solar and atmospheric studies was soon followed by reactor and accelerator-based experiments that have not only provided independent confirmation, but also improved parameter determination. Current experimental data mainly converge towards a consistent global picture -- the three-neutrino paradigm -- in which the oscillation parameters are determined as shown in figure~\ref{fig:osc}.

The top-three panels show that two mixing angles are fairly large, at odds with the corresponding mixing angles observed in the quark sector. In fact, it is the smallest lepton mixing angle, $\theta_{13}$, that lies intriguingly close in magnitude to the largest of the quark mixing angles, i.e. the Cabibbo angle. Currently we have no explanation for this fact.

\begin{figure}[!h]
\centering
\includegraphics[width=1.0\textwidth]{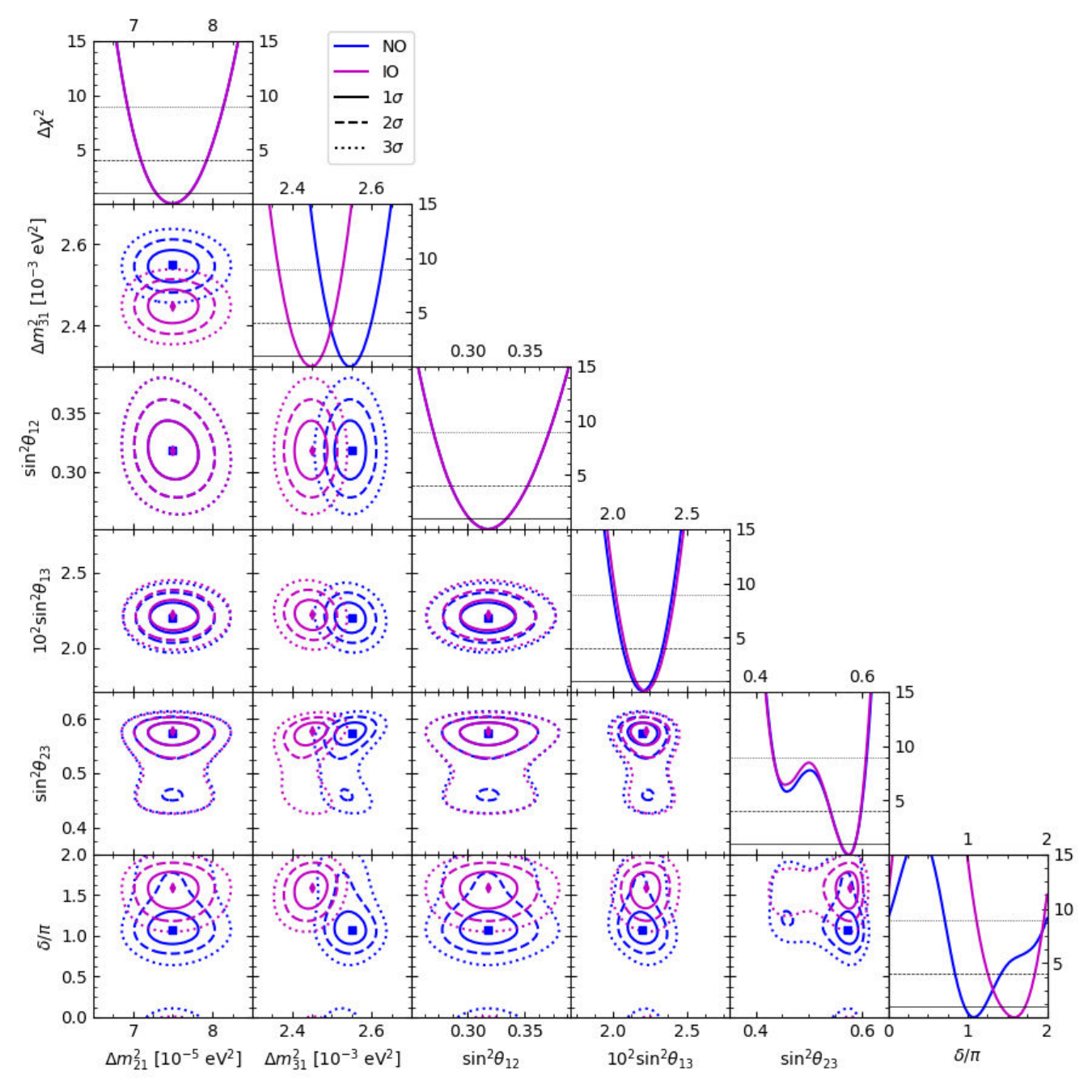}
\caption{Neutrino oscillation parameters pair-correlations. As before $\delta\equiv\delta_{CP}$. From~\cite{deSalas:2020pgw,10.5281/zenodo.4726908}.}
\label{fig:osc2}
\end{figure}

The global oscillation parameter determinations can also be shown as the ``matrix'' in figure~\ref{fig:osc2}. Besides the individual oscillation parameter determinations, given by the ``diagonal'' entries, the ``off-diagonal'' entries of figure~\ref{fig:osc2} show all pairwise
parameter correlations~\footnote{Numerical tables for the relevant $\chi^2$ profiles can be downloaded from Ref.~\cite{10.5281/zenodo.4726908}, including all pairwise correlations. These tables can be used to test various neutrino mixing patterns predicted by different flavour theories, such as those in Refs.~\cite{Hirsch:2010ru,Ma:2001dn,Babu:2002dz,Hirsch:2003dr,Ma:2006sk,Hirsch:2009mx,Hirsch:2010ru,Ding:2011gt,King:2011zj,Shimizu:2011xg,Boucenna:2011tj,deMedeirosVarzielas:2012apl,Morisi:2013eca,King:2013hj,Morisi:2013qna,Hirsch:2005mc,Hirsch:2007kh,Altarelli:2005yp,Altarelli:2008bg,Csaki:2008qq,Chen:2009gy,Burrows:2009pi,Kadosh:2010rm,Kadosh:2011id,Kadosh:2013nra,Chen:2015jta,Chen:2020udk,delAguila:2010vg,Hagedorn:2011un,Ding:2013eca,Hagedorn:2011pw,CarcamoHernandez:2015iao}.}, very useful for model-builders. Indeed, any flavour model leads to predictions for various entries of the above triangular matrix. The zenodo website in~\cite{10.5281/zenodo.4726908} will be used extensively in this review in order to examine the allowed parameter regions in different theoretical setups.

At this point one should stress that, although the overall picture provided by the ``three-neutrino paradigm'' is mostly robust, there are still loose ends. As already mentioned, the determination of the neutrino spectrum and the octant of the atmospheric angle is not fully robust yet, while the precise value of the leptonic CP phase also remains an open challenge. A robust CP determination will be a key objective of the Deep Underground Neutrino Experiment (DUNE)~\cite{DUNE:2015lol}. The experiment will have two detector systems placed along Fermilab's Long Baseline Neutrino Facility (LBNF) beam. One of them will be near the beam source, while a much larger one will be placed underground 1300 km away at the Sanford Underground Research Laboratory in South Dakota, in the same mine where Ray Davis pioneered neutrino oscillation studies in the sixties. Further improvements are expected at the ambitious T2HK proposal in Japan~\cite{Hyper-Kamiokande:2018ofw}. Reaching a precise CP phase determination will be a challenge for the coming years. Likewise, underpinning the mass ordering~\cite{JUNO:2015zny}.
Octant resolution may be harder, if the preferred $\theta_{23}$ value lies close to maximality.
See for example Refs.~\cite{Chatterjee:2017irl,Srivastava:2018ser} for details and related references. Altogether, octant discrimination, underpinning the mass-ordering and performing a precise CP determination constitute the target of the next generation of oscillation searches~\cite{Athar:2021xsd}.

Last, but not least, we mention that there are also some experimental anomalies in neutrino physics that challenge the simple picture provided by the ``three-neutrino paradigm'' suggesting, perhaps, a four-neutrino oscillation scenario~\cite{Maltoni:2004ei}. They will not be discussed in this review, the interested readers can see Refs.~\cite{Drewes:2016upu,Giunti:2019aiy,Boser:2019rta,Dasgupta:2021ies}.

\subsection{Neutrinoless double beta decay }
\label{sec:znbb}

Neutrinoless double-beta decay (or \znbb for short)  is the prime lepton number violating process~\cite{Jones:2021cga,Cirigliano:2022oqy,Dolinski:2019nrj,Vergados:2012xy}. Given that neutrinos are observed to be massive fermions, and expected to be Majorana-type~\cite{Schechter:1980gr}, it follows that \znbb decay should take place, as a consequence of neutrino exchange.
In this case, the effective mass parameter characterizing the amplitude is given as
\begin{equation}
\label{eq:mbb}
|m_{\beta\beta}|  = \left|\sum_{j=1}^3 U_{ej}^2 m_j\right|=\left|c^{\ell\,2}_{12}c^{\ell\,2}_{13} m_1 + s^{\ell\,2}
_{12}c^{\ell\,2}_{13} m_2 e^{-2i\phi_{12} }+ s^{\ell\,2}_{13} m_3 e^{-2i\phi_{13}}\right|~.
\end{equation}
Here we note that, in contrast with the parametrization adopted by the PDG, the symmetrical form of the lepton mixing matrix provides a conceptually transparent description of \znbb in which,
as it should, only Majorana phases appear~\cite{Rodejohann:2011vc}. Altogether, the original symmetrical form of the lepton mixing matrix, Eq.~(\ref{eq:lepton-mixing}), provides an insightful description both for the Dirac phase, Eq.~(\ref{eq:dell}), as well as the Majorana phases, Eq.~(\ref{eq:mbb}).

\begin{figure}[t!]
\centering
\includegraphics[width=0.60\textwidth]{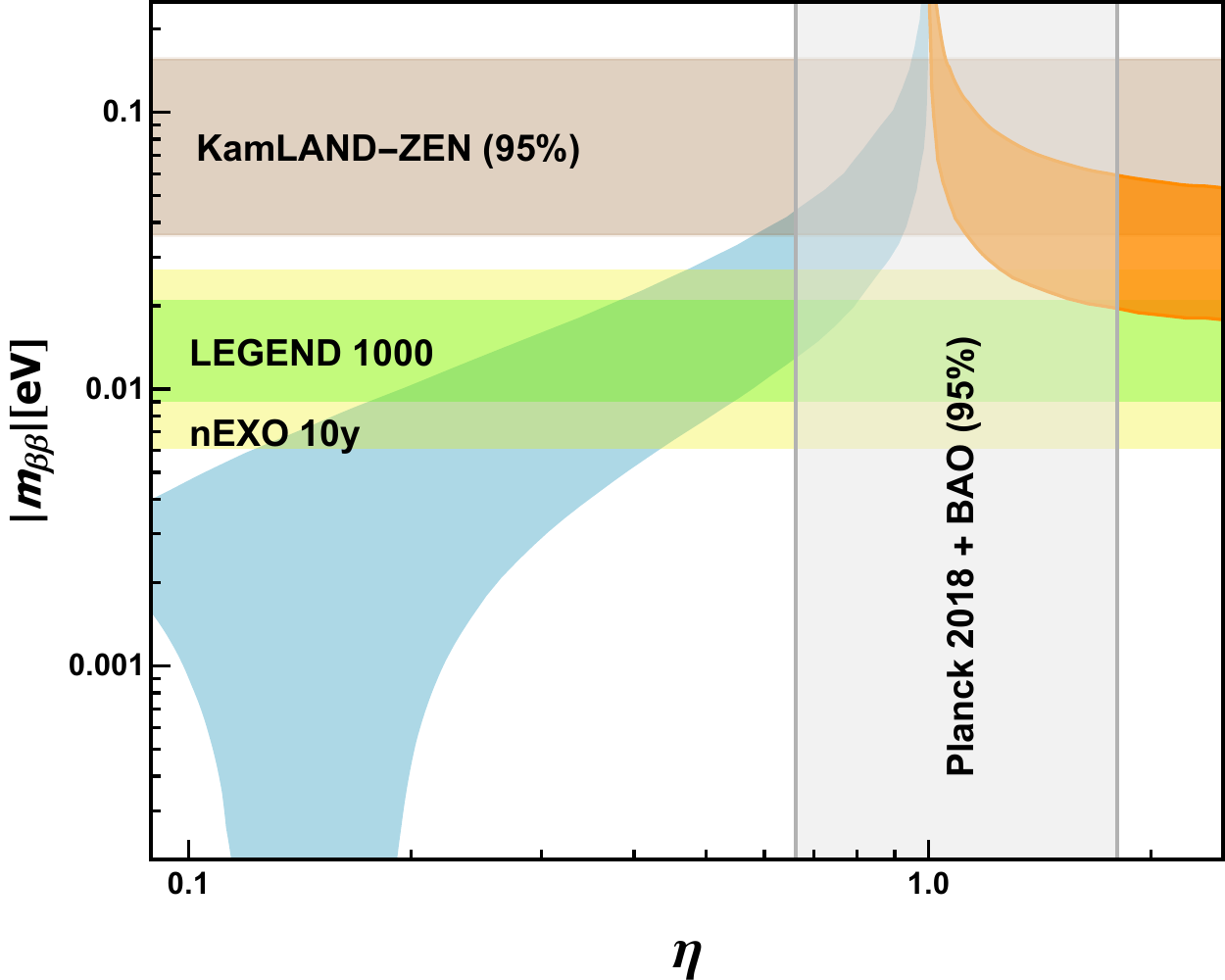}
\caption{The \znbb decay amplitude in a generic three-neutrino scheme versus the degeneracy parameter $\eta$. The curved bands are the normal and inverted ordering branches allowed by neutrino oscillations respectively~\cite{Lattanzi:2020iik}. The current experimental bound $|m_{\beta\beta}|<(36-156)\,$meV at $90\%$ confidence level (C.L.) from KamLAND-Zen~\cite{KamLAND-Zen:2022tow} and the future sensitivity ranges $|m_{\beta\beta}|<(9-21)\,$ meV from LEGEND-1000~\cite{LEGEND:2021bnm} and $|m_{\beta\beta}|<(6-27)\,$ meV from nEXO~\cite{nEXO:2021ujk} are indicated by light brown, light yellow and light green horizontal bands respectively. The vertical grey band is excluded by the $95\%$ C.L. limit $\Sigma_{i}m_{i}<0.120\,\text{eV}$ from Planck~\cite{Planck:2018vyg,Gerbino:2022nvz}.}
\label{fig:dbd1}
\end{figure}

An important point to notice from Eq.~(\ref{eq:mbb}) is that, thanks to the Majorana phases, the \znbb amplitude can vanish due to possible destructive interference among the three individual neutrino amplitudes. This is clearly seen in the blue branch of figure~\ref{fig:dbd1}, taken from Ref.~\cite{Lattanzi:2020iik}, which shows the \znbb decay amplitude versus the degeneracy parameter $\eta$. When $\eta \to 1$ neutrino masses become degenerate, and the two bands correspond to the two possible mass orderings. The most favorable case for \znbb decay detectability happens when neutrinos are nearly degenerate, as predicted in some UV-complete flavour theories~\cite{Ma:2001dn,Babu:2002dz,Hirsch:2003dr}. The $\eta\equiv 1$ case corresponds to the idealized limit where neutrinos would be strictly degenerate. In order to generate oscillations neutrino masses must deviate from exact degeneracy. This can happen in two ways, corresponding to the two curved branches seen in the figure. The normal-ordered (NO) neutrino region is indicated in the left (blue) band, while inverted ordering (IO) gives the upper-right (orange) branch. One sees that, thanks to the presence of the Majorana phases~\cite{Schechter:1980gr}, the \znbb amplitude can vanish for normal ordering (but not for inverted). The horizontal band denotes the current KamLAND-Zen limit~\cite{KamLAND-Zen:2022tow}, while the vertical one is excluded~\cite{Lattanzi:2020iik} by cosmological observations~\cite{Planck:2018vyg,Gerbino:2022nvz} e.g. by the Planck collaboration. Altogether, one sees current data strongly disfavour nearly degenerate neutrinos.

\begin{figure}[h]
\centering
\includegraphics[width=0.60\textwidth]{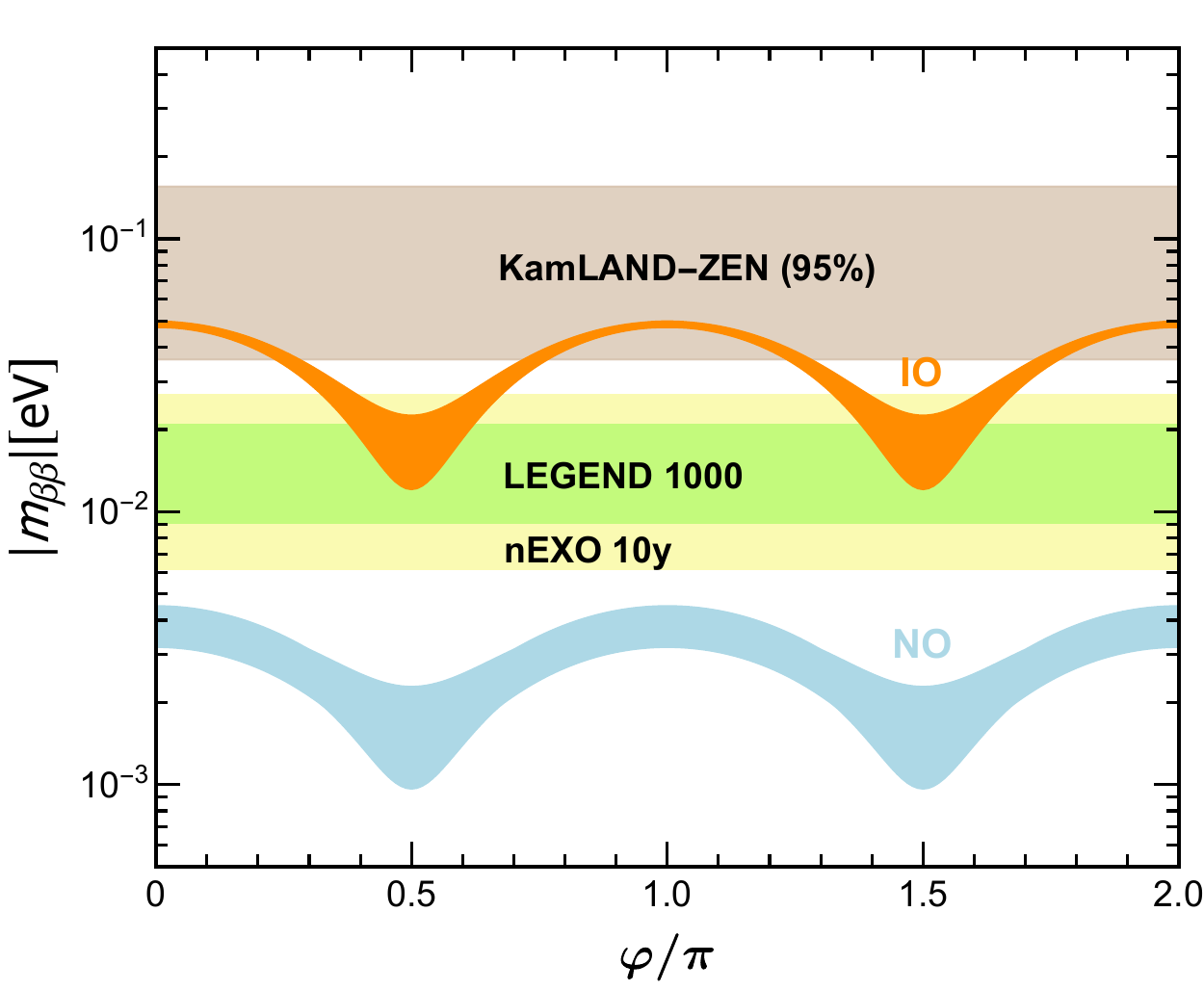}
\caption{\znbb decay amplitude when one neutrino is massless. The light-blue and light-orange bands are the current 3$\sigma$ C.L. regions for normal and inverted mass-ordering respectively. The current bound from KamLAND-ZEN~\cite{KamLAND-Zen:2022tow} and projected sensitivities of LEGEND 1000~\cite{LEGEND:2021bnm} and nEXO~\cite{nEXO:2021ujk} are indicated. }
\label{fig:dbd2}
\end{figure}

Another interesting situation happens if one (or two) of the three neutrinos is massless or nearly so, as in the ``missing partner'' seesaw mechanism~\cite{Schechter:1980gr} and other ``incomplete-multiplet'' schemes~\footnote{A massless neutrino may also arise from the presence of anti-symmetric Yukawa couplings, see e.g.~\cite{Leite:2019grf}.}. The missing partner seesaw mechanism also holds when supersymmetry is the origin of neutrino mass through the bilinear violation of R-parity~\cite{Hirsch:2000ef,Diaz:2003as,Hirsch:2004he}. The minimal viable tree-level type-I seesaw has only two right-handed neutrinos~\cite{King:1999mb,Frampton:2002qc,Raidal:2002xf}. This situation also arises within sequential right-handed neutrino dominance schemes, where the third right-handed neutrino is too heavy and decouples~\cite{King:1999cm,King:1999mb,King:2002nf,King:2013iva}. In such a scheme no cancellation is possible, even for normal-ordering~\cite{Reig:2018ztc,Barreiros:2018bju,Rojas:2018wym,Aranda:2018lif,Mandal:2021yph,Avila:2019hhv}. The resulting regions are the two periodic bands seen in figure~\ref{fig:dbd2}, which are expressed in terms of the only free parameter available, namely, the relative neutrino Majorana phase $\varphi$~\footnote{For vanishing lightest neutrino mass the relevant Majorana phase is $\varphi=\phi_{12}-\phi_{13}$ for NO and $\varphi=\phi_{12}$ for IO.}. The colored horizontal bands show current experimental limits of KamLAND-ZEN and the expected sensitivities of LEGEND 1000 and nEXO, where the width of the bands reflects nuclear matrix element uncertainties~\cite{Dolinski:2019nrj,Vergados:2012xy}. Notice in this case the existence of a predicted theoretical lower bound on $|m_{\beta\beta}|$. Taking into account the sensitivities expected at upcoming \znbb experiments, one sees that this lower bound for NO lies below detectability in the foreseeable future.

In contrast, inverse mass-ordering provides a lower bound that lies higher than the one generically expected for the IO three-massive-neutrino case. This implies a guaranteed discovery in the next round of experiments~\cite{GERDA:2019ivs,Agostini:2022zub}. In fact, for this case, the recent KamLAND-Zen limit~\cite{KamLAND-Zen:2022tow} already probes the Majorana phase, as seen by the magenta band in figure~\ref{fig:dbd2}. In short, for the one-massless-neutrino case there is a chance, perhaps, that one could be able to extract the value of the relevant Majorana phase from experiment.

\begin{figure}[h]
\centering
\includegraphics[width=0.90\textwidth]{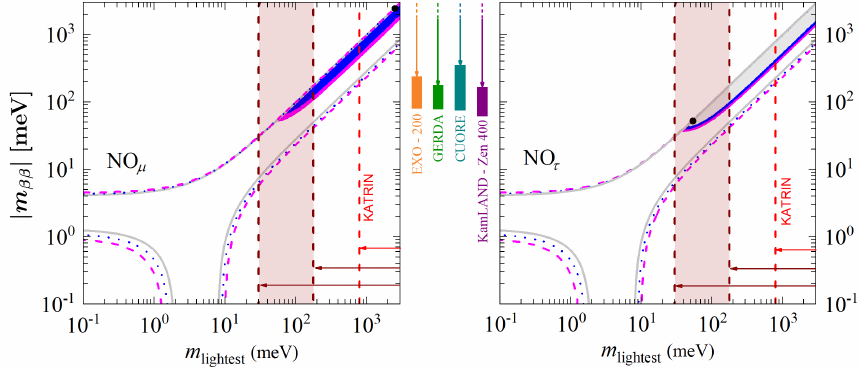}
\caption{The solid (dotted) [dashed] lines delimit the $1\sigma$ ($2\sigma$) [$3\sigma$] $|m_{\beta\beta}|$ regions allowed by oscillations.
Predictions of two normal-ordered $\mathcal{Z}_8$ schemes are given~\cite{Barreiros:2020gxu}. Vertical bars in mid-panels indicate current 95\%CL $|m_{\beta\beta}|$ upper bounds from KamLAND-Zen~400~\cite{KamLAND-Zen:2016pfg}, GERDA~\cite{GERDA:2020xhi}, CUORE~\cite{CUORE:2019yfd} and EXO-200~\cite{EXO-200:2019rkq}. }
\label{fig:dbd3}
\end{figure}

We now turn to the general three-massive-neutrino case. Although there is no guaranteed minimum value for $|m_{\beta\beta}|$ in this case, there can still be a lower-bound, even for normal mass-ordering, provided the cancellation of the amplitudes is prevented by the structure of the leptonic weak interaction vertex~\cite{King:2013hj,Hirsch:2005mc,Hirsch:2007kh,Dorame:2011eb,Dorame:2012zv,deAnda:2019jxw,deAnda:2020pti}. This can happen as a result of the imposition of a family symmetry to account for the mixing pattern seen in oscillations. As an example, figure~\ref{fig:dbd3} shows the predictions of a $\mathcal{Z}_8$ family symmetry scheme. The hollow solid (dotted) [dashed] lines delimit the $1\sigma$ ($2\sigma$) [$3\sigma$] $|m_{\beta\beta}|$ regions allowed in the general three-neutrino case. The sub-regions shown in gray, blue and magenta, respectively, are the flavour-model predictions. The black dots correspond to best-fits.  The vertical shaded band indicates the current sensitivity of cosmological data~\cite{Planck:2018vyg}. The vertical dashed red line corresponds to the KATRIN tritium beta decay~\cite{Formaggio:2021nfz} upper limit $m_{\beta}<0.8$~eV (90\% CL)~\cite{KATRIN:2021uub,KATRIN:2019yun}. The heights of the bars shown in the mid-part of figure~\ref{fig:dbd3} reflect the uncertainties in the nuclear matrix elements relevant for the computation of the decay rates. The flavour predictions for $|m_{\beta\beta}|$ illustrate our point. For example, one sees how the preferred flavour-predicted point in the right panel sits right inside the cosmologically interesting band, and close to the current \znbb limits, as indicated in between the panels.
Similar predictions for \znbb amplitudes occur in other family symmetry models, some of which will be discussed in this review, see for instance discussions given in Chapters~\ref{sec:benchmark-models} and \ref{sec:family-symmetry-cp}.

\begin{figure}[h]
\centering
\includegraphics[height=4.5cm,width=0.5\textwidth]{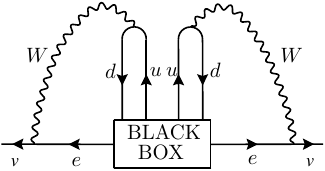}
\caption{The observation of \znbb decay implies that at least one neutrino is a Majorana fermion~\cite{Schechter:1981bd}.}
\label{fig:dbd-bb}
\end{figure}

In short, oscillations have left an important imprint upon neutrinoless double beta decay studies.
The results of the negative searches conducted so-far imply that we must consider both the possibilities of Dirac or Majorana neutrinos, as we do in this review. Nonetheless, there is a reasonable chance that, perhaps, \znbb could be seen in the coming round of experiments. This would constitute a major breakthrough. Indeed, a positive \znbb decay detection would imply, as a consequence of the black box theorem, that at least one of the neutrinos has Majorana nature~\cite{Schechter:1981bd}. The argument is illustrated in figure~\ref{fig:dbd-bb}. Note that the black-box argument holds irrespective of the underlying physics responsible for generating the process~\cite{Duerr:2011zd,Graf:2022lhj}.

In some cases, as we saw in figure~\ref{fig:dbd2}, the discovery of \znbb decay might allow us to underpin also the value of the relevant Majorana phase. Note however that, although a positive \znbb discovery would imply that at least one of the neutrinos is a Majorana particle, a negative result would not imply that neutrinos are Dirac-type, as the amplitude can be suppressed even for Majorana-type neutrinos, due to the effect of the Majorana phases. It has been argued that if a null \znbb decay signal was accompanied by a positive \zn4b quadruple beta decay signal~\cite{Heeck:2013rpa,NEMO-3:2017gmi}, then at least one neutrino should be a Dirac particle~\cite{Hirsch:2017col}.

\clearpage

\section{ Origin of neutrino masses and flavour puzzle}                    
\label{sec:orig-neutr-mass}                                                

Despite its amazing success in describing three out of the four known fundamental forces of nature, there are many drawbacks and open issues in the Standard Model. Altogether, they imply that the theory of particle physics needs a completion beyond its current form. Here we start with one of the most important issues, i.e. the lack of neutrino masses.

\subsection{Effective neutrino masses }
\label{sec:drawbacks}

\begin{figure}[hptb]
\begin{center}
\includegraphics[width=0.3\linewidth]{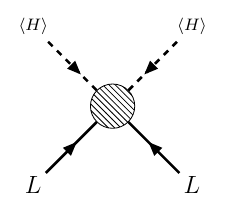}
\caption{\label{fig:neutrino-mass-Majorana} Majorana neutrino mass generation from the dimension-five operator. }
\end{center}
\end{figure}

Although the \sm lacks neutrino masses, these can arise effectively from a unique dimension-five operator characterizing lepton number non-conservation~\cite{Weinberg:1979sa}, as illustrated in figure~\ref{fig:neutrino-mass-Majorana}. In this case neutrinos are Majorana fermions, as generally expected in gauge theories. Indeed, on general grounds it was argued in~\cite{Schechter:1980gr}, within the \SM setup, that neutrinos are expected to be Majorana fermions.

However, neutrinos could also be Dirac fermions, this is currently an open experimental question. Dirac neutrinos can indeed emerge in the presence of extra symmetries, that could be discrete or continuous, global or local. For example, the imposition of $U(1)_{B-L}$ symmetry, where $B$ and $L$ are baryon and lepton numbers respectively, would forbid the Majorana mass terms of the right-handed neutrinos~\cite{Davidson:1978pm,Marshak:1979fm,Mohapatra:1980qe,Wetterich:1981bx,Leite:2020wjl}. Alternatively, as mentioned below, the Dirac nature of neutrinos could result from the presence of an underlying Peccei-Quinn symmetry~\cite{Peinado:2019mrn,Dias:2020kbj}, from the scotogenic mechanism~\cite{Farzan:2012sa,Bonilla:2016diq} or be associated with dark matter stability~\cite{Farzan:2012sa,CentellesChulia:2016rms,Bonilla:2016diq,Srivastava:2017sno,CentellesChulia:2017koy,Reig:2018mdk,Bonilla:2018ynb,Alvarado:2021fbw}. Small Dirac-type neutrino masses as in Eq.~(\ref{eq:mdir}), could arise from effective dimension-five~\cite{Roncadelli:1983ty,Gu:2006dc,Yao:2018ekp,CentellesChulia:2018gwr} as well as dimension-six operators~\cite{Yao:2017vtm,CentellesChulia:2018bkz}, that have by now been classified.

\subsection{The seesaw paradigm }
\label{sec:seesaw-paradigm-}

An attractive ultraviolet completion of the dimension-five operator is provided by the seesaw mechanism. It is specially interesting, as it gives a simple \textit{dynamical} understanding of small neutrino masses by minimizing the Higgs potential, as well as the possibility of having a stable electroweak vacuum~\cite{Elias-Miro:2011sqh,Bonilla:2015eha,Bonilla:2015kna,Mandal:2019ndp,Mandal:2020lhl,Mandal:2021acg}.

\begin{figure}[h]
\begin{center}
\includegraphics[height=5cm,width=0.98\linewidth]{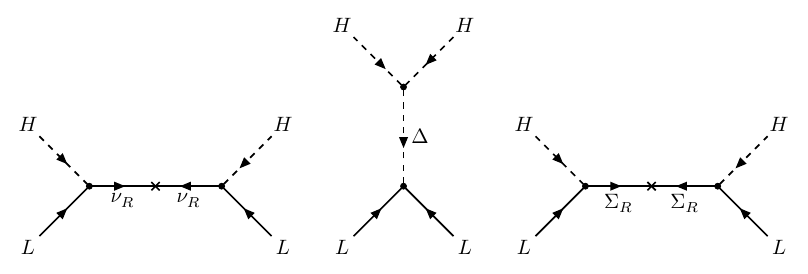}
\caption{\label{fig:t1seesaw} Feynman diagrams for the conventional seesaw types, where the mediators $\nu_{R}$ and $\Sigma_R$ are SM singlet and triplet fermions respectively, while $\Delta$ is a SM triplet scalar. }
\end{center}
\end{figure}

It has become usual to distinguish three main seesaw varieties, namely type-I~\cite{Minkowski:1977sc,Gell-Mann:1979vob,Mohapatra:1979ia,Schechter:1980gr,Cheng:1980qt,Schechter:1981cv}, type-II~\cite{Magg:1980ut,Schechter:1980gr,Schechter:1981cv,Lazarides:1980nt,Mohapatra:1980yp,Cheng:1980qt} and type-III~\cite{Foot:1988aq}, illustrated in figure~\ref{fig:t1seesaw}~\footnote{In models with extra Higgs doublets, the neutrino mass can arise from a similar diagram replacing $H$ by another doublet Higgs field~\cite{Hernandez-Garcia:2019uof}.}. These generate neutrino masses at tree level from the mediation of new heavy singlet fermion ($\nu_R$), triplet scalar ($\Delta$) and triplet fermion ($\Sigma_R$), respectively. In the simplest seesaw neutrinos acquire masses through the exchange of heavy scalars, called type-I in~\cite{Schechter:1980gr} and today called type-II. Such seesaw realization allows one to reconstruct the parameters characterizing the neutrino sector from various experiments~\cite{Mandal:2022zmy}, in particular those from high energy colliders~\cite{FileviezPerez:2008jbu,Cai:2017mow}. Moreover, the type-II seesaw opens the really tantalizing possibility of probing the absolute neutrino mass and mass ordering in collider experiments~\cite{Mandal:2022ysp}.

Here we focus on the type-I seesaw mechanism, where neutrinos acquire masses through the exchange of heavy gauge-singlet fermions, as illustrated in the left panel of figure~\ref{fig:t1seesaw}. Seesaw mediators were originally thought to lie at a high mass scale, associated to SO(10) unification or left-right symmetry. The associated physics has been covered in several textbooks~\cite{Valle:2015pba,Fukugita:2003en,Giunti:2007ry,Xing:2011zza} and reviews~\cite{Valle:2006vb}. Note that, following Ref.~\cite{Schechter:1980gr}, here we do not assume left-right symmetry in the seesaw mechanism, but simply the minimal well-tested \SM gauge structure. Notice that this seesaw description of neutrino mass generation can be made fully dynamical in the presence of a singlet vacuum expectation value driving the spontaneous violation of lepton number symmetry. The theory is then accompanied by a Nambu-Goldstone boson, dubbed Majoron~\cite{Chikashige:1980ui,Schechter:1981cv}.

Notice that in the most general case, including implementations of Weinberg's dimension five operator, and its simplest seesaw realizations, neutrinos are naturally expected to be Majorana. However, in the presence of adequate extra symmetries. Indeed, there may be good reasons for neutrinos to be Dirac-type. For example, the Dirac nature of neutrinos could also be associated to the existence of an underlying Peccei-Quinn symmetry~\cite{Peinado:2019mrn,Dias:2020kbj}. Moreover, Dirac-type neutrinos could signal the stability of dark matter~\cite{Farzan:2012sa,CentellesChulia:2016rms,Bonilla:2016diq,Srivastava:2017sno,CentellesChulia:2017koy,Reig:2018mdk,Bonilla:2018ynb,Alvarado:2021fbw}.
In the presence of extended gauge symmetries Dirac neutrinos can also emerge from the seesaw~\cite{Valle:2016kyz}. Finally, there are interesting scenarios where the Dirac nature of neutrinos is associated with a family symmetry~\cite{Aranda:2013gga,CentellesChulia:2016rms,CentellesChulia:2016fxr,Borah:2017dmk,Bonilla:2017ekt}.

\subsection{Dark matter as the source of neutrino mass}
\label{sec:neutrino-mass-and}

It is well-known that about 85\% of the matter in the universe is ``dark'' in the sense that it does not appear to interact with the electromagnetic field and is, as a consequence, very hard to detect. It is not our purpose to provide a comprehensive discussion of the observational aspects of cosmological dark matter; for a review see Ref.~\cite{Bertone:2004pz}. The first point we note is that the existence of such cosmological Dark matter is totally unaccounted for within the Standard Model. A very interesting possibility is that dark matter is made up of a novel weakly interacting massive particle, dubbed WIMP, typically present in the case of supersymmetric theories with conserved R-parity~\cite{Jungman:1995df}.

Underpinning the mechanism yielding small neutrino masses and/or dark matter is of paramount importance in particle physics, as it may select which new physics to expect as the next step. This is a very broad subject which we will not try to review. Rather, in this review we just comment on the issue of particle dark matter candidates, and how they may be simply related to the mechanism of neutrino mass generation. We focus on the very interesting possibility that the dark matter sector mediates neutrino mass generation, as postulated within the scotogenic picture~\cite{Tao:1996vb,Ma:2006km,Hirsch:2013ola,Merle:2016scw,Diaz:2016udz,Choubey:2017yyn,Restrepo:2019ilz}. An interesting twist is to imagine the existence of a universal ``hidden'' dark matter sector that \textit{seeds} neutrino mass generation, which proceeds \textit{a la seesaw}, as in the dark inverse seesaw~\cite{Mandal:2019oth} mechanism. Alternatively, one can envisage a dark linear seesaw mechanism~\cite{CarcamoHernandez:2023atk,Batra:2023bqj}.

Both dark inverse and dark linear seesaw realizations will be described below. In either case dark matter will be WIMP-like. Instead of being related to supersymmetry, WIMP dark matter in these models is closely related to neutrino physics, either as mediator or as seed of neutrino mass generation. Flavored scotogenic dark matter may also be implemented~\cite{Ding:2020vud}, providing an attractive way to reconcile low-scale radiative neutrino mass generation with dark matter, while addressing the flavour problem at the same time, see Sec.~\ref{sec:flavor-puzzle}.

Last, but not least, we mention that particle physics theories where neutrino masses arise from the spontaneous breaking of a continuous global lepton number symmetry have a natural dark matter candidate~\cite{Berezinsky:1993fm,Lattanzi:2014mia}, namely the associated Nambu-Goldstone boson, dubbed majoron~\cite{Chikashige:1980ui,Schechter:1981cv}. Indeed, the majoron is likely to pick up a mass from gravitational instanton effects, that explicitly violate global symmetries~\cite{Coleman:1988tj}. The majoron necessarily decays to neutrinos, with an amplitude proportional to their tiny mass, which typically gives it cosmologically long lifetimes~\cite{Schechter:1981cv}. The associated restrictions on the decaying warm dark matter picture coming from the CMB~\cite{Lattanzi:2007ux} as well as mono-energetic photon emission in astrophysics have been examined in detail~\cite{Lattanzi:2013uza}. Using N-body simulations it has also been been shown that the warm majoron dark matter model provides a viable alternative to the $\Lambda$CDM scenario, with predictions that can differ substantially on small scales~\cite{Kuo:2018fgw}.

\subsection{Missing partner seesaw and dark matter: the scoto-seesaw }
\label{sec:n-m}

In contrast to a left-right symmetric SO(10)-based seesaw mechanism, where the number of ``left'' and ``right'' neutrinos must match as a consequence of gauge invariance, in the most general SM-based seesaw mechanism~\cite{Schechter:1980gr} one can have any number ($m$) of ``right-handed'' neutrino mediators, since they are gauge singlets. Theories with $m<3$ in the classification of~\cite{Schechter:1980gr} could simplify substantially the form of the lepton mixing matrix~\footnote{For an early example, with lepton number symmetry conservation, see ~\cite{Schechter:1979bn}.}.

Such ``missing partner'' seesaw schemes with less ``right'' than  ``left'' neutrinos, imply that some of the active ``left'' neutrinos remain massless~\cite{Schechter:1980gr,King:1999mb,Frampton:2002qc}. The (3,2) choice corresponds to the minimal viable type-I seesaw, in which there are only two mass parameters corresponding to the experimentally measured solar and atmospheric splittings. The lightest neutrino is massless, leading to the \znbb prediction discussed in Sec.~\ref{sec:znbb}.

On the other hand, the minimal (3,1) seesaw leads to the same \znbb prediction as (3,2), but has only one neutrino mass parameter~\cite{Schechter:1980gr}. Although such setup is not consistent with current neutrino data, as it lacks solar neutrino oscillations, the later may arise from a completion in which the degeneracy between the lowest-mass neutrinos is lifted by some other mechanism, for example, as a result of radiative corrections.

Indeed, the (3,1) setup offers a template to reconcile the seesaw paradigm and the WIMP dark matter paradigm within a minimal hybrid construction called ``scoto-seesaw'' mechanism.
Such simplest (3,1) ``scoto-seesaw'' scheme provides a comprehensive theory of neutrino mass generation as well as WIMP dark matter, in which the relative magnitudes of solar and atmospheric oscillation lengths are explained due to a loop factor~\cite{Rojas:2018wym,Aranda:2018lif,Mandal:2021yph}.

\subsection{The low-scale inverse and linear seesaw mechanisms}
\label{sec:low-scale}

As a final example, one may consider having more ``right-'' than ``left-handed'' neutrinos in the seesaw mechanism, for example, two isosinglets for each ``active'' isodoublet family. This can be implemented with explicit~\cite{Mohapatra:1986bd} as well as spontaneous violation of lepton number~\cite{Gonzalez-Garcia:1988okv}. In the lepton-number-conserving limit one finds that the three light neutrinos remain massless, as in the Standard Model~\cite{Bernabeu:1987gr,Branco:1989bn,Rius:1989gk}. In contrast to the Standard Model, however, lepton flavour and lepton CP symmetries can be substantially violated~\footnote{General discussions of leptonic flavour and CP violation are given in~\cite{Nunokawa:2007qh,deGouvea:2002gf,Branco:2011zb,Vicente:2015cka,Abada:2021zcm}.}.
This shows that flavour and CP violation can exist in the leptonic weak interaction despite the masslessness of neutrinos, implying that such processes need not be suppressed by the small neutrino masses, and hence can have large rates~\cite{Bernabeu:1987gr,Branco:1989bn,Rius:1989gk,Deppisch:2004fa,Deppisch:2005zm}~\footnote{For generic references on \lfv and seesaw schemes see, for example~\cite{Ilakovac:1994kj,Arganda:2007jw,Abada:2014cca,Abada:2014vea,Abada:2015oba}.}. Such ``(3,6)'' setup~\cite{Schechter:1980gr}, where the 3 doublet neutrinos are accompanied by 6 heavy singlet neutral leptons, provides the template for building genuine ``low-scale'' realizations of the seesaw mechanism in which lepton number is restored at low values of the \lnv scale. The models are natural in {\it t'Hooft} sense, leading to small, symmetry-protected neutrino masses. Realizations of such genuine ``low-scale'' seesaw mechanisms include the inverse~\cite{Gonzalez-Garcia:1988okv,Mohapatra:1986bd} as well as the linear seesaw~\cite{Akhmedov:1995vm,Akhmedov:1995ip,Malinsky:2005bi}. Last, but not least, we notice that a dynamical realization of the seesaw mechanism involving the spontaneous violation of lepton number symmetry
can improve the consistency properties of the electroweak vacuum~\cite{Elias-Miro:2011sqh,Bonilla:2015eha,Bonilla:2015kna,Mandal:2019ndp,Mandal:2020lhl,Mandal:2021acg}.

If realized at low-scale, the type-I seesaw mechanism may also lead to signals at high-energy colliders. Already back in the LEP days, it was suggested that the neutrino mass mediators could be produced at high energy colliders~\cite{Keung:1983uu,Dittmar:1989yg,Gonzalez-Garcia:1990sbd,Aguilar-Saavedra:2012dga,Das:2012ii,Deppisch:2013cya}.
This proposal was indeed taken up by subsequent experiments, for example the ATLAS and CMS experiments at the LHC~\cite{ATLAS:2019kpx,CMS:2018iaf,CMS:2022nty}
as well as future proposals~\cite{FCC:2018evy,FCC:2018byv,Feng:2022inv,Abdullahi:2022jlv}.
These have also taken into account the possibility of having displaced vertices~\cite{Helo:2013esa,Drewes:2019fou,Beltran:2021hpq}~\footnote{Other neutrino mass mediator searches can also lead to displaced vertices~\cite{Datta:1999xq,Porod:2000hv,deCampos:2005ri}.} coming from the fact that the couplings responsible for the mediator decays can be neutrino-mass-suppressed.

\subsection{Dark inverse and linear seesaw mechanisms}
\label{sec:dark-low-scale}

The symmetry protection provided by the above schemes can be upgraded into a  ``double protection'', by having the seed of \lnv to arise radiatively. Indeed, a very interesting possibility has been suggested, namely the dark inverse~\cite{Mandal:2019oth} and also the dark linear seesaw mechanism~\cite{CarcamoHernandez:2023atk,Batra:2023bqj}. In the former case there is a universal gauge-singlet or ``hidden'' dark matter sector that \textit{seeds} neutrino mass generation, which proceeds \textit{a la seesaw}. The same idea can be used to promote the linear seesaw mechanism into a mechanism sourced by a dark sector. The latter is not unique, interesting examples were given
in~\cite{CarcamoHernandez:2023atk} and~\cite{Batra:2023bqj}.

\begin{figure}[h!]
\begin{center}
\includegraphics[width=0.65\linewidth]{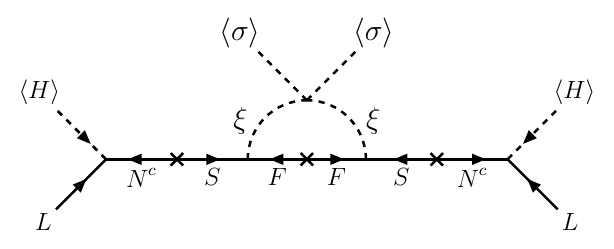}\\
\includegraphics[width=0.55\linewidth]{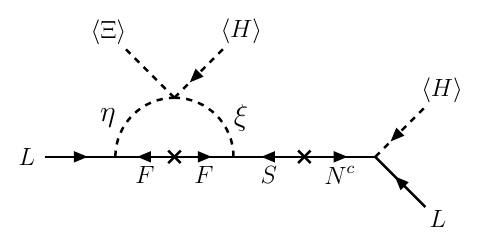}
\caption{\label{fig:darkseesaw} Feynman diagrams for ``dark'' inverse seesaw (upper panel)~\cite{Mandal:2019oth}, and the ``dark'' linear seesaw mechanism~\cite{CarcamoHernandez:2023atk} in the lower panel, which should also include the symmetrized diagram.
For another realization of the linear seesaw mechanism see Ref.~\cite{Batra:2023bqj}.}
\end{center}
\end{figure}

The dark inverse seesaw is illustrated in the upper panel of figure~\ref{fig:darkseesaw}. The \lnv loop that ``seeds'' neutrino mass is mediated by the ``dark'' gauge singlet fermion $F$ and dark singlet scalar $\xi$~\cite{Mandal:2019oth}, where $\sigma$ is a complex scalar singlet, and the new fermions $N^c$ and $S$ are SM singlets as well. This provides an elegant way to reconcile the seesaw and dark matter paradigms, providing an interesting dynamical seesaw theory where dark matter and neutrino mass generation are closely inter-connected.
Scenarios have also been proposed where a dark sector seeds neutrino mass generation radiatively within the linear seesaw mechanism~\cite{Batra:2023bqj,CarcamoHernandez:2023atk}, as illustrated in the lower panel. The dark sector contains the dark fermion $F$ which is SM singlet, a doublet dark scalar $\eta$, and a dark singlet scalar $\xi$. The field $\Xi$ is a complex scalar isotriplet, and its vacuum expectation value is restricted to be small $\langle\Xi\rangle\leq 3$ GeV by precision electroweak measurements. Neutrino masses are also calculable, since tree-level contributions are forbidden by symmetry. By having the seesaw realized at low-scale~\cite{Gonzalez-Garcia:1988okv,Mohapatra:1986bd}, as indicated in figure~\ref{fig:darkseesaw}, in either case one has charged lepton flavour violating processes e.g. $\mu \to e \gamma$ with accessible rates~\cite{Mandal:2019oth}.
We stress that these can be large, despite the tiny neutrino masses. Interesting dark-matter and collider physics implications have also been discussed.

\subsection{Neutrinos and the flavour puzzle}
\label{sec:flavor-puzzle}

We now turn to the issue of understanding the pattern of the weak interactions of quarks and leptons from first principles. Such ``flavour problem'' constitutes a major challenge in modern particle physics. Why three families of quarks and leptons? How to explain their mass hierarchies e.g. why, though otherwise so similar, the muon is about 200 times heavier than the electron? Why is the top quark mass so large compared with the masses of the other fermions? On the other hand, we also face the question of explaining why do the fermion mixing matrices follow the peculiar pattern observed? In particular, why do leptons mix so differently from the way quarks do? All of these shortcomings pose a real challenge on unified descriptions of nature.

As it stands, the Standard Model of particle physics lacks an organizing principle in terms of which to understand the flavour problem. Hence it can not be a complete theory of nature.
Although the \sm suffers from many other drawbacks, in this review we focus mainly on explaining the fermion mixing pattern and CP violation, though we also discuss some ideas to account for the fermion mass hierarchies. A ``flavour completion'' of the SM would have potentially important tests in the laboratory. Moreover, a deep understanding of the flavour problem and CP violation may prove crucial in understanding the baryon asymmetry of the universe~\cite{Sakharov:1967dj,Fukugita:1986hr}.

\begin{figure}[hptb]
\begin{center}
\includegraphics[width=0.6\textwidth]{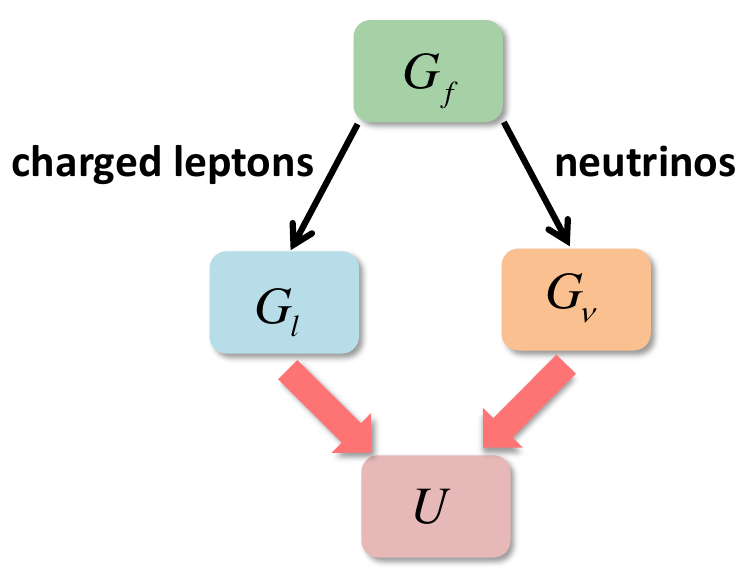}
\caption{\label{fig:flasy6} Predicting lepton flavour mixing from the flavour group $G_f$ breaking to different subgroups. }
\end{center}
\end{figure}

In this review we cover some recent attempts to account for the pattern of neutrino mixing indicated by oscillation experiments. Our key ingredient in formulating a theory of flavour is the imposition of an extra symmetry $G_f$ relating the families. The leading Lagrangian for the lepton masses should be invariant under both the SM gauge symmetry as well as the flavour symmetry, and it can be generally written as
\begin{equation}
\mathcal{L}^l_{m}=-[y_{l}(\Phi_l)]_{ij}\overline{l^{i}_R}H^{\dagger}L^{j}_L-\frac{1}{2\Lambda}[y_{\nu}(\Phi_{\nu})]_{ij}\left(\overline{(L^{i}_L)^c}\;i\tau_2H\right) \left(H^Ti\tau_2 L^j_{L}\right)+\text{h.c.}\,,
\end{equation}
where $y_l(\Phi_l)$ and $y_{\nu}(\Phi_{\nu})$ generically denote the Yukawa couplings, which are determined as functions of the flavons, and the light neutrino masses are described by the Weinberg operator~\cite{Weinberg:1979sa}. Here $\Phi_l$ and $\Phi_\nu$ denote the flavon fields with vacuum expectation values that break $G_f$ down to the residual subgroups $G_l$ and $G_{\nu}$ in the charged lepton and neutrino sectors respectively. In concrete models the higher-order terms can lead to subleading corrections.
As a result, the mismatch of the residual symmetries $G_l$ and $G_{\nu}$ allows one to make model-independent symmetry predictions for the lepton mixing matrix $U = U^{\dagger}_{l}U_{\nu}$, as illustrated in figure~\ref{fig:flasy6}.

Within the bottom-up approach one may consider remnant flavour symmetries and/or remnant CP symmetries, such as generalizations of the $\mu-\tau$ reflection symmetry~\cite{Harrison:2002kp,Harrison:2002et,Babu:2002dz,Grimus:2003yn,Harrison:2004he,Farzan:2006vj,Chen:2015siy,Chen:2018lsv}, as a way to restrict the lepton mass matrices, irrespective of the details of the underlying flavour symmetry. This can be used in turn to constrain the lepton flavour mixing parameters, such as mixing angles~\cite{deAdelhartToorop:2011re,Holthausen:2012wt,Fonseca:2014koa,Talbert:2014bda,Yao:2015dwa} and especially the CP phases~\cite{Feruglio:2012cw,Chen:2014wxa,Everett:2015oka,Chen:2015nha,Everett:2016jsk}.
In what follows we call this approach ``residual symmetry method''~\cite{Lam:2008rs,Lam:2007qc,Lam:2011ag,Lam:2011zm}. We notice that such $\mu-\tau$ reflection symmetry can emerge within UV-complete theories, e.g.~\cite{Babu:2002dz,Feruglio:2012cw}. Indeed, one may seek to constrain the flavour mixing parameters within a complete flavour theory.
Attaining a satisfactory theory that explains both quark and lepton sectors together remains an open challenge~\cite{Antusch:2014poa,Bjorkeroth:2015ora}, especially within a quark-lepton unified framework~\cite{Georgi:1974sy,Georgi:1974yf,Fritzsch:1974nn,Georgi:1979df,Gell-Mann:1979vob,Dimopoulos:1981zb,Reig:2017nrz}. This difficulty~\cite{Altarelli:2010gt,Morisi:2012fg,King:2014nza,King:2017guk,Feruglio:2019ybq,Ishimori:2010au,King:2013eh,Xing:2020ijf,Chauhan:2023faf} follows mainly due to the disparity between quark and lepton mixing angles. A possible idea is that lepton mixing angles are large with respect to quark mixing angles because neutrino masses are degenerate in the family symmetry limit~\cite{Babu:2002dz}. However, nearly degenerate neutrinos are now strongly disfavored by cosmological restrictions~\cite{Lattanzi:2020iik}.

There have been many efforts to approach the flavour problem in the quark sector in terms of an underlying family symmetry, either within a bottom-up or top-down approach~\cite{Froggatt:1978nt,Barbieri:1995uv,Blum:2007jz,Holthausen:2013vba,Araki:2013rkf,deMedeirosVarzielas:2016fqq,Li:2017abz,Lu:2018oxc,Lu:2019gqp,Hagedorn:2018gpw,Hagedorn:2018bzo}. For example, the Cabibbo mixing angle can be described by a flavour symmetry~\cite{Yao:2015dwa,Lam:2007qc,Blum:2007jz,deMedeirosVarzielas:2016fqq,Reig:2018ocz}. Similarly, quark mixing angles and CP violation may also be accounted for in the framework of generalized CP symmetries~\cite{Li:2017abz,Lu:2018oxc,Lu:2019gqp,Hagedorn:2018gpw,Hagedorn:2018bzo}.

Turning now to mass predictions, we note that some quark-lepton relations may emerge naturally from the imposition of family symmetries, even in the absence of a genuine unification group. A prime example is the ``golden'' formula,
\begin{equation}
\frac{m_\tau}{\sqrt{m_\mu m_e}}\approx \frac{m_b}{\sqrt{m_s m_d}}\,,
\end{equation}
relating charged-lepton and down-type quark masses. This successful relation may constitute, perhaps, part of any complete theory of flavour. The golden formula could be associated to an underlying Peccei-Quinn symmetry~\cite{Reig:2018ocz}. Alternatively, it could arise within UV-complete family-symmetry-based theories such as~\cite{Morisi:2011pt,Bonilla:2014xla,Bonilla:2017ekt} or~\cite{Morisi:2013eca,King:2013hj}.
Finally, we note that distinct variants of the golden formula can arise from modular symmetries~\cite{Chen:2023mwt}.

We stress that the golden-type relations involve mass ratios, stable under renormalization-group evolution, and allow us to relate quarks and leptons without invoking a genuine unified gauge group such as SU(5) or SO(10). More ambitious approaches to the flavour problem have been proposed within extra spacetime dimensions. For example, 5-D warped schemes were proposed in which mass hierarchies are accounted for by adequate choices of the
bulk mass parameters, while quark and lepton mixing angles are restricted by the imposition of a flavour symmetry~\cite{Altarelli:2005yp,Altarelli:2008bg,Csaki:2008qq,Chen:2009gy,Burrows:2009pi,Kadosh:2010rm,Kadosh:2011id,Kadosh:2013nra,Chen:2015jta,Chen:2020udk,delAguila:2010vg,Hagedorn:2011un,Ding:2013eca,Hagedorn:2011pw,CarcamoHernandez:2015iao}. Predictions for neutrino mixing, CP violation and \znbb decay, as well as a good global fit of flavour observables emerge in warped flavordynamics~\cite{Chen:2015jta,Chen:2020udk}.

On the other hand, 6-dimensional orbifold compactification has been suggested as a promising way to determine the structure of the family symmetry in four dimensions.
In this approach the 4-dimensional flavour group emerges from the symmetries between the branes in extra dimensions~\cite{deAnda:2018oik,deAnda:2018yfp}. Interesting flavour and CP predictions have been obtained from 6-dimensional orbifold compactification schemes, including neutrino oscillation and \znbb decay predictions, as well as a successful global description of the flavour observables~\cite{deAnda:2019jxw,deAnda:2020pti,deAnda:2020ssl,deAnda:2021jzc}.

\clearpage

\section{Lepton mixing patterns}                                           
\label{sec:flavor_symm}                                                    

The most striking implications of the oscillation experiments is that leptons mix quite differently from quarks. Here we consider the interpretation of the oscillation results in terms of phenomenological neutrino mixing patterns. Several of these were suggested, largely motivated by the desire to shed light on the neutrino oscillation parameters. The most salient features of the oscillation phenomenon are captured by the so called tri-bimaximal mixing (TBM) pattern~\cite{Harrison:2002er,Xing:2002sw,He:2003rm}. However there are other interesting mixing patterns, some of which we survey below~\footnote{ Note that, within a UV-complete theory framework, there can be radiative corrections to the TBM predictions~\cite{Luo:2005fc,Plentinger:2005kx,Hirsch:2006je,Wilina:2022pzy}. This issue lies beyond the scope of the model-independent approach followed here.}.

\subsection{Tri-bimaximal mixing }
\label{sec:tri-bimax-mixing}

The Tri-bimaximal lepton mixing pattern embodies bi-maximal mixing of atmospheric neutrino, tri-maximal solar mixing, no reactor mixing and hence no CP violation. It has been taken very seriously until the Daya Bay collaboration~\cite{DayaBay:2012fng,DayaBay:2012yjv,DayaBay:2013yxg} provided a robust measurement of the angle $\theta_{13}$ in 2012. The latest and most precise measurement of Daya Bay gives $\sin^2 2\theta_{13}=0.0851 \pm 0.0024$~\cite{Daya2022,DayaBay:2022orm}. Nevertheless, the TBM pattern still remains as an interesting first step towards a full description of neutrino mixing. Its mixing matrix is given by
\begin{equation}
\label{eq:UTBM}U_{TBM}=\begin{pmatrix}
\sqrt{\frac{2}{3}}   ~&~  \frac{1}{\sqrt{3}}  ~&~  0 \\
-\frac{1}{\sqrt{6}}  ~&~  \frac{1}{\sqrt{3}}  ~&~  \frac{1}{\sqrt{2}} \\
-\frac{1}{\sqrt{6}}  ~&~  \frac{1}{\sqrt{3}}  ~&~  -\frac{1}{\sqrt{2}}
\end{pmatrix}\,,
\end{equation}
which gives
\begin{equation}
\label{eq:TBM-angles}\theta_{23}=45^{\circ}, \quad \theta_{12}=\arctan\frac{1}{\sqrt{2}}\simeq35.26^{\circ},\quad \theta_{13}=0^{\circ}\,.
\end{equation}
Under the assumption of Majorana neutrinos, the most general form of the associated neutrino mass matrix in the charged lepton diagonal basis is
\begin{eqnarray}
\nonumber m^{TBM}_{\nu}&=&U_{TBM}\text{diag}(m_1, m_2, m_3)U^{T}_{TBM}\\
&=&\frac{1}{6}
\left(
\begin{array}{ccc}
4m_1+2m_2 ~&~ -2 m_1+2m_2 ~&~ -2m_1+2m_2 \\
-2m_1+2m_2 ~&~ m_1+2 m_2+3 m_3 ~&~ m_1+2 m_2-3 m_3 \\
-2m_1+2m_2 ~&~ m_1+2 m_2-3 m_3 ~&~ m_1+2 m_2+3 m_3
\end{array}
\right)\,.
\end{eqnarray}
It is easy to check that the above neutrino mass matrix is invariant under the following residual flavour transformations
\begin{equation}
\hskip-0.05cm G^{TBM}_{1}=\frac{1}{3}\left(
\begin{array}{ccc}
 1 ~& -2 ~& -2 \\
 -2 ~& -2 ~& 1 \\
 -2 ~& 1 ~& -2
\end{array}
\right),~ G^{TBM}_{2}=\frac{1}{3}\left(
\begin{array}{ccc}
 -1 ~& 2 ~& 2 \\
 2 ~& -1 ~& 2 \\
 2 ~& 2 ~& -1
\end{array}
\right),~ G^{TBM}_{3}=-\left(
\begin{array}{ccc}
1  ~&  0   ~&  0  \\
0  ~&  0  ~&  1  \\
0  ~&  1  ~&  0
\end{array}
\right)\,,
\end{equation}
up to an overall sign. In other words, the neutrino mass matrix $m^{TBM}_{\nu}$ fulfills
\begin{equation}
(G^{TBM}_{i})^{T}m^{TBM}_{\nu}G^{TBM}_{i}=m^{TBM}_{\nu}\,.
\end{equation}
The symmetry transformation $G^{TBM}_{3}$ exchanges the second and third columns as well as the second and third rows of the neutrino mass matrix, it is the well-known
$\mu-\tau$ permutation symmetry~\cite{Fukuyama:1997ky,Ma:2001mr,Balaji:2001ex,Lam:2001fb,Grimus:2001ex}, and it enforces maximal $\theta_{23}$ and vanishing
$\theta_{13}$. This symmetry can emerge in some UV-complete theories, e.g.~\cite{Babu:2002dz}.

The second transformation $G^{TBM}_{2}$ requires that the sum of the entries in each row and column of the light neutrino mass matrix are identical,
the so-called magic symmetry~\cite{Harrison:2004he,Lam:2006wy}. The invariance under $G^{TBM}_{2}$ determines one column of the lepton mixing matrix to be $(1, 1, 1)^{T}/\sqrt{3}$. The residual symmetry transformation $G_{e}$ of the charged lepton mass matrix is a generic diagonal phase matrix in the flavour basis~\footnote{If $G_{e}$ is a non-abelian subgroup, the charged lepton mass spectrum would be completely or partially degenerate. Thus $G_e$ should be a cyclic group $Z_n$ with the index $n\geq3$ or a product of cyclic groups such as $Z_2\times Z_2$ in order to distinguish among the generations.}.
The simplest choice for $G_{e}$ which can distinguish the three charged leptons is
\begin{equation}
G^{TBM}_{e}=\begin{pmatrix}
1  ~&~  0  ~&~ 0 \\
0  ~&~  \omega^2  ~&~  0 \\
0  ~&~  0   ~&~  \omega
\end{pmatrix}
\end{equation}
where  $\omega$ is the cubic root of one, i.e. $\omega=e^{2\pi i/3}$.

The group generated by $G^{TBM}_1$, $G^{TBM}_2$, $G^{TBM}_3$ and $G^{TBM}_{e}$ turns out to be $S_4$ group. Consequently, the minimal flavour symmetry group capable of yielding the tri-bimaximal mixing pattern is $S_4$~\cite{Lam:2008rs,Lam:2007qc}. However, in the absence of symmetry breaking by flavons in the $\mathbf{1}'$ and $\mathbf{1}''$ representations the $A_4$ flavor symmetry can also give rise to the tri-bimaximal mixing pattern~\cite{Ma:2001dn,Babu:2002dz,Altarelli:2005yp,Altarelli:2005yx,Altarelli:2006kg,Hirsch:2009mx,Altarelli:2012ss}. In concrete models, the tri-bimaximal mixing pattern can also be achieved through the vacuum alignment enforced by other finite non-abelian subgroups of $SO(3)$ or $SU(3)$~\cite{deMedeirosVarzielas:2005qg,deMedeirosVarzielas:2006fc,King:2009tj,Cao:2010mp,Ding:2011qt,Chen:2014wiw}.

\subsection{Generalizations of tri-bimaximal mixing}
\label{sec:gener-tri-bimax}

Soon after the tri-bimaximal mixing ansatz was proposed, possible deviations were considered. A simple generalization is obtained by assuming that just one given row or column of the lepton mixing matrix takes the same form as for tri-bimaximal~\cite{Harrison:2002kp}. If the first column of the tri-bimaximal mixing is retained, this is known as the TM1 mixing pattern~\cite{Albright:2008rp,Albright:2010ap,He:2011gb}, and the associated mixing matrix can be parameterized as
\begin{equation}
\label{eq:TM1} U_{TM1}=\frac{1}{\sqrt{6}}\begin{pmatrix}
2   ~&  \sqrt{2}\,\cos\theta    ~&  \sqrt{2}\,e^{-i\delta}\sin\theta \\
-1  ~&  \sqrt{2}\,\cos\theta-\sqrt{3}\,e^{i\delta}\sin\theta  ~&  \sqrt{3}\,\cos\theta+\sqrt{2}\,e^{-i\delta}\sin\theta  \\
-1  ~&  \sqrt{2}\,\cos\theta+\sqrt{3}\,e^{i\delta}\sin\theta   ~&  -\sqrt{3}\,\cos\theta+\sqrt{2}\,e^{-i\delta}\sin\theta
\end{pmatrix}\,.
\end{equation}
The predictions for the lepton mixing angles and leptonic Jarlskog invariant are expressed in terms of just two free parameters $\delta$ and $\theta$ as
\begin{eqnarray}
\nonumber&&\sin^2\theta_{13}=\frac{1}{3}\sin^2\theta,\qquad \sin^2\theta_{12}=\frac{1+\cos2\theta}{5+\cos2\theta},\\ &&\sin^2\theta_{23}=\frac{1}{2}+\frac{\sqrt{6}\,\cos\delta\sin2\theta}{5+\cos2\theta},\quad J_{CP}=\frac{1}{6\sqrt{6}}\sin\delta\sin2\theta\,.
\end{eqnarray}
One finds that the lepton mixing angles and Dirac CP phases are correlated as follows
\begin{equation}
3\cos^2\theta_{12}\cos^2\theta_{13}=2,\quad \tan2\theta_{23}\cos\delta^{\ell}=\frac{5\sin^2\theta_{13}-1}{2\sin\theta_{13}\sqrt{2-6\sin^2\theta_{13}}}\,.
\end{equation}
We notice that this TM1 pattern emerges as a prediction of the UV-complete models proposed in Refs.~\cite{deMedeirosVarzielas:2012apl,Chen:2020udk,Luhn:2013lkn,Li:2013jya}.

On the other hand, by retaining the second column of the tri-bimaximal mixing matrix one obtains the TM2~\cite{Albright:2008rp,Albright:2010ap,He:2011gb} pattern, a particular case of trimaximal mixing~\cite{Harrison:2002kp,Bjorken:2005rm,He:2006qd,Grimus:2008tt}, given as
\begin{equation}
\label{eq:TM2} U_{TM2}=\frac{1}{\sqrt{6}}\begin{pmatrix}
2\cos\theta  ~&~  \sqrt{2}   ~&~  2e^{-i\delta}\sin\theta \\
-\cos\theta-\sqrt{3}\,e^{i\delta}\sin\theta  ~&~  \sqrt{2}   ~&~  \sqrt{3}\,\cos\theta-e^{-i\delta}\sin\theta  \\
-\cos\theta+\sqrt{3}\,e^{i\delta}\sin\theta  ~&~ \sqrt{2}   ~&~  -\sqrt{3}\,\cos\theta-e^{-i\delta}\sin\theta
\end{pmatrix}\,.
\end{equation}
All the lepton mixing parameters depend on just two free parameters $\delta$ and $\theta$,
\begin{eqnarray}
\nonumber&&\sin^2\theta_{13}=\frac{2}{3}\sin^2\theta,~~~~\sin^2\theta_{12}=\frac{1}{2+\cos2\theta}\,,\\
&&\sin^2\theta_{23}=\frac{1}{2}-\frac{\sqrt{3}\cos\delta\sin2\theta}{2(2+\cos2\theta)}\,,~~~~J_{CP}=\frac{1}{6\sqrt{3}}\sin\delta\sin2\theta\,,
\end{eqnarray}
which implies the following correlations
\begin{equation}
3\sin^2\theta_{12}\cos^2\theta_{13}=1,\qquad \tan2\theta_{23}\cos \delta^{\ell}=\frac{\cos2\theta_{13}}{\sin\theta_{13}\sqrt{2-3\sin^2\theta_{13}}}\,.
\end{equation}
Here we note also that this TM2 neutrino mixing pattern emerges within the UV-complete models described in Refs.~\cite{King:2011zj,Shimizu:2011xg,Chen:2015jta,Ding:2020vud,Grimus:2008tt,Grimus:2009xw,Ding:2013hpa}.

\subsection{Golden ratio mixing pattern}
\label{sec:golden-ratio-mixing}

Within the standard golden ratio (GR) mixing pattern~\footnote{Another proposal for the golden ratio mixing has the solar mixing angle given as $\cos\theta_{12}=\phi_g/2$~\cite{Rodejohann:2008ir,Adulpravitchai:2009bg}.}, the lepton mixing angles are given by $\theta_{23}=45^{\circ}$, $\theta_{13}=0^{\circ}$ and $\cot\theta_{12}=\phi_{g}$,
where $\phi_{g}=(1+\sqrt{5})/2$ is the golden ratio~\cite{Datta:2003qg,Kajiyama:2007gx}. The lepton mixing matrix is of the form
\begin{equation}
\label{eq:real-GR}U_{GR}=\begin{pmatrix}
c_{12}   ~&~  s_{12}   ~&~  0  \\
-\frac{s_{12}}{\sqrt{2}}  ~ &~  \frac{c_{12}}{\sqrt{2}}  ~&~  \frac{1}{\sqrt{2}}\\
-\frac{s_{12}}{\sqrt{2}}   ~&~  \frac{c_{12}}{\sqrt{2}}  ~&~  -\frac{1}{\sqrt{2}}
\end{pmatrix}\equiv \frac{1}{\sqrt{2\sqrt{5}\phi_g}}\begin{pmatrix}
\sqrt{2}\phi_g  ~&~  \sqrt{2}   ~&~  0  \\
-1   ~&~  \phi_g  ~&~  \sqrt{\sqrt{5}\phi_g} \\
-1   ~&~  \phi_g  ~&~  -\sqrt{\sqrt{5}\phi_g}
\end{pmatrix}\,.
\end{equation}
In this case, the neutrino mass matrix is given by
\begin{eqnarray}
\label{5}\hskip-0.2cm m^{GR}_{\nu}=\frac{m_1}{2\sqrt{5}}\left(\begin{array}{ccc}
2\phi_g ~& -\sqrt{2}  ~& -\sqrt{2} \\
-\sqrt{2}  ~&  -1/\phi_g ~&  -1/\phi_g\\
-\sqrt{2}  ~& -1/\phi_g  ~& -1/\phi_g
\end{array}\right)
+\frac{m_2}{2\sqrt{5}}\left(\begin{array}{ccc}
2/\phi_g ~& \sqrt{2}   ~& \sqrt{2}\\
\sqrt{2}   ~& \phi_g   ~& \phi_g \\
\sqrt{2}   ~& \phi_g   ~& \phi_g
\end{array}\right)+\frac{m_3}{2}\left(\begin{array}{ccc} 0&0&0\\
0~&\;1~&\!\!-1\\
0~&\!\!-1&\;1
\end{array}\right)\,.
\end{eqnarray}
The residual flavour symmetry transformations of $m^{GR}_{\nu}$ take the form
\begin{eqnarray}
\nonumber&&
G^{GR}_1=\frac{1}{\sqrt{5}}\left(\begin{array}{ccc}
1  ~& -\sqrt{2} ~& -\sqrt{2} \\
-\sqrt{2}  ~& -\phi_g  ~&  1/\phi_g\\
-\sqrt{2}  ~& 1/\phi_g ~& -\phi_g
\end{array}\right),\;\;
G^{GR}_2=\frac{1}{\sqrt{5}}\left(\begin{array}{ccc}
-1 ~& \sqrt{2}  ~& \sqrt{2}\\
\sqrt{2}  ~& -1/\phi_g  ~& \phi_g\\
\sqrt{2}  ~& \phi_g  ~& -1/\phi_g
\end{array}\right)\,,\\
\label{7}&&
G^{GR}_3=-\left(\begin{array}{ccc}
1 ~&~ 0 ~&~ 0\\
0 ~&~ 0 ~&~ 1\\
0~&~ 1 ~&~ 0
\end{array}\right)\,.
\end{eqnarray}
The minimal flavour symmetry that can produce the golden ratio mixing pattern is the $A_5$ group~\cite{Everett:2008et,Feruglio:2011qq,Ding:2011cm}. Accordingly, a finite $Z_5$ subgroup is preserved by the charged lepton mass term with
\begin{equation}
G_e=\begin{pmatrix}
1  ~&  0  ~&  0  \\
0  ~&  e^{2\pi i/5}  ~& 0  \\
0  ~&  0  ~& e^{-2\pi i/5}
\end{pmatrix}\,.
\end{equation}
In order to be phenomenologically viable, the golden-ratio pattern would certainly require a revamping, along the lines considered in section~\ref{sec:revamp-lept-mixing}.

\subsection{Bi-maximal mixing pattern}
\label{subsec:bi-maximal-mixing}

For the bi-maximal mixing, both solar angle and atmospheric mixing angles are maximal $\theta_{12}=\theta_{23}=45^{\circ}$ while the reactor angle is vanishing~\cite{Barger:1998ta}. In the basis where the charged lepton mass matrix is diagonal, the corresponding neutrino mass matrix is given by
\begin{equation}
m^{BM}_{\nu}=\frac{1}{4}\left( \begin{array}{ccc}
2\left(m_1+m_2\right) ~&~ \sqrt{2} \left(m_2-m_1\right) ~&~ \sqrt{2}
\left(m_2-m_1\right) \\
\sqrt{2} \left(m_2-m_1\right) ~&~ m_1+m_2+2 m_3 ~&~
m_1+m_2-2 m_3 \\
\sqrt{2} \left(m_2-m_1\right) ~&~ m_1+m_2-2 m_3 ~&~
m_1+m_2+2 m_3 \\
\end{array}
\right)\,.
\end{equation}
The residual flavour symmetry transformations of the above neutrino mass matrix are \begin{eqnarray}
G^{BM}_{1}=\frac{1}{2}\left(
\begin{array}{ccc}
 0 & -\sqrt{2} & -\sqrt{2} \\
 -\sqrt{2} & -1 & 1 \\
 -\sqrt{2} & 1 & -1 \\
\end{array}
\right),~ G^{BM}_{2}=\frac{1}{2}\left(
\begin{array}{ccc}
 0 & \sqrt{2} & \sqrt{2} \\
 \sqrt{2} ~& -1 & 1 \\
 \sqrt{2} & 1 ~& -1 \\
\end{array}
\right),~ G^{BM}_{3}=-\left(
\begin{array}{ccc}
1  ~&  0   ~&  0  \\
0  ~&  0  ~&  1  \\
0  ~&  1  ~&  0
\end{array}
\right)\,.
\end{eqnarray}
We can choose the residual flavour symmetry of the charged lepton mass matrix to be a $Z_4$ subgroup with
\begin{equation}
G^{BM}_{e}=\begin{pmatrix}
1  ~&  0  ~&  0 \\
0  ~&  i  ~&  0  \\
0  ~&  0  ~&  -i
\end{pmatrix}\,.
\end{equation}
The group generated by $G^{BM}_1$, $G^{BM}_2$, $G^{BM}_3$ and $G^{BM}_e$ is also the $S_4$ group~\cite{Altarelli:2009gn,Li:2014eia}. Therefore the $S_4$ flavour symmetry can also be used to produce the bimaximal mixing.

Note, however, that the solar angle $\theta_{12}$ and reactor angle $\theta_{13}$ would have to undergo very large corrections in order to be compatible with current neutrino oscillation data~\cite{deSalas:2020pgw,10.5281/zenodo.4726908}, making this pattern very problematic. As a result, Bi-maximal mixing should be discarded or generalized in a radical way. One of its possible radical generalizations is the Bi-large mixing pattern discussed in section~\ref{sec:bi-large-mixing}.

Besides the tri-bimaximal, golden ratio and bi-maximal mixing patterns, there are other constant mixing patterns compatible with experimental data with nonzero $\theta_{13}$~\cite{deAdelhartToorop:2011nfg,deAdelhartToorop:2011re,Ding:2012xx,Hagedorn:2013nra}. They could be derived from the breaking of large flavor symmetry groups such as $\Delta(96)$, $\Delta(384)$ etc. No sizable corrections are necessary for these mixing patterns, however the required vacuum configuration and symmetry breaking are more complicated.

As shown above, the neutrino mass matrices corresponding to the simple tri-bimaximal, golden ratio and bi-maximal mixing patterns obey certain residual flavour symmetries. In the following section, we will show that the neutrino and charged lepton mass matrices generally can have both residual flavour and residual CP symmetries, and also that a residual flavour symmetry can be generated by a residual CP symmetry. In particular, residual symmetries can provide a method to revamp all mixing patterns discussed above. Indeed, in the next section we will show how to revamp them in a systematic manner by using the residual symmetry method. The generalized patterns are not only phenomenologically viable, but also predictive, since the form of the resulting lepton mixing matrices can be restricted.

\clearpage

\section{Flavour and CP symmetries from the bottom-up}                     
\label{sec:flavor_CP_bottom-up}                                            

In this section, we will examine quark and lepton mass matrices that have both remnant flavour and CP symmetry. Remnant flavour symmetries can be generated by performing two remnant CP transformations in succession, with explicit forms of the remnant CP symmetry
derived from the experimentally measured mixing matrix. On the other hand, the fermion mixing matrices can be constructed from the postulated residual CP transformations of the quark and lepton mass matrices. In the following, we present the remnant flavour and CP symmetries of the quark and lepton mass matrices, their parametrization and the master formula to construct the mixing matrix from the remnant CP symmetry.

\subsection{Residual symmetries of leptons}
\label{subsec:residual_sym_lepton_sector}

In the absence of a fundamental theory of flavour we study the effect of possible remnant symmetries $G_\nu$ and $G_l$ of the neutrino and charged lepton mass terms. Their existence may provide a model-independent approach towards underpinning the ultimate nature of the underlying theory. Let us now focus on this point.

The lepton masses terms are given in Eq.~\eqref{eq:charge-lepton-neutrino-dirac-mass} and Eq.~\eqref{eq:charge-lepton-neutrino-maj-mass} for Dirac and Majorana neutrinos respectively. Since in the SM the only physical mixing matrix relates to left-handed fermions, we are concerned with the Hermitian mass matrix $m^{\dagger}_{l}m_{l}$ connecting left-handed charged leptons on both sides. Under generic unitary transformations of the left-handed lepton fields $l_L$ and $\nu_L$,
\begin{equation}
l_{L}\to G_{l}l_{L},\quad\nu_{L}\to G_{\nu}\nu_{L}\,,
\end{equation}
where both $G_{l}$ and $G_{\nu}$ are three-dimensional unitary matrices, the charged lepton and neutrino mass matrices transform as
\begin{align}
\label{eq:mldag_ml_transf}&m^{\dagger}_{l}m_{l}\to G^{\dagger}_{l}m^{\dagger}_{l}m_{l}G_{l}\,,\\
\label{eq:mnu_transf}&m_{\nu}\to G^{T}_{\nu}m_{\nu}G_{\nu},~~~~\quad\quad \text{for Majorana neutrinos}\,,\\
\label{eq:mnudag_mnu_transf}&m^{\dagger}_{\nu}m_{\nu}\to G^{\dagger}_{\nu}m^{\dagger}_{\nu}m_{\nu}G_{\nu},\quad \text{for Dirac neutrinos}\,.
\end{align}
In order for the symmetry to hold, the mass matrices must satisfy
\begin{align}
\label{eq:mldag_ml_sysRQ}&G^{\dagger}_{l}m^{\dagger}_{l}m_{l}G_{l}=m^{\dagger}_{l}m_{l}\,,\\
\label{eq:mnu_sysRQ}     & G^{T}_{\nu}m_{\nu}G_{\nu}=m_{\nu},~~~~\quad\quad\text{for Majorana neutrinos}\,,\\
\label{eq:mnudag_mnu_sysRQ}&G^{\dagger}_{\nu}m^{\dagger}_{\nu}m_{\nu}G_{\nu}=m^{\dagger}_{\nu}m_{\nu},\quad \text{for Dirac neutrinos}\,.
\end{align}
Applying these invariance conditions to Eqs.~(\ref{eq:mldag_ml_general}, \ref{eq:mnu_general}, \ref{eq:mnudag_mnu_general}) and assuming no mass eigenvalue vanishes we obtain
\begin{align}
\label{eq:Ul_Cons}&U^{\dagger}_{l}G_{l}U_{l}=\textrm{diag}\left(e^{i\alpha_{e}}, e^{i\alpha_{\mu}}, e^{i\alpha_{\tau}}\right)\,,\\
\label{eq:Unu_Cons_Maj}& U^{\dagger}_{\nu}G_{\nu}U_{\nu}=\text{diag}\left(\pm1, \pm1, \pm1\right),~~~~\quad\quad \text{for Majorana neutrinos}\,,\\
\label{eq:Unu_Cons_Dirac}&U^{\dagger}_{\nu}G_{\nu}U_{\nu}=\textrm{diag}\left(e^{i\alpha_{\nu_e}}, e^{i\alpha_{\nu_{\mu}}}, e^{i\alpha_{\nu_{\tau}}}\right),\quad \text{for Dirac neutrinos}\,,
\end{align}
where $\alpha_{e, \mu, \tau}$ and $\alpha_{\nu_{e}, \nu_{\mu}, \nu_{\tau}}$, are arbitrary real parameters. It follows that the residual flavour symmetry transformations $G_{l}$ and $G_{\nu}$ are of the following form~\cite{Yao:2015dwa,Chen:2014wxa}:
\begin{align}
\label{eq:Gl_remnant}&G_{l}=U_{l}\textrm{diag}\left(e^{i\alpha_{e}}, e^{i\alpha_{\mu}}, e^{i\alpha_{\tau}}\right)U^{\dagger}_{l}\,,\\
\label{eq:Gnu_remnant_Maj}&G_{\nu}=U_{\nu}\textrm{diag}\left(\pm1, \pm1, \pm1\right)U^{\dagger}_{\nu},~~~~\quad \text{for Majorana neutrinos}\,,\\
\label{eq:Gnu_remnant_Dirac}&G_{\nu}=U_{\nu}\textrm{diag}\left(e^{i\alpha_{\nu_e}}, e^{i\alpha_{\nu_{\mu}}}, e^{i\alpha_{\nu_{\tau}}}\right)U^{\dagger}_{\nu},\quad \text{for Dirac neutrinos}\,.
\end{align}
One sees that the charged lepton mass term generically admits a $U(1)\times U(1)\times U(1)$ remnant flavour symmetry. For the neutrino mass matrix the remnant flavour symmetry depends on the nature of neutrinos. For the case of Dirac neutrinos it is also $U(1)\times U(1)\times U(1)$~\cite{Yao:2015dwa}. For Majorana neutrinos, the eight possible choices of $G_{\nu}$ in Eq.~\eqref{eq:Gnu_remnant_Maj} correspond to $Z_2\times Z_2\times Z_2$. Notice that $G_{\nu}$ and $-G_{\nu}$ should be identified as the same residual flavour symmetry transformation, since the minus sign can be absorbed as a neutrino field redefinition.
Indeed, they both lead to the same constraint on the neutrino mass matrix. We are therefore left with four possible solutions for $G_{\nu}$, which can be chosen as~\cite{Lam:2008rs,Lam:2007qc,Lam:2011ag,Lam:2011zm},
\begin{equation}
G_{i}=U_{\nu}\;d_{i}U^{\dagger}_{\nu},\qquad i=1,2,3,4\,,
\end{equation}
where the $d_i$ are given as
\begin{eqnarray}
\nonumber&&d_1=\text{diag}\left(1,-1,-1\right),\qquad d_2=\text{diag}\left(-1,1,-1\right),\\
\label{eq:d1234}&&d_3=\text{diag}\left(-1,-1,1\right),\qquad d_4=\text{diag}\left(1,1,1\right)\,.
\end{eqnarray}
One sees that $G_{4}$ is simply the trivial identity matrix, and we can further check that
\begin{equation}
G^2_{i}=1,\qquad G_{i}G_{j}=G_{j}G_{i}=G_{k}~~\text{with}~~i\neq j\neq k\neq4.
\end{equation}
It follows that the residual flavour symmetry of the Majorana neutrino mass matrix is a Klein group isomorphic to $Z_2\times Z_2$. In the flavour basis where the charged lepton mass matrix $m_{l}$ is diagonal, $U_{l}$ would be trivial, so that the lepton mixing comes just from the neutrino sector, i.e. $U_{}=U_{\nu}$. Hence the residual symmetries of the neutrino and charged lepton mass matrices are determined in terms of the mixing angles and CP violation phases contained in the mixing matrix.

An important observation is that, besides the above residual flavour symmetry, the lepton mass matrices can have residual CP (charge-conjugation and parity) symmetry. The CP transformation properties of the left-handed neutrino and charged lepton fields are given by
\begin{equation}
l_{L}(x)\stackrel{CP}{\longmapsto}iX_{l}\gamma^{0}C\bar{l}^{\,T}_{L}(\mathcal{P}x),\quad \nu_{L}(x)\stackrel{CP}{\longmapsto}iX_{\nu}\gamma^{0}C\bar{\nu}^{T}_{L}(\mathcal{P}x)\,,
\end{equation}
where $\mathcal{P}x=(t,-\vec{x})$, $C$ is the charge-conjugation matrix~\cite{Peskin:1995ev,Burgess:2006hbd,Donoghue:1992dd}, $X_{l}$ and $X_{\nu}$ are $3\times3$ unitary matrices acting in flavor space. Notice that the matrices $X_{l}$ and  $X_{\nu}$ generalize the conventional CP transformation prescription. As a result, they are referred to in the literature as generalized CP transformations~\cite{Feruglio:2012cw,Ecker:1981wv,Ecker:1983hz,Bernabeu:1986fc,Ecker:1987qp,Neufeld:1987wa,Holthausen:2012dk}.

Requiring that $X_{l}$ and $X_{\nu}$ are symmetries of the lepton mass terms in Eqs.~(\ref{eq:charge-lepton-neutrino-dirac-mass}, \ref{eq:charge-lepton-neutrino-maj-mass}) implies that the lepton mass matrices $m_{l}$ and $m_{\nu}$ should satisfy~\cite{Chen:2014wxa,Yao:2016zev}
\begin{align}
\label{eq:mldag_ml_CPRQ}&X^{\dagger}_{l}m^{\dagger}_{l}m_{l}X_{l}=\left(m^{\dagger}_{l}m_{l}\right)^{*}\,,\\
\label{eq:mnu_CPRQ}& X^{T}_{\nu}m_{\nu}X_{\nu}=m^{*}_{\nu},~~~~~~~~~~~~\quad \text{for Majorana neutrinos}\,,\\
\label{eq:mnudag_mnu_CPRQ}&X^{\dagger}_{\nu}m^{\dagger}_{\nu}m_{\nu}X_{\nu}=\left(m^{\dagger}_{\nu}m_{\nu}\right)^{*},\quad \text{for Dirac neutrinos}\,.
\end{align}
Plugging Eqs.~(\ref{eq:mldag_ml_general}, \ref{eq:mnu_general}, \ref{eq:mnudag_mnu_general}) into the above invariance conditions, one sees that the unitary transformations $U_{l}$ and $U_{\nu}$ must be subject to the following conditions
\begin{align}
\label{eq:Ul_Cons_CP}&U^{\dagger}_{l}X_{l}U^{*}_{l}=\textrm{diag}\left(e^{i\beta_{e}}, e^{i\beta_{\mu}}, e^{i\beta_{\tau}}\right)\,,\\
\label{eq:Unu_Cons_Maj_CP}& U^{\dagger}_{\nu}X_{\nu}U^{*}_{\nu}=\text{diag}\left(\pm1, \pm1, \pm1\right),~~~~~\quad \text{for Majorana neutrinos}\,,\\
\label{eq:Unu_Cons_Dirac_CP}&U^{\dagger}_{\nu}X_{\nu}U^{*}_{\nu}=\textrm{diag}\left(e^{i\beta_{\nu_e}}, e^{i\beta_{\nu_{\mu}}}, e^{i\beta_{\nu_{\tau}}}\right),\quad \text{for Dirac neutrinos}\,,
\end{align}
where $\beta_{e, \mu, \tau}$ and $\beta_{\nu_e, \nu_{\mu}, \nu_{\tau}}$ are real free parameters. Hence the residual CP transformations $X_{l}$ and $X_{\nu}$ should be of the form~\cite{Yao:2015dwa,Feruglio:2012cw,Chen:2014wxa,Everett:2015oka,Chen:2015nha,Everett:2016jsk}
\begin{align}
\label{eq:Xl_remnant}&X_{l}=U_{l}\textrm{diag}\left(e^{i\beta_{e}}, e^{i\beta_{\mu}}, e^{i\beta_{\tau}}\right)U^{T}_{l}\,,\\
\label{eq:Xnu_remnant_Maj}&X_{\nu}=U_{\nu}\textrm{diag}\left(\pm1, \pm1, \pm1\right)U^{T}_{\nu},~~~~~\quad \text{for Majorana neutrinos}\,,\\
\label{eq:Xnu_remnant_Dirac}&X_{\nu}=U_{\nu}\textrm{diag}\left(e^{i\beta_{\nu_e}}, e^{i\beta_{\nu_{\mu}}}, e^{i\beta_{\nu_{\tau}}}\right)U^{T}_{\nu},\quad \text{for Dirac neutrinos}\,.
\end{align}
Clearly, both $X_{l}$ and $X_{\nu}$ are unitary and symmetric matrices~\cite{Feruglio:2012cw,Chen:2014wxa}
\begin{equation}
X_{l}=X^{T}_{l},\quad X_{\nu}=X^{T}_{\nu},\quad X_{l}X^{\dagger}_{l}=X_{\nu}X^{\dagger}_{\nu}=1\,.
\end{equation}
From the expressions of remnant flavor symmetry in Eqs.~(\ref{eq:Gl_remnant}, \ref{eq:Gnu_remnant_Maj}, \ref{eq:Gnu_remnant_Dirac}) and remnant CP transformations in Eqs.~(\ref{eq:Xl_remnant}, \ref{eq:Xnu_remnant_Maj}, \ref{eq:Xnu_remnant_Dirac}), one can check that the residual flavour and CP symmetries satisfy the following restricted consistency conditions
\begin{eqnarray}
\nonumber&& X_{l}G^{*}_{l}X^{-1}_{l}=G^{-1}_{l}\,,\\
\nonumber && X_{\nu}G^{*}_{\nu}X^{-1}_{\nu}=G_{\nu},~\quad \text{for Majorana neutrinos}\,,
\\
&& X_{\nu}G^{*}_{\nu}X^{-1}_{\nu}=G^{-1}_{\nu},\quad \text{for Dirac neutrinos}\,.
\end{eqnarray}
If we successively perform two CP transformations on the left-handed charged lepton fields, characterized by
$X_{l}=U_{l}\textrm{diag}\left(e^{i\beta_{e}}, e^{i\beta_{\mu}}, e^{i\beta_{\tau}}\right)U^{T}_{l}$ and
$X'_{l}=U_{l}\textrm{diag}\left(e^{i\beta'_{e}}, e^{i\beta'_{\mu}}, e^{i\beta'_{\tau}}\right)U^{T}_{l}$
we obtain~\footnote{Notice that an overall minus sign is dropped in the last step, as it can be absorbed into the lepton field.}
\begin{equation}
l_{L}(x)\stackrel{CP}{\longmapsto}iX_{l}\gamma^{0}C\bar{l}^{\,T}_{L}(\mathcal{P}x)\stackrel{CP}{\longmapsto}X_{l}X'^{*}_{l}l_{L}(x)\,,
\end{equation}
with
\begin{equation}
X_{l}X'^{*}_{l}=U_{l}\textrm{diag}\left(e^{i\left(\beta_{e}-\beta'_{e}\right)}, e^{i\left(\beta_{\mu}-\beta'_{\mu}\right)}, e^{i\left(\beta_{\tau}-\beta'_{\tau}\right)}\right)U^{\dagger}_{l}\,.
\end{equation}
Moreover, from the invariance condition of $m^{\dagger}_{l}m_{l}$ under the residual CP symmetry, Eq.~\eqref{eq:mldag_ml_CPRQ}, it is easy to show that
\begin{equation}
X'^{T}_{l}X^{\dagger}_{l}m^{\dagger}_{l}m_{l}X_{l}X'^{*}_{l}=m^{\dagger}_{l}m_{l}\,.
\end{equation}
This means that performing two CP transformations in succession is equivalent to a flavour symmetry transformation $X_lX'^{\ast}_l\equiv G_{l}$~\cite{Chen:2014wxa}. The same conclusion also holds true for neutrinos if they are Dirac particles.

For the case of Majorana neutrinos, there are eight possibilities for $X_{\nu}$. In this case one should take $X_{\nu}$ and $-X_{\nu}$ as the same CP transformation, since the minus sign can be absorbed into the neutrino fields and the mass term involves a product of two neutrino fields. As a result only four of them are relevant. Without loss of generality they can be chosen to be~\cite{Chen:2014wxa,Everett:2015oka,Chen:2015nha}
\begin{equation}
X_{i}=U_{\nu}d_i U^{T}_{\nu},~~ i=1,2,3,4\,,\label{eq:residual-CP-nu}
\end{equation}
with $d_i$ given in Eq.~\eqref{eq:d1234}. The remaining four can be obtained from the above by multiplying an overall $-1$ factor. It is straightforward to check that the neutrino mass matrix $m_{\nu}$ satisfies
\begin{equation}
X^{\dagger}_jX_{i}^Tm_{\nu}X_iX_j^{\ast}=m_{\nu}\,.
\end{equation}
As a result, remnant flavour symmetries can be generated by remnant CP symmetries as well.
Explicitly, we have the following relations~\cite{Chen:2014wxa}:
\begin{equation}
\label{eq:CP_flavor_relations}
\begin{split}
&X_2X^{*}_3=X_3X^{*}_2= X_4X^{*}_1=X_1X^{*}_4=G_1,\\
&X_1X^{*}_3=X_3X^{*}_1=X_4X^{*}_2=X_2X^{*}_4=G_2,\\
&X_1X^{*}_2=X_2X^{*}_1=X_4X^{*}_3=X_3X^{*}_4=G_3,\\
&X_1X^{*}_1=X_2X^{*}_2=X_3X^{*}_3=X_4X^{*}_4=G_4=1\,.
\end{split}
\end{equation}
As a result, once we impose a set of generalized CP transformations, there is always an associated flavour symmetry. Furthermore, Eq.~\eqref{eq:CP_flavor_relations} implies that any residual CP transformation can be expressed in terms of the remaining ones as follows~\cite{Chen:2014wxa},
\begin{equation}
\label{eq:CP_3_indep}X_{i}=X_{j}X^{\ast}_{m}X_{n},\qquad i\neq j\neq m\neq n\,.
\end{equation}
In other words, only three of the four remnant CP transformations are independent. In the flavour basis where $m_{l}$ is diagonal, the residual CP transformation $X_{l}$ is an arbitrary diagonal phase matrix, while $X_{\nu}=U_{}\textrm{diag}\left(\pm1, \pm1, \pm1\right)U^{T}_{}$. Hence the remnant CP symmetry can be constructed from the neutrino mixing matrix, and its explicit form can be determined more precisely with the improved measurement of the mixing angles and CP phases~\cite{deSalas:2020pgw,10.5281/zenodo.4726908}.

Here we have so far assumed that the three light neutrino masses are non-vanishing. If the lightest neutrino is massless one can analyze the residual symmetry of the neutrino mass matrix in the same way. This interesting situation occurs generically in the ``missing partner'' seesaw mechanism~\cite{Schechter:1980gr} in which there are less ``right'' than ``left''-handed neutrinos~\footnote{The present neutrino data allows the possibility that the lightest neutrino is massless, consequently at least two right-handed neutrinos are necessary in type-I seesaw~\cite{King:1999mb,Frampton:2002qc,Raidal:2002xf}.} (a massless neutrino also emerges in theories where anti-symmetric Yukawa couplings are involved in generating neutrino mass~\cite{Leite:2019grf}). Note that an incomplete fermion multiplet structure can also be interesting in cosmological models of dark matter~\cite{Reig:2018ztc,Barreiros:2018bju,Rojas:2018wym,Aranda:2018lif,Mandal:2021yph,Avila:2019hhv} as well as leptogenesis (for a discussion of seesaw leptogenesis with two and three right-handed neutrinos see~\cite{Abada:2018oly,Drewes:2021nqr}.)

With one neutrino massless one has $m_1=0$ for normal ordered (NO) neutrino mass spectrum, and $m_3=0$ for the inverted ordered (IO) spectrum. If neutrinos have Dirac nature, one finds that the residual CP transformation $X_{\nu}$ would still be given by Eq.~\eqref{eq:Xnu_remnant_Dirac}. On the other hand, if neutrinos are Majorana particles, one must use Eq.~\eqref{eq:Xnu_remnant_Maj} replacing the diagonal entry ``$\pm1$'' in the position (11) for NO and (33) for IO with an arbitrary phase factor~\cite{Li:2017zmk}, i.e.
\begin{equation}
X_{\nu}=\left\{
\begin{array}{lc}
U_{\nu}\textrm{diag}\left(e^{i\beta}, \pm1, \pm1\right)U^{T}_{\nu},\quad \text{for NO}\,,\\
U_{\nu}\textrm{diag}\left(\pm1, \pm1, e^{i\beta}\right)U^{T}_{\nu},\quad \text{for IO}\,,
\end{array}
\right.
\end{equation}
where $\beta$ is real. Similarly the residual flavour symmetry $G_{\nu}$ becomes~\cite{Chen:2014wxa,Li:2017zmk}
\begin{equation}
G_{\nu}=\left\{
\begin{array}{lc}
U_{\nu}\textrm{diag}\left(e^{i\alpha}, \pm1, \pm1\right)U^{\dagger}_{\nu},\quad \text{for NO}\,,\\
U_{\nu}\textrm{diag}\left(\pm1, \pm1, e^{i\alpha}\right)U^{\dagger}_{\nu},\quad \text{for IO}\,,
\end{array}
\right.
\end{equation}
with real $\alpha$. Hence the generalized CP and residual flavour symmetry groups have the structure $U(1)\times Z_2$ (instead of $Z_2\times Z_2$) modulo a possible overall factor $-1$ in this case. On the other hand, this kind of residual symmetry can enforce a massless Majorana neutrino. The breaking of finite discrete groups into this form of $G_{\nu}$ in the neutrino sector was analyzed in~\cite{Joshipura:2013pga,Joshipura:2014pqa,King:2016pgv}, where the angle $\alpha$ was a rational multiple of $\pi$.

\subsection{Reconstructing lepton mixing from remnant CP symmetry}
\label{sec:constructing-mixing-res-sym}

As shown in above, residual CP symmetries can be derived from the mixing matrix, and conversely, the lepton mixing matrix can be constructed from the remnant CP symmetries in the neutrino and the charged lepton sectors. In concrete models, we can start from a set of CP transformations $\mathcal{X_{CP}}$ respected by the Lagrangian at some high energy scale.
Subsequently $\mathcal{X_{CP}}$ is spontaneously broken by some scalar fields into different remnant symmetries in the neutrino and the charged lepton sectors. The misalignment between the two remnant symmetries is responsible for the mismatch of the rotations which diagonalize the neutrino and charged lepton matrices, leading to the lepton mixing matrix.
We now present the general parametrization for the remnant CP symmetries of the neutrino and charged lepton sector, and the corresponding restrictions on the unitary transformations $U_{\nu}$ and $U_l$.

We start from the simplest nontrivial case in which a single remnant CP transformation $X_R$ is preserved by the neutrino mass matrix. As shown in section~\ref{subsec:residual_sym_lepton_sector}, $X_R$ should be a symmetric unitary matrix, otherwise the light neutrino masses would be degenerate. Thus $X_R$ can be parameterized as follows~\cite{Chen:2014wxa}:
\begin{equation}
\label{eq:X_one}X_R=e^{i\kappa_1}v_{1}v^{T}_{1}+e^{i\kappa_2}v_{2}v^{T}_{2}+e^{i\kappa_3}v_{3}v^{T}_{3}\,,
\end{equation}
where the phases $\kappa_1$, $\kappa_2$ and $\kappa_3$ can be taken in the range of $0$ and $2\pi$ without loss of generality, $v_1$, $v_2$ and $v_3$ are mutually orthogonal vectors with
\begin{eqnarray}
\nonumber&&v_1=\begin{pmatrix}
\cos\varphi            \\
\sin\varphi\cos\phi    \\
\sin\varphi\sin\phi
\end{pmatrix}\,,~~~~v_2=\begin{pmatrix}
\sin\varphi\cos\rho   \\
-\sin\phi\sin\rho-\cos\varphi\cos\phi\cos\rho   \\
\cos\phi\sin\rho-\cos\varphi\sin\phi\cos\rho
\end{pmatrix},\\
&&v_3=\begin{pmatrix}
\sin\varphi\sin\rho     \\
\sin\phi\cos\rho-\cos\varphi\cos\phi\sin\rho    \\
-\cos\phi\cos\rho-\cos\varphi\sin\phi\sin\rho
\end{pmatrix}\,.
\end{eqnarray}
Invariance of the neutrino mass matrix under $X_R$ implies that $U_{\nu}$ should be subject to the constraint in Eq.~\eqref{eq:Unu_Cons_Maj_CP} and Eq.~\eqref{eq:Unu_Cons_Dirac_CP}
for Majorana and Dirac neutrinos. As a consequence, $U_{\nu}$ is fixed to be~\cite{Chen:2014wxa}
\begin{equation}
\label{eq:Unu-one-CP}U_{\nu}=\left(v_1, v_2, v_3\right)\text{diag}\big(e^{i\frac{\kappa_1}{2}}, e^{i\frac{\kappa_2}{2}}, e^{i\frac{\kappa_3}{2}}\big)O_{3}(\theta_1, \theta_2, \theta_3)Q_{\nu}\,,
\end{equation}
where $O_3$ is a generic real orthogonal matrix,
\begin{equation}
\label{eq:orthogonal-matrix}
O_{3}(\theta_1, \theta_2, \theta_3)=\begin{pmatrix}
1 ~& 0 ~& 0 \\
0 ~& \cos\theta_1   ~&   \sin\theta_1 \\
0 ~& -\sin\theta_1  ~&   \cos\theta_1
\end{pmatrix}
\begin{pmatrix}
\cos\theta_2   ~&   0    ~&    \sin\theta_2 \\
0   ~&   1   ~&   0 \\
-\sin\theta_2   ~&   0   ~&    \cos\theta_2
\end{pmatrix}
\begin{pmatrix}
\cos\theta_3     ~&    \sin\theta_3    ~&    0 \\
-\sin\theta_3    ~&    \cos\theta_3    ~& 0   \\
0    ~&    0     ~&    1
\end{pmatrix}\,,
\end{equation}
where the real rotation angles $\theta_{1,2,3}\in[0, 2\pi)$ are free. The matrix $Q_{\nu}$ is diagonal, and it entries are $\pm1$ and $\pm i$ which encode the CP parity of the neutrinos, while it is unphysical for Dirac neutrinos.

If two remnant CP transformations $X_{R1}$ and $X_{R2}$ out of the original CP symmetry are preserved in the neutrino sector, they can generally be written as~\cite{Chen:2014wxa}
\begin{eqnarray}
\nonumber&&X_{R1}=e^{i\kappa_1}v_1v^T_1+e^{i\kappa_2}v_2v^T_2+ e^{i\kappa_3}v_3v^T_3\,,\\
&&X_{R2}=e^{i\kappa_1}v_1v^T_1-e^{i\kappa_2}v_2v^T_2-e^{i\kappa_3}v_3v^T_3\,,
\end{eqnarray}
for Majorana neutrinos. A remnant flavour transformation $G_R$ can be obtained from $X_{R1}$ and $X_{R2}$ as~\footnote{For Dirac neutrinos, the neutrino mixing matrix $U_{\nu}$ would be completely fixed up to column permutations, if the order of the remnant flavour transformation $G_{R}=X_{R1}X^{*}_{R2}=X_{R2}X^{*}_{R1}$ is greater than or equal to three, so as to distinguish the three families.}
\begin{equation}
\label{eq:GR}G_{R}=X_{R1}X^{*}_{R2}=X_{R2}X^{*}_{R1}=2v_1v^{T}_1-1\,,
\end{equation}
which satisfies $G^2_R=1$. Hence a remnant $Z_2$ flavor symmetry generated by $G_R$ is induced and it fixes one column of $U_{\nu}$ to be $v_1$. Besides the parameters characterizing the remnant CP symmetry, $U_{\nu}$ is determined just by a free rotation angle $\theta$~\cite{Chen:2014wxa},
\begin{equation}
\label{eq:Unu-two-CP}U_{\nu}=\left(v_1, v_2, v_3\right)\text{diag}\big(e^{i\frac{\kappa_1}{2}}, e^{i\frac{\kappa_2}{2}}, e^{i\frac{\kappa_3}{2}}\big)R_{23}(\theta)P_{\nu}Q_{\nu}\,,
\end{equation}
where $R_{23}(\theta)$ denotes a rotation matrix through an angle $\theta$ in the (23)-plane with $0\leq\theta<\pi$,
\begin{equation}
\label{eq:R23-matrix}R_{23}(\theta)=\left(
\begin{array}{ccc}
 1 ~& 0 ~& 0 \\
 0 ~& \cos\theta  ~& \sin\theta  \\
 0 ~& -\sin\theta  ~& \cos\theta
\end{array}
\right)\,.
\end{equation}
Since the remnant symmetry can not constrain the ordering of the light neutrino mass eigenvalues, $U_{\nu}$ is determined up to independent column permutations, and consequently $P_{\nu}$ is a generic permutation matrix which can take six possible forms $1$, $P_{12}$, $P_{13}$, $P_{23}$, $P_{23}P_{12}$, $P_{23}P_{13}$ with
\begin{equation}
P_{12}=\begin{pmatrix}
0  ~&~ 1  ~&~  0 \\
1  ~&~  0 ~&~  0 \\
0  ~&~  0  ~&~  1
\end{pmatrix},~~~P_{13}=\begin{pmatrix}
0 ~&~  0 ~&~  1 \\
0 ~&~  1  ~&~  0 \\
1 ~&~  0  ~&~ 0
\end{pmatrix},~~~P_{23}=\begin{pmatrix}
1 ~&~  0  ~&~  0  \\
0  ~&~ 0  ~&~  1 \\
0  ~&~  1 ~&~ 0
\end{pmatrix}\,.
\end{equation}
We see that the unitary transformations $U_{\nu}$ for the two permutations $P_{\nu}$ and $P_{23}P_{\nu}$ are related by the redefinitions $\theta\to\theta-\pi/2$ and $Q_{\nu}\to P^{T}_{\nu} \text{diag}(1,1,-1)P_{\nu}Q_{\nu}$. Hence only three inequivalent permutations of the columns are relevant in this case. In other words, the fixed vector $\left(\cos\varphi, \sin\varphi\cos\phi, \sin\varphi\sin\phi\right)^{T}$ can be the first column, the second column or the third column of the matrix $U_{\nu}$.

For the case of Majorana neutrinos, we consider the scenario that all independent remnant CP transformations are preserved by the neutrino mass matrix. The remnant CP transformations can be parameterized as~\cite{Chen:2015nha}
\begin{eqnarray}
\nonumber&&X_{R1}= e^{i\kappa_1}v_1v^T_1+e^{i\kappa_2}v_2v^T_2+e^{i\kappa_3}v_3v^T_3\,,\\
\nonumber&&X_{R2}=e^{i\lambda_1}v_1v^T_1+e^{i\lambda_2}w_2w^T_2+e^{i\lambda_3}w_3w^T_3\,,\\
\nonumber&& X_{R3}=e^{i\lambda_1}v_1v^T_1-e^{i\lambda_2}w_2w^T_2-e^{i\lambda_3}w_3w^T_3\,,\\
&& X_{R4}=e^{i\kappa_1}v_1v^T_1-e^{i\kappa_2}v_2v^T_2-e^{i\kappa_3}v_3v^T_3\,,
\end{eqnarray}
where $v_1$, $w_2$ and $w_3$ also form another set of real orthonormal vectors with
\begin{equation}
w_2=\cos\xi v_2-\sin\xi v_3,\qquad w_3=\sin\xi v_2+\cos\xi v_3\,,
\end{equation}
and the phases $e^{i\lambda_1}$, $e^{i\lambda_2}$ and $e^{i\lambda_3}$ are given by
\begin{equation}
e^{i\lambda_1}=-e^{i\kappa_1},\quad e^{i\lambda_2}=-\frac{e^{i\kappa_2}\cos^2\xi+e^{i\kappa_3}\sin^2\xi}{\left|e^{i\kappa_2}\cos^2\xi+ e^{i\kappa_3}\sin^2\xi\right|},\quad e^{i\lambda_3}=\frac{e^{i\kappa_2}\sin^2\xi+e^{i\kappa_3}\cos^2\xi}{\left|e^{i\kappa_2}\sin^2\xi+e^{i\kappa_3}\cos^2\xi\right|}\,.
\end{equation}
A remnant Klein four flavour symmetry $K_4\equiv\left\{1, G_{R1}, G_{R2}, G_{R3}\right\}$ can be generated by performing two CP transformations,
and the three nontrivial residual flavour symmetry transformations $G_{Ri}$ for $i=1, 2, 3$ can be expressed as
\begin{eqnarray}
\nonumber&&G_{R1}=X_{R1}X^{*}_{R4}=v_1v_1^T-v_2v_2^T-v_3 v_3^T\,,\\
\nonumber&&G_{R2}=X_{R1}X^{*}_{R3}=-v_1v^T_1-c_{22}v_2v_2^T-c_{33}v_3v_3^T-c_{23}v_2v_3^T-c_{32}v_3v_2^T,\\
&&G_{R3}=X_{R1}X^{*}_{R2}=-v_1v^T_1+c_{22}v_2v_2^T+c_{33}v_3v_3^T+c_{23}v_2v_3^T+c_{32}v_3v_2^T,
\end{eqnarray}
with
\begin{equation}
c_{22}=-c_{33}=-\frac{\cos2\xi}{\left|e^{i\kappa_2}\cos^2\xi+e^{i\kappa_3}\sin^2\xi\right| }\,,~~c_{23}=c^{*}_{32}=\frac{\cos\left(\frac{\kappa_2-\kappa_3}{2}\right)e^{i\frac{\kappa_2-\kappa_3}{2}} \sin2\xi }{\left|e^{i\kappa_2}\cos^2\xi + e^{i\kappa_3}\sin^2\xi \right|}\,.
\end{equation}
In this case, the neutrino mixing matrix $U_{\nu}$ is completely determined by the remnant CP transformations without additional free parameters~\cite{Chen:2015nha},
\begin{equation}
\label{eq:Unu-four-CP}U_{\nu}=\left(v_1, v_2, v_3\right)\text{diag}\big(e^{i\frac{\kappa_1}{2}}, e^{i\frac{\kappa_2}{2}}, e^{i\frac{\kappa_3}{2}}\big)R_{23}(\chi)P_{\nu}Q_{\nu}\,,
\end{equation}
where the angle $\chi$ fulfills
\begin{equation}
\tan2\chi=\cos\left(\frac{\kappa_2-\kappa_3}{2}\right)\tan2\xi\,.
\end{equation}
Notice that the lepton mixing angles and the three CP violating phases are completely fixed by the remnant CP symmetry in this case. Moreover, note that the master \textit{formulae} of Eqs.~(\ref{eq:Unu-one-CP}, \ref{eq:Unu-two-CP}, \ref{eq:Unu-four-CP}) for $U_{\nu}$ hold irrespective of how the remnant CP symmetry is dynamically realized.

\subsection{Residual symmetries of quarks}
\label{subsec:residual_sym_quark_sector}

In this section, we turn to the residual flavour and CP symmetries of the quark mass matrices. The Lagrangian for the quark masses is given in Eq.~\eqref{eq:quark_mass_ci_dirac}. One can easily check that the Hermitian combinations $m^{\dagger}_{U}m_{U}$ and $m^{\dagger}_{D}m_{D}$ are invariant under the following unitary transformations~\cite{Yao:2015dwa},
\begin{equation}
U_{L}\to  G_{u}U_{L},\quad D_{L}\to  G_{d}D_{L}\,,
\end{equation}
with
\begin{equation}
\label{eq:Gu_Gd}G_{u}=V_{u}\textrm{diag}\left(e^{i\alpha_{u}}, e^{i\alpha_{c}}, e^{i\alpha_{t}}\right)V^{\dagger}_{u},\quad G_{d}=V_{d}\textrm{diag}\left(e^{i\alpha_{d}}, e^{i\alpha_{s}}, e^{i\alpha_{b}}\right)V^{\dagger}_{d}\,,
\end{equation}
where $\alpha_{q}$ ($q=u, d, c, s, t, b$) are arbitrary phase parameters. Hence the following equalities are satisfied,
\begin{equation}
G^{\dagger}_{u}m^{\dagger}_{U}m_{U}G_{u}=m^{\dagger}_{U}m_{U},\quad
G^{\dagger}_{d}m^{\dagger}_{D}m_{D}G_{d}=m^{\dagger}_{D}m_{D}\,.
\end{equation}
In other words, both the up-quark mass matrix $m_U$ and down-quark mass matrix $m_{D}$ have a residual $U(1)\times U(1)\times U(1)$ flavour symmetry~\cite{Yao:2015dwa}. Notice that the same conclusion holds true for any Dirac fermion mass matrix, hence it also applies for the charged lepton mass term.

Let us now turn to the discussion of residual CP symmetries of the quark mass term. Assuming the left-handed quarks $U_{L}$ and $D_{L}$ transform as
\begin{equation}
U_{L}(x)\stackrel{CP}{\longmapsto}iX_{u}\gamma^{0}C\overline{U}^{\,T}_{L}(\mathcal{P}x),\quad D_{L}(x)\stackrel{CP}{\longmapsto}iX_{d}\gamma^{0}C\overline{D}^{T}_{L}(\mathcal{P}x)\,,
\end{equation}
this is a symmetry of the quark mass matrices $m_{U}$ and $m_{D}$ if and only if $m^{\dagger}_{U}m_{U}$ and $m^{\dagger}_{D}m_{D}$ fulfill the conditions
\begin{equation}
X^{\dagger}_{u}m^{\dagger}_{U}m_{U}X_{u}=(m^{\dagger}_{U}m_{U})^{\star},\quad
X^{\dagger}_{d}m^{\dagger}_{D}m_{D}X_{d}=(m^{\dagger}_{D}m_{D})^{\star}\,.
\end{equation}
Similarly to the charged lepton sector, one can show that $X_{u}$ and $X_{d}$ must take the following form
\begin{equation}
\label{eq:Xu-Xd}X_{u}=V_{u}\textrm{diag}\left(e^{i\beta_{u}}, e^{i\beta_{c}}, e^{i\beta_{t}}\right)V^{T}_{u},\quad X_{d}=V_{d}\textrm{diag}\left(e^{i\beta_{d}}, e^{i\beta_{s}}, e^{i\beta_{b}}\right)V^{T}_{d}\,,
\end{equation}
where $\beta_{q}$ ($q=u, d, c, s, t, b$) are real. It is easy to see that both $X_{u}$ and $X_{d}$ are unitary symmetric matrices, and that residual flavour symmetries $G_{u}$ and $G_{d}$ will be generated by $X_{u}$ and $X_{d}$, respectively. In the basis where the up-quark mass matrix is diagonal, $V_{u}$ is diagonal, so that $V_d$ coincides with the CKM matrix, $V_d=V_{CKM}$. As a result, the explicit form of the remnant flavour symmetries $G_{u}$, $G_{d}$ and the remnant CP symmetries $X_{u}$, $X_{d}$ can be fixed in terms of the measured values of the CKM matrix elements~\cite{Workman:2022ynf}. As in section~\ref{sec:constructing-mixing-res-sym} for lepton sector, the quark mixing matrix can also be fixed by the remnant CP symmetry.

So far we have shown that lepton and quark mass terms generically admit residual flavour and CP symmetries. These are associated to the charged current flavour mixing matrix. One may ask about the origin of these residual symmetries. It is well known that the spontaneous breaking of the \SM gauge symmetry preserves the $U(1)$ gauge symmetry associated to electromagnetism. Likewise, one may conjecture that at some high energy scale the true underlying theory has a flavour and CP symmetry which is subsequently broken down to the residual symmetry or a subgroup of it. This gives a further motivation for introducing flavour and CP symmetries, i.e. to explain flavour mixing and CP violation in a model-independent manner. We also saw that the residual flavour symmetry of the charged leptons and quarks should be contained in $U(1)\times U(1)\times U(1)$. The same holds for neutrinos if they are Dirac particles. In contrast, the residual flavour symmetry of the neutrino sector should be a subgroup of $Z_2\times Z_2\times Z_2$ if they are Majorana fermions.

In a pioneering work~\cite{Froggatt:1978nt}, Froggatt and Nielsen originally took the $G_f=U(1)$ flavor symmetry in order to explain the quark mass ratios and the CKM mixing angles, which are expressed as powers of small $G_f$ breaking parameters. This is the so-called FN mechanism. The magnitudes of the quark masses and the CKM matrix can be reproduced qualitatively for appropriate choice of horizontal charges. However, typically this does not allow quantitative predictions. The Froggatt-Nielsen $U(1)$ flavor symmetry can be gauged, and the anomaly-free charge assignments can be found in~\cite{Froggatt:1998he,Allanach:2018vjg,Costa:2019zzy}. Anomaly-free inverted Froggatt-Nielsen models
were studied in Ref.~\cite{Smolkovic:2019jow}. Concerning the lepton sector, broken flavor symmetries based on non-abelian discrete groups were found to reproduce certain interesting mixing patterns, such as TBM as a first order approximation~\cite{Ma:2001dn,Babu:2002dz,Altarelli:2005yp,Altarelli:2005yx}. Continuous Lie groups $U(2)$~\cite{Barbieri:1995uv,Barbieri:1996ww,Barbieri:1997tu,Linster:2018avp}, $SO(3)$~\cite{King:2005bj,King:2006me,Reig:2018ocz} and $SU(3)$~\cite{King:2001uz,King:2003rf,deMedeirosVarzielas:2005ax,Antusch:2007re,Bazzocchi:2008rz,deAnda:2018yfp} as global flavor symmetries were also studied to address the flavor puzzles. Typically, they must be strongly broken. In order to avoid the massless Goldstone bosons, these continuous non-abelian flavor symmetry groups should be a local gauge symmetry~\cite{Reig:2018ocz,Grinstein:2010ve}. The cancellation of gauge anomalies automatically leads to the addition of fermions whose masses are inversely proportional to the known fermion masses, thus the flavor violating effects for the light generations are naturally suppressed. Moreover, the scale of flavor physics could be as low as a TeV, while avoiding all flavor and precision electroweak bounds, yet within reach of the LHC and future colliders~\cite{Grinstein:2010ve,Alonso:2016onw}.

Besides the use of full-fledged flavor symmetries, there are more phenomenological approaches to the flavor puzzle involving, e.g., the use of texture zeros in the fermion mass matrices~\cite{Weinberg:1977hb,Fritzsch:1977za,Fritzsch:1977vd,Frampton:2002yf}. Alternatively the assumption of lepton mixing ``anarchy''~\cite{Hall:1999sn,deGouvea:2012ac}. In the present review, we shall focus on the approach of discrete flavor symmetry as well as generalized CP symmetry.

\clearpage

\section{Viable lepton mixing patterns }                                   
\label{sec:revamp-lept-mixing}                                             

The ``constant'' lepton mixing patterns discussed in section~\ref{sec:flavor_symm}, such as the tri-bimaximal, golden ratio and bi-maximal mixing, characterized by numerical predictions for the mixing angles and phase, are all ruled out by neutrino oscillation data~\cite{deSalas:2020pgw,10.5281/zenodo.4726908}, especially by the precise measurement of the ``reactor angle'' $\theta_{13}$~\cite{Daya2022,DayaBay:2022orm}. They all need to be ``revamped'' in order to be compatible with experimental data and to enable meaningful theoretical predictions for CP violation.

In this section we assume that neutrinos are Majorana particles and show how the imposition of residual flavour and CP symmetries~\cite{Chen:2015siy,Chen:2018lsv,Chen:2014wxa,Chen:2016ica,Chen:2018zbq,CentellesChulia:2019ldn} can be used to produce systematic generalizations of the patterns discussed in the section~\ref{sec:flavor_symm}. Indeed, imposing the residual symmetries $G_i$ in Sec.~\ref{sec:flavor_CP_bottom-up} fixes the $i$-th column of the mixing matrix. In this way one can obtain generalized patterns which can be not only viable, but also predictive, in which the mixing matrix is described by just a few parameters. This model-independent approach of predicting new mixing patterns holds irrespective of how the relevant mass matrices arise from first principles. We now describe some examples.

\subsection{Revamped TBM mixing}
\label{subsec:revamp_TBM}

Working in the charged lepton diagonal basis, we start our discussion with the ``complex TBM'' matrix (cTBM)~\cite{Chen:2018eou}, which is given by
\begin{equation}
 U_{cTBM}=\begin{pmatrix}
\sqrt{\frac{2}{3}}  ~&~ \frac{ e^{-i \rho } }{\sqrt{3}}    ~&~ 0                               \\
-\frac{e^{i \rho }}{\sqrt{6}}    ~&~ \frac{1}{\sqrt{3}}   ~&~ \frac{e^{-i\sigma}}{\sqrt{2}}    \\
\frac{e^{i (\rho + \sigma)}}{\sqrt{6}}  ~&~  -\frac{e^{i\sigma}}{\sqrt{3}}   ~&~ \frac{1}{\sqrt{2}}
\end{pmatrix}\,.
\label{eq:ctbm}
\end{equation}
This cTBM mixing matrix predicts the same mixing angles as the usual real TBM pattern in Eq.~\eqref{eq:TBM-angles}, though the Majorana phases are non-vanishing. Within the symmetrical parametrization of the lepton mixing matrix in Eqs.~(\ref{eq:lepton-mixing}) and (\ref{eq:dell}) they are given by
\begin{equation}
\phi_{12}=\rho,~~~~~\phi_{23}= \sigma\,.
\end{equation}
The TBM matrix in Eq.~\eqref{eq:UTBM} corresponds to zero Majorana phases $\rho=\sigma=0$ and hence we call it real TBM~\footnote{The minus sign of the third row is absorbed into the charged leptons.}. From Eq.~\eqref{eq:residual-CP-nu}, we know the four CP symmetry matrices $X_{1,2,3,4}$ associated with the cTBM mixing pattern are
\begin{equation}
X_i = U_{cTBM} d_i U_{cTBM}^T\,,
\end{equation}
where $d_{1,2,3,4}$ are diagonal matrices with entries $\pm1$ given in Eq.~\eqref{eq:d1234}. Thus the above four CP symmetries are given in matrix form as
\begin{eqnarray}
\nonumber X_1 & = & \frac{1}{6}
\left(\begin{array}{ccc}
 4-2 e^{-2i\rho} ~& -2 e^{-i\rho}-2 e^{i\rho} ~& 2 e^{i(\rho+\sigma)}+2e^{-i(\rho-\sigma)} \\
 -2e^{-i\rho}-2e^{i\rho} ~& -2+e^{2i\rho}-3 e^{-2i\sigma} ~& -3e^{-i\sigma}-e^{i(2\rho+\sigma)}+2 e^{i\sigma} \\
 2 e^{i(\rho+\sigma)}+2e^{-i(\rho-\sigma)} ~&~ -3e^{-i\sigma}-e^{i(2\rho+\sigma)}+2 e^{i\sigma} ~&~ -3+e^{2i(\rho+\sigma)} -2 e^{2i\sigma} \\
\end{array}\right)\,,\\
\nonumber X_2 & = & \frac{1}{6}
\left(\begin{array}{ccc}
 -4+2 e^{-2i\rho} ~& 2 e^{-i\rho}+2 e^{i\rho} ~& -2 e^{i (\rho+\sigma)  }-2 e^{-i (\rho-\sigma) } \\
 2 e^{-i\rho}+2 e^{i\rho} ~& 2-e^{2 i\rho}-3 e^{-2 i\sigma} ~& -3 e^{-i\sigma}+e^{i (2\rho+\sigma) }-2 e^{i\sigma} \\
 -2 e^{i (\rho+\sigma)  }-2 e^{-i (\rho-\sigma) } ~& -3 e^{-i\sigma}+e^{i (2\rho+\sigma) }-2 e^{i\sigma} ~&
 -3  - e^{2 i(\rho+\sigma) } +2 e^{2 i\sigma} \\
\end{array}\right)\,,\\
\nonumber X_3 & = & \frac{1}{6}
\left(\begin{array}{ccc}
 -4-2 e^{-2 i\rho} ~& -2 e^{-i\rho}+2 e^{i\rho} ~& -2 e^{i(\rho+\sigma)}+2 e^{-i (\rho-\sigma) } \\
 -2 e^{-i\rho}+2 e^{i\rho} ~& -2-e^{2 i\rho}+3 e^{-2 i\sigma} ~& 3 e^{-i\sigma}+e^{i (2\rho+\sigma) }+2 e^{i\sigma} \\
 -2 e^{i (\rho+\sigma)  }+2 e^{-i (\rho-\sigma)} ~&~ 3 e^{-i\sigma}+e^{i (2\rho+\sigma)  }+2 e^{i\sigma} ~&~ 3-e^{2 i (\rho+\sigma)  }-2 e^{2 i\sigma} \\
\end{array}\right)\,,\\
\label{eq:X0} X_4 & = & \frac{1}{6}
\left(\begin{array}{ccc}
 4+2 e^{-2 i\rho} ~& 2 e^{-i\rho}-2 e^{i\rho} ~& 2 e^{i (\rho+\sigma) }-2 e^{-i (\rho-\sigma) } \\
 2 e^{-i\rho}-2 e^{i\rho} ~& 2+e^{2 i\rho}+3 e^{-2 i\sigma} ~& 3 e^{-i\sigma}-e^{i (2\rho+\sigma) }-2 e^{i\sigma} \\
 2 e^{i (\rho+\sigma) }-2 e^{-i (\rho-\sigma)} ~& 3 e^{-i\sigma}-e^{i (2\rho+\sigma) }-2 e^{i\sigma} ~& 3+e^{2i(\rho+\sigma)}+2 e^{2i\sigma} \\
\end{array}\right)\,.
\label{eq:ccp-sym}
\end{eqnarray}
The CP symmetries corresponding to the ``standard'' real TBM matrix of Eq.~\eqref{eq:UTBM} are obtained simply by taking the limit of $\rho, \sigma \to 0$ in Eq.~\eqref{eq:ccp-sym}. These CP symmetries are therefore given by
\begin{eqnarray}
\nonumber &&X_1=\frac{1}{3}\left(
\begin{array}{ccc}
 1 ~&~  -2  ~&~ 2 \\
-2  ~&~  -2  ~&~ -1 \\
2 ~&~ -1 ~&~ -2
\end{array}\right),~~~~~X_2=\frac{1}{3}\left(
\begin{array}{ccc}
 -1 ~&~  2 ~& -2 \\
 2 ~&~ -1 ~&~ -2 \\
-2 ~&~ -2 ~&~ -1
\end{array}\right)\,,\\
\label{eq:resid-CP}&&X_3=\left(\begin{array}{ccc}
-1 ~&~ 0 ~&~ 0 \\
0 ~&~ 0 ~&~ 1 \\
0 ~&~ 1 ~&~ 0
\end{array}\right),~~~~~X_4=\left(\begin{array}{ccc}
 1 ~&~ 0 ~&~ 0 \\
 0 ~&~ 1 ~&~ 0 \\
0 ~&~ 0 ~&~ 1
\end{array}\right)\,.
\end{eqnarray}
As shown in Eq.~\eqref{eq:CP_flavor_relations}, the residual flavour symmetry can be generated by the CP transformations,
\begin{eqnarray}
\nonumber&&\hskip-0.2in G_1 = X_2X^{*}_3=X_3X^{*}_2=X_4X^{*}_1=X_1X^{*}_4\,,~~~ G_2 = X_1X^{*}_3=X_3X^{*}_1=X_4X^{*}_2=X_2X^{*}_4\,,\\
\label{eq:res-CP-flavor}&&\hskip-0.2in G_3 = X_1X^{*}_2=X_2X^{*}_1=X_4X^{*}_3=X_3X^{*}_4\,,~~~ G_4 = X_1X^{*}_1=X_2X^{*}_2=X_3X^{*}_3=X_4X^{*}_4\,.
\end{eqnarray}
It is our goal here to obtain generalized but restricted forms for the mixing matrices starting from the ``original'' ones by exploiting residual flavour and CP symmetries.
Notice that only three of the four CP and flavour symmetries are really independent~\cite{Chen:2014wxa, Chen:2015nha}. If any three of the four CP symmetries in Eq.~\eqref{eq:X0} are imposed simultaneously, the neutrino mixing matrix would be the cTBM matrix in Eq.~\eqref{eq:ctbm} with $\theta_{13}=0$. Therefore, we will impose only two or only one of these CP symmetries, so that realistic mixing patterns with non-vanishing $\theta_{13}$ and CP violation are obtained.

\subsubsection{Case a: $G_1$ flavour and $X_1, X_4$ CP symmetries}
\label{sec:g_1-flavor-x_1}

The requirement that the CP transformations $X_1$ and $X_4$ are symmetries of the neutrino mass matrix $m_{\nu}$ implies that the $G_1$ flavour symmetry is preserved and $m_{\nu}$ satisfies
\begin{equation}
X_1^T m_\nu X_1= m_\nu^{\ast},~~~~
X_4^T m_\nu X_4= m_\nu^{\ast}\,.
\end{equation}
Consequently the light neutrino mass matrix is of the following form
\begin{equation}
m'_{\nu}=U_{cTBM}^T m_{\nu}U_{cTBM} =
\begin{pmatrix}
 m_1 ~& 0 ~& 0 \\
 0 ~& m_2 ~& \delta m \\
 0 ~& \delta m ~& m_3
\end{pmatrix}\,,
\end{equation}
where the parameters $m_1$, $m_2$, $m_3$ and $\delta m$ are real. The mass matrix $m'_{\nu}$ can be diagonalized by a real orthogonal matrix $R_{23}(\theta)$ given by
\begin{equation}
R_{23}(\theta)=
\left(\begin{array}{ccc}
1 ~& 0 ~& 0 \\
0 ~& \cos\theta ~& \sin\theta \\
0 ~& -\sin\theta ~& \cos\theta
\end{array}\right)~~\text{with}~~
\tan2\theta=\frac{2\,\delta m}{m_3-m_2}\,.
\end{equation}
As a result, in this case the lepton mixing matrix is given as
\begin{eqnarray}
\nonumber U_{} &=& U_{cTBM}\,R_{23}\,Q_{\nu} \\ \label{eq:UPMNS-RTBM-A}&=&\frac{1}{\sqrt{6}}\begin{pmatrix}
2 ~& \sqrt{2}e^{-i \rho} \cos \theta ~&  \sqrt{2}e^{-i \rho} \sin \theta \\
-e^{i \rho} ~& \sqrt{2}\cos \theta -\sqrt{3}e^{-i \sigma}\sin \theta  ~&
\sqrt{2}\sin \theta+\sqrt{3}e^{-i \sigma} \cos \theta      \\
e^{i (\rho + \sigma)}  ~& -\sqrt{3}\sin \theta-\sqrt{2}e^{i \sigma} \cos \theta  ~&  \sqrt{3}\cos \theta-\sqrt{2}e^{i \sigma} \sin \theta
\end{pmatrix}Q_{\nu}\,,
\end{eqnarray}
where $Q_{\nu}=\text{diag} (e^{ik_1\pi/2}, e^{ik_2\pi/2} , e^{ik_3\pi/2})$ is a diagonal unitary matrix with $k_{1,2,3}=0, 1, 2, 3$. The entries $\pm1$ and $\pm i$ encode the CP parities of the neutrino states and render the neutrino mass eigenvalues non-negative.
From Eq.~\eqref{eq:UPMNS-RTBM-A} one  can then extract the expressions of lepton mixing angles and CP violating phases as follows,
\begin{eqnarray}
\nonumber&&\sin^2\theta_{13}=\frac{\sin^2\theta}{3},~~\sin^2\theta_{12}=\frac{\cos^2\theta}{\cos^2\theta+2},~~\sin^2\theta_{23}=\frac{1}{2}+\frac{\sqrt{6} \sin  2\theta \cos \sigma}{2\cos^2\theta+4}\,,\\
\nonumber&&\sin\delta^{\ell}=-\frac{\text{sign}(\sin2\theta)(\cos^2\theta+2) \sin \sigma}{\sqrt{(\cos^2\theta+2)^2-6\sin^2 2\theta \cos^2\sigma}}\,,~~\tan\delta^{\ell} = \frac{2+\cos^2\theta}{2-5\cos^2\theta}\tan\sigma\,,\\
\label{eq:angles-phases-RTBM-A}&&
\phi_{12}=\rho + \frac{(k_1-k_2)\pi}{2}\,,\qquad\phi_{13}=\rho+\frac{(k_1-k_3)\pi}{2}\,.
\end{eqnarray}
Notice that in the symmetric parametrization the CP violating phase characterizing neutrino oscillations is given by the invariant combination $\delta^{\ell}=\phi_{13}-\phi_{12}-\phi_{23}$~\cite{Rodejohann:2011vc}, see Eq.~\eqref{eq:dell}. We see that the first column of the lepton mixing matrix in Eq.~\eqref{eq:UPMNS-RTBM-A} is $(2, -e^{i\rho}, e^{i(\rho+\sigma)})^{T}/\sqrt{6}$ which is in common with that of the cTBM mixing pattern. This arises from the preserved $G_1$ symmetry. Eliminating the parameters $\theta$ and $\sigma$ in Eq.~\eqref{eq:angles-phases-RTBM-A}, we see that the lepton mixing angles and CP phases are correlated with each other according to
\begin{equation}
\label{eq:cor-RTBM-A}\cos^2\theta_{12}\cos^2\theta_{13}=\frac{2}{3}\,,~~~~
\tan2\theta_{23}\cos\delta^{\ell}=\frac{5\sin^2\theta_{13}-1}{2\sin\theta_{13}\sqrt{2-6\sin^2\theta_{13}}}\,.
\end{equation}

\begin{figure}[h!]
\begin{center}
\includegraphics[width=0.95\linewidth]{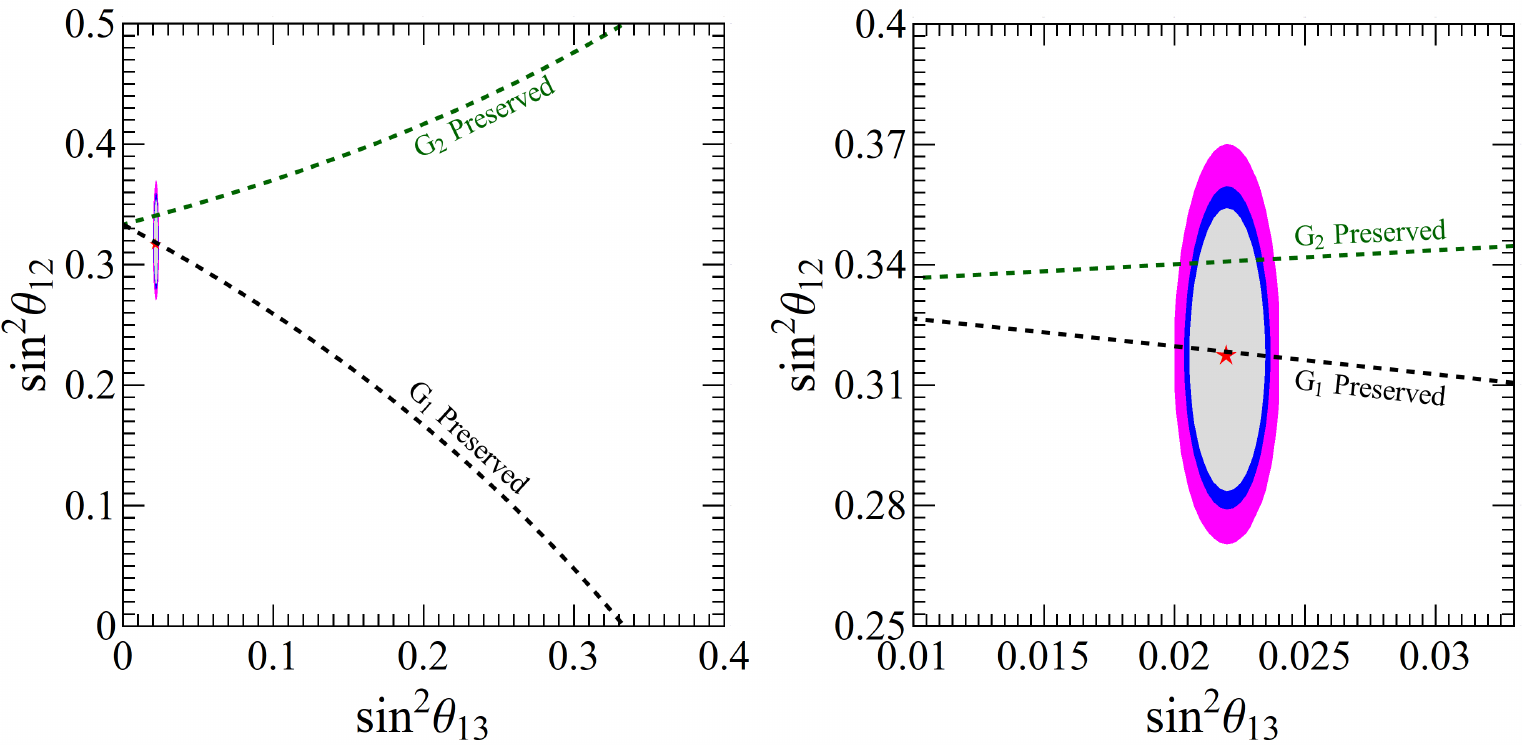}
\caption{\label{fig:g1L} Predicted correlation between $\sin^2\theta_{12}$ and $\sin^2\theta_{13}$ in the revamped TBM scheme. The black-dashed line corresponds to the case where $G_1$ is preserved by the neutrino sector (Eq.~\eqref{eq:cor-RTBM-A}, left), while the green-dashed one refers to the case where $G_2$ is preserved (Eq.~\eqref{eq:cor-RTBM-C}, left). The right panel is a zoom of the left one. The global fit regions correspond to $90\%$, $95\%$ and $99\%$ confidence levels~\cite{deSalas:2020pgw,10.5281/zenodo.4726908}. }
\end{center}
\end{figure}

\begin{figure}[h!]
\begin{center}
\includegraphics[width=0.95\linewidth]{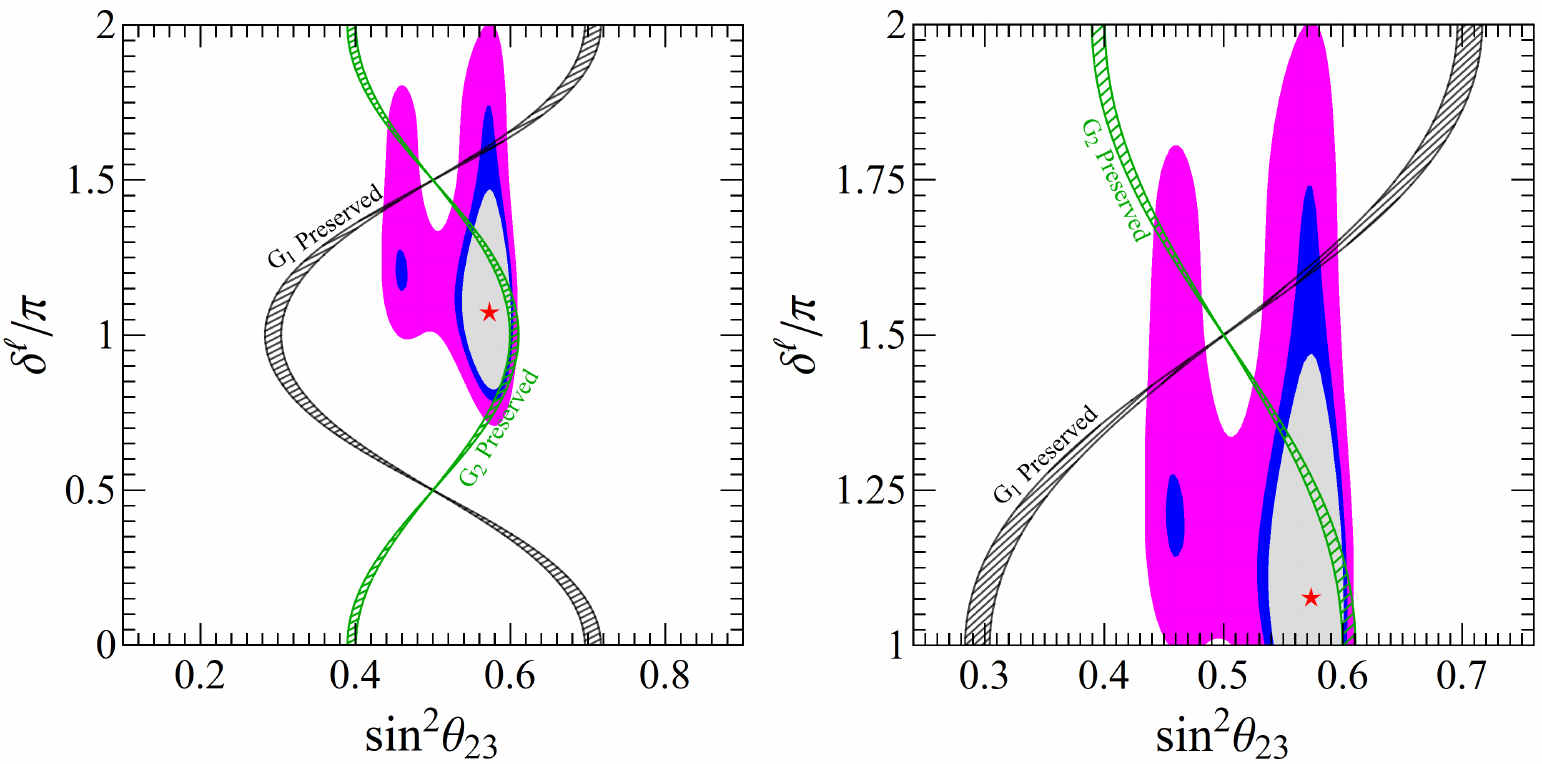}
\caption{\label{fig:g1R}
Predicted correlation between $\delta^{\ell}$ and $\sin^2\theta_{23}$ in the revamped TBM scheme. The black region corresponds to the case where $G_1$ is preserved by the neutrino sector, as given by Eq.~\eqref{eq:cor-RTBM-A}, right. The green region corresponds to the case where $G_2$ is preserved, as given by Eq.~\eqref{eq:cor-RTBM-C}, right. The right panel is a zoom of the left one. The global fit regions correspond to $90\%$, $95\%$ and $99\%$ confidence levels~\cite{deSalas:2020pgw,10.5281/zenodo.4726908}.  }
\end{center}
\end{figure}

The first equation in Eq.~\eqref{eq:cor-RTBM-A} relates the solar and the reactor angles while, for given values of the latter, the second equation correlates the CP phase $\delta^{\ell}$ and the atmospheric angle. These correlations can be used to test the mixing matrix of Eq.~\eqref{eq:UPMNS-RTBM-A} at current and future oscillation experiments.
Notice that these correlations are a generic feature of mass matrices which preserve the $G_1$ symmetry. These are displayed in figure~\ref{fig:g1L} and figure~\ref{fig:g1R}.
In the limit of $\rho,\sigma\to 0$, we see that the mixing angles $\theta_{12}$ and $\theta_{13}$ remain the same, while the Dirac CP phase vanishes $\sin \delta^{\ell} \to 0$, so that CP would be conserved in neutrino oscillations. Notice that both Majorana phases~\cite{Schechter:1980gk} become some integer multiples of $\pi/2$ and therefore they correspond to just CP signs~\cite{Schechter:1981hw,Wolfenstein:1981rk}.

When the two CP symmetries $X_2$ and $X_3$ are imposed, the neutrino mass matrix preserves the flavour symmetry $G_1= X_2X^{\ast}_3=X_3X^{\ast}_2$ as well. The resulting predictions for lepton mixing parameters are obtained from Eq.~\eqref{eq:angles-phases-RTBM-A} by redefining $\rho\to\rho+\pi/2 $ and $\sigma\to \sigma-\pi/2$.

\subsubsection{Case b: $G_2$ flavour and $X_2, X_4$ CP symmetries }
\label{sec:g_2-flavor-x_2}

The combination of $X_2$ and $X_4$ leads to the conservation of $G_2 = X_2X^{*}_4=X_4X^{*}_2$ flavour symmetry. The invariance of neutrino mass matrix under the action of $X_2$ and $X_4$ requires
\begin{equation}
X_2^T m_\nu X_2=m_\nu^{*}\,,~~~~X_4^T m_\nu X_4=m_\nu^{*}\,,
\end{equation}
from which we can determine the neutrino mass matrix to be of the following form
\begin{equation}
m'_{\nu}=U_{cTBM}^T m_{\nu}U_{cTBM} =
\left(\begin{array}{ccc}
 m_1 & 0 & \delta m \\
 0 & m_2 & 0 \\
 \delta m & 0 & m_3
\end{array}\right)\,,
\end{equation}
where $m_{1,2,3}$ and $\delta m$ are generic real parameters. The matrix $m'_{\nu}$ can be diagonalized by a rotation matrix $R_{13}(\theta)$ in the (13)-plane,
\begin{equation}
R_{13}(\theta)=\begin{pmatrix}
\cos\theta ~& 0 ~& \sin\theta \\
0 ~& 1 ~& 0 \\
-\sin\theta ~& 0 ~& \cos\theta
\end{pmatrix}
~~\text{with}~~\tan2\theta=\frac{2\,\delta m}{m_3-m_1}\,.
\end{equation}
Consequently the residual CP symmetries $X_2$ and $X_4$ fix the lepton mixing matrix to be
\begin{eqnarray}
\nonumber U&=&U_{cTBM}R_{13}(\theta) Q_{\nu}\\
\label{eq:U-RTBM-C}&=&\frac{1}{\sqrt{6}}\begin{pmatrix}
2 \cos \theta  ~& \sqrt{2}e^{-i \rho} ~ & 2\sin\theta      \\
-e^{i \rho} \cos \theta - \sqrt{3}e^{-i \sigma} \sin \theta  ~& \sqrt{2}
~& -e^{i \rho} \sin \theta +\sqrt{3}e^{-i \sigma} \cos \theta     \\
e^{i (\rho + \sigma)} \cos \theta -\sqrt{3}\sin \theta  ~& -\sqrt{2}e^{i\sigma}  ~& e^{i (\rho + \sigma)} \sin \theta +\sqrt{3}\cos\theta
\end{pmatrix}\,Q_{\nu}\,.
\end{eqnarray}
Note that the second column of the mixing matrix is $(e^{-i \rho}, 1, -e^{i\sigma})^{T}/\sqrt{3}$ which matches with that of the cTBM mixing pattern. We can extract the mixing angles and the predicted CP violation phases from Eq.~\eqref{eq:U-RTBM-C} in the usual way, leading to
\begin{eqnarray}
\nonumber&&\sin^2\theta_{13}=\frac{2 \sin^2\theta}{3}\,,~~~ \sin^2\theta_{12}=\frac{1}{2\cos^2\theta+1}\,,~~~\sin^2\theta_{23}=\frac{1}{2}-\frac{\sqrt{3} \sin  2\theta \cos(\rho + \sigma)}{4\cos^2\theta+2}\,, \\
\nonumber&&\sin\delta^{\ell}=-\frac{\text{sign}(\sin2\theta) (2\cos^2\theta+1)\sin(\rho + \sigma)}{\sqrt{(2\cos^2\theta+1)^2-3 \cos^2(\rho + \sigma) \sin^2 2\theta}}\,, ~~~
\tan\delta^{\ell}=\frac{(2\cos^2\theta+1)\tan(\rho + \sigma)}{1-4\cos^2\theta}\,, \\
\label{eq:angles-phases-RTBM-C}&& \phi_{12}=\rho + \frac{(k_1-k_2)\pi}{2}\,,~~~~ \phi_{13}=\frac{(k_1-k_3)\pi}{2}\,.
\end{eqnarray}
The mixing parameters are again correlated with each other, as follows
\begin{equation}
\label{eq:cor-RTBM-C}\sin^2\theta_{12}\cos^2\theta_{13}=\frac{1}{3}\,,
\qquad \tan2\theta_{23} \cos\delta^{\ell}=\frac{\cos2\theta_{13}}{\sin\theta_{13}\sqrt{2-3\sin^2\theta_{13}}}\,.
\end{equation}
These correlations lead to predictions for the oscillations parameters, given in figures~\ref{fig:g1L} and \ref{fig:g1R}.

For the case that the CP symmetries $X_1$ and $X_3$ are preserved, the flavour symmetry $G_2=X_1X^{*}_3=X_3X^{*}_1$ would be preserved as well~\footnote{The imposition of $G_3$ is uninteresting here, as it leads to $\theta_{13}=0$.}. The resulting predictions for lepton mixing matrix and mixing parameters can be obtained from Eqs.~(\ref{eq:U-RTBM-C}, \ref{eq:angles-phases-RTBM-C}) by redefining $\rho\to\rho+\pi/2$, $\sigma\to\sigma-\pi$.\\[-.3cm]

Finally, if a single CP symmetry is preserved by the neutrino mass matrix, the lepton mixing matrix is determined up to a three dimensional orthogonal matrix. The resulting lepton flavour mixing predictions can be analyzed in a similar fashion~\cite{Chen:2018zbq}. The simple TBM mixing matrix can also be revamped by exploiting the generalized CP symmetries of the charged lepton mass matrix~\cite{Chen:2019fgb}.

\subsection{Revamped golden ratio mixing scheme}
\label{sec:revamp-gold-ratio}

In a way analogous to what we did for the TBM mixing matrix, we can also revamp the golden ratio mixing pattern. We start from the complex golden ratio (cGR) mixing matrix, in the charged lepton diagonal basis,
\begin{equation}
U_{cGR}=\frac{1}{\sqrt{2\sqrt{5}\,\phi_g}}\begin{pmatrix}
\sqrt{2}\phi_g  ~&~  \sqrt{2} e^{-i\rho}  ~&~  0  \\
-e^{i\rho}   ~&~  \phi_g  ~&~  \sqrt{\sqrt{5}\,\phi_g}\,e^{-i\sigma} \\
-e^{i(\rho+\sigma)}   ~&~  \phi_g e^{i\sigma} ~&~  -\sqrt{\sqrt{5}\,\phi_g}
\end{pmatrix}\,,
\end{equation}
which reduces to the real GR mixing matrix of Eq.~\eqref{eq:real-GR} in the limit of $\rho=\sigma=0$. The four CP symmetry matrices $X_i=U_{cGR}\;d_i\; U_{cGR}^T$ associated with the cGR mixing pattern are of the following form,
\begin{eqnarray}
\nonumber X_1&=&\frac{1}{\sqrt{5}}\begin{pmatrix}
 1+\frac{2i\sin\rho}{\phi_g}e^{-i\rho} ~&~ -\sqrt{2}\cos\rho ~&~ -\sqrt{2} \, e^{i \sigma } \cos\rho \\
 -\sqrt{2} \cos\rho ~&~
 -\frac{1+\sqrt{5} e^{-2i\sigma }}{2}+\frac{i\sin\rho}{\phi_g}e^{i\rho} ~&~
\frac{-e^{i\sigma}+\sqrt{5}e^{-i\sigma}}{2}+\frac{i\sin\rho}{\phi_g}e^{i(\rho+\sigma)}\\
 -\sqrt{2} e^{i \sigma } \cos\rho  ~&~ \frac{-e^{i\sigma}+\sqrt{5}e^{-i\sigma}}{2}+\frac{i\sin\rho}{\phi_g}e^{i(\rho+\sigma)} ~&~  -\frac{e^{2i\sigma}+\sqrt{5}}{2}+\frac{i\sin\rho}{\phi_g}e^{i(\rho+2\sigma)} \\
\end{pmatrix}\,,\\
\nonumber X_2&=&\frac{1}{\sqrt{5}}\begin{pmatrix}
 -1-\frac{2i\sin\rho}{\phi_g}e^{-i\rho} ~&~ \sqrt{2}\cos\rho  ~&~ \sqrt{2}\, e^{i \sigma } \cos\rho \\
 \sqrt{2}\, \cos\rho  ~&~  \frac{1-\sqrt{5}\, e^{-2i\sigma}}{2}-\frac{i\sin\rho}{\phi_g}e^{i\rho}  ~&~ \frac{e^{i\sigma}+\sqrt{5}\,e^{-i\sigma}}{2}-\frac{i\sin\rho}{\phi_g}e^{i(\rho+\sigma)} \\
 \sqrt{2}\, e^{i \sigma } \cos\rho ~&~ \frac{e^{i\sigma}+\sqrt{5}\,e^{-i\sigma}}{2}-\frac{i\sin\rho}{\phi_g}e^{i(\rho+\sigma)} ~&~\frac{e^{2i\sigma}-\sqrt{5}}{2}-\frac{i\sin\rho}{\phi_g}e^{i(\rho+2\sigma)}\\
\end{pmatrix}\,,\\
\nonumber X_3&=& \frac{1}{\sqrt{5}}\begin{pmatrix}
 -1-\frac{2\cos\rho}{\phi_g}e^{-i\rho} ~&~ \sqrt{2}\, i\sin\rho ~&~ \sqrt{2}\, i  e^{i\sigma}\sin\rho\\
 \sqrt{2}\, i\sin\rho ~&~ \frac{-1+\sqrt{5}\,e^{-2i\sigma}}{2}-\frac{\cos\rho}{\phi_g}e^{i\rho}  ~&~
 -\frac{e^{i\sigma}+\sqrt{5}\,e^{-i\sigma}}{2}-\frac{\cos\rho}{\phi_g}e^{i(\rho+\sigma)} \\
 \sqrt{2}\, ie^{i\sigma}\sin\rho   ~&~ -\frac{e^{i\sigma}+\sqrt{5}\,e^{-i\sigma}}{2}-\frac{\cos\rho}{\phi_g}e^{i(\rho+\sigma)} ~&~ \frac{-e^{2i\sigma}+\sqrt{5}}{2}-\frac{\cos\rho}{\phi_g}e^{i(\rho+2\sigma)} \\
\end{pmatrix}\,,\\
X_4&=&\frac{1}{\sqrt{5}}\begin{pmatrix}
 1+\frac{2\cos\rho}{\phi_g}e^{-i\rho} ~&~ -\sqrt{2}\,i\sin\rho ~&~ -\sqrt{2}\,i e^{i\sigma}\sin\rho \\
 -\sqrt{2}\, i\sin\rho ~&~ \frac{1+\sqrt{5}\,e^{-2i\sigma}}{2}+\frac{\cos\rho}{\phi_g}e^{i\rho}  ~&~ \frac{e^{i\sigma}-\sqrt{5}\,e^{-i\sigma}}{2}+\frac{\cos\rho}{\phi_g}e^{i(\rho+\sigma)}\\
 -\sqrt{2}\,i e^{i\sigma} \sin\rho   ~&~ \frac{e^{i\sigma}-\sqrt{5}\,e^{-i\sigma}}{2}+\frac{\cos\rho}{\phi_g}e^{i(\rho+\sigma)} ~&~ \frac{e^{2i\sigma}+\sqrt{5}}{2}+\frac{\cos\rho}{\phi_g}e^{i(\rho+2\sigma)}\\
\end{pmatrix}\,.
\label{eq:res-cp-sym-GR}
\end{eqnarray}
Taking the limit of $\rho, \sigma \rightarrow 0$, we obtain
\begin{eqnarray}
\nonumber && X_1=\frac{1}{\sqrt{5}}
\begin{pmatrix}
 1 ~&~ -\sqrt{2} ~&~ -\sqrt{2} \\
 -\sqrt{2} ~&~ -\phi_g ~&~ 1/\phi_g \\
 -\sqrt{2} ~&~ 1/\phi_g ~&~ -\phi_g
\end{pmatrix} \,,~~~
X_2=\frac{1}{\sqrt{5}}
\begin{pmatrix}
 -1 ~&~ \sqrt{2} ~&~ \sqrt{2} \\
 \sqrt{2} ~&~ -1/\phi_g ~&~ \phi_g \\
 \sqrt{2} ~&~ \phi_g ~&~ -1/\phi_g
\end{pmatrix} \,,\\
&&X_3=-\begin{pmatrix}
 1 ~&~  0 ~&~ 0 \\
 0 ~&~ 0  ~&~ 1 \\
 0 ~&~ 1 ~&~ 0
\end{pmatrix}\,,~~~~~
X_4=\begin{pmatrix}
 1 ~&~ 0 ~&~ 0 \\
 0 ~&~ 1 ~&~ 0 \\
 0 ~&~ 0 ~&~ 1
\end{pmatrix}\,.
\end{eqnarray}
The relations between residual flavour and CP symmetries in Eq.~\eqref{eq:res-CP-flavor} are fulfilled. If all the four remnant CP transformations in Eq.~\eqref{eq:res-cp-sym-GR} are preserved by the neutrino mass matrix, the complex GR mixing pattern with vanishing $\theta_{13}$ would be produced.

Similarly to section~\ref{subsec:revamp_TBM}, we consider the scenario of partially preserved remnant CP symmetries. If the CP symmetries $X_1, X_4$ or $X_2, X_3$ are preserved in the neutrino sector, the remnant flavour symmetry $G_1=X_1X^{\ast}_4=X_4X^{\ast}_1 = X_2X^{\ast}_3=X_3X^{\ast}_2$ would be conserved as well. As a consequence, the first column of the lepton mixing matrix would be determined to be $(\sqrt{2}\phi_g, -e^{i\rho}, -e^{i(\rho+\sigma)})^{T}/\sqrt{2\sqrt{5}\,\phi_g}$, the same as in the cGR mixing. It follows that the relation $\cos^2\theta_{12}\cos^2\theta_{13}=\frac{\phi_g}{\sqrt{5}}$ is fulfilled. Using the $3\sigma$ allowed range $2.000\times10^{-2}\leq\sin^2\theta_{13}\leq2.405\times10^{-2}$~\cite{deSalas:2020pgw,10.5281/zenodo.4726908}, we find the solar mixing angle must lie in the region $0.2586\leq\sin^2\theta_{12}\leq0.2616$ which doesn't overlap with the experimental $3\sigma$ range of $\theta_{12}$~\cite{deSalas:2020pgw,10.5281/zenodo.4726908}.

\begin{figure}[h!]
\begin{center}
\includegraphics[width=0.95\linewidth]{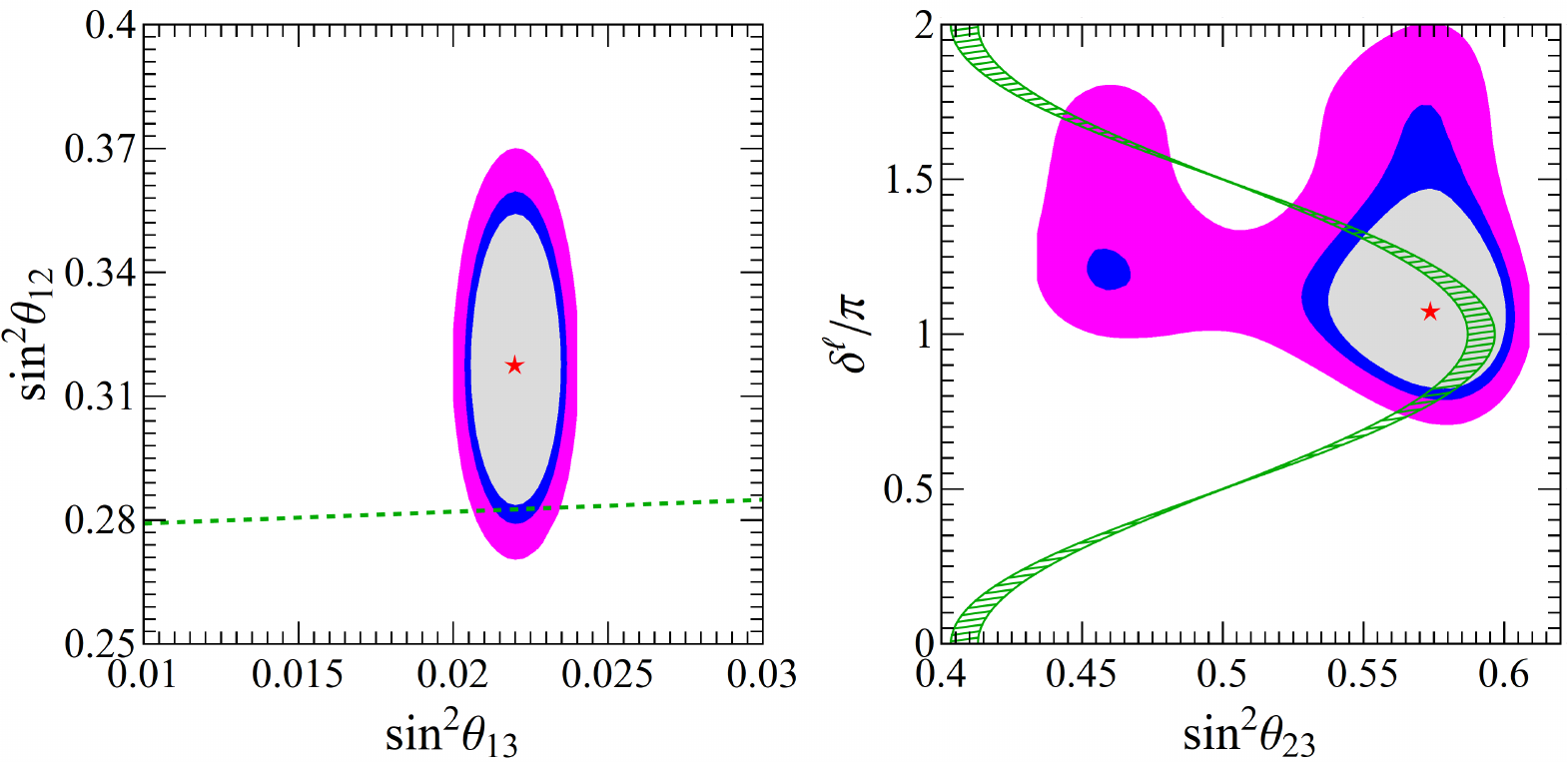}
\caption{\label{fig:revamp-GR}  Predicted oscillation parameter correlations in the revamped GR scheme. The ``generic'' global-fit allowed regions corresponding to $90\%$, $95\%$ and $99\%$ confidence level are displayed~\cite{deSalas:2020pgw,10.5281/zenodo.4726908}.}
\end{center}
\end{figure}

The phenomenologically viable case is found when the CP symmetries $X_2, X_4$ or $X_1, X_3$ are preserved by the neutrino mass matrix. The mixing parameters for the latter CP transformation can be obtained from those of the former by redefining $\rho\to\rho+\pi/2, \sigma\to \sigma-\pi$. Without loss of generality, we shall focus on preserved CP transformations $X_2, X_4$ which leads to conservation of the flavour symmetry generated by the $G_2=X_2X^{*}_4=X_4X^{*}_2$ transformation~\footnote{Notice again that imposing $G_3$ is uninteresting here, as it leads to $\theta_{13}=0$.}. The lepton mixing matrix is found to be
\begin{small}
\begin{eqnarray}
\nonumber &&\hskip-0.5in U=U_{cGR}R_{13}(\theta) Q_{\nu}\\
&&\hskip-0.3in=\frac{1}{\sqrt{2\sqrt{5}\phi_g}}\begin{pmatrix}
\sqrt{2}\phi_g\cos\theta   ~&  \sqrt{2} e^{-i \rho }  ~& \sqrt{2}\phi_g\sin \theta \\
-e^{i\rho}\cos\theta-5^{1/4}e^{-i \sigma}\sqrt{\phi_g}\sin\theta ~& \phi_g  ~&-e^{i\rho}\sin\theta+5^{1/4}e^{-i\sigma}\sqrt{\phi_g}\cos\theta\\
-e^{i(\rho+\sigma)}\cos\theta+5^{1/4}\sqrt{\phi_g}\sin\theta  ~&  e^{i\sigma}\phi_g  ~&  -e^{i(\rho+\sigma)}\sin\theta-5^{1/4}\sqrt{\phi_g}\cos\theta
\end{pmatrix}Q_{\nu}\,.
\end{eqnarray}
\end{small}
The mixing angles and CP violating phases read as
\begin{eqnarray}
\nonumber&&\sin^2\theta_{13}=\frac{\phi_g\sin^2\theta}{\sqrt{5}}\,,
~~~\sin^2\theta_{12}=\frac{1}{1+\phi_g^2\cos^2\theta}\,,
~~~\sin^2\theta_{23}=\frac{1}{2}-\frac{\sqrt{\sqrt{5}\phi_g}\sin2\theta\cos(\rho+\sigma)}{2\phi_g^2\cos^2\theta+2}\,,\\
\nonumber&&\sin\delta^{\ell}=
-\frac{\text{sign}(\sin2\theta)(\phi_g^2\cos^2\theta+1)\sin(\rho+\sigma)}
{\sqrt{(\phi_g^2\cos^2\theta+1)^2-\sqrt{5}\phi_g\sin^22\theta\cos^2(\rho+\sigma)}}\,,\\
&&\tan\delta^{\ell} =-\frac{(\phi_g^2\cos^2\theta+1)\tan(\rho+\sigma)}{\cos2\theta+\phi_g^2\cos^2\theta}\,,~~~\phi_{12}=\rho+\frac{(k_1-k_2)\pi}{2}\,,
~~~ \phi_{13}=\frac{(k_1-k_3)\pi}{2}\,.
\end{eqnarray}
As a consequence, we can derive the following exact relations among the mixing parameters
\begin{equation}
\label{eq:corr-revamp-cGR}\sin^2\theta_{12}\cos^2\theta_{13}=\frac{1}{\sqrt{5}\phi_g}\,,
~~~~ \tan2\theta_{23} \cos \delta^{\ell}=
\frac{\phi_g^2\cot^2\theta_{13}-2}{2\sqrt{\phi_g^2\cot^2\theta_{13}-1}}\,.
\end{equation}
For the best fit value $\sin^2\theta_{13}=2.200\times 10^{-2}$~\cite{deSalas:2020pgw,10.5281/zenodo.4726908}, we find the solar mixing angle $\sin^2\theta_{12}=0.2826$ which is within the $3\sigma$ range. The correlations of Eq.~\eqref{eq:corr-revamp-cGR} are displayed in figure~\ref{fig:revamp-GR}.

\subsection{Bi-large mixing}
\label{sec:bi-large-mixing}

It is unlikely that any revamping procedure can make the bi-maximal mixing pattern of section~\ref{subsec:bi-maximal-mixing} consistent with the oscillation data, since the measured value of the solar angle $\theta_{12}$ deviates too much from the maximal value~\cite{deSalas:2020pgw,10.5281/zenodo.4726908}. The bi-large pattern is phenomenologically motivated by the fact that the smallest lepton mixing angle $\theta_{13}$ is similar in magnitude to the largest of the elements of the quark mixing matrix, the Cabibbo angle. It suggests that the latter may act as the universal seed for quark and lepton mixings~\cite{Boucenna:2012xb,Ding:2012wh,Roy:2014nua,Chen:2019egu,Roy:2020vtm}.
Bi-large neutrino mixing implies that the lepton and quark sectors may be related with each other, a possible new strategy in the quest for quark-lepton symmetry and unification.

Within the simplest bi-large mixing hypothesis, the solar and atmospheric mixing angles are expressed as~\cite{Boucenna:2012xb},
\begin{equation}
\sin\theta_{13}=\lambda,~~~\sin\theta_{12}=s\lambda,~~~\sin\theta_{23}=a\lambda\,,
\end{equation}
where the small parameter $\lambda$ is the reactor angle, $s$ and $a$ are free parameters of order one. Using the best fit values of the mixing angles~\cite{deSalas:2020pgw,10.5281/zenodo.4726908}, one finds $\lambda\simeq0.148$, $s\simeq3.802$ and $a\simeq5.108$ for normal neutrino mass ordering. Since the bi-large approach describes the structure of the lepton mixing matrix in terms of $\theta_{13}$ as input, no revamping is needed.

\subsubsection{Bi-large mixing from abelian family symmetry}
\label{sec:bi-large-mixing-1}

As shown above, the bi-large mixing ansatz assumes that the three lepton mixing angles are of the same order of magnitude, to first approximation,
\begin{equation}
\label{eq:BL1}\texttt{BL}_1:~~ \sin\theta_{12}\sim\lambda_c,~~~\sin\theta_{13}\sim\lambda_c,~~~ \sin\theta_{23}\sim\lambda_c\,,
\end{equation}
where $\lambda_c\simeq0.23$ is the Cabibbo angle~\cite{Workman:2022ynf}, and the notation ''$\sim$'' implies that the above relations contain unknown factors of order one. The freedom in these factors can be used to obtain an adequate description of neutrino mixing. Although global analyses of neutrino oscillation data show preference for the second octant of the atmospheric mixing angle $\theta_{23}$~\cite{deSalas:2020pgw,10.5281/zenodo.4726908}, the significance is not yet overwhelming. For $\theta_{23}$ in the preferred higher octant, it has been proposed~\cite{Ding:2012wh} that the bi-large mixing ansatz could be
\begin{equation}
\label{eq:BL2}\texttt{BL}_2:~~ \sin\theta_{12}\sim\lambda_c,~~~\sin\theta_{13}\sim\lambda_c,~~~ \sin\theta_{23}\sim1-\lambda_c\,.
\end{equation}
Here we will show how the above variants of the bi-large mixing patterns can be achieved in a Froggatt-Nielsen-type scenario~\cite{Froggatt:1978nt} with an Abelian flavour symmetry.
Assuming the presence of two Higgs doublets $H_{u,d}$, and that neutrino masses are described by an effective Weinberg operator we generalize the well-known $U(1)$ flavour symmetry to a larger $U(1)\times Z_m\times Z_n\subset U(1)\times U(1)'\times U(1)''$ family symmetry. In order to break the $U(1)$, $Z_m$ and $Z_n$ flavour groups we assume three flavons $\Theta_1$, $\Theta_2$ and $\Theta_3$ with horizontal charges $\Theta_1: (-1,0,0)$, $\Theta_2: (0,-1,0)$, $\Theta_3: (0,0,-1)$, respectively, and their VEVs $\vev{\Theta_{1,2,3}}/\Lambda$ scaled by the cutoff $\Lambda$ are of order $\lambda_c$. We assume the horizontal charges of $H_{u,d}$ to be zero. Fermion masses are then described by the following effective Yukawa couplings~\cite{Ding:2012wh},
\begin{eqnarray}
\nonumber&& \mathcal{W}=(y_u)_{ij}Q_iU^c_jH_u\left(\frac{\Theta_1}{\Lambda}\right)^{F(Q_i)+F(U^c_j)}\left(\frac{\Theta_2}{\Lambda}\right)^{\left[Z_m(Q_i)+Z_m(U^c_j)\right]}\left(\frac{\Theta_3}{\Lambda}\right)^{\left[Z_n(Q_i)+Z_n(U^c_j)\right]}\\
\nonumber&&\qquad +(y_d)_{ij}Q_iD^c_jH_d\left(\frac{\Theta_1}{\Lambda}\right)^{F(Q_i)+F(D^c_j)}\left(\frac{\Theta_2}{\Lambda}\right)^{\left[Z_m(Q_i)+Z_m(D^c_j)\right]}\left(\frac{\Theta_3}{\Lambda}\right)^{\left[Z_n(Q_i)+Z_n(D^c_j)\right]}\\
\nonumber&&\qquad +(y_e)_{ij}L_iE^c_jH_d\left(\frac{\Theta_1}{\Lambda}\right)^{F(L_i)+F(E^c_j)}\left(\frac{\Theta_2}{\Lambda}\right)^{\left[Z_m(L_i)+Z_m(E^c_j)\right]}\left(\frac{\Theta_3}{\Lambda}\right)^{\left[Z_n(L_i)+Z_n(E^c_j)\right]}\\
\label{eq:U(1)_extended}&&\qquad +(y_{\nu})_{ij}\frac{1}{\Lambda}L_{i}L_{j}H_{u}H_{u}\left(\frac{\Theta_1}{\Lambda}\right)^{F(L_i)+F(L_j)}\left(\frac{\Theta_2}{\Lambda}\right)^{\left[Z_m(L_i)+Z_m(L^{}_j)\right]}
\left(\frac{\Theta_3}{\Lambda}\right)^{\left[Z_n(L_i)+Z_n(L^{}_j)\right]}\,,
\end{eqnarray}
where $F(\psi)$ denotes the $U(1)$ charge of the field $\psi$,  $Z_{m,n}(\psi)$ stand for the $Z_{m,n}$ charge of $\psi$, and the brackets $[\ldots]$ around the exponents denote that we are modding out by $m$ ($n$) according to the $Z_m$ ($Z_n$) addition rule, i.e., $\left[Z_m(Q_i)+Z_m(U^c_j)\right]=Z_m(Q_i)+Z_m(U^c_j)~(\text{mod}\; m)$. Hence fermion mass matrices are expressed in terms of the horizontal charges as follows,
\begin{eqnarray}
\nonumber&&(M_u)_{ij}=(y_{u})_{ij}\;\lambda_c^{F(Q_i)+F(U^c_j)+\left[Z_m(Q_i)+Z_m(U^c_j)\right]+\left[Z_n(Q_i)+Z_n(U^c_j)\right]}\;v_u\,, \\
\nonumber&&(M_d)_{ij}=(y_{d})_{ij}\;\lambda_c^{F(Q_i)+F(D^c_j)+\left[Z_m(Q_i)+Z_m(D^c_j)\right]+\left[Z_n(Q_i)+Z_n(D^c_j)\right]}\;v_d\,,  \\
\nonumber&&(M_{e})_{ij}=(y_{e})_{ij}\;\lambda_c^{F(L_i)+F(E^c_j)+\left[Z_m(L_i)+Z_m(E^c_j)\right]+\left[Z_n(L_i)+Z_n(E^c_j)\right]}\;v_d\,,  \\
\label{eq:mass-matrices-U1}&&(M_{\nu})_{ij}=(y_{\nu})_{ij}\;\lambda_c^{F(L_i)+F(L_j)+\left[Z_m(L_i)+Z_m(L^{}_j)\right]+\left[Z_n(L_i)+Z_n(L^{}_j)\right]}\;\frac{v^2_u}{\Lambda}\,.
\end{eqnarray}
If all the horizontal charges are positive, the hierarchical structure of the mass matrices allows a simple order-of-magnitude estimate of the various mass ratios and mixing angles.
For instance, the entries of the CKM matrix are estimated to be,
\begin{equation}
(V_{CKM})_{ij}\sim\lambda_c^{F_{\rm eff}(Q_i)-F_{\rm eff}(Q_j)\pm \alpha m\pm \beta n}\,,
\end{equation}
where $F_{\rm eff}(\psi)=F(\psi)+Z_m(\psi)+Z_n(\psi)$, and $\alpha,\beta=0, 1$ depends on the charge assignment under $Z_m$ and $Z_n$. Likewise for the lepton sector, one obtains
\begin{equation}
\sin\theta_{ij}\sim\lambda_c^{F_{\rm eff}(L_i)-F_{\rm eff}(L_j)\pm \alpha m\pm \beta n}\,.
\end{equation}
Notice that the mixing angles can be enhanced or suppressed by $\lambda^{\pm m\pm n}$ relative to the scaling predictions obtained with the continuous $U(1)\times U(1)' \times U(1)''$ family symmetry. Moreover, $U(1)\times Z_m\times Z_n$ reduces to $U(1)\times Z_m$ if $n=1$, and to the usual $U(1)$ if $m=n=1$.

\begin{itemize}[labelindent=-0.8em, leftmargin=1.2em]
\item{ Model for $\texttt{BL}_1$ mixing }

The family symmetry group is $U(1)\times Z_3\times Z_4$, and we assign the lepton fields to transform under the flavour symmetry as follows~\cite{Ding:2012wh},
\begin{eqnarray}
\nonumber&&L_{1}:~(4,1,3),~~\quad~~ L_{2}:~(3,2,2),~~\quad~~ L_{3}:~(1,1,1), \\
\label{eq:ass-leptons-BL1}&& E^c_{1}:~(3,2,2),~~\quad~~ E^c_{2}:~(1,2,2),~~\quad~~ E^c_{3}:~(0,0,0)\,.
\end{eqnarray}
One can then read out the pattern of charged lepton and neutrino mass matrices,
\begin{equation}
\label{eq:cmm}M_{e}\sim\left(\begin{array}{ccc}
\lambda^{8}_c ~& \lambda^6_c   ~& \lambda^8_c \\
\lambda^{7}_c ~& \lambda^5_c   ~& \lambda^{7}_c \\
\lambda^7_c ~&  \lambda^5_c   ~& \lambda^3_c
\end{array}\right)v_d,~~\quad~~M_{\nu}\sim\left(
\begin{array}{ccc}
\lambda^{12}_c  ~&  \lambda^8_c  ~& \lambda^7_c  \\
\lambda^8_c  ~&  \lambda^7_c  ~& \lambda^7_c \\
\lambda^7_c  ~&  \lambda^7_c  ~& \lambda^6_c
\end{array}
\right)\frac{v^2_u}{\Lambda}\,.
\end{equation}
These give rise to the following mass ratios and lepton mixing angles,
\begin{eqnarray}
\nonumber&&~~\qquad~~~\frac{m_e}{m_{\mu}}\sim\lambda^3_c,\qquad  \frac{m_{\mu}}{m_{\tau}}\sim\lambda^2_c\,,\\
\nonumber&&m_{1}\sim\lambda^8_c\,\frac{v^2_u}{\Lambda},~~~ m_{2}\sim\lambda^7_c\,\frac{v^2_u}{\Lambda},~~~ m_{3}\sim\lambda^6_c\,\frac{v^2_u}{\Lambda}\,,\\
&&\sin\theta_{12}\sim\lambda_c,~\quad~ \sin\theta_{13}\sim\lambda_c,~\quad~  \sin\theta_{23}\sim\lambda_c\,.
\end{eqnarray}
This way we obtain the $\texttt{BL}_1$ mixing pattern. A simple numerical analysis with different seed procedures for the order-one Yukawa coefficients, leads to the
$\theta_{23}$ distributions given in figure~\ref{fig:theta23-his-BL1}. One sees that $\sin^2\theta_{23}<1/2$ (first octant) is preferred in this case.

\begin{figure}[h!]
\begin{center}
\includegraphics[width=1.0\linewidth]{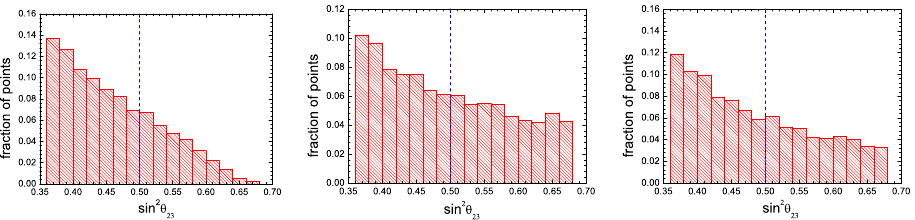}
\caption{\label{fig:theta23-his-BL1}
Distributions of the atmospheric neutrino mixing angle $\theta_{23}$ in the $\mathrm{\tt BL_1}$ model. The left, middle and right panels are obtained with flat, exponential and Gaussian seed procedures. From~\cite{Ding:2012wh}. }
\end{center}
\end{figure}

We now extend the model to the quark sector along the lines of $SU(5)$ unification.
Each family of standard quarks and leptons is embedded in $SU(5)$ multiplets $\mathbf{10}=(Q, U^c, E^c)$ and $\bar{\mathbf{5}}=(D^c, L)$. The fields within a single SU(5) multiplet transform in the same way under the family symmetry. Therefore the lepton assignment in Eq.~\eqref{eq:ass-leptons-BL1} implies that the quark charges under the flavour group $U(1)\times Z_3\times Z_4$ should be
\begin{eqnarray}
\nonumber&& Q_{1}:~(3,2,2),~~\quad~~ Q_{2}:~(1,2,2), ~~\quad~~ Q_{3}:~(0,0,0)\,, \\
\nonumber&& U^c_{1}:~(3,2,2),~~\quad~~ U^c_{2}:~(1,2,2),~~\quad~~ U^c_{3}:~(0,0,0)\,,   \\
\label{eq:qs}&& D^c_{1}:~(4,1,3),~~\quad~~ D^c_{2}:~(3,2,2),~~\quad~~ D^c_{3}:~(1,1,1)\,.
\end{eqnarray}
Consequently, the up-type and down-type quark mass matrices are given as
\begin{eqnarray}
\label{eq:quark_matrices-BL1}M_{u}\sim\begin{pmatrix}
 \lambda^7_c ~& \lambda^5_c  ~&  \lambda^7_c \\
 \lambda^5_c ~& \lambda^3_c  ~&  \lambda^5_c \\
 \lambda^7_c ~& \lambda^5_c  ~&  1
\end{pmatrix}v_u  , \qquad  M_d\sim
\begin{pmatrix}
 \lambda^8_c  ~&  \lambda^7_c  ~& \lambda^7_c \\
 \lambda^6_c  ~&  \lambda^5_c  ~& \lambda^5_c \\
 \lambda^8_c  ~&  \lambda^7_c  ~& \lambda^3_c
\end{pmatrix} v_d\,,
\end{eqnarray}
The resulting quark masses and CKM mixing matrix are determined to follow the pattern
\begin{eqnarray}
\nonumber&& m_u\sim \lambda^7_cv_u,~~\quad~~  m_c\sim \lambda^3_cv_u,~~\quad~~ m_t\sim v_u, \\
\nonumber&&m_d\sim \lambda^8_c v_d,~~\quad~~  m_s\sim \lambda^5_c v_d,~~\quad~~  m_b\sim \lambda^3_c v_d \,,\\
&&|V_{us}|\sim\lambda^2_c,~~\quad~~ |V_{cb}|\sim\lambda^2_c,~~\quad~~ |V_{ub}|\sim\lambda^4_c\,,
\end{eqnarray}
which are in agreement with experimental data except $|V_{us}|$, for which an accidental enhancement of $\mathcal{O}(\lambda^{-1})$ amongst the free order-one coefficients is needed
so as to reproduce the correct Cabibbo angle.

\item{Model for $\texttt{BL}_2$ mixing }

Here the flavour symmetry is $U(1)\times Z_2$, and the charge assignments for the quarks and leptons are taken as~\cite{Ding:2012wh}
\begin{eqnarray}
\nonumber&& ~~\quad~~D^c_1, L_1 :~(3,0),\quad\quad  D^c_2, L_2 :~(3,1),\quad\quad D^c_3, L_3 :~(2,0),\\
&&Q_1, U^c_1, E^c_1 :~ (4,0),~~\quad~~  Q_2, U^c_2, E^c_2 :~ (2,1),~~\quad~~  Q_3, U^c_3, E^c_3 :~ (0,1)\,.
\end{eqnarray}
Consequently the neutrino and charged lepton mass matrices take the following form
\begin{equation}
M_{e}\sim\begin{pmatrix}
\lambda^{7}_c ~& \lambda^6_c   ~& \lambda^4_c \\
\lambda^{8}_c ~& \lambda^5_c   ~& \lambda^{3}_c \\
\lambda^6_c  ~&  \lambda^5_c  ~& \lambda^3_c
\end{pmatrix}v_d,~~~~\quad~~
M_{\nu}\sim\begin{pmatrix}
\lambda^{6}_c  ~&  \lambda^7_c  ~& \lambda^5_c  \\
\lambda^7_c  ~&  \lambda^6_c  ~& \lambda^6_c \\
\lambda^5_c  ~&  \lambda^6_c  ~& \lambda^4_c
\end{pmatrix} \frac{v^2_u}{\Lambda}\,.
\end{equation}
The charged lepton mass matrix $M_e$ has a ``lopsided'' structure, and can give the correct order-of-magnitude for the charged lepton masses and a large 2-3 mixing. Combining neutrino and charged lepton diagonalization matrices, the resulting $\texttt{BL}_2$ lepton mixing pattern is determined as,
\begin{equation}
\label{eq:mixing_BL2}\sin\theta_{12}\sim\lambda_c,~~\quad~~\sin\theta_{13}\sim\lambda_c\,~~\quad~~ \sin\theta_{23}\sim1\,.
\end{equation}
Using the master formula of Eq.~\eqref{eq:mass-matrices-U1} we can easily read off the quark mass matrices as
\begin{eqnarray}
M_{u}\sim \begin{pmatrix}
 \lambda^8_c ~& \lambda^7_c   ~& \lambda^5_c \\
 \lambda^7_c ~&  \lambda^4_c  ~& \lambda^2_c \\
 \lambda^5_c  ~ & \lambda^2_c  & 1
\end{pmatrix} v_u,  ~~\quad~~   M_d\sim
\begin{pmatrix}
 \lambda^7_c  ~&  \lambda^8_c  ~ &  \lambda^6_c \\
 \lambda^6_c  ~&   \lambda^5_c  ~& \lambda^5_c \\
 \lambda^4_c  ~&  \lambda^3_c ~& \lambda^3_c
\end{pmatrix} v_d\,.
\end{eqnarray}
This leads to the following pattern of CKM matrix elements and quark mass ratios,
\begin{eqnarray}
\nonumber &|V_{us}|\sim\lambda_c, ~\quad~ |V_{cb}|\sim\lambda^2_c, ~\quad~ |V_{ub}|\sim\lambda^3_c\,, \\
&\frac{m_u}{m_c}\sim\lambda^4_c, ~\quad~ \frac{m_c}{m_t}\sim\lambda^4_c, ~\quad~ \frac{m_d}{m_s}\sim\lambda^2_c,  ~~\quad~~ \frac{m_s}{m_b}\sim\lambda^2_c, ~\quad~ \frac{m_b}{m_t}\sim\lambda^3_c\,,
\end{eqnarray}
in very good qualitative agreement with observed values.
\end{itemize}

\subsubsection{Confronting bi-large mixing with oscillation data}

We now make the bi-large mixing ansatz more predictive. We assume a CP conserving neutrino diagonalization matrix $U_\nu$, with its three angles related to the Cabibbo angle in a simple manner, and a CKM-like charged lepton diagonalization. Under these assumptions we illustrate the predictive power of bi-large mixing~\cite{Chen:2019egu}.

\begin{itemize}[labelindent=-0.8em, leftmargin=1.2em]

\item{Constraining bi-large mixing: pattern I }

Here the neutrino mixing angles are assumed to be related to the Cabibbo angle as follows~\cite{Chen:2019egu}
\begin{equation}
\sin\theta_{13}=\lambda_c,~~~\sin\theta_{12}=2\lambda_c, ~~~
\sin\theta_{23}=1-\lambda_c \,.
\end{equation}
The Dirac CP phase is taken as $\delta^{\nu}_{CP}=\pi$ with vanishing Majorana phases. The resulting neutrino diagonalization matrix is given by
\begin{equation}
\label{eq:unu1}U_{\nu}\simeq\begin{pmatrix}
1-\frac{5\lambda^2_c}{2} ~& 2\lambda_c ~ &-\lambda_c \\
\lambda_c-2\sqrt{2}\lambda^{3/2}_c ~& \sqrt{2\lambda_c}-\frac{\lambda^{3/2}_c}{2\sqrt{2}} ~ & 1-\lambda_c-\frac{\lambda^2_c}{2} \\
2\lambda_c+\sqrt{2}\lambda^{3/2}_c  ~& -1+\lambda_c ~& \sqrt{2\lambda_c}-\frac{\lambda^{3/2}_c}{2\sqrt{2}}
\end{pmatrix}\,.
\end{equation}
Motivated by $SO(10)$, we take a CKM-type charged lepton diagonalization matrix
\begin{equation}
U_{l}=R_{23}(\theta^{CKM}_{23})\,\Phi\,R_{12}(\theta^{CKM}_{12})\,\Phi^\dagger\simeq\begin{pmatrix}
1-\frac{\lambda^2_c}{2} ~&~ \lambda_c e^{-i\phi} ~&~ 0 \\
-\lambda_c e^{i\phi}  ~&~  1-\frac{\lambda^2_c}{2}  ~ &~  A\lambda^2_c \\
A\lambda^3_ce^{i\phi} ~ &~  -A\lambda^2_c  ~&~ 1
\end{pmatrix}\,,
\end{equation}
where $\sin\theta^{\rm CKM}_{12}=\lambda_c$, $\sin\theta^{\rm CKM}_{23}=A\lambda^{2}_c$, $\lambda_c=0.22453\pm0.00044$ and $A=0.836\pm0.015$ are the Wolfenstein parameters~\cite{Workman:2022ynf}, and $\Phi={\rm diag}(e^{-i\phi/2}, e^{i\phi/2},1)$ where $\phi$ is a free phase parameter. One sees that the lepton mixing matrix $U=U^{\dagger}_lU_{\nu}$ only depends on a single free phase parameter $\phi$, which can in general take values between $-\pi$ and $\pi$. The leptonic mixing angles and Jarlskog invariant are found to be
\begin{eqnarray}
\nonumber \sin^2\theta_{13} &\simeq& 4\lambda^2_c (1-\lambda_c)\cos^2\frac{\phi}{2}\,, \\
\nonumber \sin^2\theta_{12} &\simeq & 2\lambda^2_c\big(2-2\sqrt{2\lambda_c}\cos\phi+\lambda_c\big)\,,\\
\nonumber \sin^2\theta_{23} &\simeq& (1-\lambda_c)^2-2\sqrt{2}A\lambda_c^\frac{5}{2}-2\lambda^3_c(1+2\cos\phi)\,,\\
J_{CP}&\simeq& - 2\left(\sqrt{2}+\sqrt{\lambda}\right)\lambda^{5/2}_c \sin\phi\,.
\end{eqnarray}
We show these correlations in figure~\ref{fig:cor_Cons-BL1}. They show how the precise measurement of the reactor angle can be promoted to sharp predictions for solar and atmospheric mixing angles. Similarly, the Dirac CP phase is also predicted, up to a two-fold degeneracy.

\begin{figure}[t!]
\begin{center}
\begin{tabular}{cc}
\hskip-0.5cm\includegraphics[width=0.47\textwidth]{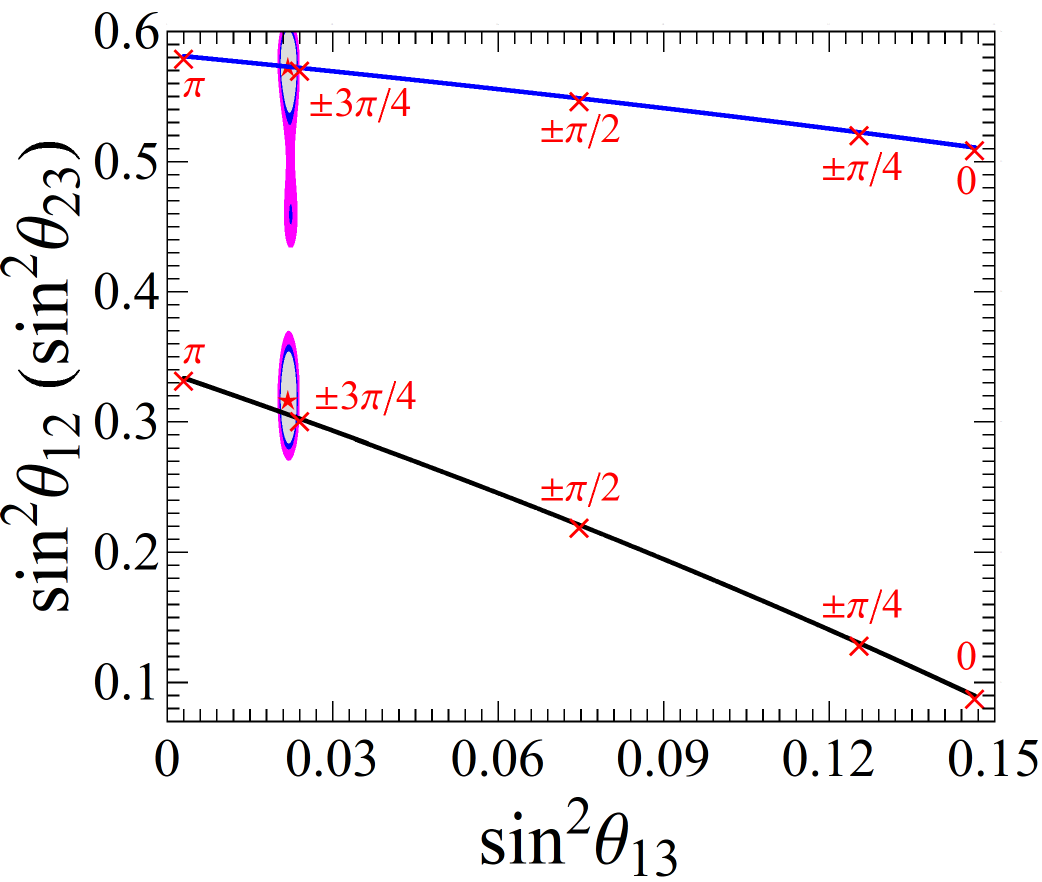}   & \hskip-0.2cm  \includegraphics[width=0.47\textwidth]{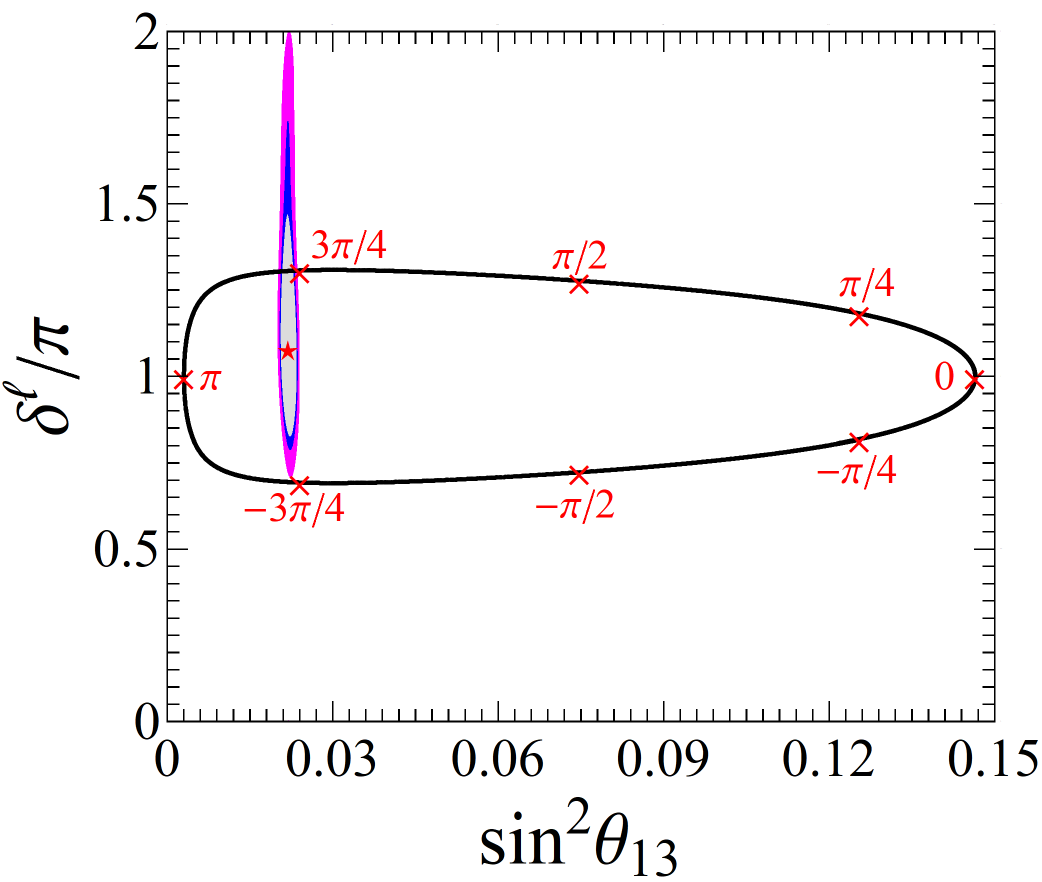}
\end{tabular}
\caption{\label{fig:cor_Cons-BL1}Predicting the solar and atmospheric mixing parameters $\sin^2\theta_{12}$ (lower, black line) and $\sin^2\theta_{23}$ (upper, blue line) versus $\sin^2\theta_{13}$ in the first kind of constrained bi-large mixing scheme. The right panel shows $\delta^{\ell}$ versus $\sin^2\theta_{13}$. The allowed $90\%$, $95\%$ and $99\%$ CL regions of the global oscillation fit are displayed~\cite{deSalas:2020pgw,10.5281/zenodo.4726908}. The crosses correspond to $\phi=0, \pi, \pm\pi/4, \pm\pi/2, \pm3\pi/4$. }
\end{center}
\end{figure}

\item{Constraining bi-large mixing: pattern II }

In this case the bi-large ansatz for the neutrino mixing angles is~\cite{Chen:2019egu}
\begin{equation}
\sin\theta^\nu_{13}=\lambda_c\,, ~~~ \sin\theta^\nu_{12} = 2\lambda_c\,, ~~~ \sin\theta^\nu_{23}=3\lambda_c\,,
\end{equation}
with $\delta^{\nu}_{CP}=\pi$ and vanishing Majorana phases. Motivated by $SU(5)$ unification, we take the charged-lepton diagonalization matrix to be of the form
\begin{equation}\label{eq:Ul2}
U_{l} = \Phi^{\dagger} R^{T}_{12}(\,\theta_{12}^{CKM}\,)\Phi R^{T}_{23}(\,\theta_{23}^{CKM}\,)\simeq
\begin{pmatrix}
1-\frac{1}{2}\lambda^2_c ~&~ -\lambda_c\, e^{i \phi} ~&~ A \lambda^3_c e^{i \phi} \\
\lambda_c\, e^{- i \phi} ~&~ 1- \frac{1}{2}\lambda^2_c ~&~ - A\lambda^2_c \\
0 ~&~ A \lambda^2_c ~&~ 1
\end{pmatrix}\,.
\end{equation}
Taking into account both charged and neutral diagonalizations we can extract the following results for the lepton mixing angles and leptonic Jarlskog invariant,
\begin{eqnarray}
\nonumber\sin^{2}\theta_{13}& \simeq& \lambda^{2}_c- 6 \lambda^3_c \cos\phi+8\lambda^4_c \,, \\
\nonumber \sin^{2}\theta_{12} &\simeq & \lambda^{2}_c(5 + 4 \cos\phi )  - 2\lambda^4_c(8+13\cos\phi) \,, \\
\nonumber\sin^{2}\theta_{23} & \simeq& 9\lambda^{2}_c+6\lambda^{3}_c(A
+ \cos\phi)-\lambda^4_c(8-2A\cos\phi-A^2)\,, \\
J_{CP} &\simeq& -\left[3+(16+A)\lambda_c\right]\lambda^{3}_c\sin\phi \,.
\end{eqnarray}
The resulting correlations between mixing parameters are displayed in figure~\ref{fig:cor_Cons-BL2}. As before, requiring $\sin^2\theta_{13}$ in the 3$\sigma$ range~\cite{deSalas:2020pgw,10.5281/zenodo.4726908}, we find that the other oscillation parameters $\theta_{12}$, $\theta_{23}$ and $\delta_{CP}$ vary in the following regions
\begin{eqnarray}
\nonumber&&0.02000\leq\sin^2\theta_{13}\leq0.02405\,,~~~~~0.314824\leq\sin^2\theta_{12}\leq0.322459\,,\\
&&0.511755\leq\sin^2\theta_{23}\leq0.513969\,,~~~~~ 1.26643\leq\delta_{CP}/\pi\leq1.27402\,.
\end{eqnarray}

\begin{figure}[t!]
\begin{center}
\begin{tabular}{cc}
\hskip-0.5cm\includegraphics[width=0.46\textwidth]{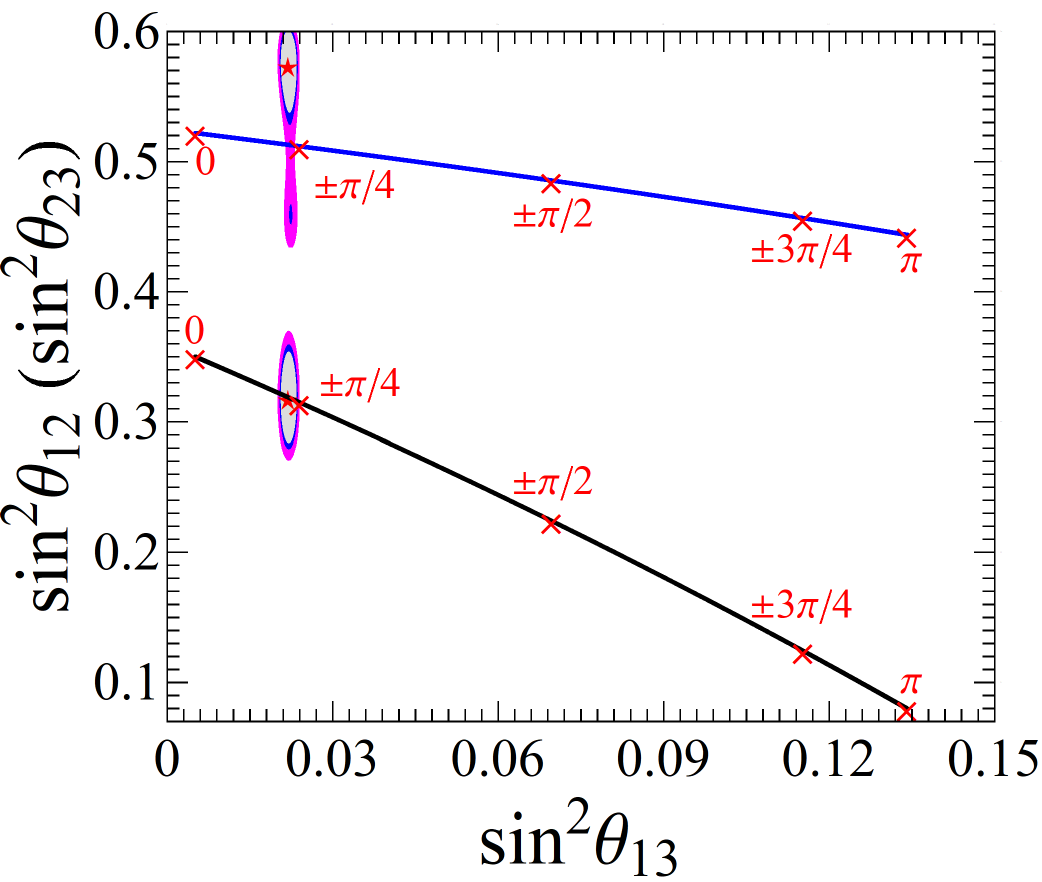}   & \hskip-0.2cm \includegraphics[width=0.46\textwidth]{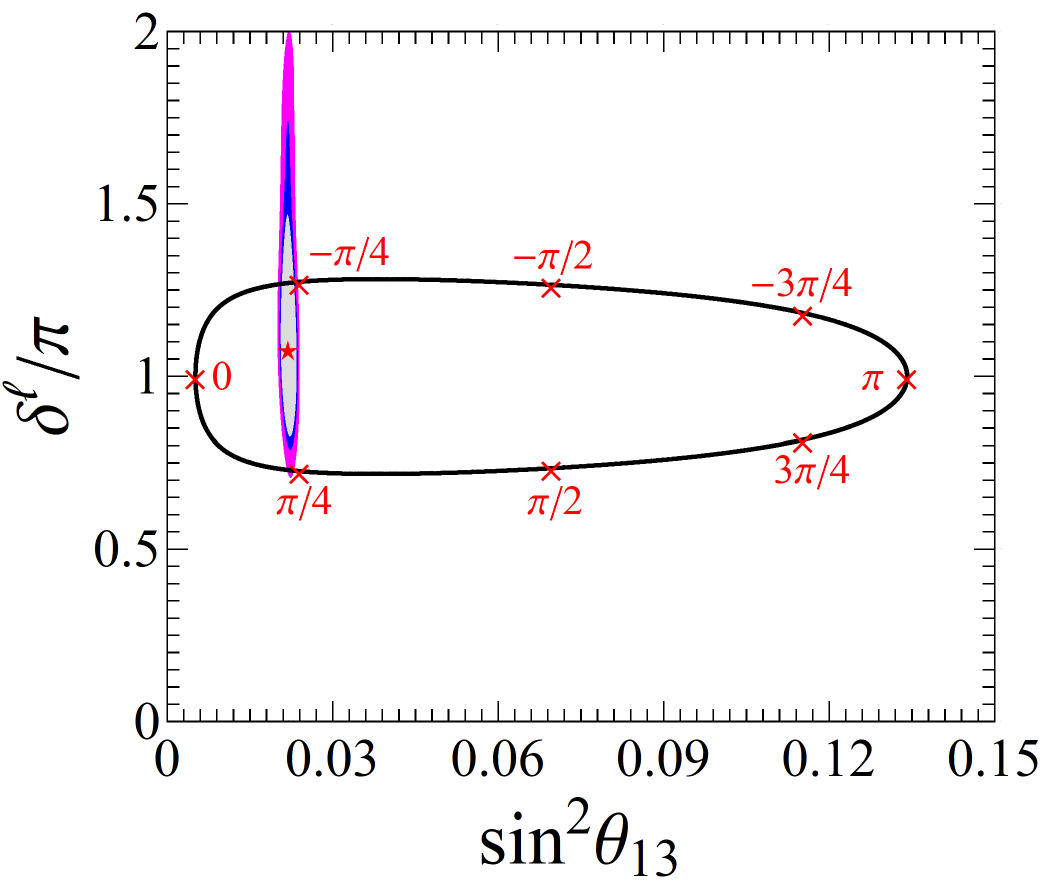}
\end{tabular}
\caption{\label{fig:cor_Cons-BL2}
Predicting the solar and atmospheric mixing angles and Dirac phase in the second constrained bi-large mixing scheme. We adopt the same convention as in figure~\ref{fig:cor_Cons-BL1}. }
\end{center}
\end{figure}

\end{itemize}

\clearpage

\section{Lepton mixing from flavour and CP symmetry}                       
\label{sec:lepton_flavor_CP_symm}                                          

Symmetries have been widely used to address the family puzzle. Early attempts employed a $U(1)$ flavour symmetry~\cite{Froggatt:1978nt}, spontaneously broken by the vacuum expectation value of a singlet scalar field. This would qualitatively account for the small quark mass ratios, as well as mixing angles. With the discovery of neutrino oscillations the idea of flavour symmetry was substantially extended, with many symmetry groups and breaking patterns studied. It has been found that finite discrete flavour groups~\cite{Ishimori:2012zz} are particularly suitable to reproduce the large lepton mixing angles and provide non-trivial predictions. The basic theory is assumed to be invariant under a flavour group $G_f$, but $G_f$ is spontaneously broken into different subgroups $G_{\nu}$ and $G_{l}$
in the neutrino and charged lepton sectors respectively. This is achieved by the VEVs of a set of scalar fields, and the misalignment between neutrino and charged lepton mass matrices
arises from the non-trivial breaking pattern of the flavour symmetry.

Neutrino model building using non-abelian discrete flavor symmetries has been surveyed in a number of dedicated reviews~\cite{Ishimori:2010au,Altarelli:2010gt,Morisi:2012fg,King:2013eh,King:2014nza,King:2017guk,Feruglio:2019ybq,Xing:2020ijf,Chauhan:2023faf}.
A prime non-abelian discrete flavour symmetry is $A_4$~\cite{Ma:2001dn,Babu:2002dz,Altarelli:2005yp,Altarelli:2005yx} which, under some circumstances, leads to the celebrated tri-bimaximal mixing pattern. Although no longer compatible with the experimental measurement of the reactor angle $\theta_{13}\simeq8.5^{\circ}$~\cite{DayaBay:2012fng,RENO:2012mkc,Daya2022,DayaBay:2022orm,DoubleChooz:2012gmf,DayaBay:2012yjv}, the simplest $A_4$ symmetry can, as we saw, be revamped so as to produce viable and predictive patterns of neutrino mixing.

All in all, the idea that lepton mixing emerges from the mismatch of the embedding of the two residual subgroups $G_{\nu}$ and $G_{l}$ into of the flavour group $G_f$ is still viable, interesting and predictive. In particular, if combining flavour symmetry with the generalized CP symmetry, one can not only accommodate lepton mixing angles but also predict leptonic CP violating phases in terms of few free parameters. In what follows, we shall review the possible schemes of predicting lepton mixing from residual symmetry, emphasizing the role of generalized CP symmetry. The results only depend on the assumed symmetry breaking pattern and are independent of the details of the residual symmetry and the particle content of the flavor symmetry breaking sector, or possible additional symmetries of the theory.

\subsection{Lepton mixing from flavour symmetry alone}
\label{subsec:lepton-mixing-from-3}

\begin{figure}[t!]
\centering
\includegraphics[width=0.98\textwidth]{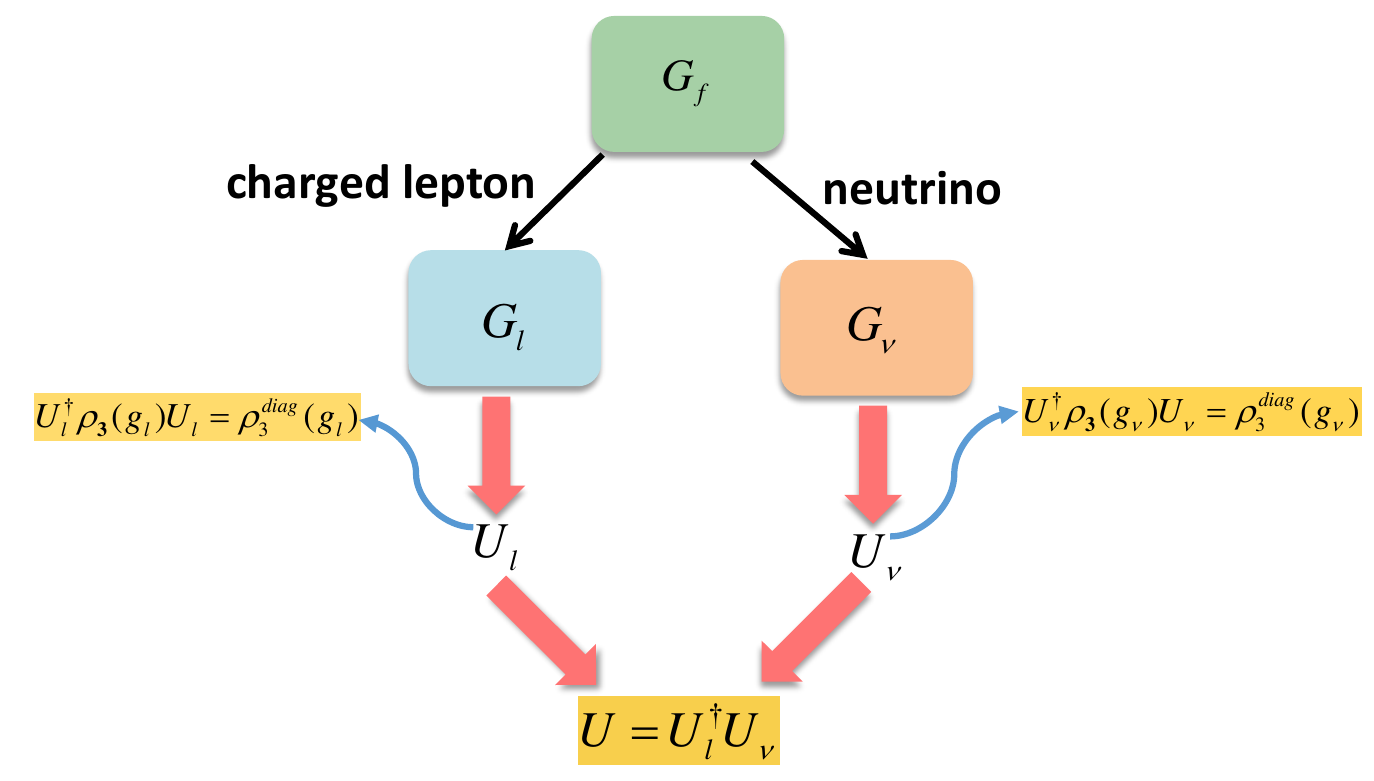}
\caption{\label{fig:Gl-Gnu-flavor}Model-independent predictions for the lepton mixing matrix from the flavour symmetry $G_f$ broken to different residual subgroups $G_{l}$ and $G_{\nu}$ in the neutrino and charged lepton sectors. Here $g_{l}$ and $g_{\nu}$ are $G_{l}$ and $G_{\nu}$ generators, respectively. See text for details. }
\end{figure}

Within a top-down approach, one imposes a certain flavour symmetry group $G_f$ at some high energy scale. The full Lagrangian is invariant under $G_f$, which is subsequently broken down to different subgroups $G_{\nu}$ and $G_{l}$ in the neutrino and charged lepton sectors respectively. The residual symmetry $G_{l}$ can be any abelian subgroup of $G_{f}$, and the same holds for $G_{\nu}$ if neutrinos are Dirac particles. On the other hand $G_{\nu}$ can only be the Klein group $K_4$ (or a subgroup) for the case of Majorana neutrinos. Throughout this review, we shall focus on the scenario in which $G_{f}$ is a non-abelian finite discrete group. In general, the three families of left-handed lepton doublets are assigned to an irreducible faithful three-dimensional representation $\rho_{\mathbf{3}}$ of $G_{f}$.
We assume that the charged lepton and neutrino mass matrices are invariant under the action of the elements of $G_{l}$ and $G_{\nu}$, i.e.
\begin{equation}
\label{eq:inv_mc}\rho^{\dagger}_{\mathbf{3}}(g_l)m^{\dagger}_{l}m_{l}\rho_{\mathbf{3}}(g_l)=m^{\dagger}_{l}m_{l},\qquad g_{l}\in G_{l}\,,
\end{equation}
and
\begin{eqnarray}
\label{eq:6.2}
\nonumber&&\rho^{\dagger}_{\mathbf{3}}(g_{\nu})m^{\dagger}_{\nu}m_{\nu}\rho_{\mathbf{3}}(g_{\nu})=m^{\dagger}_{\nu}m_{\nu}, \quad g_{\nu}\in G_{\nu},\quad \text{for Dirac neutrinos}\,,\\
\label{eq:inv_mnu}&&\rho^{T}_{\mathbf{3}}(g_{\nu})m_{\nu}\rho_{\mathbf{3}}(g_{\nu})=m_{\nu}, \quad g_{\nu}\in G_{\nu},\quad ~~~~~~~~\text{for Majorana neutrinos}\,.
\end{eqnarray}
Notice that it is sufficient to impose the conditions in Eqs.~(\ref{eq:inv_mc}) and (\ref{eq:inv_mnu}) on the generators of $G_l$ and $G_{\nu}$. Moreover, we see that these conditions imply
\begin{equation}
\left[\rho_{\mathbf{3}}(g_l), m^{\dagger}_{l}m_{l}\right]=0,\qquad \left[\rho_{\mathbf{3}}(g_{\nu}), m^{\dagger}_{\nu}m_{\nu}\right]=0\,.
\end{equation}
As $\rho_{\mathbf{3}}(g_l)$ and $m^{\dagger}_{l}m_{l}$ commute with each other, the unitary transformation $U_{l}$ that diagonalizes $m^{\dagger}_{l}m_{l}$ also diagonalizes
$\rho_{\mathbf{3}}(g_l)$ up to permutations and phases of columns
\begin{equation}
\label{eq:Ul_diag}U^{\dagger}_{l}\rho_{\mathbf{3}}(g_l)U_{l}=\text{diag}(e^{i\alpha_e}, e^{i\alpha_\mu}, e^{i\alpha_\tau})\,.
\end{equation}
Since $g_{l}$ is an element of the discrete flavour symmetry group $G_{f}$, the $e^{i\alpha_{e,\mu,\tau}}$ are all roots of unity. Analogously $\rho_{\mathbf{3}}(g_{\nu})$ and $m^{\dagger}_{\nu}m_{\nu}$ (or $m_{\nu}$) are diagonalized by the same matrix $U_{\nu}$,
\begin{equation}
U^{\dagger}_{\nu}\rho_{\mathbf{3}}(g_{\nu})U_{\nu}=\left\{\begin{array}{cc}
\text{diag}(e^{i\beta_e}, e^{i\beta_\mu}, e^{i\beta_\tau}),& \text{for Dirac neutrinos} ,\\[0.1in]
\text{diag}(\pm 1,\pm 1,\pm 1),& ~~~\text{for Majorana neutrinos} \,,
\end{array}
\right.
\end{equation}
where $\beta_{e, \mu, \tau}$ are rational multiples of $\pi$. Then the lepton mixing matrix $U_{}$ is determined as
\begin{equation}
U_{}=U^{\dagger}_{l}U_{\nu}
\end{equation}
up to independent row and column permutations. In short, given a family symmetry group $G_{f}$ and the residual subgroups $G_{l}$ and $G_{\nu}$, the unitary transformations $U_{l}$ and $U_{\nu}$ as well as the lepton matrix $U_{\text{}}$ can be obtained by diagonalizing the representation matrices of the generators of $G_{l}$ and $G_{\nu}$, as illustrated in the figure~\ref{fig:Gl-Gnu-flavor}. In practice, we only need to find the eigenvectors of $\rho_{\mathbf{3}}(g_l)$ and $\rho_{\mathbf{3}}(g_{\nu})$ that form the column vectors of $U_{l}$ and $U_{\nu}$. In this approach, it is not necessary to construct the explicit form of the mass matrices $m^{\dagger}_{l}m_{l}$ and $m^{\dagger}_{\nu}m_{\nu}$ (or $m_{\nu}$)
although this can be accomplished in a straightforward manner. Since the lepton masses cannot be predicted in this framework, in particular the neutrino mass spectrum can have either normal ordering or inverted ordering, the unitary matrices $U_{l}$ and $U_{\nu}$ are uniquely fixed up to permutations and phases of their column vectors. As a consequence, the lepton mixing matrix $U$ is determined up to independent row and column permutations
and arbitrary phase matrices multiplied from the left and right sides. Therefore the Majorana CP phases are not constrained by the residual flavour symmetry.

Notice that if we switch the roles of the subgroups $G_{\nu}$ and $G_{l}$, the lepton matrix $U_{}$ would become its Hermitian conjugate. If two pairs of subgroups $\{G_{l}, G_{\nu}\}$ and $\{G'_{l}, G'_{\nu}\}$ are conjugate~\footnote{The two pairs of groups are conjugate if their generators $g_e$, $g_{\nu}$ and $g'_e$, $g'_{\nu}$ are related by group conjugacy, i.e., $g'_l=hg_l h^{-1}$, $g'_{\nu}=hg_{\nu} h^{-1}$, where $h$ is an element of $G_f$.}, both residual symmetries would lead to the same result for the lepton mixing matrix. The reason is that if the generators of $G_{l}$ and $G_{\nu}$ are diagonalized by $U_{l}$ and $U_{\nu}$ respectively, $\rho_{\mathbf{3}}(h)U_{l}$ and $\rho_{\mathbf{3}}(h)U_{\nu}$ would diagonalize those of $G'_{l}$ and $G'_{\nu}$. On the other hand, if the left-handed lepton fields are assigned to the complex-conjugate of the triplet representation, the lepton mixing matrix would be complex conjugated, so that the predictions for the lepton mixing angles are unchanged and the signs of the CP violation phases are inverted.

\subsubsection{Fully preserved residual symmetry $G_{\nu}=K_4$ }
\label{sec:lepton-mixing-from}

As already mentioned, if the flavour group $G_{f}$ is of finite order and the residual symmetries $G_{l}$ and $G_{\nu}$ distinguish the three families of charged leptons and neutrinos then the lepton mixing matrix would be completely determined by the residual symmetries, regardless of the lepton masses or any other parameter of the underlying theory~\cite{deAdelhartToorop:2011re,Holthausen:2012wt,Fonseca:2014koa,Talbert:2014bda,Yao:2015dwa}. This is the so-called direct model building approach~\cite{King:2009ap,King:2013eh}. This happens, for instance, if $G_{\nu}$ is the Klein group $K_4$ and $G_{l}$ is a cyclic group $Z_n$ with $n\geq3$ or the product of several cyclic groups. A complete classification of all resulting mixing matrices has been performed by using theorems on roots of unity~\cite{Fonseca:2014koa}. One finds that the lepton mixing matrix can take 17 sporadic patterns plus one infinite series, and only the latter could be compatible with the experimental data, given as~\cite{Fonseca:2014koa,Yao:2015dwa}
\begin{equation}
\label{eq:trimaximal_flavor}U=\frac{1}{\sqrt{3}}\begin{pmatrix}
\sqrt{2}\cos\theta ~&~ 1 ~&~ -\sqrt{2}\sin\theta \\
-\sqrt{2}\cos\left(\theta-\frac{\pi}{3}\right) ~&~ 1 ~&~ \sqrt{2}\sin\left(\theta-\frac{\pi}{3}\right) \\
-\sqrt{2}\cos\left(\theta+\frac{\pi}{3}\right) ~&~ 1 ~&~ \sqrt{2}\sin\left(\theta+\frac{\pi}{3}\right)
\end{pmatrix}\,,
\end{equation}
where $\theta$ is a rational multiple of $\pi$, with its exact value completely determined by group-theoretical considerations. This result is consistent with comprehensive scanning over the finite groups and possible symmetry breaking patterns~\cite{King:2013vna,Talbert:2014bda,Yao:2015dwa}. We obtain the following expressions for the lepton mixing angles
\begin{equation}
\sin^2\theta_{13}=\frac{2}{3}\sin^2\theta,~~~\sin^2\theta_{12}=\frac{1}{1+2\cos^2\theta},~~~\sin^2\theta_{23}=\frac{2\sin^2(\theta-\frac{\pi}{3})}{1+2\cos^2\theta}\,.
\end{equation}
Combining the above equations one sees that the three mixing angles depend on a single parameter $\theta$, consequently the following relations must be satisfied,
\begin{equation}
\label{eq:correlation_trimaximal}3\sin^2\theta_{12}\cos^2\theta_{13}=1,\qquad \sin^2\theta_{23}=\frac{1}{2}\pm\frac{1}{2}\tan\theta_{13}\sqrt{2-\tan^2\theta_{13}}\,.
\end{equation}
Using the best fit value of the reactor mixing angle $\sin^2\theta_{13}\simeq 0.022$ for NO~\cite{deSalas:2020pgw,10.5281/zenodo.4726908} we get
\begin{equation}
\sin^2\theta_{12}\simeq0.341,\quad \sin^2\theta_{23}\simeq 0.605 ~\mathrm{or}~0.395\,.
\end{equation}
The value of solar mixing angle is within the experimentally preferred $2\sigma$ region, the atmospheric angle $\sin^2\theta_{23}\simeq 0.605$ lies in the $3\sigma$ region,
while another possible value $\sin^2\theta_{23}\simeq 0.395$ is disfavored by the present data~\cite{deSalas:2020pgw,10.5281/zenodo.4726908}. As regards the CP violation phases, the Majorana phases are unconstrained~\footnote{One can predict the values of Majorana phases by including the generalized CP symmetry. For instance, for the flavor symmetry $\Delta(6n^2)=\left(Z_n\times Z_n\right)\rtimes S_3$ in combination with CP, the Majorana phase $\phi_{12}$ can take several discrete values for each $n$, and $\phi_{13}$ is a multiple of $\pi/2$~\cite{King:2014rwa} if residual $K_4$ flavor symmetry and CP symmetry are preserved by neutrino mass matrix.} and the Dirac CP phase vanishes
\begin{equation}
\sin\delta_{CP}=0\,.
\end{equation}
Notice that current measurements still do not provide a fully robust global CP determination~\cite{deSalas:2020pgw,10.5281/zenodo.4726908}.

In order to reproduce the experimentally favored mixing angles the minimal flavour symmetry group is $G_f=(Z_{18}\times Z_6)\rtimes S_3$~\cite{Holthausen:2012wt,Yao:2015dwa}
for Majorana neutrinos, while for Dirac neutrinos the minimal flavour group is $G_f=(Z_{9}\times Z_3)\rtimes S_3$ ~\cite{Yao:2015dwa}. Hence the order of the flavour symmetry group should be at least 648 and 162 for Majorana and Dirac neutrinos, respectively, leading to a realistic value of $\theta=\pi/18$. In concrete models, the residual symmetries $G_{l}$ and $G_{\nu}$ are spoiled by higher order terms involving flavon fields. These induce corrections suppressed by the flavon VEVs with respect to the flavor scale.

\subsubsection{Partially preserved residual symmetry $G_{\nu}=Z_2$}
\label{sec:lepton-mixing-from-1}

The idea of partially preserved residual symmetry was proposed~\cite{Ge:2011ih,Ge:2011qn,Hernandez:2012ra,Hernandez:2012sk} in order to accommodate a non-zero Dirac CP violation phase $\delta^{\ell}$  and degrade the order of the flavour symmetry. In this approach, part of the residual symmetry of the neutrino mass matrix arises from the assumed flavour symmetry at high the energy scale. A widely studied scenario is that a $Z_2$ (instead of $K_4$) subgroup is preserved in the neutrino sector, i.e. the residual group is
$G_{\nu}=Z^{g_{\nu}}_2$, where $g_{\nu}$ refers to the generator of $G_{\nu}$. Note that the presentation rule of the cyclic group $Z^{g}_n$ is $Z^{g}_n\equiv\left\{1, g, g^2,\ldots, g^{n-1}\right\}$.

\begin{figure}[t!]
\begin{center}
\begin{tabular}{cc}
\includegraphics[width=0.49\linewidth]{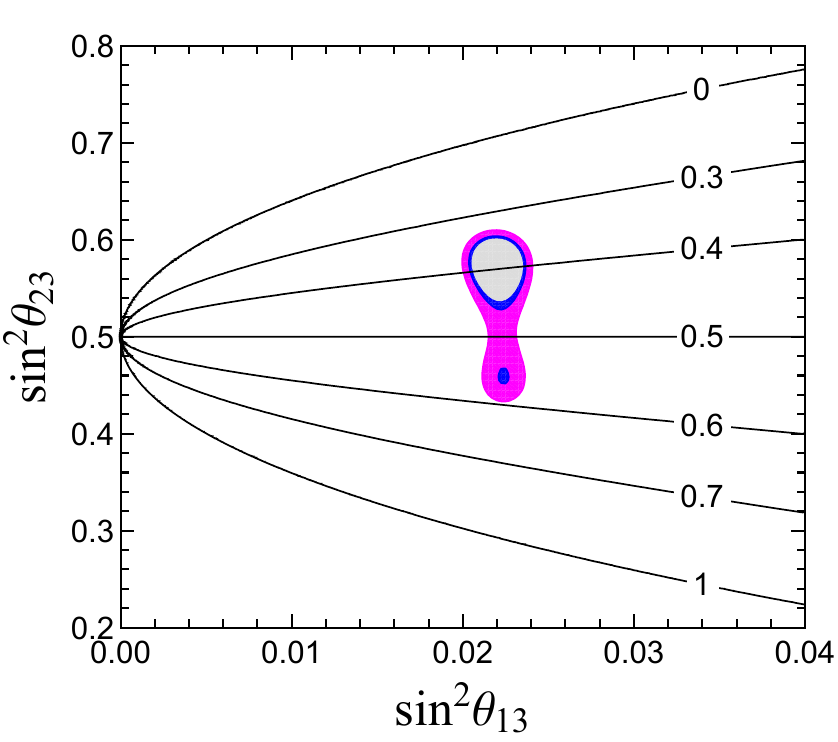}  &
\includegraphics[width=0.49\linewidth]{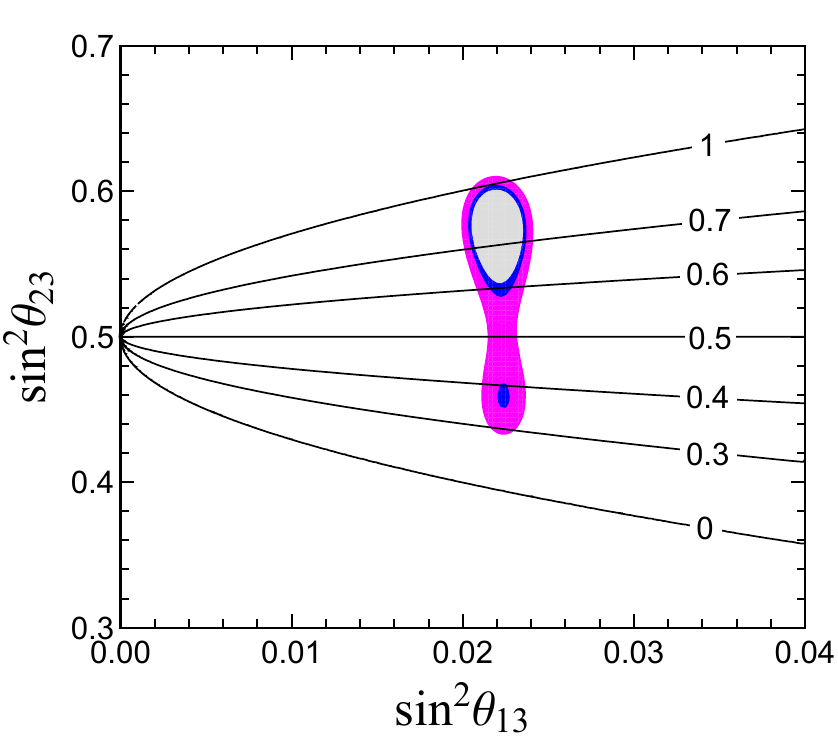}
\end{tabular}
\caption{\label{fig:deltaCP_contour_Z2} Contours of $|\delta_{CP}/\pi|$ in the $\sin^2\theta_{23}-\sin^2\theta_{13}$ plane. The left panel is for the case that the fixed column is $(2/\sqrt{6}, -1/\sqrt{6}, -1/\sqrt{6})^{T}$ (TM1), and the right panel is for $(1/\sqrt{3}, 1/\sqrt{3}, 1/\sqrt{3})^{T}$ (TM2). The global-fit allowed oscillation regions corresponding to $90\%$, $95\%$ and $99\%$ confidence level are shown~\cite{deSalas:2020pgw,10.5281/zenodo.4726908}}
\end{center}
\end{figure}

The invariance of the neutrino mass matrix under $Z^{g_{\nu}}_2$ requires that Eq.~\eqref{eq:6.2} holds. Consequently the residual flavour symmetry imposes the following restriction on the unitary transformation
\begin{equation}
U^{\dagger}_{\nu}\rho_{\mathbf{3}}(g_{\nu})U_{\nu}=\pm P^{T}_{\nu}\text{diag}(1, -1,-1)P_{\nu}\,,
\end{equation}
where $P_{\nu}$ is a permutation matrix, and we have taken into account that the eigenvalues of $\rho_{\mathbf{3}}(g_{\nu})$ is $+1$ or $-1$, since $g_{\nu}$ is of order two.
Let us denote $U_{0}$ as a diagonalization matrix of $\rho_{\mathbf{3}}(g_{\nu})$ with
\begin{equation}
U^{\dagger}_{0}\rho_{\mathbf{3}}(g_{\nu})U_{0}=\pm \text{diag}(1, -1,-1)\,,
\end{equation}
Then the unitary transformation $U_{\nu}$ would be of the form
\begin{equation}
U_{\nu}=U_{0}U_{23}(\theta, \delta)P_{\nu}\,,
\end{equation}
where $U_{23}(\theta, \delta)$ is a block diagonal complex unitary rotation,
\begin{equation}
U_{23}(\theta, \delta)=\begin{pmatrix}
1 &  0  &  0  \\
0  &  \cos\theta  &  \sin\theta e^{-i\delta}  \\
0  &  -\sin\theta e^{i\delta}   &  \cos\theta
\end{pmatrix}\,.
\end{equation}
Notice that, since the residual $Z_2$ flavour symmetry can not fully distinguish the three neutrino families in this case, only one column of $U_{\nu}$ is numerically fixed.
In the charged lepton sector, a residual subgroup $G_{l}$ is preserved, so that the unitary transformation $U_{l}$ obeys the condition $U^{\dagger}_{l}\rho_{\mathbf{3}}(g_l)U_{l}=\text{diag}(e^{i\alpha_e}, e^{i\alpha_\mu}, e^{i\alpha_\tau})$ as shown in Eq.~\eqref{eq:Ul_diag}, where $\alpha_{e, \mu, \tau}$ are rational multiples of $\pi$. Thus the assumed residual symmetry allows us to determine the lepton mixing matrix in terms of two free parameters as
\begin{equation}
U_{}=U^{\dagger}_{l}U_{0}U_{23}(\theta, \delta)P_{\nu}\,.
\end{equation}
Hence only one column is fixed by residual symmetry in this case, and this is dubbed semi-direct approach in~\cite{King:2013eh}. For example, if the flavour group is $S_4$ and the residual symmetries are chosen as $G_{\nu}=Z^{SU}_2$ and $G_{l}=Z^{T}_3$, then $U_0=U_{TBM}$ and the lepton mixing is
\begin{equation}
U=\frac{1}{\sqrt{6}}\begin{pmatrix}
2   ~&  \sqrt{2}\,\cos\theta    ~&  \sqrt{2}\,e^{-i\delta}\sin\theta \\
-1  ~&  \sqrt{2}\,\cos\theta-\sqrt{3}\,e^{i\delta}\sin\theta  ~&  \sqrt{3}\,\cos\theta+\sqrt{2}\,e^{-i\delta}\sin\theta  \\
-1  ~&  \sqrt{2}\,\cos\theta+\sqrt{3}\,e^{i\delta}\sin\theta   ~&  -\sqrt{3}\,\cos\theta+\sqrt{2}\,e^{-i\delta}\sin\theta
\end{pmatrix}\,.
\end{equation}
We see that the first column is $(2/\sqrt{6}, -1/\sqrt{6}, -1/\sqrt{6})^{T}$ which is in common with the TBM mixing pattern~\cite{Harrison:2002er,Xing:2002sw,He:2003rm}. In fact, this is exactly the TM1 lepton mixing matrix in Eq.~\eqref{eq:TM1}~\cite{Albright:2008rp,Albright:2010ap,He:2011gb}. In figure~\ref{fig:deltaCP_contour_Z2} we plot the contour of $|\delta_{CP}|$ in the plane $\sin^2\theta_{23}$ versus $\sin\theta_{13}$.
The value of the Dirac CP phase $\delta_{CP}$ is not too constrained, as long as the atmospheric mixing angle $\theta_{23}$ remains poorly measured.

On the other hand, if the residual groups are $G_{\nu}=Z^{S}_2$ and $G_{l}=Z^{T}_3$ for either $A_4$ or $S_4$ flavour symmetry, one column of the lepton mixing matrix is enforced to be $(1/\sqrt{3}, 1/\sqrt{3}, 1/\sqrt{3})^{T}$ with,
\begin{equation}
U=\frac{1}{\sqrt{6}}\begin{pmatrix}
2\cos\theta  ~&~  \sqrt{2}   ~&~  2e^{-i\delta}\sin\theta \\
-\cos\theta-\sqrt{3}\,e^{i\delta}\sin\theta  ~&~  \sqrt{2}   ~&~  \sqrt{3}\,\cos\theta-e^{-i\delta}\sin\theta  \\
-\cos\theta+\sqrt{3}\,e^{i\delta}\sin\theta  ~&~ \sqrt{2}   ~&~  -\sqrt{3}\,\cos\theta-e^{-i\delta}\sin\theta
\end{pmatrix}\,,
\end{equation}
which is exactly the tri-maximal mixing pattern TM2 in Eq.~\eqref{eq:TM2}~\cite{Harrison:2002kp,Bjorken:2005rm,He:2006qd,Grimus:2008tt}. The predictions for the Dirac phase $\delta_{CP}$ are shown in figure~\ref{fig:deltaCP_contour_Z2}. The figure refers to the case of normal mass ordering, with very similar results for the case of inverse-ordered masses.

\subsection{Combining flavour and CP symmetry}
\label{sec:combining-flavour-cp}

\begin{figure}[t!]\centering
\begin{tabular}{cc}
\includegraphics[width=0.6\linewidth]{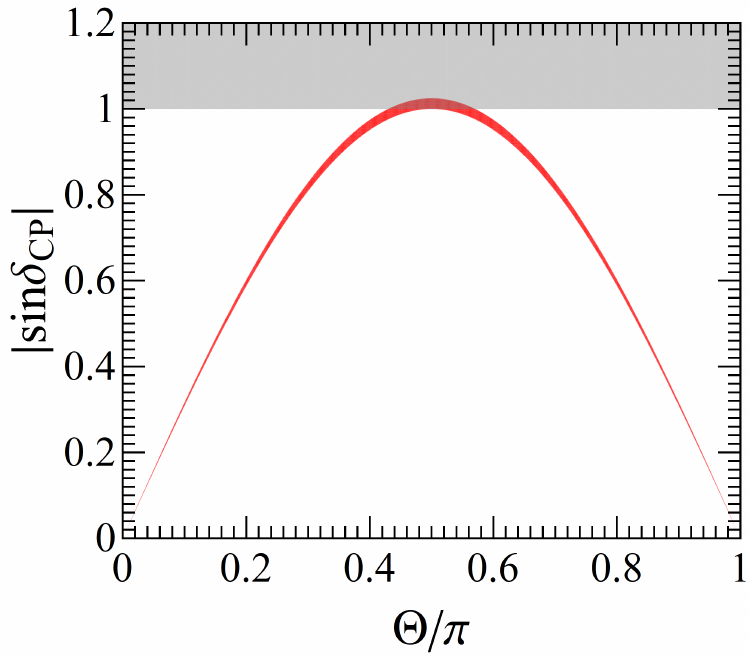}
\end{tabular}
\caption{\label{fig:sinDeltaCP_Theta} $|\sin\delta_{CP}|$ regions versus $\Theta$ characterizing the generalized $\mu-\tau$ reflection, where $\theta_{23}$ is required to lie in its $3\sigma$ allowed range~\cite{deSalas:2020pgw,10.5281/zenodo.4726908}.}
\end{figure}

We saw how a non-vanishing Dirac CP phase can be obtained if a $Z_2$ rather than the Klein group $K_4$ is the residual flavour symmetry preserved by the neutrino mass matrix. However, $\delta_{CP}$ can vary within a wide region. Moreover, the Majorana phases can not be predicted just from the flavour symmetry. In order to understand the CP violating phases, one can impose a generalized CP (gCP) symmetry~\cite{Bernabeu:1986fc,Ecker:1987qp,Neufeld:1987wa}, see~\cite{Feruglio:2012cw,Grimus:1995zi,Holthausen:2012dk} for discussions of gCP to the lepton flavor mixing problem. A simple example is the $\mu-\tau$ reflection symmetry~\cite{Harrison:2002kp,Harrison:2002et,Babu:2002dz,Grimus:2003yn,Harrison:2004he,Farzan:2006vj,King:2018kka,King:2019tbt}, see Ref.~\cite{Xing:2015fdg} for review. The $\mu-\tau$ reflection symmetry is a combination of the canonical CP transformation with the $\mu-\tau$ exchange symmetry, which exchanges a muon (tau) neutrino with a tau (muon) antineutrino,
\begin{equation}
\begin{pmatrix}
\nu_e \\
\nu_{\mu} \\
\nu_{\tau}
\end{pmatrix}\mapsto \begin{pmatrix}
\nu^{c}_e \\
\nu^{c}_{\tau} \\
\nu^{c}_{\mu}
\end{pmatrix}=\begin{pmatrix}
1  ~&  0   ~&  0  \\
0 ~&  0   ~&  1  \\
0  ~&  1   ~&  0
\end{pmatrix}\begin{pmatrix}
\nu^{c}_e \\
\nu^{c}_{\mu} \\
\nu^{c}_{\tau}
\end{pmatrix}\,.
\end{equation}
Notice that the generalized CP transformation matrix is not diagonal in family space.
In the charged lepton mass basis, the neutrino mass matrix invariant under $\mu-\tau$ reflection has the following form
\begin{equation}
m_{\nu}=\begin{pmatrix}
a   ~&   b  ~&   b^{*}  \\
b   ~&   c  ~&  d  \\
b^{*}  ~&   d  ~&  c^{*}
\end{pmatrix}\,,
\end{equation}
where $a$ and $d$ are real, $b$ and $c$ are complex numbers. As a result, both atmospheric mixing angle $\theta_{23}$ and Dirac CP phase would be maximal, while Majorana phases take CP conserving values. These should be contrasted with present neutrino oscillation data~\cite{deSalas:2020pgw,10.5281/zenodo.4726908}. Deviations of $\theta_{23}$ and $\delta_{CP}$ from maximal values can easily arise from a more general $\mu-\tau$ reflection symmetry acting on the neutrino sector. Such CP transformation is of the following form~\cite{Chen:2015siy},
\begin{equation}
\label{eq:X_mutau-gener}
\begin{pmatrix}
\nu_e \\
\nu_{\mu} \\
\nu_{\tau}
\end{pmatrix}\mapsto \begin{pmatrix}
\nu^{c}_e \\
\cos\Theta\,\nu^c_{\mu}+i\sin\Theta\,\nu^{c}_{\tau} \\
\cos\Theta\,\nu^c_{\tau}+i\sin\Theta\,\nu^{c}_{\mu}
\end{pmatrix}=\begin{pmatrix}
1  ~&  0   ~&  0  \\
0 ~&  \cos \Theta  ~&  i\sin\Theta  \\
0  ~&   i\sin\Theta   ~&  \cos \Theta
\end{pmatrix}\begin{pmatrix}
\nu^{c}_e \\
\nu^{c}_{\mu} \\
\nu^{c}_{\tau}
\end{pmatrix}\,,
\end{equation}
where the angle $\Theta$ characterizes the CP transformation. It reduces to the $\mu-\tau$ reflection symmetry in the limit of $\Theta=\pi/2$. The atmospheric mixing angle $\theta_{23}$ and Dirac phase $\delta_{CP}$ are predicted to be strongly correlated as follows~\cite{Chen:2015siy},
\begin{equation}
\label{eq:corre_theta23_deltaCP}\sin^{2} \delta_{\rm CP} \sin^22\theta_{23}= \sin^{2} \Theta \,,
\end{equation}
while the Majorana phases take on trivial CP conserving values.  The correlation in Eq.~\eqref{eq:corre_theta23_deltaCP} allows us to predict the range of the Dirac CP violating phase $|\sin\delta_{CP}|$ as a function of the parameter $\Theta$ as shown in figure~\ref{fig:sinDeltaCP_Theta}. Note that the sign of $\delta_{CP}$ can not be fixed.

\subsubsection{Mathematical consistency}
\label{sec:math-prel}

Family symmetries can be generated if one successively performs two generalized CP transformations. Hence generalized CP symmetries can be thought of as associated to some underlying flavour symmetry. A convenient strategy for defining such CP transformations is to start from a flavour symmetry group $G_f$ and find all possible CP transformations $H_{CP}$ which can generate the given flavour group transformations. The purpose of this section is to determine the restricted lepton mixing matrices that can be obtained from discrete flavour and CP symmetries.

It is highly non-trivial to define a CP transformation consistently in the presence of a family symmetry $G_f$~\cite{Feruglio:2012cw,Grimus:1995zi,Holthausen:2012dk,Chen:2014tpa}.
Let us consider a set of fields $\varphi$ in a generic irreducible representation $\mathbf{r}$ of $G_f$, transforming under the action of $G_f$ as
\begin{equation}
\varphi(x)\stackrel{G_f}{\longrightarrow} \rho_{\mathbf{r}}(g) \varphi(x),
\qquad g \in G_f\ ,
\end{equation}
where $\rho_{\mathbf{r}}(g)$ denotes the representation matrix for any element $g$ in the irreducible representation $\mathbf{r}$. The generalized CP acts on $\varphi$ as
\begin{equation}
\varphi(x)\stackrel{\mathcal{CP}}{\longrightarrow}X_{\mathbf{r}}\,\varphi^{*}(x_{\mathcal{P}})\,,
\end{equation}
where $x_{\mathcal{P}}=(t,-\vec{x})$, and $X_{\mathbf{r}}$ is the CP transformation matrix in flavor space, assumed to be a unitary matrix, so as to leave the kinetic term invariant~\footnote{When $\varphi$ denotes a spinor, the obvious action of CP on the spinor indices is understood.}. A physical CP transformation should map each field $\varphi(x)$ in any irreducible representation $\mathbf{r}$ of $G_f$ into its complex conjugate
$\varphi^{*}(x_{\mathcal{P}})$ in the complex representation $\mathbf{r}^{*}$~\cite{Chen:2014tpa}. If we first perform a CP transformation, then apply a flavour symmetry transformation, and subsequently an inverse CP transformation we obtain
\begin{equation}
\varphi(x)\stackrel{\mathcal{CP}}{\longrightarrow}X_{\mathbf{r}}\,\varphi^{*}(x_{\mathcal{P}})\stackrel{G_f}{\longrightarrow}X_{\mathbf{r}}\rho^{*}_{\mathbf{r}}(g)\varphi^{*}(x_{\mathcal{P}})
\stackrel{\mathcal{CP}^{-1}}{\longrightarrow}X_{\mathbf{r}}\rho^{*}_{\mathbf{r}}(g)X^{-1}_{\mathbf{r}}\varphi(x)\,.
\end{equation}

\begin{figure}[t!]
\centering
\includegraphics[width=0.68\textwidth]{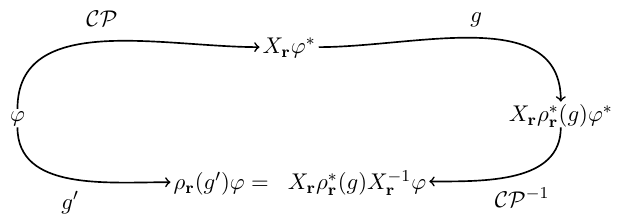}
\caption{\label{fig:consistency-flavor-CP} The consistency condition for flavour and generalized CP symmetries.  }
\end{figure}

As shown in figure~\ref{fig:consistency-flavor-CP}, the theory should still be invariant since it is invariant under each transformation individually. To make the theory consistent the resulting net transformation should be equivalent to a flavour symmetry transformation $\rho_{\mathbf{r}}(g')$ of some flavour group element $g'$, i.e.
\begin{equation}
\label{eq:Cons-Cond}X_{\mathbf{r}}\rho^{*}_{\mathbf{r}}(g)X^{-1}_{\mathbf{r}}=\rho_{\mathbf{r}}(g'),
~~~~~ g, g' \in G_f\,
\end{equation}
where the elements $g$ and $g'$ are independent of the representation $\mathbf{r}$. Eq.~\eqref{eq:Cons-Cond} is the important consistency condition which must be fulfilled for all irreducible representations of $G_f$ in order to ensure generalized CP and flavour symmetry invariance simultaneously. If the condition Eq.~\eqref{eq:Cons-Cond} is not fulfilled, the group $G_f$ is not the full symmetry group of the Lagrangian, and one would have to consider a larger group, which closes under CP transformations. The allowed form of the generalized CP transformations is strongly restricted by the consistency condition of Eq.~\eqref{eq:Cons-Cond}.

In practice, it suffices to consider the consistency conditions for the generators of $G_f$.
It is remarkable that both $e^{i\theta}X_{\mathbf{r}}$ and $\rho_{\mathbf{r}}(h)X_{\mathbf{r}}$ satisfy the consistency condition Eq.~\eqref{eq:Cons-Cond}
for a generalized CP transformation $X_{\mathbf{r}}$, where $\theta$ is real and $h$ is any element of $G_f$. For a well-defined CP transformation $X_{\mathbf{r}}$, $\rho_{\mathbf{r}}(h)X_{\mathbf{r}}$ is also a viable CP transformation for any $h\in G_f$. The two CP transformations $X_{\mathbf{r}}$ and $\rho_{\mathbf{r}}(h)X_{\mathbf{r}}$ differ by a flavour symmetry transformation $\rho_{\mathbf{r}}(h)$. Since the latter is certainly a symmetry of the Lagrangian, these two CP transformations are indistinguishable. Moreover, Eq.~\eqref{eq:Cons-Cond} implies that the generalized CP transformation $X_{\mathbf{r}}$ maps the group element $g$ into $g'$ and the flavour group multiplication is preserved
under this mapping, i.e. $X_{\mathbf{r}}\rho^{*}_{\mathbf{r}}(g_1g_2)X^{-1}_{\mathbf{r}}=X_{\mathbf{r}}\rho^{*}_{\mathbf{r}}(g_1)X^{-1}_{\mathbf{r}}X_{\mathbf{r}}\rho^{*}_{\mathbf{r}}(g_2)X^{-1}_{\mathbf{r}}$. Therefore the CP transformation $X_{\mathbf{r}}$ defines a homomorphism $\mathfrak{u}: g\to g',\; g, g'\in G_f$ of the family symmetry group $G_f$. Note that the homomorphism $\mathfrak{u}'$ associated with the CP transformation $\rho_{\mathbf{r}}(h)X_{\mathbf{r}}$ for any $h\in G_f$ is related to $\mathfrak{u}$ by conjugation:
\begin{equation}
\mathfrak{u}'(g)=h\mathfrak{u}(g)h^{-1},~~~~\forall~h,g\in G_f\,.
\end{equation}
Notice that when $\rho_{\mathbf{r}}$ is a faithful representation, the elements $g$ and $g'$ have the same order, the mapping defined in Eq.~\eqref{eq:Cons-Cond} is bijective,
and thus the associated CP transformation becomes an automorphism, see Ref.~\cite{Holthausen:2012dk} for a more formal treatment. Furthermore, taking trace on both sides of the consistency condition in Eq.~\eqref{eq:Cons-Cond}, we find that the group characters $\chi_{\mathbf{r}}$ fulfill
\begin{equation}
\chi_{\mathbf{r}}(g')=\text{tr}[\rho_{\mathbf{r}}(g')]=\text{tr}[X_{\mathbf{r}}\rho^{*}_{\mathbf{r}}(g)X^{-1}_{\mathbf{r}}]=\text{tr}[\rho^{*}_{\mathbf{r}}(g)]=\text{tr}[\rho^{\dagger}_{\mathbf{r}}(g)]=\chi_{\mathbf{r}}(g^{-1})\,.
\end{equation}
Hence $g'$ and $g^{-1}$ should be in the same conjugacy class, that is to say the CP transformation corresponds to a class-inverting automorphism of $G_f$. As a consequence, when determining the generalized CP transformation compatible with a flavour symmetry group $G_f$ it is sufficient to focus on the class-inverting automorphisms. Because $g$ and $g'$ in Eq.~\eqref{eq:Cons-Cond} are generally different group elements, flavour transformations and CP transformations in general do not commute. Hence the mathematical structure of the group comprising $G_f$ and generalized CP is in general a semi-direct product, and the
full symmetry is $G_f\rtimes H_{CP}$~\cite{Feruglio:2012cw}, where $H_{CP}$ is the group generated by generalized CP transformations. The semi-direct product would reduce to a direct product in the case of $g=g'$.

\subsubsection{Implications of residual flavour and CP symmetries}
\label{sec:general-paradigm-}

The presence of generalized CP allows for more symmetry breaking patterns than flavor symmetry alone. In this approach, the parent flavour and CP symmetries are broken down to different residual subgroups $G_l\rtimes H^l_{CP}$ and $G_{\nu}\rtimes H^{\nu}_{CP}$ in the charged lepton and neutrino sectors respectively. The mismatch between the residual symmetries $G_l\rtimes H^l_{CP}$ and $G_{\nu}\rtimes H^{\nu}_{CP}$ gives rise to a certain lepton mixing pattern. It is remarkable that the lepton flavour mixing is fixed by the group structure of $G_{f}\rtimes H_{CP}$ and the residual symmetries~\cite{Chen:2014wxa,Chen:2015nha}. The details of the breaking mechanisms realizing the assumed residual symmetries are irrelevant. In the following, we present the formalism to determine the lepton mixing matrix from the residual symmetries of the charged lepton and neutrino mass matrices. We shall assume neutrinos to be Majorana particles, thus the residual flavour symmetry $G_{\nu}$ is a $Z_2$ or Klein subgroup $K_4$ (Dirac neutrinos can be discussed in a very similar way). As usual the three generations of left-handed leptons are assumed to transform as a faithful irreducible triplet $\mathbf{3}$ of the $G_f$ symmetry.

For the residual symmetry $G_{l}\rtimes H^{l}_{CP}$ to hold, the Hermitian combination $m^{\dagger}_{l}m_{l}$ of the charged lepton mass matrix should be invariant under the action of $G_{l}\rtimes H^{l}_{CP}$, i.e.,
\begin{align}
\label{eq:flavor_cons_Charg}  &\rho^{\dagger}_{\mathbf{3}}(g_{l})m^{\dagger}_{l}m_{l}\rho_{\mathbf{3}}(g_{l})=m^{\dagger}_{l}m_{l},  \qquad g_{l}\in G_{l}\,, \\
\label{eq:CP_cons_Charg}  &X^{\dagger}_{l\mathbf{3}}m^{\dagger}_{l}m_{l}X_{l\mathbf{3}}=(m^{\dagger}_{l}m_{l})^{*},  \qquad X_{l\mathbf{3}} \in H^{l}_{CP}\,,
\end{align}
where the charged lepton mass matrix $m_l$ is given in the convention in which the left-handed (right-handed) fields are on the right-hand (left-hand) side of $m_l$. We denote the unitary diagonalization matrix of $m^{\dagger}_{l}m_{l}$ as $U_{l}$ which satisfies $U^{\dagger}_{l}m^{\dagger}_{l}m_{l}U_{l}=\text{diag}(m^2_e, m^2_{\mu}, m^2_{\tau})$.
Once the residual symmetry $G_{l}\rtimes H^{l}_{CP}$ and the triplet representation $\mathbf{3}$ are specified, the explicit form of $m^{\dagger}_{l}m_{l}$ can be constructed
from Eqs.~(\ref{eq:flavor_cons_Charg},~\ref{eq:CP_cons_Charg}) in a straightforward way, and thus $U_{l}$ can be determined. In fact, one can also directly extract the restrictions on $U_{l}$ from Eqs.~(\ref{eq:flavor_cons_Charg},\ref{eq:CP_cons_Charg})
without working out the explicit form of $m^{\dagger}_{l}m_{l}$ as follows
\begin{equation}
\label{eq:Gl-HlCP-constrain}U_l^{\dagger}\rho_{\mathbf{3}}(g_l)U_l=\rho^{diag}_{\mathbf{3}}(g_l)\,,~~~~~~ U_l^{\dagger}X_{l\mathbf{3}}U_l^*=X^{diag}_{l\mathbf{3}}\,,
\end{equation}
where both $\rho^{diag}_{\mathbf{3}}(g_l)$ and $X^{diag}_{l\mathbf{3}}$ are diagonal phase matrices.

The first identity in Eq.~\eqref{eq:Gl-HlCP-constrain} comes from imposing the residual flavour symmetry $G_l$, and the latter arises from the residual CP symmetry $H^{l}_{CP}$. We see that both $\rho_{\mathbf{3}}(g_l)$ and $m^{\dagger}_{l}m_{l}$ can be diagonalized by the same unitary matrix $U_{l}$, and the residual CP transformation $X_{l\mathbf{3}}=U_{l}X^{diag}_{l\mathbf{3}}U^{T}_{l}$ should be a symmetric unitary matrix. Moreover, Eq.~\eqref{eq:Gl-HlCP-constrain} implies that the residual flavour and CP symmetries should satisfy the following restricted consistency conditions:
\begin{equation}
\label{eq:Gch-Xch-conSconD}X_{l\mathbf{r}}\rho^{*}_{\mathbf{r}}(g_{l})X^{-1}_{l\mathbf{r}}=\rho_{\mathbf{r}}(g^{-1}_{l}),\qquad
g_{l}\in G_{l}\,,
\end{equation}
If the residual flavour symmetry $G_{l}$ distinguishes the three families, the inclusion of generalized CP $H^l_{CP}$ would not add any information as far as lepton mixing is concerned~\cite{Ding:2013bpa,Li:2014eia}. Likewise, the requirement that $G_{\nu}\rtimes H^{\nu}_{CP}$ is preserved in the neutrino sector implies that the neutrino mass matrix $m_{\nu}$
must be invariant under the action of $G_{\nu}\rtimes H^{\nu}_{CP}$, i.e.,
\begin{equation}
\label{eq:mnu-Gnu}
\begin{aligned}
&\rho^T_{\mathbf{3}}(g_\nu)m_\nu\rho_{\mathbf{3}}(g_\nu)=m_\nu,~~~~ g_{\nu}\in G_{\nu}\,, \\
&X_{\nu\mathbf{3}}^Tm_\nu X_{\nu\mathbf{3}}=m_\nu^*,~~~~ X_{\nu\mathbf{3}} \in H^{\nu}_{CP}\,.
\end{aligned}
\end{equation}
Since $G_{\nu}$ is a $Z_2$ or $K_4$ subgroup for Majorana neutrinos, the eigenvalues of the representation matrix $\rho_{\mathbf{3}}(g_{\nu})$ can only be $+1$ or $-1$. The diagonalization matrix of $m_{\nu}$ is denoted as $U_{\nu}$ and it satisfies $U_\nu^Tm_\nu U_\nu=\text{diag}(m_1, m_2, m_3)$. From Eq.~\eqref{eq:mnu-Gnu}, we find that the residual symmetry $G_{\nu}\rtimes H^{\nu}_{CP}$ leads to the following restrictions on the unitary transformation $U_{\nu}$,
\begin{equation}
\label{eq:cons-nu-Z2CP}U_{\nu}^{\dagger}\rho_{\mathbf{3}}(g_\nu)U_\nu=\text{diag}(\pm 1,\pm 1,\pm 1)\,,~~~~~~U_\nu^{\dagger}X_{\nu\mathbf{3}}U_{\nu}^*=\text{diag}(\pm 1,\pm 1,\pm 1)\,.
\end{equation}
Hence the residual CP transformation $X_{\nu\mathbf{3}}$ should be a symmetric matrix~\cite{Feruglio:2012cw,Chen:2014wxa,Chen:2015nha} and the restricted consistency condition in the neutrino sector is
\begin{equation}
\label{eq:Gnu-Xnu-cond}X_{\nu\mathbf{r}}\rho^*_{\mathbf{r}}(g_\nu)X_{\nu\mathbf{r}}^{-1}=\rho_{\mathbf{r}}(g_\nu),\quad
g_{\nu}\in G_{\nu},~~X_{\nu\mathbf{r}}\in H^{\nu}_{CP}\,,
\end{equation}
which implies that the residual flavour symmetry $G_{\nu}$ and residual CP symmetry $H^{\nu}_{CP}$ in the neutrino sector commute with each other. Consequently, for Majorana neutrinos the mathematical structure of the residual symmetry is a direct product $G_{\nu}\times{H}^{\nu}_{CP}$. Solving the constraints Eqs.~(\ref{eq:Gl-HlCP-constrain}, \ref{eq:cons-nu-Z2CP}) imposed by the residual symmetry, one can fix the unitary transformations $U_{l}$, $U_{\nu}$ and hence the lepton mixing matrix, as
\begin{equation}
U=U^{\dagger}_lU_{\nu}\,.
\end{equation}
In the following, we present several different choices for the residual scheme and the corresponding predictions for the lepton mixing matrix.

\subsection{Predictive scenarios with CP symmetry}

Assuming a generalized CP symmetry, the Majorana CP violation phases can be predicted and we have more choices for the possible residual symmetries. In particular, some scenarios are quite predictive, with the lepton mixing matrix elements depending only on few free parameters, as shown in table~\ref{tab:residual-sym-Nump}. For notational simplicity, we denote $\mathcal{G}_l\equiv G_l\rtimes H^l_{CP}$ and $\mathcal{G}_{\nu}\equiv G_{\nu}\rtimes H^{\nu}_{CP}$ which are the residual symmetries of the charged lepton and neutrino sectors respectively~\footnote{In concrete model building, one could possibly have more branches of residual symmetry~\cite{King:2015dvf,Ding:2018fyz,Ding:2018tuj,Chen:2019oey}.}.
We have assumed that  the residual flavour symmetry $Z_n$ with $n\geq3$ is sufficient to distinguish the three generations of charged leptons, otherwise it can be taken to be a product of cyclic groups. The unitary transformations $\Sigma_{l}$ and $\Sigma_{\nu}$ are the Takagi factorizations of the residual CP transformations $H^l_{CP}$ and $H^{\nu}_{CP}$ respectively, in the presence of residual CP. They should also diagonalize the residual flavour symmetry transformations $G_l$ and $G_{\nu}$ respectively. Moreover, $R_{23}(\theta)$ is a rotation matrix in the (23)-plane, Eq.~\eqref{eq:R23-matrix}, while $O_{3}(\theta_1, \theta_2, \theta_3)$ is a general three-dimensional rotation matrix, Eq.~\eqref{eq:orthogonal-matrix}. Both $P_{l}$ and $P_{\nu}$ are permutation matrices since the neutrino and charged lepton masses are not constrained by the residual symmetry. Furthermore, $Q_{l}$ and $Q_{\nu}$ are diagonal phase matrices, $Q_{l}$ can be absorbed into the charged lepton fields, the non-zero elements of $Q_{\nu}$ encoding the CP parity of neutrinos are equal to $\pm1$ and $\pm i$ and they can shift the Majorana phases $\phi_{12}$ and $\phi_{13}$ by $\pm\pi/2$ or $\pm\pi$. In what follows, we present two predictive scenarios for illustration.

\begin{table}[h!]
\begin{center}
\begin{tabular}{|c|c|c|c|}
\hline
$ \mathcal{G}_{l}$   &   $\mathcal{G}_{\nu}$   ~&~   $U$  &  number of parameters \\ \hline

$Z_n$  & $K_4\times CP$   & $Q^{\dagger}_lP^{T}_l\Sigma^{\dagger}_l\Sigma_{\nu}P_{\nu}Q_{\nu}$   &  0 \\ \hline

$Z_n$   &  $Z_2\times CP$  & $Q_l^{\dagger}P_l^T\Sigma_l^{\dagger}\Sigma_{\nu} R_{23}(\theta)P_{\nu}Q_{\nu}$ & \multirow{2}{*}{1}  \\ \cline{1-3}

$Z_2\times CP$  & $K_4\times CP'$   & $Q^{\dagger}_lP^{T}_lR^{T}_{23}(\theta_l)\Sigma^{\dagger}_l\Sigma_{\nu}P_{\nu}Q_{\nu}$ & \\ \hline

$Z_2\times CP$   &  $Z_2\times CP'$  &  $Q^{\dagger}_lP^{T}_lR^{T}_{23}(\theta_l)\Sigma^{\dagger}_l\Sigma_{\nu}R_{23}(\theta_{\nu})P_{\nu}Q_{\nu}$ & \multirow{2}{*}{2}  \\ \cline{1-3}

$Z_2$   &  $K_4\times CP$  &  $Q^{\dagger}_lP^{T}_lU^{\dagger}_{23}(\theta_l, \delta_l)\Sigma^{\dagger}_l\Sigma_{\nu}P_{\nu}Q_{\nu}$  &  \\ \hline

$Z_n$  &  $CP$  &  $Q_l^{\dagger}P_l^T\Sigma_l^{\dagger}\Sigma_{\nu} O_{3}(\theta_1, \theta_2, \theta_3)Q_{\nu}$ & \multirow{3}{*}{3}  \\ \cline{1-3}

$CP$  &  $K_4\times CP'$  &  $Q_l^{\dagger}O^{T}_{3}(\theta_1, \theta_2, \theta_3)\Sigma_l^{\dagger}\Sigma_{\nu} Q_{\nu}$ & \\ \cline{1-3}

$Z_2$  &  $Z_2\times CP$  &  $Q_l^{\dagger}P_l^TU^{\dagger}_{23}(\theta_l, \delta_l)\Sigma_l^{\dagger}\Sigma_{\nu}R_{23}(\theta_{\nu})P_{\nu}Q_{\nu}$  & \\ \hline

\end{tabular}
\caption{\label{tab:residual-sym-Nump}{Possible choices of residual symmetries and corresponding predictions for lepton mixing matrix, when the latter depends on up to three free parameters. If the residual CP is absent from $\mathcal{G}_{\nu}$, the Majorana CP phases are not restricted.}}
\end{center}
\end{table}

\subsubsection{$\mathcal{G}_{l}=Z_n\; (n\geq3)$, $\mathcal{G}_{\nu}=Z_2\times CP$}
\label{sec:abelian-subgroup-g_l}

We consider the scenario where the three families of left-handed leptons transform inequivalently as one-dimensional representations under the residual flavour symmetry $G_l=Z_n$ with $n\geq3$, so that they are distinguished by the abelian subgroup $G_l$. The representation matrix $\rho_3(g_l)$ can be diagonalized by a unitary matrix $\Sigma_{l}$ fulfilling $\Sigma_l^{\dagger}\rho_{\mathbf{3}}(g_l)\Sigma_l=\rho^{diag}_{\mathbf{3}}(g_l)$,
where the three columns of $\Sigma_l$ are formed by the three eigenvectors of $\rho_{\mathbf{3}}(g_l)$ and $\Sigma_l$ is determined up to an arbitrary diagonal unitary matrix $Q_l$ and a permutation matrix $P_l$. Eq.~\eqref{eq:Gl-HlCP-constrain} implies that the charged lepton diagonalization matrix $U_{l}$ coincides with $\Sigma_l$, i.e.,
\begin{equation}
\label{eq:Ul1}U_l=\Sigma_lP_lQ_l\,.
\end{equation}
Concerning the neutrino sector, since $X_{\nu\mathbf{3}}$ is a symmetric and unitary matrix, by performing the Takagi factorization $X_{\nu\mathbf{3}}$ can be written as
\begin{equation}
\label{eq:XnuSigma}
 X_{\nu\mathbf{3}}=\Sigma_{\nu}\Sigma_{\nu}^T\,,~~~\text{with}~~~\Sigma_{\nu}^{\dagger}\rho_{\mathbf{3}}(g_{\nu})\Sigma_{\nu}=\pm\text{diag}(1, -1, -1)\,.
\end{equation}
The procedure of obtaining the unitary matrix $\Sigma_{\nu}$ has been given in~\cite{Yao:2016zev}. The neutrino diagonalization matrix $U_{\nu}$ satisfying the conditions in Eq.~\eqref{eq:cons-nu-Z2CP} is determined to be of the following form~\cite{Feruglio:2012cw,Yao:2016zev},
\begin{equation}
\label{eq:Unu-Z2CP}U_{\nu}=\Sigma_{\nu} R_{23}(\theta)P_{\nu}Q_{\nu}\,,
\end{equation}
where $R_{23}(\theta)$ stands for a rotation matrix through an angle $\theta$ in the (23)-plane and it takes the form of Eq.~\eqref{eq:R23-matrix}, and $P_{\nu}$ is a generic permutation matrix. Here $Q_{\nu}$ in Eq.~\eqref{eq:Unu-Z2CP} is a diagonal matrix with elements equal to $\pm1$ and $\pm i$. Hence, without loss of generality it can be given as,
\begin{equation}
Q_{\nu}=\begin{pmatrix}
1 ~& 0 ~& 0 \\
1 ~& i^{k_1} & 0 \\
0 ~& 0 ~& i^{k_2}
\end{pmatrix}
\end{equation}
with $k_{1,2}=0, 1, 2, 3$. Hence the lepton mixing matrix $U_{}$ is of the form~\cite{Feruglio:2012cw,Chen:2014wxa,Yao:2016zev}
\begin{equation}
\label{eq:Upmns-Zn-Z2CP}
U_{}=U_l^{\dagger}U_{\nu}=Q_l^{\dagger}P_l^T\Sigma_l^{\dagger}\Sigma_{\nu} R_{23}(\theta)P_{\nu}Q_{\nu}\,.
\end{equation}
Notice that, as usual, the phase matrix $Q_{l}$ can be absorbed into the charged lepton fields~\cite{Schechter:1980gr}, while the effect of $Q_{\nu}$ is to shift the Majorana CP phases $\phi_{12}$ and $\phi_{13}$ by integral multiples of $\pi/2$. It is remarkable that the lepton mixing matrix is constrained to only depend on a single free parameter $\theta$, whose value can be fixed by the precisely measured reactor mixing angle $\theta_{13}$. We can therefore predict the values of the other two mixing angles $\theta_{12}$, $\theta_{23}$ as well as the CP violation phases. The lepton mixing matrix is determined up to possible permutations of rows and columns, since both neutrino and charged lepton masses are not constrained in this approach. Moreover, the $Z_2$ residual flavour symmetry can only distinguish one neutrino family from the other two generations, so that only one column of the lepton mixing matrix is fixed through the choice of residual symmetry, as can be seen from Eq.~\eqref{eq:Upmns-Zn-Z2CP}. Comparing with the oscillation data in Eqs.~(\ref{eq:U-NO}, \ref{eq:U-IO}), one can determine the phenomenologically allowed permutation matrices $P_{l}$ and $P_{\nu}$.

\begin{figure}[t!]
\centering
\includegraphics[width=0.98\textwidth]{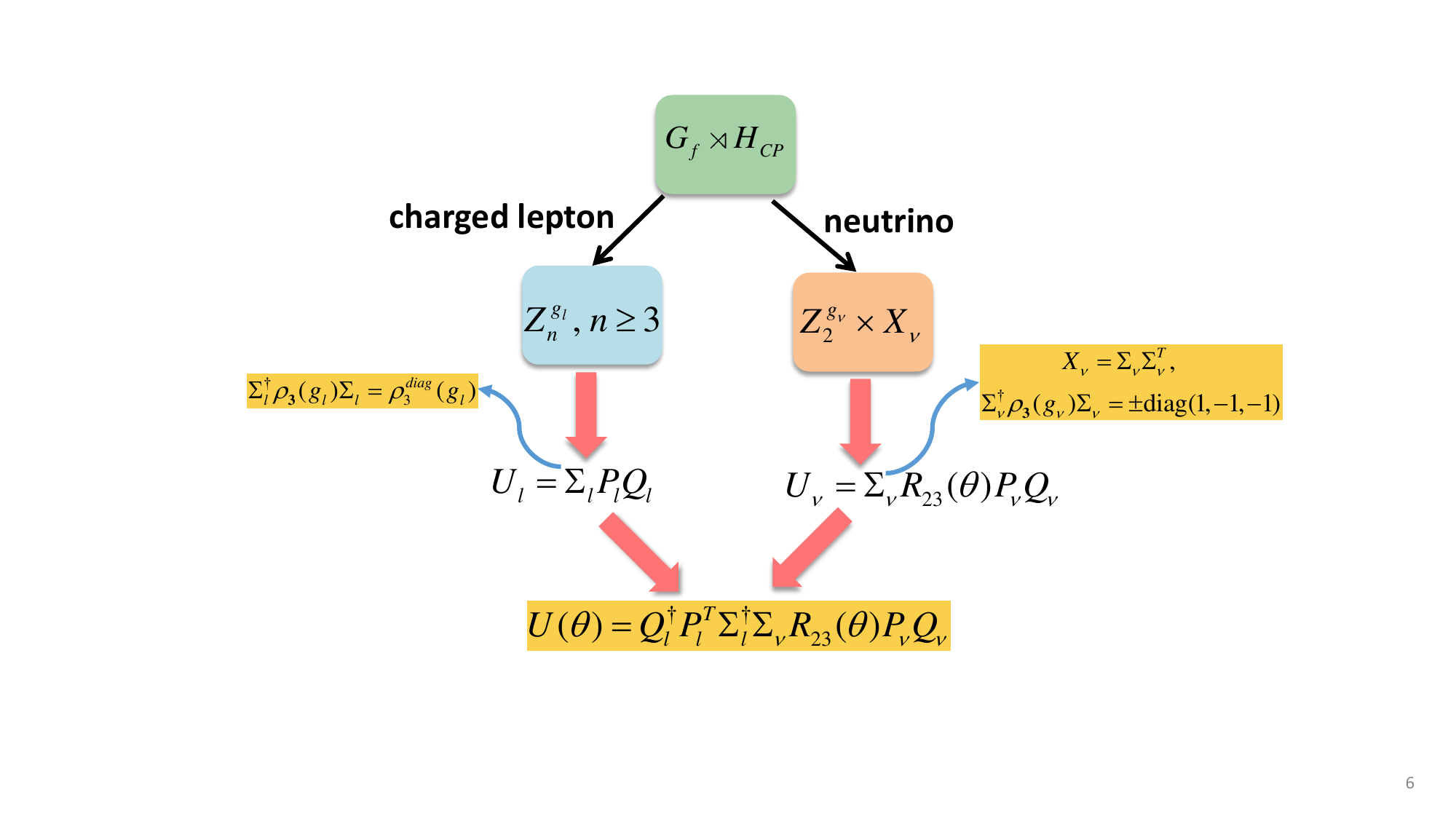}
\caption{\label{fig:Zn-z2xCP-charge-leptons-nu}The model independent predictions for lepton mixing matrix for the preserved residual symmetry abelian subgroup $G_l$ and $Z_2\times CP$ in the  charged lepton and neutrino sectors respectively.  }
\end{figure}

It follows from the above that, for any postulated residual symmetry subgroups $G_l$ and $Z_2\times CP$, the lepton mixing matrix can be extracted by using Eq.~\eqref{eq:Upmns-Zn-Z2CP} in a simple manner, the recipe is summarized in figure~\ref{fig:Zn-z2xCP-charge-leptons-nu}. The predicted lepton mixing matrix only depends on the structure of the symmetry group $G_{f}\rtimes H_{CP}$ and the assumed residual symmetry, regardless of the underlying dynamics which breaks $G_{f}\rtimes H_{CP}$ down to the residual subgroup.

As an illustration, we consider the $S_4$ flavor symmetry in combination with the generalized CP symmetry. The group theory and representation of $S_4$ are given in~\ref{sec:ap_S4Group}. The outer automorphism group of $S_4$ is trivial~\cite{Holthausen:2012dk,Li:2013jya}, all the automorphisms of $S_4$ are inner automorphisms which are group conjugations. Without loss of generality it is sufficient to consider the representative automorphism $\mathfrak{u}:(s, t, u)\to(s, t^{-1}, u)$, where $s$, $t$ and $u$ are the generators of $S_4$, the corresponding generalized CP transformation $X^{0}_{\mathbf{r}}$ is determined by the following consistency equations
\begin{eqnarray}
\nonumber&&X^{0}_{\mathbf{r}}\rho_{\mathbf{r}}^{*}(s)\left(X_{\mathbf{r}}^{0}\right)^{-1}=\rho_{\mathbf{r}}(\mathfrak{u}(s))=\rho_{\mathbf{r}}(s)\,,\\
\nonumber&&X^{0}_{\mathbf{r}}\rho_{\mathbf{r}}^{*}(t)\left(X_{\mathbf{r}}^{0}\right)^{-1}=\rho_{\mathbf{r}}(\mathfrak{u}(t))=\rho_{\mathbf{r}}(t^{2})\,,\\
&&X^{0}_{\mathbf{r}}\rho_{\mathbf{r}}^{*}(u)\left(X_{\mathbf{r}}^{0}\right)^{-1}=\rho_{\mathbf{r}}(\mathfrak{u}(u))=\rho_{\mathbf{r}}(u)\,.
\end{eqnarray}
From the explicit form of the representation matrices in Eqs.~(\ref{eq:S4-rep-singlet}, \ref{eq:S4-rep-doublet}, \ref{eq:S4-rep-triplet}), it follows that $X^{0}_{\mathbf{r}}$ is the unit matrix up to an arbitrary overall phase,
\begin{equation}
X^{0}_{\mathbf{r}}=1\,.
\end{equation}
Including the family symmetry transformation, the generalized CP transformation consistent with the $S_4$ symmetry is given by~\cite{Ding:2013hpa,Li:2013jya,Li:2014eia},
\begin{equation}
X_{\mathbf{r}}=\rho_{\mathbf{r}}(g)X^{0}_{\mathbf{r}}=\rho_{\mathbf{r}}(g),\quad g\in S_4\,.
\end{equation}
It was found that three sets of residual symmetries are compatible with the experimental data~\cite{Feruglio:2012cw}. Notice that a fourth residual symmetry with $G_{l}=Z^{t}_{3}$, $G_{\nu}=Z^{su}_2$, $X_{\nu}=1$ is not phenomenologically viable, since the reactor and atmospheric angles can not be accommodated simultaneously.

\begin{description}[labelindent=-0.6em, leftmargin=0.3em]
\item[(\romannumeral1)]{$G_{l}=Z^{t}_{3}$, $G_{\nu}=Z^{s}_2$, $X_{\nu}=1$ }

In this case, the unitary transformation $\Sigma_{l}$ is the unit matrix since the representation matrix $\rho_{\mathbf{3}}(t)$ is diagonal, and the Takagi factorization matrix $\Sigma_{\nu}$ is found to be
\begin{equation}
\Sigma_{\nu}=\frac{1}{\sqrt{6}}\left(
\begin{array}{ccc}
 \sqrt{2} ~& 2 ~& 0 \\
 \sqrt{2} ~& -1 ~& \sqrt{3} \\
 \sqrt{2} ~& -1 ~& -\sqrt{3}
\end{array}
\right)\,.
\end{equation}
Hence the lepton mixing matrix is determined to be
\begin{equation}
\label{UPMNS-S4CP-I}U=\frac{1}{\sqrt{6}}\left(
\begin{array}{ccc}
2\cos\theta ~&~ \sqrt{2} ~&~ 2 \sin \theta \\
-\cos\theta-\sqrt{3}\sin\theta ~&~ \sqrt{2} ~&~ \sqrt{3}\cos\theta-\sin\theta\\
-\cos\theta+\sqrt{3}\sin\theta ~&~ \sqrt{2} ~&~ -\sqrt{3}\cos\theta-\sin\theta
\end{array}
\right)Q_{\nu}\,,
\end{equation}
where we have taken $P_{\nu}=P_{12}$ such that $(1, 1, 1)^{T}/\sqrt{3}$ is in the second column, in order to be consistent with experimental data. We can extract the lepton mixing parameters as follows,
\begin{eqnarray}
\nonumber&&\sin^2\theta_{13}=\frac{2}{3}\sin^2\theta\,,~~\sin^2\theta_{23}=\frac{1}{2}-\frac{\sqrt{3}\sin2\theta}{2(2+\cos2\theta)}=\frac{1}{2}\pm\frac{1}{2}\tan\theta_{13}\sqrt{2-\tan^2\theta_{13}}\,,\\
\label{mixing-pars-S4CP-I}&&\sin^2\theta_{12}=\frac{1}{2+\cos2\theta}=\frac{1}{3\cos^2\theta_{13}},~~~\sin2\phi_{12}=\sin2\phi_{13}=\sin\delta_{CP}=0\,.
\end{eqnarray}
Using the experimental best fit value $(\sin^2\theta_{13})^{\text{bf}}=0.022$~\cite{deSalas:2020pgw,10.5281/zenodo.4726908}, we find the solar mixing angle $\sin^2\theta_{12}\simeq0.341$ and the
atmospheric angle
\begin{equation}
\sin^2\theta_{23}\simeq\left\{\begin{array}{cc}
0.426  & ~\text{for}~  \theta<\pi/2  \\
0.574  & ~\text{for}~   \theta>\pi/2
\end{array}
\right.\,.
\end{equation}
The Dirac and Majorana CP phases are determined to take on CP conserving values in this case.

\item[(\romannumeral2)]{$G_{l}=Z^{t}_{3}$, $G_{\nu}=Z^{s}_2$, $X_{\nu}=u$ }

This case differs from the previous one in the residual CP $X_{\nu}$. The unitary matrix $\Sigma_{\nu}$ is
\begin{equation}
\Sigma_{\nu}=\frac{1}{\sqrt{6}}\left(
\begin{array}{ccc}
  \sqrt{2}\,i ~& 2 i ~& 0 \\
 \sqrt{2}\,i ~& -i ~& \sqrt{3} \\
  \sqrt{2}\,i ~& -i ~& -\sqrt{3}
\end{array}
\right)\,.
\end{equation}
Consequently, for $P_{\nu}=P_{12}$, the lepton mixing matrix is fixed to be
\begin{equation}
\label{UPMNS-S4CP-II}U=\frac{1}{\sqrt{6}}\left(
\begin{array}{ccc}
2 \cos \theta ~& \sqrt{2} ~& 2 \sin \theta \\
-\cos \theta+i \sqrt{3} \sin \theta ~& \sqrt{2} ~& -\sin\theta-i\sqrt{3}\cos\theta\\
-\cos \theta-i \sqrt{3} \sin \theta ~& \sqrt{2} ~& -\sin\theta+i\sqrt{3}\cos\theta
\end{array}
\right)Q_{\nu}\,.
\end{equation}
We can extract the lepton mixing parameters as
\begin{eqnarray}
\nonumber&&\sin^2\theta_{13}=\frac{2}{3}\sin^2\theta\,,~~~\sin^2\theta_{12}=\frac{1}{2+\cos2\theta}=\frac{1}{3\cos^2\theta_{13}}\,,\\
\label{mixing-pars-S4CP-II}&&\sin^2\theta_{23}=\frac{1}{2}\,,~~~|\sin\delta_{CP}|=1\,,~~~\sin2\phi_{12}=\sin2\phi_{13}=0\,.
\end{eqnarray}
Because the $\mu-\tau$ reflection symmetry $X_{\nu}=u$ is imposed on the neutrino mass matrix~\cite{Harrison:2002kp,Harrison:2002et,Harrison:2004he,Grimus:2003yn},
both atmospheric angle $\theta_{23}$ and Dirac CP phase $\delta_{CP}$ are maximal, while the Majorana CP phases $\phi_{12}$ and $\phi_{13}$ are trivial.

\item[(\romannumeral3)]{$G_{l}=Z^{t}_{3}$, $G_{\nu}=Z^{su}_2$, $X_{\nu}=u$}

The Takagi factorization matrix $\Sigma_{\nu}$ is determined to be
\begin{equation}
\Sigma_{\nu}=\frac{1}{\sqrt{6}}\left(
\begin{array}{ccc}
 2 i ~& \sqrt{2}\,i  ~& 0 \\
 -i ~&  \sqrt{2}\,i  ~& -\sqrt{3} \\
 -i ~&  \sqrt{2}\,i  ~& \sqrt{3} \\
\end{array}
\right)\,.
\end{equation}
Using the master formula Eq.~\eqref{eq:Upmns-Zn-Z2CP}, we obtain the lepton mixing matrix as
\begin{equation}
\label{UPMNS-S4CP-IV}U=\frac{1}{\sqrt{6}}\left(
\begin{array}{ccc}
 2 ~& \sqrt{2} \cos \theta ~& \sqrt{2} \sin \theta \\
 -1 ~& \sqrt{2} \cos \theta-i \sqrt{3} \sin \theta ~&
 \sqrt{2}\sin \theta+i \sqrt{3} \cos \theta \\
 -1 ~& \sqrt{2} \cos \theta+i \sqrt{3} \sin \theta ~&
   \sqrt{2} \sin \theta-i \sqrt{3} \cos \theta \\
\end{array}
\right)Q_{\nu}\,.
\end{equation}
The predictions for mixing angles and CP violation phases are
\begin{eqnarray}
\nonumber&&\sin^2\theta_{13}=\frac{1}{3}\sin^2\theta\,,~~~\sin^2\theta_{12}=\frac{1+\cos2\theta}{5+\cos2\theta}=\frac{1-3\sin^2\theta_{13}}{3\cos^2\theta_{13}}\,,\\
\label{eq:caseIV-Z2xCP-S4}&&\sin^2\theta_{23}=\frac{1}{2}\,,~~~|\sin\delta_{CP}|=1\,,~~~\sin2\phi_{12}=\sin2\phi_{13}=0\,.
\end{eqnarray}
For $(\sin^2\theta_{13})^{\text{bf}}=0.022$~\cite{deSalas:2020pgw,10.5281/zenodo.4726908} we find, from Eq.~\eqref{eq:caseIV-Z2xCP-S4}, the solar mixing angle as $\sin^2\theta_{12}\simeq0.318$. Similar to case {\romannumeral2}, the $\mu-\tau$ reflection symmetry $X_{\nu}=u$ implies maximal values of $\theta_{23}$ and $\delta_{CP}$, and CP conserving values of $\phi_{12}$,$\phi_{13}$.
\end{description}

\begin{figure}[!h]
\centering
\begin{tabular}{cc}
\includegraphics[width=0.44\textwidth]{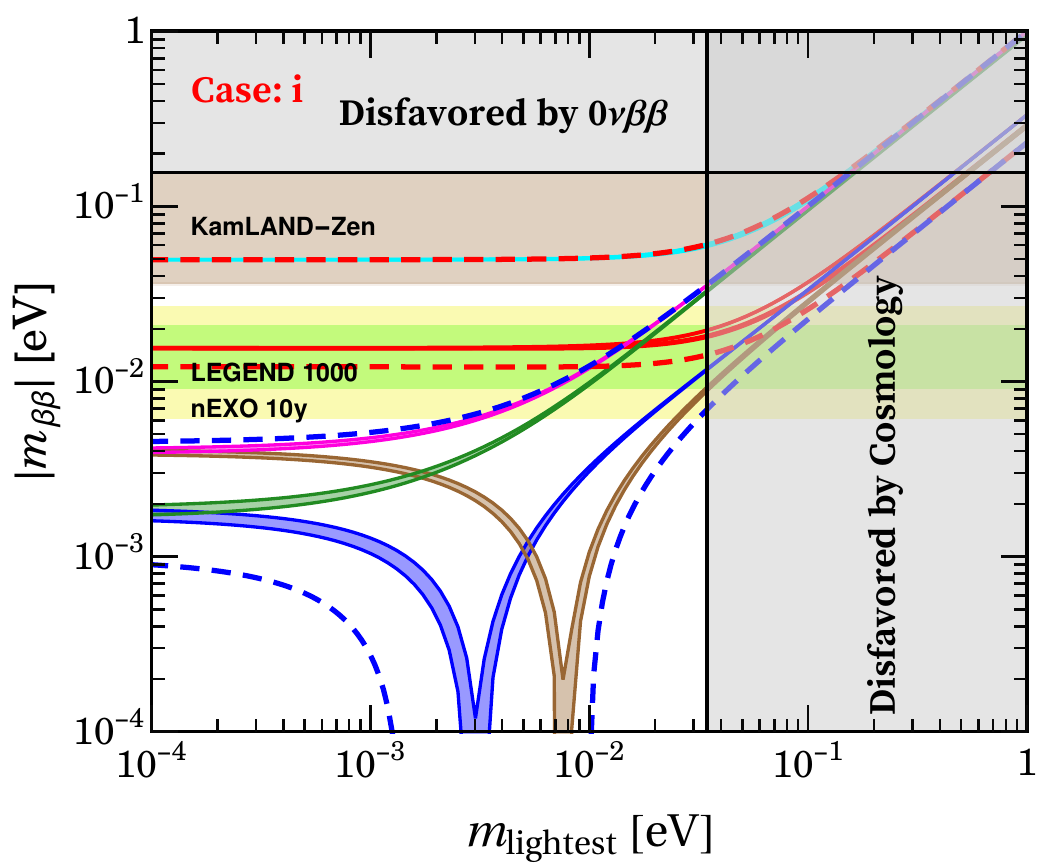} &
\includegraphics[width=0.44\textwidth]{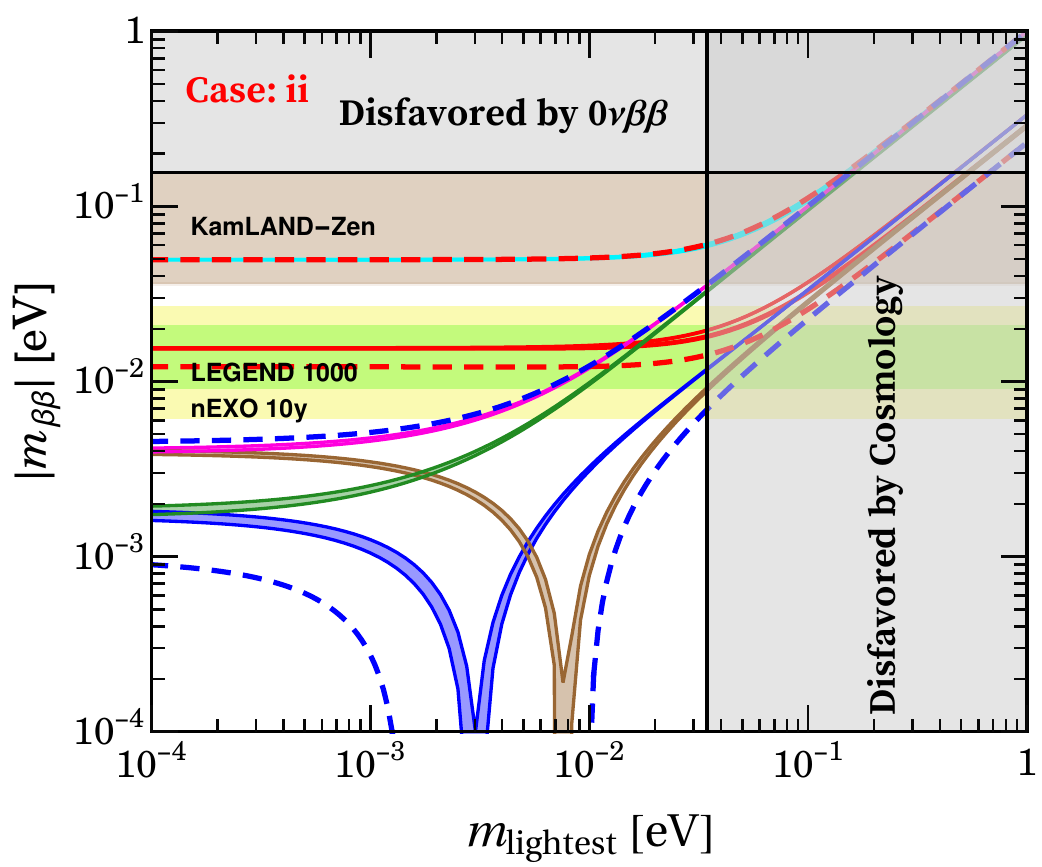}\\
\includegraphics[width=0.44\textwidth]{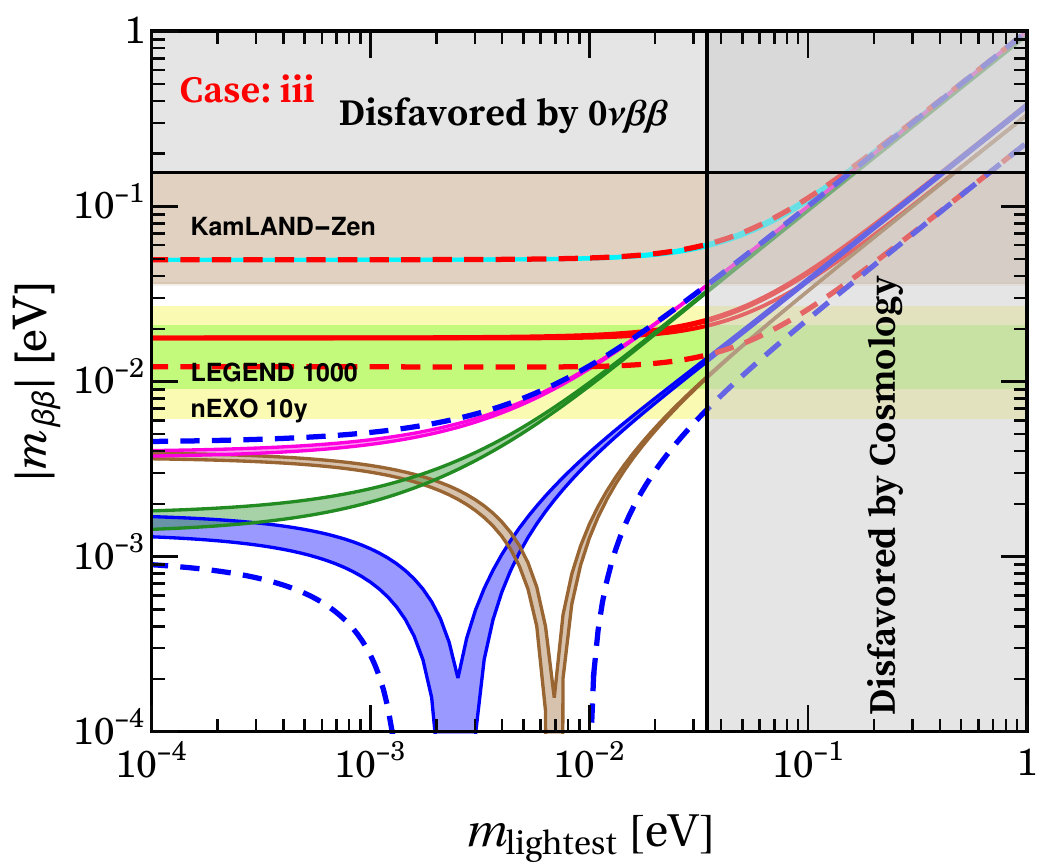}&~~~~~~~~~
\multirow{1}[30]{0.4\linewidth}[5.00cm]{\includegraphics[width=0.9\linewidth]{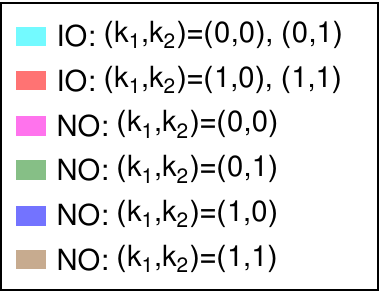}}
\end{tabular}
\caption{\label{fig:onbb-Z2xCP-S4}Predicted effective Majorana neutrino mass $|m_{\beta\beta}|$ for three viable mixing patterns when the $S_4$ flavour symmetry and CP symmetry are broken to an Abelian subgroup and $Z_2\times CP$ in the charged lepton sector and neutrino sector respectively. The red (blue) dashed lines indicate the most general allowed regions for IO (NO) neutrino mass ordering obtained by varying the mixing parameters in their $3\sigma$ ranges~\cite{deSalas:2020pgw,10.5281/zenodo.4726908}. The current experimental bound $|m_{\beta\beta}|<(36-156)\,$meV at $90\%$ C.L. from KamLAND-Zen~\cite{KamLAND-Zen:2022tow} and the future sensitivity ranges $|m_{\beta\beta}|<(9-21)\,$ meV from LEGEND-1000~\cite{LEGEND:2021bnm} and $|m_{\beta\beta}|<(6-27)\,$ meV from nEXO~\cite{nEXO:2021ujk} are indicated by light brown, light yellow and light green horizontal bands respectively. The vertical grey exclusion band represents the bound $\sum_im_i<0.120$ eV from Planck at $95\%$ C.L.~\cite{Planck:2018vyg,Gerbino:2022nvz}.}
\end{figure}

The values of the effective $|m_{\beta\beta}|$ parameter characterizing the \znbb decay amplitude can be determined for the viable cases associated to the lepton mixing matrices in Eqs.~(\ref{UPMNS-S4CP-I}, \ref{UPMNS-S4CP-II}, \ref{UPMNS-S4CP-IV}). This effective mass parameter depends on the CP parity encoded in $Q_{\nu}$. Since the rotation angle $\theta$ is strongly constrained by the reactor angle $\theta_{13}$ and the Majorana phases can be predicted, the effective mass $|m_{\beta\beta}|$ is severely restricted, as shown in figure~\ref{fig:onbb-Z2xCP-S4}. The narrow width of each band comes from varying $\theta_{13}$ and the neutrino mass squared differences over their allowed $3\sigma$  ranges~\cite{deSalas:2020pgw}.

For inverted neutrino mass ordering, the term proportional to $m_3$ in $|m_{\beta\beta}|$ is suppressed by the small values of $m_3$ and $\sin^2\theta_{13}$, consequently the value of $k_2$ is almost irrelevant. For $(k_1, k_2)=(0, 0), (0, 1)$, the effective mass $m_{\beta\beta}$ is close to the upper boundary of the IO region obtained by using the $3\sigma$ global data, while $|m_{\beta\beta}|$ is close to the lower boundary of IO for $(k_1, k_2)=(1, 0), (1, 1)$.  \\[-.4cm]

Symmetry breaking patterns of abelian subgroups in the charged lepton sector and $Z_2\times CP$ in the neutrino sector have also been studied for many flavour symmetry groups combined with generalized CP, such as $A_4$~\cite{Feruglio:2012cw,Ding:2013bpa,Li:2016nap}, $S_4$~\cite{Feruglio:2012cw,Li:2013jya,Ding:2013hpa,Li:2014eia,Feruglio:2013hia,Lu:2016jit,Penedo:2017vtf}, $\Delta(27)$~\cite{Branco:2015hea,Branco:2015gna}, $\Delta(48)$~\cite{Ding:2013nsa,Ding:2014hva}, $A_5$~\cite{Li:2015jxa,DiIura:2015kfa,Ballett:2015wia,Turner:2015uta,DiIura:2018fnk}, $\Delta(96)$~\cite{Ding:2014ssa} and the infinite group series $\Delta(3n^2)=(Z_n\times Z_n)\rtimes Z_3$~\cite{Hagedorn:2014wha,Ding:2015rwa}, $\Delta(6n^2)=(Z_n\times Z_n)\rtimes S_3$~\cite{Hagedorn:2014wha,Ding:2014ora} and $D^{(1)}_{9n, 3n}=(Z_{9n}\times Z_{3n})\rtimes S_3$~\cite{Li:2016ppt}. For popular flavour symmetries $A_4$, $S_4$ and $A_5$, the Dirac CP phase is predicted to take simple values $\delta_{CP}=0, \pi, \pm\pi/2$, and the Majorana phases take on CP conserving values for the experimentally viable mixing patterns.  Moreover, the values of the Dirac CP phase and atmospheric mixing angle are correlated in flavor groups $A_4$, $S_4$ and $A_5$, the atmospheric angle is maximal for $\delta_{CP}=\pm\pi/2$ and non-maximal for $\delta_{CP}=0, \pi$. Both Dirac and Majorana CP phases can depend nontrivially on the parameter $\theta$ for larger flavour symmetry groups such as~$\Delta(96)$~\cite{Ding:2014ssa}. A systematical classification of the possible mixing patterns resulting from the pair of residual symmetry subgroups $\{G_l, Z_2\times CP\}$ has been performed~\cite{Yao:2016zev}. The neutrino mixing patterns compatible with oscillation data are given as follows~\cite{Yao:2016zev}:
\begin{eqnarray}
\nonumber&&U^{I(a)}=\frac{1}{\sqrt{3}}
\begin{pmatrix}
\sqrt{2} \sin \varphi _1 ~&~ e^{i \varphi _2} ~&~ \sqrt{2}\cos\varphi_1 \\
\sqrt{2}\cos\left(\varphi _1-\frac{\pi }{6}\right) ~&~ -e^{i\varphi_2} ~&~ -\sqrt{2} \sin \left(\varphi_1-\frac{\pi}{6}\right) \\
\sqrt{2}\cos\left(\varphi_1+\frac{\pi}{6}\right)   ~&~  e^{i\varphi_2} ~&~ -\sqrt{2}\sin\left(\varphi_1+\frac{\pi}{6}\right)
\end{pmatrix}R_{23}(\theta)Q_{\nu}\,,\\
\nonumber&&U^{I(b)}=\frac{1}{\sqrt{3}}
\begin{pmatrix}
\sqrt{2}\cos\varphi_1 ~&~ e^{i \varphi _2} ~&~  \sqrt{2} \sin \varphi _1 \\
-\sqrt{2} \sin \left(\varphi_1-\frac{\pi}{6}\right)  ~&~ -e^{i\varphi_2} ~&~ \sqrt{2}\cos\left(\varphi _1-\frac{\pi }{6}\right) \\
-\sqrt{2}\sin\left(\varphi_1+\frac{\pi}{6}\right) ~&~  e^{i\varphi_2} ~&~\sqrt{2}\cos\left(\varphi_1+\frac{\pi}{6}\right)
\end{pmatrix}R_{12}(\theta)Q_{\nu}\,,\\
\nonumber&&U^{II}=\frac{1}{\sqrt{3}}
\begin{pmatrix}
 e^{i \varphi _1} ~&~ 1 ~&~ e^{i \varphi _2} \\
 \omega  e^{i \varphi _1} ~&~ 1 ~&~ \omega ^2 e^{i \varphi _2} \\
 \omega ^2 e^{i \varphi _1} ~&~ 1 ~&~ \omega  e^{i \varphi _2} \\
\end{pmatrix}R_{13}(\theta)Q_{\nu}\,, \\
\nonumber&&U^{III}=\frac{1}{\sqrt{3}}
\begin{pmatrix}
 \sqrt{2} e^{i \varphi _1} \sin \varphi _2 ~&~ 1 ~&~ \sqrt{2} e^{i \varphi _1} \cos \varphi _2 \\
 \sqrt{2} e^{i \varphi _1} \cos \left(\varphi _2+\frac{\pi }{6}\right) ~&~ 1 ~&~ -\sqrt{2} e^{i \varphi _1} \sin \left(\varphi _2+\frac{\pi }{6}\right) \\
 -\sqrt{2} e^{i \varphi _1} \cos \left(\varphi _2-\frac{\pi }{6}\right) ~&~ 1 ~&~ \sqrt{2} e^{i \varphi _1} \sin \left(\varphi _2-\frac{\pi }{6}\right)
\end{pmatrix}R_{13}(\theta)Q_{\nu}\,,\\
\nonumber&&U^{IV(a)}=\frac{1}{\sqrt{2\sqrt{5}\,\phi_g}}\begin{pmatrix}
-\sqrt{2}\,\phi_g  ~&~ \sqrt{2} ~&~ 0\\
1 ~&~ \phi_g ~&~ -\sqrt{\sqrt{5}\,\phi_g} \\[1mm]
1 ~&~ \phi_g ~&~ \sqrt{\sqrt{5}\,\phi_g}
\end{pmatrix}R_{13}(\theta)Q_{\nu}\,, \\
\nonumber&&U^{IV(b)}=\frac{1}{\sqrt{2\sqrt{5}\,\phi_g}}\begin{pmatrix}
-\sqrt{2}\,\phi_g i ~&~ \sqrt{2} ~&~ 0\\
i ~&~ \phi_g ~&~ -\sqrt{\sqrt{5}\,\phi_g} \\[1mm]
i ~&~ \phi_g ~&~ \sqrt{\sqrt{5}\,\phi_g}
\end{pmatrix}R_{13}(\theta)Q_{\nu}\,, \\
\nonumber&&U^{V}=\frac{1}{2}\begin{pmatrix}
 \phi_g  ~&~ 1 ~&~ \phi_g -1 \\
 \phi_g -1 ~&~ -\phi_g  ~&~ 1 \\
 1 ~&~ 1-\phi_g  ~&~ -\phi_g
\end{pmatrix}R_{23}(\theta)Q_{\nu}\,,\\
\nonumber&&U^{VI}=\frac{1}{2\sqrt{3}}\begin{pmatrix}
(\sqrt{3}-1)e^{i \varphi } ~&~ 2 ~&~ -(\sqrt{3}+1)e^{i\left(\varphi+\frac{3\pi}{4}\right)} \\
-(\sqrt{3}+1)e^{i\varphi} ~&~ 2 ~&~ (\sqrt{3}-1)e^{i\left(\varphi+\frac{3\pi}{4}\right)} \\
2e^{i\varphi} ~&~ 2 ~&~ 2e^{i\left(\varphi+\frac{3\pi}{4}\right)}
\end{pmatrix}R_{13}(\theta)Q_{\nu}\,,\\
&&U^{VII}=\frac{1}{2\sqrt{6}}
\begin{pmatrix}
  -\frac{\sqrt{3}}{s_3} ~&~  2 \sqrt{2} ~&~  \frac{s_2-s_1}{s_1 s_2} \\[1mm]
  \frac{\sqrt{3}}{s_2} ~&~  2 \sqrt{2} ~&~ -\frac{s_1+s_3}{s_1 s_3} \\[1mm]
 \frac{\sqrt{3}}{s_1} ~&~  2 \sqrt{2} ~& \frac{s_2+s_3}{s_2 s_3}
 \end{pmatrix}R_{23}(\theta)Q_{\nu}\,,
\end{eqnarray}
up to row and column permutations, where $s_n\equiv\sin (2n\pi/7)$ with $n=1, 2, 3$ and $R_{ij}(\theta)$ is the rotation matrix through an angle $\theta$ in the $(ij)$-plane. The parameters $\varphi_1$, $\varphi_2$ and $\varphi$ are group theoretical indices characterizing the flavour group and the residual symmetry and they can only take some discrete values for a given flavour symmetry. The possible values of $\varphi_1$, $\varphi_2$ and $\varphi$ for different finite flavour group $G_f$ up to order 2000 and the corresponding predictions for lepton mixing parameters are given in~\cite{Yao:2016zev}. It is remarkable that the above mixing patterns can be obtained from the flavour groups $\Delta(6n^2)$, $D^{(1)}_{9n, 3n}$, $A_5$ and $\Sigma(168)$ in combination with generalized CP symmetry.

As shown in table~\ref{tab:residual-sym-Nump}, if the residual symmetries $K_4\times CP$ and $Z_2\times CP'$ are preserved in the neutrino and charged lepton sectors respectively, the lepton mixing matrix would only depends on a single real parameter $\theta$ as well. In this case, one row of the mixing matrix is completely fixed by residual symmetry, regardless of the free parameter $\theta$.

It turns out that only one type of mixing pattern can accommodate the oscillation data~\cite{Yao:2016zev}:
\begin{equation}
\label{eq:neutrino-K4-charlep-Z2CP}U^{VIII}=\frac{1}{2}R^T_{13}(\theta)\left(
\begin{array}{ccc}
 \sqrt{2} e^{i \varphi _1} & -\sqrt{2} e^{i \varphi _1} & 0 \\
 1 & 1 & -\sqrt{2} e^{i \varphi _2} \\
 1 & 1 & \sqrt{2} e^{i \varphi _2}
\end{array}
\right)Q_{\nu}\,,
\end{equation}
where $\varphi_1$ and $\varphi_2$ take discrete values determined by the choice of the residual symmetry and the flavour symmetry group.

\subsubsection{$\mathcal{G}_{l}=Z_2\times CP$, $\mathcal{G}_{\nu}=Z_2\times CP'$}
\label{sec:resid-symm-z_2t}

\begin{figure}[t!]
\centering
\includegraphics[width=0.98\textwidth]{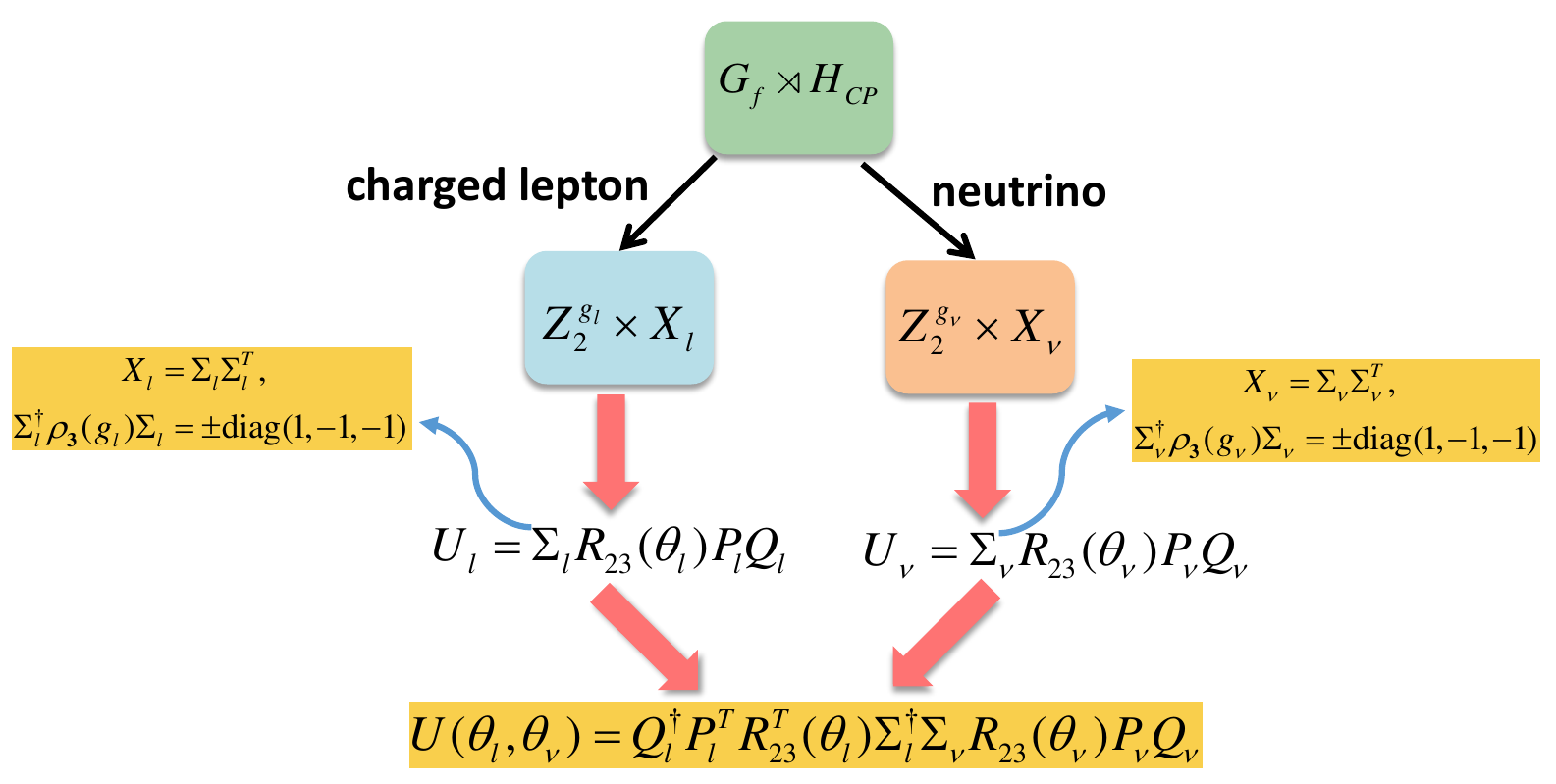}
\caption{\label{fig:z2xCP-nu-charge-leptons}
Model-independent predictions for the lepton mixing matrix when the residual symmetry has the structure $Z_2\times CP$ in both neutrino and charged lepton sectors. }
\end{figure}

The residual subgroups in both the neutrino and charged lepton sectors are of the structure $Z_2\times CP$ in this scheme~\cite{Lu:2016jit,Li:2017abz,Lu:2019gqp}. Since the neutrino sector still preserves the residual symmetry $Z^{g_\nu}_2\times X_{\nu}$, the unitary transformation $U_{\nu}$ is given by Eq.~\eqref{eq:Unu-Z2CP}.
The residual symmetry of the charged lepton sector is denoted as $Z^{g_l}_2\times X_{l}$ in this scheme, and the constrained consistency condition in Eq.~\eqref{eq:Gch-Xch-conSconD}
should be fulfilled, i.e., $X_{l\mathbf{r}}\rho^*_{\mathbf{r}}(g_l)X_{l\mathbf{r}}^{-1}=\rho_{\mathbf{r}}(g_l)$.
The three families of left-handed lepton doublets are assigned to transform as a faithful triplet $\rho_{\mathbf{3}}$ of the flavour group $G_f$,
where the representation $\rho_{\mathbf{3}}$ can be either irreducible or reducible. The residual symmetry $Z^{g_l}_2\times X_{l}$ of the charged lepton sector requires that $m^{\dagger}_lm_l$ is invariant under $Z^{g_l}_2\times X_{l}$,
where $m_{l}$ is the charged lepton mass matrix. Thus the unitary transformation $U_l$ of the charged leptons should fulfill
\begin{equation}
\label{eq:Ul-Z2xCP}U_l^{\dagger}\rho_{\mathbf{3}}(g_l)U_l=\text{diag}(\pm1, \pm1, \pm1)\,,~~~~~~ U_l^{\dagger}X_{l\mathbf{3}}U_l^*=\text{diag}\left(e^{-i\alpha_e}, e^{-i\alpha_{\mu}}, e^{-i\alpha_{\tau}} \right)\equiv Q^{\dagger2}_{l}\,,
\end{equation}
where $Q_l=\text{diag}\left(e^{i\alpha_e/2}, e^{i\alpha_{\mu}/2}, e^{i\alpha_{\tau}/2}\right)$ is a diagonal phase matrix with $\alpha_{e, \mu, \tau}$ real. Notice that the eigenvalue of $\rho(g_l)$ is $+1$ or $-1$ because the generator $g_l$ is of order 2. The residual CP transformation $X_l$ should be a symmetric matrix, otherwise the charged lepton masses would be partially degenerate. Likewise, for the neutrino sector we can perform the Takagi factorization for $X_{l}$ as follows
\begin{equation}
\label{eq:XlSigma-Z2xCP}
X_{l\mathbf{3}}=\Sigma_{l}\Sigma_{l}^T\,,~~~~\Sigma_{l}^{\dagger}\rho_{\mathbf{3}}(g_{l})\Sigma_{l}=\pm\text{diag}(1, -1, -1)\,.
\end{equation}
Then the residual symmetry constrains $U_l$ to be
\begin{equation}
U_l=\Sigma_l R_{23}(\theta_l) P_l Q_l
\end{equation}
where $P_l$ is a generic three-dimensional permutation matrix, since the charged lepton masses are not predicted in this scheme.
Hence the lepton mixing matrix is determined to be of the following form
\begin{equation}
\label{eq:U-Z2xCP-nu-ch}U=U^{\dagger}_lU_{\nu}=Q^{\dagger}_lP^{T}_lR^{T}_{23}(\theta_l)\Sigma^{\dagger}_l\Sigma_{\nu}R_{23}(\theta_{\nu})P_{\nu}Q_{\nu} \,,
\end{equation}
where the phase matrix $Q_l$ can be absorbed into the charged lepton fields.
In this scenario, the recipe for extracting the predicted lepton mixing matrix from the residual symmetry is summarized in figure~\ref{fig:z2xCP-nu-charge-leptons}. One sees that the resulting lepton mixing matrix only depends on two free rotation angles $\theta_{l}$ and $\theta_{\nu}$, lying in the range $0\leq\theta_{l, \nu}<\pi$. The precisely measured reactor angle $\theta_{13}$ and the solar angle $\theta_{12}$ can be accommodated for certain values of $\theta_{l}$ and $\theta_{\nu}$, leading to relations for the other mixing parameters. Notice that in this scheme only one element of the mixing matrix is fixed to a constant value by the residual subgroups.

A comprehensive study of the lepton mixing patterns which can arise from the breaking of $S_4$ and CP symmetries into two distinct $Z_2\times CP$ subgroups in the neutrino and charged lepton sectors leads to eighteen phenomenologically viable cases~\cite{Lu:2016jit}. In the following, we give one typical example which predicts non-trivial CP violation phases. The three families of left-handed leptons are embedded in a triplet $\mathbf{3}$ of $S_4$, and the residual symmetries of the neutrino and charged lepton mass matrices are $Z^{st^2su}_2\times X_{l}$ and $Z^{s}_2\times X_{\nu}$ respectively, with $X_{l}=t^2$ and $X_{\nu}=su$. The Takagi factorization $\Sigma_{l}$ and $\Sigma_{\nu}$ are found to be
\begin{equation}
\Sigma_{l}=\frac{1}{\sqrt{6}}\begin{pmatrix}
 2 ~& 0 ~& -\sqrt{2} \\
e^{\frac{i\pi}{3}} ~& -\sqrt{3}\, e^{\frac{i\pi}{3}} ~& \sqrt{2}\, e^{\frac{i\pi}{3}}  \\
e^{-\frac{i\pi}{3}} ~& \sqrt{3}\, e^{-\frac{i\pi}{3}} ~& \sqrt{2}\,e^{-\frac{i\pi}{3}}
\end{pmatrix},~~~~\Sigma_{\nu}=\frac{1}{\sqrt{6}}
\begin{pmatrix}
\sqrt{2}\,i  ~& 0 ~& -2 \\
\sqrt{2}\,i ~& -\sqrt{3}\,i  ~& 1 \\
\sqrt{2}\,i  ~& \sqrt{3}\,i ~& 1
\end{pmatrix}\,.
\end{equation}
The lepton mixing matrix can be easily obtained by using the master formula of Eq.~\eqref{eq:U-Z2xCP-nu-ch}, one of its elements is fixed to be $1/\sqrt{2}$. In order to be compatible with experimental data, the fixed element can be either the (23) or (33) entry, so that we can take the permutation matrices $(P_{l}, P_{\nu})=(P_{12}, P_{13})$ or $(P_{13}.P_{12}, P_{13})$. For the first case, $(P_{l}, P_{\nu})=(P_{12}, P_{13})$, we find that the lepton mixing parameters are
\begin{equation}
\sin^2\theta_{13}=\frac{1}{2}\cos^2\theta_{l},~~~\sin^2\theta_{12}=\frac{1}{2}+\frac{\left(1-3\cos2\theta_l\right)\sin2\theta_{\nu}}{6-2\cos2\theta_l},~~\sin^2\theta_{23}=\frac{1}{2-\cos^2\theta_l}\,,
\end{equation}
while the CP invariants are determined as
\begin{equation}
J_{CP}=\frac{\sin2\theta_l\cos2\theta_{\nu}}{8\sqrt{2}},~~~I_1=\frac{\left(2\sin2\theta_l-3\sin4\theta_l\right)\cos2\theta_{\nu}}{16\sqrt{2}},~~~
I_2=\frac{\sin \theta _l \cos ^3\theta _l \cos 2 \theta _{\nu}}{2 \sqrt{2}}\,.
\end{equation}
The Jarlskog invariant $J_{CP}$ of neutrino oscillations is related to $\delta^{\ell}$ in Eq.~(\ref{eq:dell}), while the other two invariants $I_1$ and $I_2$ are given in terms of the basic Majorana phases $\phi_{12},\phi_{13}$ involved in \znbb decay, see section~\ref{sec:introduction}. One finds the following expressions~\cite{Branco:2011zb,Jarlskog:1985ht,Branco:1986gr,Nieves:1987pp},
\begin{eqnarray}
\nonumber J_{CP}&=&\text{Im}\left(U_{11}U_{33}U_{13}^{*}U_{31}^{*}\right)=\frac{1}{8}\sin2\theta_{12}\sin2\theta_{13}\sin2\theta_{23}\cos\theta_{13}\sin\delta^{\ell}\,,\\
\nonumber I_{1}&=&\text{Im}\left(U_{11}^{2*}U_{12}^{2}\right)=-\sin^{2}\theta_{12}\cos^{2}\theta_{12}\cos^{4}\theta_{13}\sin2\phi_{12},\\
I_{2}&=&\text{Im}\left(U_{11}^{2*}U_{13}^{2}\right)=-\cos^{2}\theta_{12}\cos^{2}\theta_{13}\sin^{2}\theta_{13}\sin2\phi_{13}\,.
\end{eqnarray}

The lepton mixing matrices corresponding to the two kinds of permutations $(P_{l}, P_{\nu})=(P_{13}.P_{12}, P_{13})$ and $(P_{l}, P_{\nu})=(P_{12}, P_{13})$ are related to each other by the exchange of the second and third rows. Thus the atmospheric angle $\theta_{23}$ and Dirac CP violation phase $\delta^{\ell}$ become $\pi/2-\theta_{23}$ and $\delta^{\ell}+\pi$ respectively, while the other mixing angles $\theta_{12}$, $\theta_{13}$ and the Majorana CP phases $\phi_{12}$ and $\phi_{13}$ remain unchanged.

\begin{table}[!h]
\footnotesize
\begin{center}
\begin{tabular}{|c|c|c|c|c|c|c|c|c|c|}
\hline
& \multirow{2}{*}{$(P_{l}, P_{\nu})$} & \multirow{2}{*}{$\chi^{2}_{\mathrm{min}}$}  & \multirow{2}{*}{$(\theta^{\mathrm{bf}}_{l},\theta^{\mathrm{bf}} _{\nu})/\pi$} & \multirow{2}{*}{$\sin^{2}\theta_{13}$} & \multirow{2}{*}{$\sin^{2}\theta_{12}$} & \multirow{2}{*}{$\sin^{2}\theta_{23}$} & \multirow{2}{*}{$\delta^{\ell}/\pi$} & $\phi_{12}/\pi$ & $\phi_{13}/\pi$ \\
& & & & & & & &(mod 1/2) &(mod 1/2)\\
\hline
\multirow{8}{*}{NO} &\multirow{4}{*}{$(P_{12}, P_{13})$} & \multirow{4}{*}{20.078} &(0.433,0.938) & \multirow{4}{*}{0.022} & \multirow{4}{*}{0.318} & \multirow{4}{*}{0.511} & \multirow{2}{*}{0.539} & \multirow{2}{*}{0.397} & \multirow{2}{*}{0.468} \\
\cline{4-4}
 & & & (0.567,0.562) & & & & & & \\
\cline{4-4}
\cline{8-10}
& & & (0.433,0.562) & & & &\multirow{2}{*}{1.461} &\multirow{2}{*}{0.103} &\multirow{2}{*}{0.032} \\
\cline{4-4}
& & & (0.567,0.938) & & & & & & \\
\cline{2-10}
& \multirow{4}{*}{$(P_{13}.P_{12}, P_{13})$} & \multirow{4}{*}{37.057} & (0.433,0.938) & \multirow{4}{*}{0.022} & \multirow{4}{*}{0.318} & \multirow{4}{*}{0.489} &\multirow{2}{*}{1.539} &\multirow{2}{*}{0.397} &\multirow{2}{*}{0.468} \\
\cline{4-4}
& & & (0.567,0.562) & & & & & & \\
\cline{4-4}
\cline{8-10}
 & & & (0.433,0.562) & & & &\multirow{2}{*}{0.461} & \multirow{2}{*}{0.103} & \multirow{2}{*}{0.032} \\
\cline{4-4}
& & & (0.567,0.938) & & & & & & \\ \hline
\multirow{8}{*}{IO} &\multirow{4}{*}{$(P_{12}, P_{13})$} & \multirow{4}{*}{15.352} &(0.432,0.938) & \multirow{4}{*}{0.022} & \multirow{4}{*}{0.318} & \multirow{4}{*}{0.511} & \multirow{2}{*}{0.539} & \multirow{2}{*}{0.396} & \multirow{2}{*}{0.468} \\
\cline{4-4}
 & & & (0.568,0.562) & & & & & & \\
\cline{4-4}
\cline{8-10}
& & & (0.432,0.562) & & & &\multirow{2}{*}{1.461} &\multirow{2}{*}{0.104} &\multirow{2}{*}{0.032} \\
\cline{4-4}
& & & (0.568,0.938) & & & & & & \\
\cline{2-10}
& \multirow{4}{*}{$(P_{13}.P_{12}, P_{13})$} & \multirow{4}{*}{27.629} & (0.432,0.938) & \multirow{4}{*}{0.022} & \multirow{4}{*}{0.318} & \multirow{4}{*}{0.489} &\multirow{2}{*}{1.539} &\multirow{2}{*}{0.396} &\multirow{2}{*}{0.468} \\
\cline{4-4}
& & & (0.568,0.562) & & & & & & \\
\cline{4-4}
\cline{8-10}
 & & & (0.432,0.562) & & & &\multirow{2}{*}{0.461} & \multirow{2}{*}{0.104} & \multirow{2}{*}{0.032} \\
\cline{4-4}
& & & (0.568,0.938) & & & & & & \\
\hline
\end{tabular}
\caption{\label{tab:Z2xCP-nu-charlep-BF} $\chi^{2}$ analysis for the residual symmetries  $Z^{st^2su}_2\times X_{l}$ in the charged lepton sector and $Z^{s}_2\times X_{\nu}$ in the neutrino sector with $X_{l}=t^2$ and $X_{\nu}=su$. We give the best fit values $\theta^{\mathrm{bf}}_{l}$ and $\theta^{\mathrm{bf}}_{\nu}$ for $\theta_{l}$ and $\theta_{\nu}$ corresponding to $\chi^2_{\mathrm{min}}$. We also list the mixing angles and $CP$ violating phases at the best fit point. }
\end{center}
\end{table}

\begin{figure}[!h]
\centering
\includegraphics[width=0.4\textwidth]{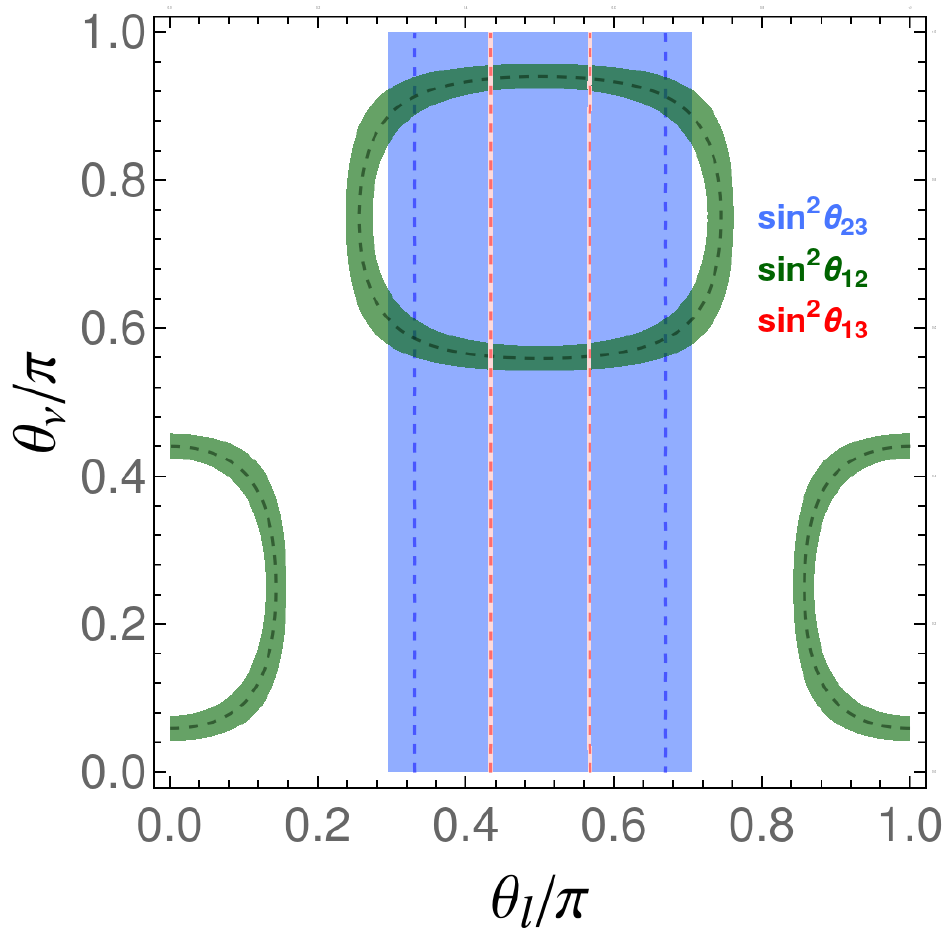}~~
\includegraphics[width=0.4\textwidth]{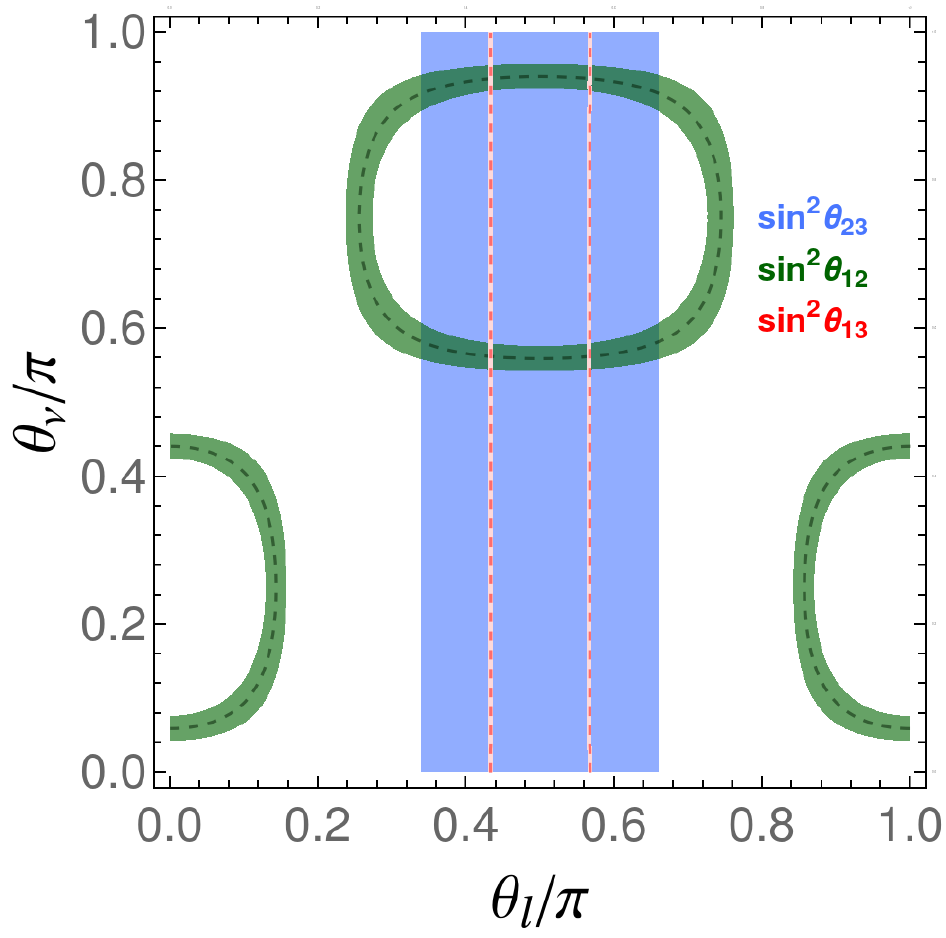}
\caption{\label{fig:Z2xCP-nu-char-S4-contour} Contour plots of $\sin^2\theta_{ij}$ in the plane $\theta_{\nu}$ versus $\theta_{l}$ for the residual symmetries $Z^{st^2su}_2\times X_{l}$ in the charged lepton sector and $Z^{s}_2\times X_{\nu}$ in the neutrino sector with $X_{l}=t^2$ and $X_{\nu}=su$. The red, blue and green areas are the $3\sigma$ regions of $\sin^{2}\theta_{13},\sin^{2}\theta_{23}$ and $\sin^{2}\theta_{12}$ respectively. The dashed lines correspond to the best fit mixing angle values taken from~\cite{deSalas:2020pgw,10.5281/zenodo.4726908}. }
\end{figure}

The results of the $\chi^2$ analysis are presented in table~\ref{tab:Z2xCP-nu-charlep-BF}. Moreover, a numerical analysis is performed, with both $\theta_{l}$ and $\theta_{\nu}$ varying freely in the range of $0$ to $\pi$, requiring all the three lepton mixing angles to lie in the experimental $3\sigma$ regions~\cite{deSalas:2020pgw,10.5281/zenodo.4726908}. In figure~\ref{fig:Z2xCP-nu-char-S4-contour} we display the $3\sigma$ contour regions for $\sin^2\theta_{12}$, $\sin^2\theta_{13}$ and $\sin^2\theta_{23}$,
as well as their experimental best fit values in the $\theta_{l}-\theta_{\nu}$ plane.  Dashed (solid) lines are the best fit mixing-angle values~\cite{deSalas:2020pgw,10.5281/zenodo.4726908}. Left (right) panels are for $(P_{l}, P_{\nu})=(P_{12}, P_{13})$ and $(P_{l}, P_{\nu})=(P_{13}.P_{12}, P_{13})$, respectively. One sees that the lepton mixing angles can be accommodated in the small regions around the best fit points.

\begin{figure}[!h]
\centering
\includegraphics[width=0.4\textwidth]{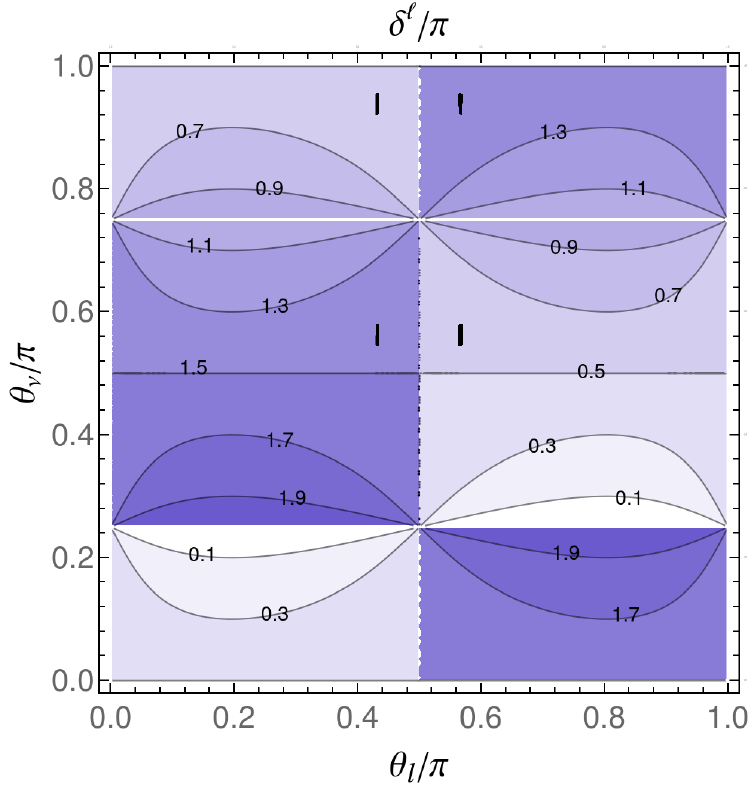}\\
\includegraphics[width=0.4\textwidth]{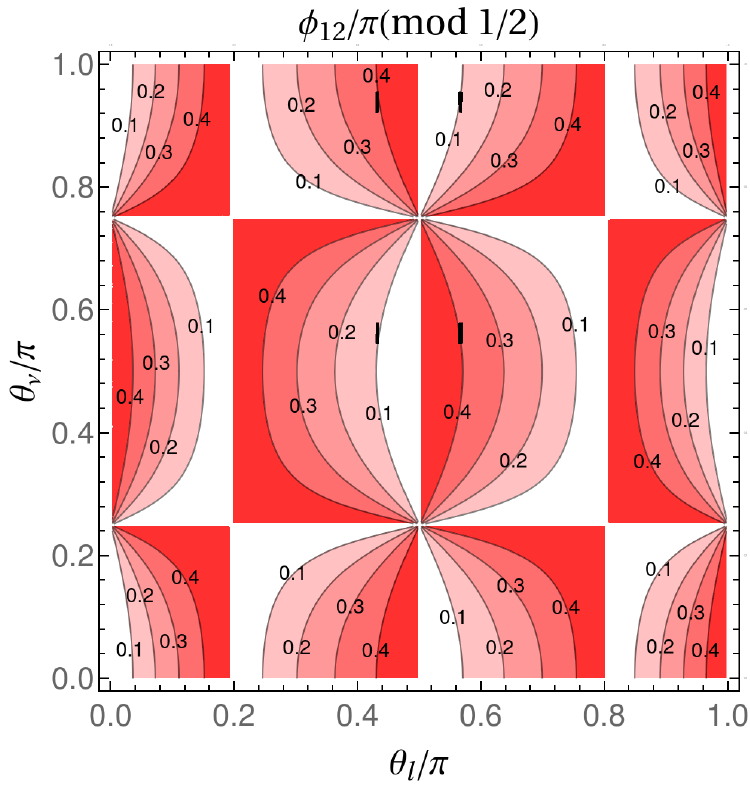}
\includegraphics[width=0.4\textwidth]{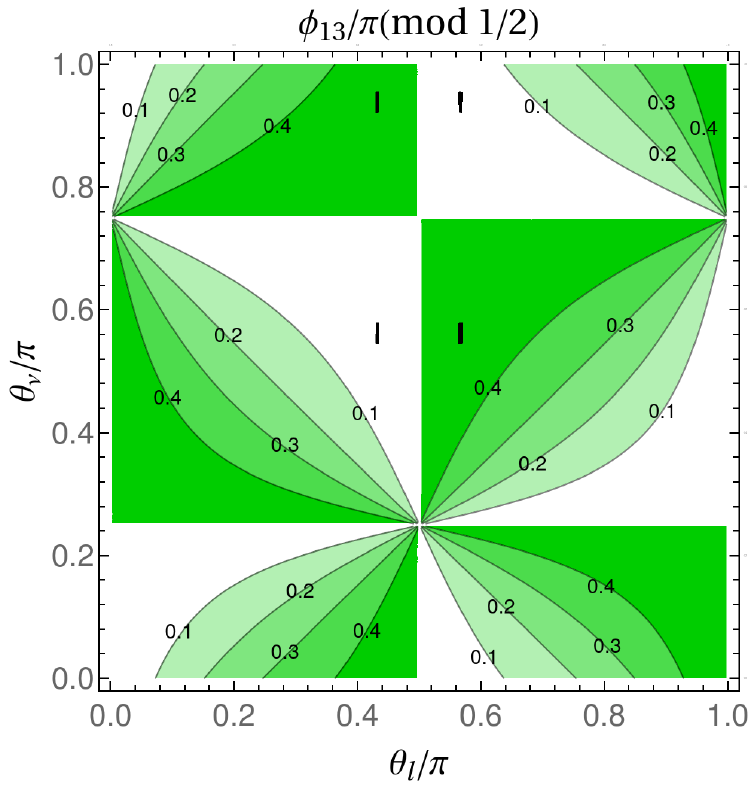}
\caption{\label{fig:contour_CP_phases} Contour plots of the $CP$ violation phases $\delta^{\ell}$, $\phi_{12}$ and $\phi_{13}$ in the $\theta_{l}-\theta_{\nu}$ plane, where the residual symmetries are $Z^{st^2su}_2\times X_{l}$ in the charged lepton sector and $Z^{s}_2\times X_{\nu}$ in the neutrino sector with $X_{l}=t^2$ and $X_{\nu}=su$.
In the black areas all three lepton mixing angles lie within their experimental $3\sigma$ ranges. Here we choose the row and column permutations $(P_{l}, P_{\nu})=(P_{12}, P_{13})$, the Dirac CP phase $\delta^{\ell}$ changes to $\delta^{\ell}+\pi$ while the Majorana phases $\phi_{12}$ and $\phi_{13}$ are invariant for $(P_{l}, P_{\nu})=(P_{13}.P_{12}, P_{13})$.  }
\end{figure}

\begin{figure}[hptb!]
\centering
\begin{tabular}{>{\centering\arraybackslash} m{7.5cm} >{\centering\arraybackslash} m{7.5cm} }
\includegraphics[width=0.5\textwidth]{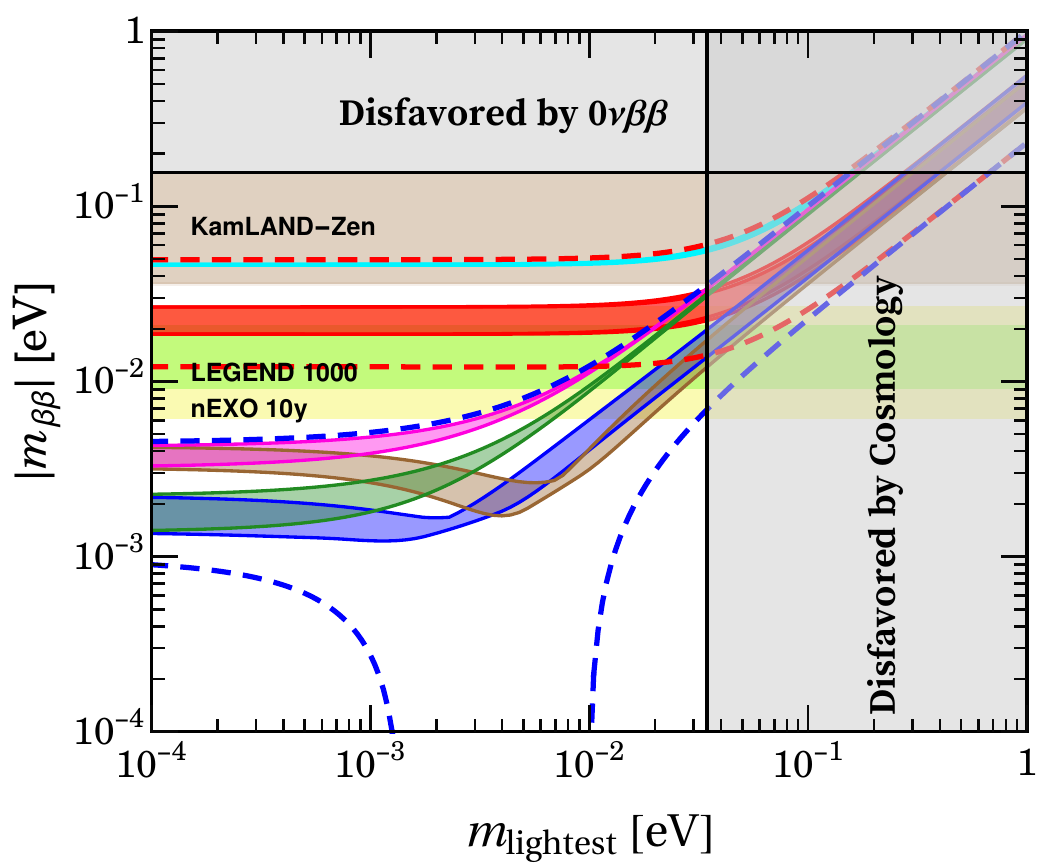}&~~
\includegraphics[width=0.30\textwidth]{fig/note_v5.pdf}
\end{tabular}
\caption{The effective Majorana neutrino mass $m_{\beta\beta}$ versus the lightest neutrino mass, for residual symmetries $Z^{st^2su}_2\times X_{l}$ in the charged lepton and $Z^{s}_2\times X_{\nu}$ in the neutrino sector, with $X_{l}=t^2$ and $X_{\nu}=su$. Here we adopt the same convention for the different bands as used in figure~\ref{fig:onbb-Z2xCP-S4}. }
\label{fig:mee-Z2xCP-nu-charg-S4}
\end{figure}

In figure~\ref{fig:contour_CP_phases} we show the contour plots of the CP violation phases $\delta^{\ell}$, $\phi_{12}$ and $\phi_{13}$ in the plane $\theta_{\nu}$ versus $\theta_{l}$.
The black areas denote the regions where the lepton mixing angles are compatible with oscillation data within $3\sigma$. These will be testable at forthcoming long baseline neutrino oscillation experiments. Since all the lepton mixing angles and CP phases are predicted to lie in narrow regions, we also have tight predictions for the effective Majorana mass of neutrinoless double beta decay, as shown in figure~\ref{fig:mee-Z2xCP-nu-charg-S4}.

\subsection{Quark and lepton mixing from a common flavour group}
\label{subsec:lepton-quark-mixing-flavor-CP}

Discrete flavour symmetries are particularly suitable to account for the large lepton mixing angles. As discussed in~\cite{deAdelhartToorop:2011re,Lam:2007qc,Blum:2007jz,Holthausen:2013vba,Araki:2013rkf,Yao:2015dwa,deMedeirosVarzielas:2016fqq} they can also address the quark mixing pattern. Assuming that the flavour group of quarks is broken down to different residual subgroups in the up- and down-quark sectors, only the Cabibbo mixing between the first two quark families can be generated with only flavour symmetry~\cite{Lam:2007qc,Blum:2007jz,Araki:2013rkf,Yao:2015dwa,deMedeirosVarzielas:2016fqq}. It is remarkable that the hierarchical quark mixing angles and CP violation can be explained if the flavour symmetry is extended with CP symmetry and the residual subgroups of the up- and down-quark sectors are $Z^{g_u}_2\times X_u$ and $Z^{g_d}_2\times X_d$ respectively~\cite{Li:2017abz,Lu:2019gqp}, where both CP transformations $X_u$ and $X_d$ are unitary and symmetric. The diagonalization matrices $U_{u}$ and $U_d$ would be restricted to be of the following form
\begin{equation}
U_u=\Sigma_u R_{23}(\theta_u) P_u Q_u\,,~~~U_d=\Sigma_d R_{23}(\theta_d) P_d Q_d
\end{equation}
where $P_{u,d}$ are three dimensional permutation matrices, $Q_{u,d}$ are generic phase matrices, and $\Sigma_{u,d}$ are the Takagi factorizations of $X_{u,d}$, fulfilling
\begin{eqnarray}
\nonumber&&X_{u}=\Sigma_{u}\Sigma_{u}^T\,,~~~~\Sigma_{u}^{\dagger}\rho(g_{u})\Sigma_{u}=\pm\text{diag}(1, -1, -1)\,,\\
&&X_{d}=\Sigma_{d}\Sigma_{d}^T\,,~~~~\Sigma_{d}^{\dagger}\rho(g_{d})\Sigma_{d}=\pm\text{diag}(1,  -1, -1)\,.
\end{eqnarray}
As a consequence, the CKM mixing matrix is predicted as
\begin{equation}
\label{eq:U-Z2xCP-up-down}V_{CKM}=U^{\dagger}_{u}U_{d}=Q^{\dagger}_uP^{T}_uR^{T}_{23}(\theta_u)\Sigma^{\dagger}_u\Sigma_{d}R_{23}(\theta_{d})P_{d}Q_{d} \,,
\end{equation}
which only depends on two free rotation angles $\theta_u$ and $\theta_d$, limited in the range $0\leq\theta_{u,d}<\pi$. Notice that both $Q_{u}$ and $Q_d$ are unphysical, as they can be absorbed into quark fields. We take the flavour symmetry as the dihedral group $D_n$ which has only one- and two-dimensional irreducible representations, as shown in~\ref{sec:dihedral-Group}.

Since the top quark is much heavier than the others, the first two families of left-handed quarks are assigned to a doublet of $D_n$, while the third generation is a $D_n$ singlet. Without loss of generality, we can take
\begin{equation}
\begin{pmatrix}
Q_1\\
Q_2
\end{pmatrix}\sim\mathbf{2}_1,~~~Q_3\sim\mathbf{1}_1\,,
\end{equation}
where $Q_1\equiv\left(u_L, d_L\right)^{T}$, $Q_2\equiv\left(c_L, s_L\right)^{T}$, and $Q_3\equiv\left(t_L, b_L\right)^{T}$. The dihedral group and CP symmetry are broken down to $Z_2^{SR^{z_u}}\times X_u$ in up-quark sector and $Z_2^{SR^{z_d}}\times X_d$ in down-quark sector with
$X_u=R^{-z_u+x_u}$ and $X_d=R^{-z_d+x_d}$, where $z_{u}, z_{d}=0,1,\dots,n-1$, $x_u=x_d=0$ for odd $n$ and $x_u, x_d=0, n/2$ if the group index $n$ is even. The Takagi factorization of the residual symmetry $Z_{2}^{SR^{z}}\times R^{-z+x}$ is determined to be
\begin{equation}
\Sigma=\frac{i^{2x/n}}{\sqrt{2}}\begin{pmatrix}
-e^{-\frac{i\pi z }{n}}  ~& 0 ~&  e^{-\frac{i\pi z }{n}} \\
e^{\frac{i\pi z }{n}}   ~& 0  ~&  e^{\frac{i\pi z }{n}} \\
0 ~& \sqrt{2}\;i^{-2x/n} ~& 0
\end{pmatrix}\,.
\end{equation}
From our master formula in Eq.~\eqref{eq:U-Z2xCP-up-down}, we find that the quark mixing matrix takes the following form,
\begin{equation}
\label{eq:VCKM_Z2CP-updown}
V_{CKM}=\begin{pmatrix}
\cos\varphi_{1} &~ -c_{d}\sin\varphi_{1} &~ s_{d}\sin\varphi_{1} \\
c_{u}\sin\varphi_{1} &~ s_{d}s_{u}e^{i\varphi_{2}}+c_{d}c_{u}\cos\varphi_{1} &~ c_{d}s_{u}e^{i\varphi_{2}}-s_{d}c_{u}\cos\varphi_{1}\\
-s_{u}\sin\varphi_{1} &~ s_{d}c_{u}e^{i\varphi_{2}}-c_{d}s_{u}\cos\varphi_{1} &~ c_{d}c_{u}e^{i\varphi_{2}}+s_{d}s_{u}\cos\varphi_{1}
\end{pmatrix}\,,
\end{equation}
up to row and column permutations with
\begin{equation}
\varphi_1=\frac{(z_u-z_d)\pi}{n},~~~\varphi_2=\frac{(x_u-x_d)\pi}{n}
\end{equation}
and $c_{d}\equiv \cos\theta_{d}$, $s_{d}\equiv \sin\theta_{d}$, $c_{u}\equiv\cos\theta_{u}$, $s_{u}\equiv\sin\theta_{u}$.
The parameters $\varphi_1$ and $\varphi_2$ depend on the choice of residual symmetry, and they can take the following discrete values
\begin{eqnarray}
\nonumber&&\varphi_{1}~(\text{mod}~2\pi)=0, \frac{1}{n}\pi, \frac{2}{n}\pi, \dots, \frac{2n-1}{n}\pi\,,\\
&&\varphi_{2}~(\text{mod}~2\pi)=0, \frac{1}{2}\pi, \frac{3}{2}\pi\,.
\end{eqnarray}
One can straightforwardly extract the quark mixing parameters from Eq.~\eqref{eq:VCKM_Z2CP-updown}. Eliminating the free parameters $\theta_u$ and $\theta_d$, one obtains the following correlations among the quark mixing angles and CP phase~\cite{Lu:2019gqp},
\begin{equation}
\label{eq:quark-corr-Dn-CP}\cos^2\theta^q_{13}\cos^{2}\theta^q_{12}=\cos^2\varphi_{1},~~~\sin\delta^{q}\simeq\frac{\sin2\varphi_{1}\sin\varphi_2}{\sin2\theta^{q}_{12}\cos^2\theta^{q}_{13}\cos\theta^{q}_{23}}\,.
\end{equation}
The experimental data on the CKM matrix can be well accommodated for $\varphi_{1}=\pi/14$, $\varphi_{2}=\pi/2$, which can be achieved from the $D_{14}$ flavour group with the residual symmetry indices
$z_u=1$, $z_d=0$, $x_u=7$, $x_d=0$. The best-fit values of $\theta_{u, d}$ and mixing parameters are determined to be,
\begin{eqnarray}
\nonumber&&\theta_{u}=0.01237\pi,~~\quad \theta_{d}=0.99473\pi,~~\quad \sin\theta_{12}^{q}=0.22249\,,\\
&&\sin\theta_{13}^{q}=0.00369,\quad \sin\theta_{23}^{q}=0.04206,\quad J^{q}_{CP}=3.104\times 10^{-5}\,.
\end{eqnarray}
Notice that $\sin\theta_{13}^{q}$, $\sin\theta_{23}^{q}$ and $J^{q}_{CP}$ are consistent with the global fit results of the UTfit collaboration~\cite{UTfit}. The mixing angle $\sin\theta_{12}^{q}$ is about $1\%$ smaller than its measured value, so that higher-order corrections in a concrete model are needed to reconcile it with the data.

The $D_{14}$ flavour group can also explain the lepton flavour structure if it is broken down to $Z_{2}^{SR^{z_{l}}}\times X_l$ and $Z_{2}^{SR^{z_{\nu}}}\times X_{\nu}$ in the charged lepton and neutrino sector, respectively, where $X_{l}=R^{-z_{l}+x_l}$, $X_{\nu}=R^{-z_{\nu}+x_{\nu}}$ with $z_{l, \nu}=0,1,\dots,13$ and $x_{l, \nu}=0, 7$. The lepton mixing matrix has the same form as Eq.~\eqref{eq:VCKM_Z2CP-updown}, the rotation angles $\theta_{u}$ and $\theta_d$ should be replaced with $\theta_l$ and $\theta_{\nu}$ respectively. Choosing the residual symmetry indices $z_{l}=4$, $z_{\nu}=0$, $x_l=7$ and $x_{\nu}=0$, we have $\varphi_1=2\pi/7$ and $\varphi_2=\pi/2$. Choosing the permutations as $P_l=P_{12}P_{23}$ and $P_{\nu}=P_{13}$, the lepton mixing angles can be accommodated for certain values of the free parameters $\theta_{l, \nu}$:
\begin{small}
\begin{eqnarray}
\nonumber&&\hskip-0.4in \theta^{\mathrm{bf}}_{e}=0.439\pi,~ \theta^{\mathrm{bf}}_{\nu}=0.811\pi,~ \chi^2_{\text{min}}=4.147,~ \sin^2\theta_{13}=0.0220,~ \sin^2\theta_{12}=0.318\,, \\
&&\hskip-0.4in \sin^2\theta_{23}=0.603,~ \delta^{\ell}/\pi=1.530,~ \phi_{12}/\pi=-0.082~(\text{mod}~1/2),~ \phi_{13}/\pi=1.474~(\text{mod}~1/2)\,.
\end{eqnarray}
\end{small}
In summary, one sees how the dihedral group as well as the residual symmetry $Z_2\times CP$ provide an interesting opportunity for model building.
Indeed, we saw how the $D_{14}$ flavour symmetry can provide a unified description of flavour mixing for both quarks and leptons.

If both left-handed quarks and leptons are assigned as irreducible triplets of the flavour symmetry and the residual symmetry is $Z_2\times CP$,
one finds that $\Delta(294)$ is the minimal flavour group that can generate realistic quark and lepton flavour mixing patterns~\cite{Lu:2018oxc}.
In contrast,  the singlet plus doublet assignment seems better than the triplet assignment. Once the CP symmetry is included, the order of the flavour symmetry group can be reduced considerably,
i.e. 28 versus 294 in this scheme.

There are also other schemes to explain the mixing patterns of quarks and leptons using flavour and CP symmetries. For instance, quark and lepton mixing patterns can arise from the stepwise breaking of these symmetries to different residual subgroups in different sectors of the theory~\cite{Hagedorn:2018gpw,Hagedorn:2018bzo}, with charged fermion mass hierarchies generated by operators with different numbers of flavons. For a concrete model with $\Delta(384)$ flavour symmetry see~\cite{Hagedorn:2018bzo}.

\subsection{ Geometrical CP violation}

It is well-known that CP symmetry is broken by complex Yukawa couplings in the SM, which leads to the CP violation in charged current interactions through the CKM matrix. However, the origin of CP violation is still a mystery. Analogous to the electroweak symmetry, the CP symmetry could be spontaneously broken by the VEVs of some scalar fields~\cite{Lee:1973iz}. In models of spontaneous CP violation, the Lagrangian is invariant under the CP symmetry so that all parameters of the scalar potential are real in a certain basis. Spontaneous CP violation is achieved through complex VEVs for the Higgs multiplets which also break the gauge symmetry. Usually the phases of the fields depend on the coupling constants in the scalar potential. \par

The phases of the Higgs multiplets could have geometrical values, independently of the potential parameters if there is an additional (accidental) CP symmetry of the potential. The resulting CP breaking vacua lead to the so-called geometrical CP violation or calculable phases~\cite{Branco:1983tn}. It has been shown that more than two Higgs doublets and non-abelian symmetry relating the Higgs multiplets are necessary conditions in order to realize the geometrical CP violation. It turns out that $\Delta(27)$ and $\Delta(54)$ are the smallest groups which lead to calculable phases~\cite{Branco:1983tn,deMedeirosVarzielas:2011zw}. \par

If one assigns three Higgs doublets $H\equiv (H_1, H_2, H_3)$ to a triplet of $\Delta(27)$ or $\Delta(54)$, the scalar potential has only one relevant phase dependent term, i.e.
\begin{equation}
\label{eq:V-GCPV}V=V_0+\left[\lambda_4\sum_{i\neq j\neq k}(H^{\dagger}_iH_j)(H^{\dagger}_iH_k)+\text{h.c.}\right]\,.
\end{equation}
The traditional CP transformation $H_i\xrightarrow{\mathcal{CP}} H^{*}_i$ forces the coupling $\lambda_4$ to be real. Then one can obtain the following two possible vacua with calculable phases~\cite{Branco:1983tn}
\begin{eqnarray}
\nonumber \langle H\rangle&=&\frac{v}{\sqrt{3}}\begin{pmatrix}
1\\
\omega\\
\omega^2
\end{pmatrix},~~~~~\lambda_4<0\,,\\
\label{eq:Gphases}\langle H\rangle&=&\frac{v}{\sqrt{3}}\begin{pmatrix}
\omega^2\\
1\\
1
\end{pmatrix},~~~~~\lambda_4>0\,.
\end{eqnarray}
The same scalar potential as Eq.~\eqref{eq:V-GCPV} and calculable phases in Eq.~\eqref{eq:Gphases} can be obtained from other non-abelian symmetry groups such as $\Delta(3n^2)$~\cite{Luhn:2007uq} and $\Delta(6n^2)$~\cite{Escobar:2008vc}, where $n$ is a multiple of 3. \par

The observed fermion masses and flavor mixings could possibly be accommodated if one properly assigns the transformations of fermion fields under the symmetry group~\cite{deMedeirosVarzielas:2011zw}. In short, the geometrical CP violation arises from the correct interplay among the scalar content, non-abelian symmetry group and CP symmetry.

\clearpage

\section{Testing flavor and CP symmetries  }                               
\label{sec:test-symmetry}                                                  

As shown above, discrete flavor and generalized CP symmetries allow us to predict the lepton mixing matrix in terms of few parameters. Eliminating free parameters generally leads to correlations among the lepton mixing angles and CP violation phases. Such predictions are often called lepton mixing \textit{sum rules} in the literature, though they are not always strictly so. For example, for the residual symmetry $\mathcal{G}_{l}=Z_n\; (n\geq3)$, $\mathcal{G}_{\nu}=Z_2\times CP$ discussed in section~\ref{sec:abelian-subgroup-g_l}, the lepton mixing matrix depends on a unique real parameter $\theta$, as shown in Eq.~\eqref{eq:Upmns-Zn-Z2CP}. The parameter $\theta$ becomes determined in terms of the precisely measured reactor mixing angle $\theta_{13}$, leading to sharp predictions for the leptonic mixing angles and CP violation phases that can be tested at current and future neutrino oscillation experiments. \par

Ultimately they could be used to distinguish different symmetry-based flavor models.
It is remarkable that, besides the mixing angles, flavor symmetry in combination with generalized CP symmetry allows us to predict both the Dirac and Majorana leptonic CP violation phases. Hence the effective neutrino mass $|m_{\beta\beta}|$ is constrained to lie within narrow regions, as can be seen from figures~\ref{fig:onbb-Z2xCP-S4} and~\ref{fig:mee-Z2xCP-nu-charg-S4}.\par

As a result one could also test flavor and CP symmetries by confronting with the data of the current and forthcoming $0\nu\beta\beta$ experiments. In fact, some flavor models relate the Majorana phases to the neutrino masses, which could also lead to restrictions on the effective mass $|m_{\beta\beta}|$, and a very powerful tool to test and discriminate flavour models.

\subsection{Testing mixing predictions}

Neutrino physics has entered the precision era, providing a good opportunity for probing different flavor models. The reactor angle $\theta_{13}$ is the best-measured leptonic mixing parameter. The precise measurement of nonzero $\theta_{13}$ by Daya Bay~\cite{DayaBay:2012fng,DayaBay:2018yms}, Double Chooz~\cite{DoubleChooz:2011ymz,DoubleChooz:2014kuw}, and RENO~\cite{RENO:2012mkc,RENO:2018dro} has excluded many flavor models predicting $\theta_{13}$ close to zero. The long baseline neutrino oscillation experiments NO$\nu$A~\cite{NOvA:2004blv} and T2K~\cite{T2K:2011qtm} give measurements of the leptonic mixing angle $\theta_{23}$ and also the first hint of leptonic CP violation~\cite{T2K:2018rhz,NOvA:2019cyt,T2K:2019bcf} associated to the leptonic Dirac CP violation phase $\delta_{\text{CP}}$. However, the value of $\delta_{CP}$ has not been significantly constrained by neutrino oscillation experiments. The latest data of T2K favor near-maximal Dirac CP violation phase, the atmospheric mixing angle $\theta_{23}>45^{\circ}$ and normal mass ordering~\cite{T2K:2023smv}. The data of T2K constrain the CP phase $\delta_{CP}$ in the range $\delta_{CP}=-1.97^{+0.97}_{-0.70}$, and $\delta_{CP}=0, \pi$ is excluded at more than $90\%$ confidence level~\cite{T2K:2023smv}. In comparison with T2K,
the data of NO$\nu$A exclude the CP phase in the vicinity of $\delta_{CP}=\pi/2$ at more than $3\sigma$ for the inverted mass ordering, and the values around $\delta_{CP}=3\pi/2$ in the normal ordering are disfavored at $2\sigma$ confidence~\cite{NOvA:2021nfi}. Improved CP measurements constitute the target of upcoming experiments such as DUNE. The future long baseline neutrino oscillation experiments DUNE~\cite{DUNE:2015lol,DUNE:2020ypp} and T2HK~\cite{Hyper-KamiokandeProto-:2015xww} should measure $\theta_{23}$ and $\delta_{\rm CP}$ with very good precision. It is expected that DUNE can observe the signal of lepton CP violation with $5\sigma$ significance after about 7 years if $\delta_{CP}=-\pi/2$ and after about 10 years for $50\%$ of $\delta_{CP}$ values, and CP violation can be observed with $3\sigma$ significance for $75\%$ of $\delta_{CP}$ values after about 13 years of running~\cite{DUNE:2015lol,DUNE:2020ypp}. T2HK has shown that it can expect a discovery of CP violation over $76\% (58\%)$ of the parameter space at $3\sigma (5\sigma)$~\cite{Hyper-KamiokandeProto-:2015xww}. The forthcoming medium baseline reactor neutrino experiment JUNO can make very precise measurement of the solar angle $\theta_{12}$, and the error of $\sin^2\theta_{12}$ can be reduced to the level of $0.5\%-0.7\%$~\cite{JUNO:2015zny}.

\begin{figure}[t!]
\centering
\includegraphics[width=1.0\textwidth]{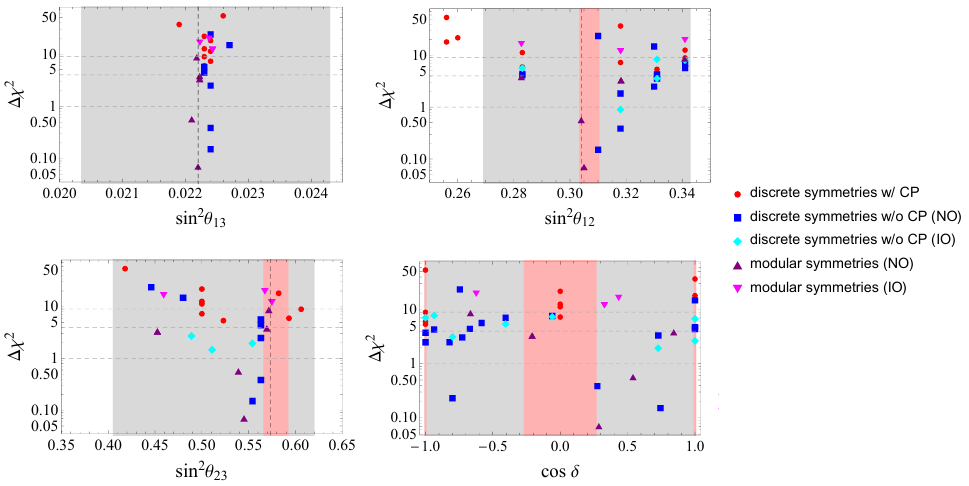}
\caption{Best-fit predictions of the models based on discrete symmetries broken to certain residual symmetries of the lepton mass matrices \cite{Petcov:2018snn,Blennow:2020snb}, and the models based on modular symmetries discussed in~\cite{Gehrlein:2020jnr,Novichkov:2020eep,Novichkov:2021evw}. The gray regions are the $3\sigma$ ranges and the dashed line is the best fit value for normal-ordered neutrino masses. The red band in the panel of $\sin^2\theta_{12}$ is the prospective $3\sigma$ range after 6 years of JUNO running~\cite{JUNO:2015zny}, and the red regions in the panels of $\sin^2\theta_{23}$ and $\cos\delta$ are the prospective $3\sigma$ ranges after 15 years of DUNE running~\cite{DUNE:2020ypp}. This figure is taken from~\cite{Gehrlein:2022nss}. }
\label{fig:bf-predictions-models-Snowmass}
\end{figure}

We show in figure~\ref{fig:bf-predictions-models-Snowmass} the best fit predictions for the mixing parameters in some typical models based on discrete flavor symmetry with or without generalized CP symmetry and modular symmetry. We see that the synergy between JUNO and long baseline neutrino experiments DUNE and T2HK  will be extremely powerful for testing the huge number of flavor symmetry models.

As an illustration, we give an extensively discussed prediction for lepton mixing angles as follows~\cite{Marzocca:2013cr,Ballett:2014dua},
\begin{equation}
\cos\delta_{CP} = \frac{\tan\theta_{23}}{\sin2\theta_{12}\, \sin\theta_{13}}
\left[\cos2\theta^\nu_{12} + \left(\sin^2\theta_{12} - \cos^2\theta^\nu_{12}\right)
\left(1 - \cot^2\theta_{23}\,\sin^2\theta_{13}\right)\right]\,,
\label{eq:solar-sum-rule}
\end{equation}
which generally arises when considering charged-lepton corrections to neutrino mixing matrices which are completely fixed by symmetry with the neutrino angle $\theta^{\nu}_{13}=0$ and charged lepton rotation angle $\theta^{e}_{13}=0$. This prediction gives $\cos\delta_{CP}$ as a function of the measured lepton mixing angles and one fixed parameter $\theta^\nu_{12}$ determined by the underlying discrete symmetry. It is noticeable that Eq.~\eqref{eq:solar-sum-rule} is specified by fixing the value of just one parameter, the angle $\theta^{\nu}_{12}$. Thus one can enumerate the viable models of this type by deriving the values of $\theta^{\nu}_{12}$ from symmetry considerations. This leads us the following well-motivated sum rules characterized by specific values of  $\theta^\nu_{12}$, namely the one based on TBM mixing with $\theta^\nu_{12}= \arcsin(1/\sqrt3) \approx 35.26^{\circ}$~\cite{Harrison:2002er,Xing:2002sw,He:2003rm}, the one based on BM mixing
with $\theta^\nu_{12}=45^{\circ}$~\cite{Barger:1998ta}, the one based on the type A golden ratio mixing (GRA) with $\theta^\nu_{12}=\arctan(1/\phi_g)\approx31.72^{\circ}$~\cite{Datta:2003qg,Kajiyama:2007gx}, the one based on the type B golden ratio mixing (GRB) with $\theta^\nu_{12}=\arccos(\phi_g/2)=36^{\circ}$~\cite{Rodejohann:2008ir,Adulpravitchai:2009bg}, and the one based on hexagonal (HG) mixing with $\theta^{\nu}_{12}=30^{\circ}$~\cite{Albright:2010ap,Kim:2010zub}. \par

Using the sum rule of Eq.~\eqref{eq:solar-sum-rule} and plugging into  the best fit values of the neutrino mixing angles for NO~\cite{deSalas:2020pgw}, one can straightforwardly determine the value of $\cos\delta_{CP}$ and the CP phase $\delta_{CP}$ for the above mentioned values of $\theta^{\nu}_{12}$,
\begin{eqnarray}
\nonumber\text{TBM}:~&&\cos\delta_{CP}\approx-0.08,~~~\delta_{CP}\approx\pm94.65^{\circ}\,,
\\
\nonumber \text{GRA}:~&&\cos\delta_{CP}\approx0.41,~~~\delta_{CP}\approx\pm66.09^{\circ}\,,\\
\nonumber\text{GRB}:~&&\cos\delta_{CP}\approx-0.18,~~~\delta_{CP}\approx\pm100.65^{\circ}\,,\\
\label{eq:cosdelta-solar-HG}\text{HG}:~&&\cos\delta_{CP}\approx0.63,~~~\delta_{CP}\approx\pm50.90^{\circ}\,,
\end{eqnarray}
Notice that there are two values of $\delta_{CP}$ of opposite sign for each value of $\cos\delta_{CP}$. The sum rule of Eq.~\eqref{eq:solar-sum-rule} for the BM case, $\theta^{\nu}_{12}=45^{\circ}$, is not compatible with the current best fit values of the lepton mixing angles~\cite{deSalas:2020pgw}, and will be dropped hereafter. We also take into account the experimental uncertainties of the three lepton mixing angles by varying $\theta_{12}$, $\theta_{23}$, and $\theta_{13}$ in their $3\sigma$ experimentally allowed regions. The ranges of the CP violation $\delta_{\mathrm{CP}}$ obtained from Eq.~\eqref{eq:solar-sum-rule} are summarized in table~\ref{tab:deltaCP-ranges-solar-sum-rule}. We see that out of all mixing angles, the $3\sigma$ allowed range of $\theta_{12}$ causes the largest uncertainty in $\delta_{CP}$ resulting from Eq.~\eqref{eq:solar-sum-rule}.

\begin{table}[t!]
\centering
\resizebox{1.0\textwidth}{!}{
\begin{tabular}{|c|cccc|}
\hline
& \multicolumn{4}{c|}{Ranges of $\delta_{CP}$ obtained by varying} \\
& $\theta_{12}$ in $3\sigma$ & $\theta_{23}$ in $3\sigma$ & $\theta_{13}$ in $3\sigma$ & $\theta_{12}\&\theta_{23}\&\theta_{13}$ in $3\sigma$ \\
\hline

TBM & $\pm[240.54^{\circ},289.19^{\circ}]$ & $\pm[264.58^{\circ},268.06^{\circ}]$   & $\pm[264.86^{\circ},265.89^{\circ}]$ & $\pm[235.27^{\circ},291.21^{\circ}]$ \\
GRA & $\pm[271.01^{\circ},322.93^{\circ}]$   & $\pm[289.72^{\circ},295.36^{\circ}]$  & $\pm[292.95^{\circ},294.82^{\circ}]$ & $\pm[270.22^{\circ},332.89^{\circ}]$\\
GRB & $\pm[233.08^{\circ},283.21^{\circ}]$  & $\pm[258.11^{\circ},263.51^{\circ}]$   & $\pm[258.57^{\circ},260.20^{\circ}]$ & $\pm[226.19^{\circ},284.41^{\circ}]$ \\
HG & $\pm[284.70^{\circ},360^{\circ}]$   & $\pm[300.64^{\circ},312.11^{\circ}]$   & $\pm[307.15^{\circ},310.98^{\circ}]$  & $\pm[283.21^{\circ},360^{\circ}]$\\
\hline
\end{tabular}}
\caption{The prediction of the sum rule Eq.~\eqref{eq:solar-sum-rule} for the allowed ranges of $|\delta_{CP}|$ due to the present $3\sigma$ uncertainties
in the values of the neutrino mixing angles. Here we vary at least one lepton mixing angle in its corresponding $3\sigma$ intervals for the NO spectrum~\cite{deSalas:2020pgw}. }
\label{tab:deltaCP-ranges-solar-sum-rule}
\end{table}

\begin{figure}[t!]
\centering
\includegraphics[width=0.60\textwidth]{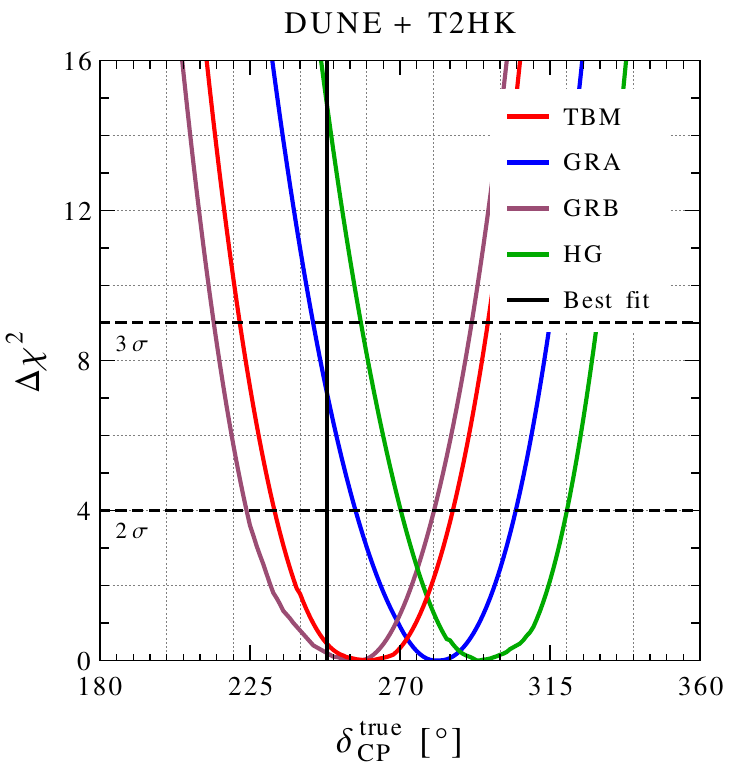}
\caption{Compatibility of the prediction in Eq.~\eqref{eq:solar-sum-rule} with any potentially true value of the Dirac CP phase $\delta_{CP}$ in the range $\delta_{CP}\in [180^{\circ}, 360^{\circ}]$. Four values of $\theta^{\nu}_{12}$ are considered: $\theta^\nu_{12}=\arcsin(1/\sqrt3) \approx 35.26^{\circ}$ for TBM,
$\theta^\nu_{12}=\arctan(1/\phi_g)\approx31.72^{\circ}$ for GRA,
$\theta^\nu_{12}=36^{\circ}$ for GRB, and $\theta^\nu_{12}=30^{\circ}$ for HG. From~\cite{Agarwalla:2017wct}. }
\label{fig:chi2-vs-deltaCP}
\end{figure}

Future long baseline experiments DUNE and T2HK will be able to make precision measurement of $\delta_{CP}$ and $\theta_{23}$. The combination of DUNE and T2HK provides better sensitivity to $\delta_{CP}$ than either of these two experiments in isolation. The prospective DUNE+T2HK data should allow one to test the predictions for $\cos\delta_{CP}$, as shown in figure~\ref{fig:chi2-vs-deltaCP}. Given the $3\sigma$ range $\delta_{CP}\in[127.80^{\circ}, 358.20^{\circ}]$ of the Dirac phase $\delta_{CP}$ from the latest global analysis~\cite{deSalas:2020pgw}, only $\delta_{CP}$ values in the interval of $180^{\circ}$ to $360^{\circ}$ are considered in figure~\ref{fig:chi2-vs-deltaCP}.  We see that a significant part of the true value of $\delta_{CP}$ gets disfavoured at more than $3\sigma$ for each symmetry predicted value of the angle $\theta^{\nu}_{12}$. Furthermore, detailed analysis showed that future facilities DUNE and T2HK in combination with JUNO could distinguish the different cases of $\theta^{\nu}_{12}$~\cite{Agarwalla:2017wct,Ballett:2014dua}.

In summary, the existence of mixing predictions is a characteristic feature of flavor symmetry models, as explicitly shown in Eqs.~(\ref{eq:cor-RTBM-A}, \ref{eq:cor-RTBM-C}, \ref{eq:corr-revamp-cGR}, \ref{eq:correlation_trimaximal}, \ref{eq:corre_theta23_deltaCP}, \ref{mixing-pars-S4CP-I}, \ref{mixing-pars-S4CP-II}, \ref{eq:caseIV-Z2xCP-S4}, \ref{eq:quark-corr-Dn-CP}). These relations highlight the testability of flavor symmetry, and there are many other possible relations among the lepton mixing parameters in the literature~\cite{Ballett:2014dua,King:2005bj,Masina:2005hf,Antusch:2005kw,Ballett:2013wya,Marzocca:2013cr,Girardi:2015vha}. The phenomenological implications of these sum rules and the prospects of testing them at precision neutrino facilities have been extensively discussed~\cite{Ballett:2013wya,Ballett:2014dua,Agarwalla:2017wct,Blennow:2020snb,Blennow:2020ncm,Ding:2019zhn,Costa:2023bxw}.

\subsection{Testing mass sum rules}

The neutrino mass matrix leading to the three neutrino masses in discrete flavor symmetry models typically involves a reduced set of parameters. Indeed, several discrete flavor-symmetry-based models yield a constrained neutrino mass matrix, leading to specific neutrino mass sum-rules. This is turn is phenomenologically interesting, as it can lead to predictions for the effective neutrinoless double beta mass parameter~\cite{Dorame:2011eb}. For example, one can have a \znbb lower bound on even for normal ordered neutrino spectrum~\cite{Dorame:2012zv}. \par

If the light neutrino mass matrix depends on two complex parameters~\cite{Gehrlein:2017ryu} one can extract a relation between the three complex neutrino mass eigenvalues, leading to a neutrino mass sum rule. Typically neutrino mass sum rules can arise from any neutrino mass generation mechanism in which the structure of the mass matrix is generated by two flavons. Prime examples neutrino mass sum rules are $2\tilde{m}_2+\tilde{m}_3=\tilde{m}_1$ and $2\tilde{m}^{-1}_2+\tilde{m}^{-1}_3=\tilde{m}^{-1}_1$ predicted by early $A_4$ models~\cite{Altarelli:2005yp,Altarelli:2005yx,Ma:2005sha}. For systematic categorization and consequences of neutrino mass sum rules in beta decay, $0\nu\beta\beta$ decay and cosmology see Refs.~\cite{Barry:2010yk,Dorame:2011eb,King:2013psa,Gehrlein:2015ena,Gehrlein:2016wlc}. A sample of the flavor models in the literature, includes the following twelve different neutrino mass sum rules~\cite{King:2013psa}:
\begin{align}
\nonumber &\tilde{m}_1+\tilde{m}_2=\tilde{m}_3, &&\tilde{m}_1+ \tilde{m}_3 =2\tilde{m}_2\,, \\
\nonumber &2\tilde{m}_2+\tilde{m}_3=\tilde{m}_1, &&\tilde{m}_1+\tilde{m}_2=2\tilde{m}_3\,,\\
\nonumber &\tilde{m}_1 + \frac{\sqrt{3}+1}{2} \tilde{m}_3 = \frac{\sqrt{3}-1}{2} \tilde{m}_2, && \tilde{m}_1^{-1}+\tilde{m}_2^{-1}=\tilde{m}_3^{-1}\,,\\
\nonumber &2\tilde{m}_2^{-1}+\tilde{m}_3^{-1}=\tilde{m}_1^{-1},&& \tilde{m}_1^{-1}+\tilde{m}_3^{-1}=2\tilde{m}_2^{-1}\,,\\
\nonumber &\tilde{m}_3^{-1} \pm 2i\tilde{m}_2^{-1}=\tilde{m}_1^{-1},&& \tilde{m}_1^{1/2}- \tilde{m}_3^{1/2}=2\tilde{m}_2^{1/2}\,,\\
\label{eq:nu-mass-sum-rule}&\tilde{m}_1^{1/2}+\tilde{m}_3^{1/2}=2\tilde{m}_2^{1/2},&& \tilde{m}_1^{-1/2}+\tilde{m}_2^{-1/2}=2\tilde{m}_3^{-1/2}\,,
\end{align}
where $\tilde{m}_i$ stand for the complex neutrino mass eigenvalues, which can be expressed in terms of the Majorana phases $\phi_i \in [0,2\pi)$ and the physical mass eigenvalues $m_i \ge 0$ as $\tilde{m}_{i}=m_i e^{-i\phi_i}$ with $\phi_3$ chosen to be unphysical. These sum rules appeared in models based on $A_4$, $S_4$, $A_5$, $T'$, $T_7$, $\Delta(54)$ and $\Delta(96)$ flavor symmetry, when the three neutrino mass eigenvalues can be described by two model parameters only, see Ref.~\cite{King:2013psa} for a very good discussion and references on models predicting these sum rules. In addition, five new mass sum rules have been identified in flavor models based on modular symmetries with residual symmetries~\cite{Gehrlein:2020jnr}. From Eq.~\eqref{eq:nu-mass-sum-rule} one sees that all neutrino mass sum rules can be parametrized in the following manner~\cite{King:2013psa, Gehrlein:2015ena}:
\begin{equation}
c_1 \left(m_1 e^{-i\phi_{1}}\right)^d
e^{i\Delta \chi_{13}}+
c_2 \left(m_2 e^{-i\phi_{2}}\right)^d
e^{i\Delta \chi_{23}}
+m_3^d~=0 \,,
\label{eq:parametrization-numass-SR}
\end{equation}
with $c_1, c_2>0$, where $\phi_1$ and $\phi_2$  are the Majorana phases. The parameters $c_1$, $c_2$, $d$, $\Delta \chi_{13}$, and $\Delta \chi_{23}$ characterize the sum rule, and they can be straightforwardly read out for any of the above twelve known sum rules.
Interpreting the complex numbers as vectors in the complex plane, the neutrino mass sum rule can be geometrically understood as a sum of three vectors which form a triangle in the complex plane~\cite{Barry:2010yk}. Since the sum rule in Eq.~\eqref{eq:parametrization-numass-SR} is a complex equation, it requires both real part and imaginary parts to be vanishing, i.e.,
\begin{eqnarray}
\nonumber&& c_1m_1^d\cos\beta+c_2m^d_2\cos\alpha+m^d_3=0\,,\\
\label{eq:mass-sum-rule-ReIm}&&c_1m_1^d\sin\beta+c_2m^d_2\sin\alpha=0\,,
\end{eqnarray}
with the angles $\alpha \equiv -d \phi_{2}+\Delta \chi_{23}$ and $\beta \equiv -d \phi_1+\Delta \chi_{13}$. Eq.~\eqref{eq:mass-sum-rule-ReIm} allows to express $\alpha$ and $\beta$ in terms of the parameters of the sum rule,
\begin{eqnarray}
\nonumber &&\cos\alpha=\frac{c_1^2 m_1^{2d}-c_ 2^2 m_2^{2d}-m_3^{2d}}{2 c_2 m_2^d m_3^d}\,,~~~\cos\beta=\frac{c_2^2 m_2^{2d}-c_ 1^2 m_1^{2d}-m_3^{2d}}{2 c_1 m_1^d m_3^d}\,,\\
\nonumber&&\sin\alpha=\pm\frac{\sqrt{4c^2_2m^{2d}_2m^{2d}_3-(c^2_1m^{2d}_1-c^2_2m^{2d}_2-m^{2d}_3)^2}}{2 c_2 m_2^d m_3^d}\,,\\
\label{eq:mass-sum-rule-cos}&&\sin\beta=\mp\frac{\sqrt{4c^2_2m^{2d}_2m^{2d}_3-(c^2_1m^{2d}_1-c^2_2m^{2d}_2-m^{2d}_3)^2}}{2 c_1 m_1^d m_3^d}\,,
\end{eqnarray}
which relate Majorana CP phases with light neutrino masses. Given that $|\cos\alpha|\leq1$, $|\sin\alpha|\leq1$, $|\cos\beta|\leq1$ and $|\sin\beta|\leq1$, it follows that the validity of the sum rule requires the three neutrino masses to satisfy the following triangle inequalities,
\begin{eqnarray}
\nonumber && |c_1m^d_1e^{i\beta}|\leq |c_2m^d_2e^{i\alpha}|+|c_3m^d_3|,\\
\nonumber &\text{and}&~~|c_2m^d_2e^{i\alpha}|\leq |c_1m^d_1e^{i\beta}|+|c_3m^d_3|,\\
\label{eq:triangle-sum-rule} &\text{and}&~~|c_3m^d_3|\leq |c_1m^d_1e^{i\beta}|+|c_2m^d_2e^{i\alpha}|\,,
\end{eqnarray}
which implies a triangle formed out of the three vectors $c_1m^d_1e^{i\beta}$, $c_2m^d_2e^{i\alpha}$ and $c_3m^d_3$ in the complex plane. Given the solar and atmospheric neutrino mass squared differences $\Delta m^2_{\text{sol}}\equiv m_2^2-m_1^2=7.50^{+0.22}_{-0.20}\times 10^{-5}~\text{eV}^2$ and $\Delta m^2_{\text{atm}}\equiv|m_3^2-m_1^2|=2.55\,(2.45)^{+0.02}_{-0.03}\times 10^{-3}~\text{eV}^2$~\cite{deSalas:2020pgw} for NO (IO), the light neutrino masses are related to the smallest mass $m_1$ ($m_3$) as follows,
\begin{eqnarray}
\nonumber&& \text{NO}:~m_2=\sqrt{\Delta m^2_{\text{sol}}+m^2_1},~~~m_3=\sqrt{\Delta m^2_{\text{atm}}+m^2_1}\,,\\
&&\text{IO}:~m_1=\sqrt{\Delta m^2_{\text{atm}}+m^2_3},~~~m_2=\sqrt{\Delta m^2_{\text{atm}}+\Delta m^2_{\text{sol}}+m^2_3}\,.
\end{eqnarray}
Thus mass rules usually lead to a lower limit on the lightest neutrino mass through the triangle inequalities in Eq.~\eqref{eq:triangle-sum-rule}, and some sum rules may only allow for a certain mass ordering. For instance, neutrino masses can only be normal ordered for the sum rule  $2\tilde{m}_2+\tilde{m}_3=\tilde{m}_1$ since the triangle inequality $m_3+m_1\geq2m_2$ cannot be fulfilled for IO. Solving the inequality $m_3\leq m_1+2m_2$ in the case of NO, one obtain the lower limit on $m_1$,
\begin{equation}
m_1\geq \sqrt{\frac{\Delta m^2_{\text{atm}}}{8}}\,(1-3r)\simeq 0.016\;\text{eV}\,,
\end{equation}
where the exact result has been expanded in terms of the small ratio $r\equiv \Delta m^2_{\text{sol}}/\Delta m^2_{\text{atm}}$. Another benchmark sum rule $2\tilde{m}^{-1}_2+\tilde{m}^{-1}_3=\tilde{m}^{-1}_1$ works for both mass orderings, and the limits on the lightest neutrino mass from the triangle inequalities are found to be
\begin{eqnarray}
\nonumber&& \text{NO}:~ 0.0043\;\text{eV}\simeq\sqrt{\dfrac{\Delta m^2_{\text{sol}}}{3}}\left(1-\dfrac{4\sqrt{3r}}{9}\right)\leq m_1 \leq\sqrt{\dfrac{\Delta m^2_{\text{sol}}}{3}}\left(1+\dfrac{4\sqrt{3r}}{9}\right)\simeq 0.0057\;\text{eV}\;,\\
&& \text{IO}:~m_3\geq\sqrt{\frac{\Delta m^2_{\text{atm}}}{8}}\,\left(1+\dfrac{r}{3}\right)\simeq 0.018\;\text{eV}\,.
\end{eqnarray}

\begin{figure}[h!]
\centering
\includegraphics[scale=0.56]{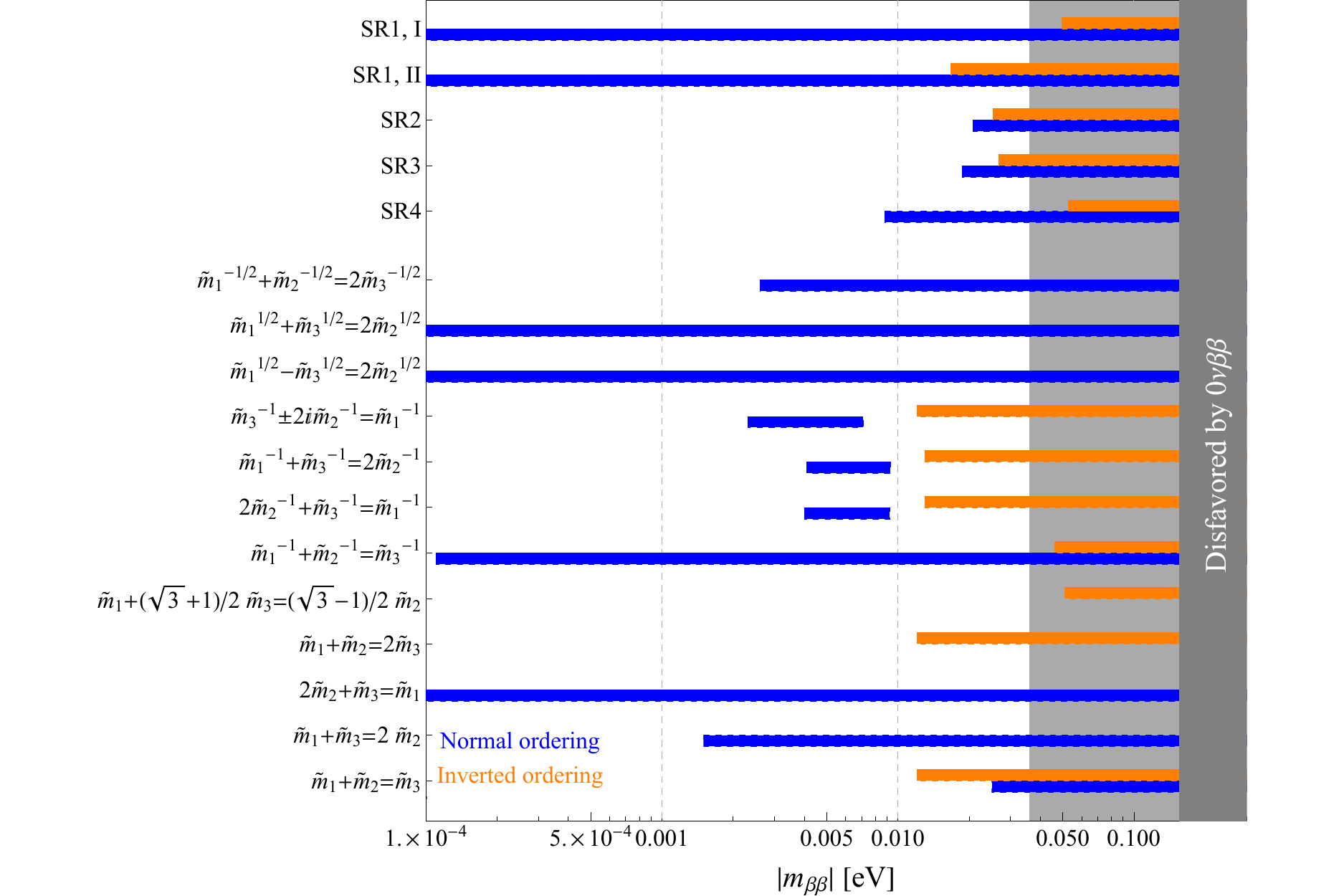}
\caption{Predictions for $|m_{\beta\beta}|$ from different neutrino mass sum rules. The upper five mass sum rules have been derived in models based on modular symmetries~\cite{Gehrlein:2020jnr}, the lower twelve mass sum rules come from models based on discrete symmetries~\cite{King:2013psa}. The  grey regions show the constraints on $|m_{\beta\beta}|$ from Ref.~\cite{KamLAND-Zen:2022tow}  using commonly-adopted nuclear matrix-element calculations. Taken from~\cite{Cirigliano:2022oqy}.}
\label{fig:mbb-sum-rules}
\end{figure}

One sees that certain mass sum rules predict a value of the lightest neutrino mass close to the current upper limit from cosmology and therefore they could be probed in the near future. Moreover, the lower bound on the lightest neutrino mass obtained in this way combined with the predictions for Majorana phases in Eq.~\eqref{eq:mass-sum-rule-cos} can be used to estimate the lower bound for the effective mass $|m_{\beta\beta}|$ from the general expression in Eq.~\eqref{eq:mbb}. A detailed description of the analysis procedure can be found in Refs.~\cite{Barry:2010yk,King:2013psa}.  The predictions for the effective Majorana neutrino mass $|m_{\beta\beta}|$ for the different sum rules are displayed in figure~\ref{fig:mbb-sum-rules}.

One sees that certain sum rules can be fully or partially probed by the next generation of $0\nu\beta\beta$ decay experiments even taking into account the large uncertainties in the nuclear matrix element calculations. The exact mass sum rule could be violated by the higher-order terms resulting from flavour symmetry breaking or by the renormalization group evolution effects~\cite{Gehrlein:2015ena}. It was shown that the predictions of the sum rules are still valid at least qualitatively~\cite{Gehrlein:2015ena,Gehrlein:2016wlc}. To sum up one can say that, precision measurements of mixing angles, an observation of neutrinoless double beta decay can also provide insights into underlying flavor symmetries.

\subsection{Flavor symmetry toolkit}

One can check the validity of flavor symmetry models and constrain their predicted mixing parameters by comparing them those extracted from global oscillation fit results~\cite{deSalas:2020pgw,10.5281/zenodo.4726908,Esteban:2020cvm,Capozzi:2021fjo}.
Nevertheless, numerical simulation of neutrino experiments and statistical analysis are often necessary to test and discriminate theoretical models in the current and future neutrino oscillation experiments such as NO$\nu$A, T2K, JUNO, DUNE and T2HK. The public available software such as \texttt{GLoBES}~\cite{Huber:2004ka,Huber:2007ji}, \texttt{Prob3++}~\cite{Calland:2013vaa}, \texttt{nuCRAFT}~\cite{Wallraff:2014qka} and \texttt{nuSQuIDS}~\cite{Arguelles:2021twb} have been widely used to simulate the experimental characteristics of neutrino oscillation experiments. The compatibility between the experimental data and the expected outcome of a given neutrino experiment is frequently evaluated by the chi-square test. The involved simulation and analysis with \texttt{GLoBES} is quite involved and time-consuming. In order to facilitate the analysis of leptonic flavour symmetry models in neutrino oscillation experiments, a dedicated package \texttt{FaSE-GLoBES} has been recently developed~\cite{Tang:2020jeh} and is available via the link~\url{https://github.com/tcwphy/FASE_GLoBES}. \texttt{FaSE-GLoBES} is a supplemental tool for \texttt{GLoBES}, written in \texttt{c/c++} language, and allows users to assign any flavour symmetry model and analyze how it can be constrained and tested by the simulated neutrino oscillation experiments.

\clearpage

\section{Benchmark UV-complete models in 4-dimensions (4-D)}               
\label{sec:benchmark-models}                                               

In addition to the model-independent approaches described in previous chapters, the flavour problem may be tackled in a more complete, UV-complete manner, by guessing the structure of the underlying family symmetry and building explicit models on a case-by-case basis, for reviews see, e.g. Refs.~\cite{Altarelli:2010gt,Morisi:2012fg,King:2013eh,King:2017guk,King:2014nza,Feruglio:2019ybq}.
In this chapter we present two simple extensions of the \sm implementing family symmetry within the renormalisable \SM gauge field theoretic framework.

\subsection{A benchmark flavour-symmetric scotogenic model }
\label{sec:model}

Here $A_4$ is used as flavour symmetry within the scotogenic picture~\cite{Ma:2006km} in which the neutrino masses are generated radiatively, and the lightest of the mediators is identified with the dark matter particle. We adopt the singlet-triplet extension~\cite{Hirsch:2013ola,Merle:2016scw,Diaz:2016udz,Choubey:2017yyn,Restrepo:2019ilz} of the original model~\cite{Ma:2006km}, making it substantially richer in the associated phenomenology~\cite{Avila:2019hhv}. We employ the Ma-Rajasekaran basis~\cite{Ma:2001dn}, the representation matrices of the generators and the Clebsch-Gordon coefficients are listed in Table~\ref{tab:A4-irr-decomp}.

We now present a flavored extension of the theory proposed in~\cite{Ding:2020vud}. The basic fields and their symmetry transformation properties are summarized in table~\ref{tab:fields-EXD-TM1}. The left-handed leptons form an $A_4$ triplet, while right-handed ones come in as inequivalent singlets. Besides the SM fields, the original singlet-triplet scotogenic model~\cite{Hirsch:2013ola} contains new weak iso-triplet and iso-singlet fermions $\Sigma$ and $F$. Together with the Higgs scalars $\phi$ and $\Omega$, these transform as triplets under the action of the family group $A_4$, while the dark scalar $\eta$ is a flavour singlet.

A characteristic prediction of this model is the lower bound for the \znbb decay amplitude, discussed in section~\ref{sec:znbb}.
This follows from its incomplete fermion multiplet nature, see figure~\ref{fig:dbd2}. In contrast to its original form, both charged lepton and neutrino mass matrices will now have a non-trivial structure, predicting trimaximal neutrino mixing. As usual, the $\mathbb{Z} _{2}$ parity is imposed in order to ensure the stability of the dark matter candidate and the radiative nature of neutrino mass generation.

\begin{table}[h!]
\centering
\begin{tabular}{|c|c|c|c|c|c|c|c| }
\hline
& \multicolumn{2}{ c |  }{~~Standard Model~~ } &  \multicolumn{2}{ c | }{ ~dark fermions~ }  & \multicolumn{3}{c|}{~ scalars~}  \\
\hline
 &~   $L$   ~& $e_R$, $\mu_R$, $\tau_R$     & ~ $\Sigma$ ~&    $F$   &~  $\phi$  ~&~ $\eta$ ~&~ $\Omega$~ \\
\hline
multiplicity &   3 &  3 &  3 &  3 &  3 &  1 & 3\\
\hline
$\rm SU(3)_c$  &   1    & 1       &   1     &    1    &  1  &    1    &   1    \\
$\rm SU(2)_L$  &  2    &  1        &    3     &   1  &    2   &    2    &   3     \\
$\rm U(1)_Y$     &   $-1$    & $-2$         &   0     &   0   &   1  & 1    &    0     \\  \hline
$\mathbb{Z} _{2}$      & $1$     & $1$    & $-1$    & $-1$  & $1$  & $-1$   &  $1$    \\
  \hline
$A_4$  &  $\mathbf{3}$  &  $\mathbf{1}, \mathbf{1'}, \mathbf{1''} $  & $\mathbf{3}$  &  $\mathbf{3}$  &  $\mathbf{3}$  &  $\mathbf{1}$  &  $\mathbf{3}$  \\
\hline
\end{tabular}
\caption{Transformation properties of the fields in the scotogenic model with $A_4$ family symmetry~\cite{Ding:2020vud}.\label{tab:fields-EXD-TM1} }
\end{table}

With the fields and symmetry assignments in table~\ref{tab:fields-EXD-TM1}, we can read out the Yukawa terms relevant to fermion masses as follows,
\begin{eqnarray}
\nonumber
\mathcal{L_Y} & \supset & -y_e (\overline{L} \phi )_{\mathbf{1}} e_R-y_{\mu} (\overline{L} \phi )_{\mathbf{1''}} \mu_R-y_{\tau} (\overline{L} \phi )_{\mathbf{1'}} \tau_R-Y_{F}\left(\overline{L}F\right)_{\mathbf{1}}\tilde{\eta} - Y_{\Sigma}(\overline{L}\tilde{\Sigma}^{c})_{\mathbf{1}}\tilde{\eta}\\
 &~& -Y_{\Omega,1}\left( {\rm Tr} \left[\left(\overline{\Sigma}\Omega\right)_{\mathbf{3}_{S}}
 \right]F\right)_{\mathbf{1}}-Y_{\Omega,2}\left( {\rm Tr}
\left[\left(\overline{\Sigma}\Omega\right)_{\mathbf{3}_{A}} \right]F\right)_{\mathbf{1}} + \text{h.c.}\,.
\label{eq:Lagrangian-fermion1}
\end{eqnarray}
In addition we have the bare mass terms
\begin{eqnarray}
 \label{eq:Lagrangian-fermion2} \mathcal{L_M} & \supset & -\frac{1}{2} M_{\Sigma} {\rm Tr} \big(( \overline{\Sigma} \tilde{\Sigma}^{c})_{\mathbf{1}} \big) - \frac{1}{2} M_F\left( \overline{F^{c}} F\right)_{\mathbf{1}} + \text{h.c.}\,,
\end{eqnarray}
where $\tilde{\eta}=i\sigma_2\eta^{*}$. The $SU(2)_L$ triplets $\Sigma$ and $\Omega$ are written in $2\times 2$ matrix notation as
\begin{equation}
\Omega=\left(\begin{array}{cc}
\Omega^{0}/\sqrt{2} &\Omega^{+}  \\
 \Omega^{-} &-\Omega^{0}/\sqrt{2}
\end{array}\right),~~~~
\Sigma=\left(\begin{array}{cc}
\Sigma^{0}/\sqrt{2} &\Sigma^{+}  \\
 \Sigma^{-} &-\Sigma^{0}/\sqrt{2}
\end{array}\right)
\end{equation}
with $\tilde{\Sigma}^{c}\equiv i\sigma_{2}\Sigma^{c}i\sigma_{2}$. The scalar triplet $\Omega$ is assumed to be real.

The $A_4$ flavour symmetry is broken by the VEVs of the scalar fields $\phi$ and $\Omega$, with the following VEV alignment in flavour space
\begin{equation}
\label{eq:vacuum-EXD1}\langle\phi \rangle=\begin{pmatrix}
1 \\
1 \\
1
\end{pmatrix} v_{\phi},~~~~
\vev{\Omega}=\begin{pmatrix}
1 \\
0 \\
0
\end{pmatrix} v_{\Omega},~~~~
\langle\eta\rangle=0\,,
\end{equation}
which can be a global minimum of the $A_4$-invariant scalar potential in certain regions of parameter space~\cite{Ding:2020vud}. Notice that the VEVs of $\phi$ and $\Omega$ break the $A_4$ flavour symmetry down to the subgroups $Z^t_3$ and $Z^s_2$ respectively, where the superscripts denote the generators of the subgroups. The $\rho$ parameter constrains the VEV $v_{\Omega}$ to be small, and the current electroweak precision tests lead to~\cite{Workman:2022ynf}
\begin{equation}
\label{eq:ewpt}
 v_{\Omega} \leq 4.5~\text{GeV}~~~{\rm at}~3\sigma ~{\rm CL}
\end{equation}

\subsubsection{Charged lepton masses}
\label{sec:charg-lept-mass}

The first three terms in the Yukawa Lagrangian are responsible for the charged lepton masses. Inserting the VEV of $\phi$ and using the multiplication law for the contraction of two triplets in table~\ref{tab:A4-irr-decomp}, one can straightforwardly read out the charged lepton mass matrix as
\begin{equation}
M_{\ell}=\begin{pmatrix}
y_e   &   y_{\mu}   & y_{\tau}  \\
y_{e} & \omega y_{\mu}  & \omega^2y_{\tau} \\
y_e  &  \omega^2 y_{\mu} & \omega y_{\tau} \\
\end{pmatrix}v_{\phi}\,.
\end{equation}
The matrix $M_{\ell}M^{\dagger}_{\ell}$ can be diagonalized to $\text{diag}(3 |y_ev_{\phi}|^2, 3|y_{\mu}v_{\phi}|^2, 3|y_{\tau}v_{\phi}|^2)$ by means of the unitary transformation
\begin{equation}
U_{\ell}=\frac{1}{\sqrt{3}} \begin{pmatrix}
1 ~&   1   ~& 1  \\
1 ~& \omega   ~& \omega^2 \\
1 ~&  \omega^2 ~& \omega  \\
\end{pmatrix}\,,
\label{eq:Ul-EXD1}
\end{equation}
which is a constant matrix.

\subsubsection{Dark fermion masses}
\label{sec:dark-fermion-masses}

With the alignment of $\Omega$ in Eq.~\eqref{eq:vacuum-EXD1}, we find the mass matrix of the dark fermions $F$ and $\Sigma^0$ is of the following form
\begin{equation}
\label{eq:M-chi}
M_\chi =\begin{pmatrix} M_\Sigma & 0 & 0 & 0 & 0 & 0 \\
0 & M_{F} & 0 & 0 & 0 & 0 \\
0 &  0 & M_\Sigma  & (Y_{\Omega,1}-Y_{\Omega,2})v_{\Omega} & 0 & 0 \\
0 &  0 & (Y_{\Omega,1}-Y_{\Omega,2})v_{\Omega} & M_{F} & 0 & 0 \\
0 &  0 & 0 & 0 & M_\Sigma  & (Y_{\Omega,1}+Y_{\Omega,2})v_{\Omega} \\
0 &  0 & 0 & 0 & (Y_{\Omega,1}+Y_{\Omega,2})v_{\Omega} & M_{F} \\
\end{pmatrix}
\end{equation}
in the convention of $-\frac{1}{2}(\overline{\Sigma^0_{1}},\overline{F^{c}_{1}},\overline{\Sigma^0_{2}},\overline{F^{c}_{3}},\overline{\Sigma^0_{3}},\overline{F^{c}_{2}})M_{\chi} \left(\Sigma_{1}^{0c}, F_{1}, \Sigma_{2}^{0c}, F_{3}, \Sigma_{3}^{0c}, F_{2}\right)^{T}$. The last block in Eq.~(\ref{eq:M-chi}) determines the masses of the dark Majorana fermions $F$ and $\Sigma$. The symmetric complex $6\times 6$ matrix $M_{\chi}$ can be diagonalized by a $6\times 6$ block-diagonal matrix $V$:
\begin{equation}
V^{T}M_{\chi}V=\text{diag}(m_{\chi^{0}_{1}},m_{\chi^{0}_{2}},m_{\chi^{0}_{3}},m_{\chi^{0}_{4}},m_{\chi^{0}_{5}},m_{\chi^{0}_{6}})\,,
\end{equation}
with
\begin{equation}
V = \left(\begin{array}{cccccc}1 ~&~ 0 &~ 0 ~& 0 & 0 & 0 \\
0 ~&~ 1 &~ 0 ~& 0 & 0 & 0 \\
0 ~&~ 0 &~ \cos\theta_1\, e^{i(\phi_{1}+\varrho_{1})/2} ~& \sin\theta_1\, e^{i(\phi_{1}+\sigma_{1})/2} ~& 0~ & ~0 \\
0 ~&~ 0 &~  -\sin\theta_1\, e^{i(-\phi_{1}+\varrho_{1})/2} ~& \cos\theta_1\, e^{i(-\phi_{1}+\sigma_{1})/2} ~& 0~ & ~0 \\
0 ~&~ 0 &~ 0 & 0 ~& ~\cos\theta_2\, e^{i(\phi_{2}+\varrho_{2})/2} ~& \sin\theta_2 \, e^{i(\phi_{2}+\sigma_{2})/2}\\
0 ~&~ 0 &~ 0 & 0 ~& ~-\sin\theta_2\, e^{i(-\phi_{2}+\varrho_{2})/2} ~& \cos\theta_2 \, e^{i(-\phi_{2}+\sigma_{2})/2} \\ \end{array}\right)\,.
\end{equation}
Following the method described in~\ref{app:diag}, one finds that the rotation angles $\theta_1$ and $\theta_2$ are given as
\begin{equation}
\tan2\theta_1=\frac{\Delta_{34}}{M_{F}^{2}-M_{\Sigma}^{2}}\,,\qquad
\tan2\theta_2=\frac{\Delta_{56}}{M_{F}^{2}-M_{\Sigma}^{2}}\,,
\end{equation}
with
\begin{eqnarray}
\nonumber&&  \Delta_{34}=2Y_{-}\sqrt{M_{\Sigma}^{2}+M_{F}^{2}+2M_{\Sigma}M_{F}\cos2\phi_{34}}\,,\\
\nonumber&& \Delta_{56}=2Y_{+}\sqrt{M_{\Sigma}^{2}+M_{F}^{2}+2M_{\Sigma}M_{F}\cos2\phi_{56}}\,,\\
\nonumber&&Y_{-}\equiv |(Y_{\Omega,1}-Y_{\Omega,2})v_{\Omega}|\,,~~~~ \phi_{34}\equiv \text{arg}((Y_{\Omega,1}-Y_{\Omega,2})v_{\Omega})\,,\\
&&Y_{+}\equiv |(Y_{\Omega,1}+Y_{\Omega,2})v_{\Omega}|\,,~~~~ \phi_{56}\equiv \text{arg}((Y_{\Omega,1}+Y_{\Omega,2})v_{\Omega})\,.
\end{eqnarray}
The eigenvalues $M_{\chi_{1,2,3,4,5,6}^{0}}$ are given by
\begin{eqnarray}\nonumber
&&m_{\chi_{1}^{0}}=M_{\Sigma}\,,~~~~~~ m_{\chi_{2}^{0}}=M_{F}\,,\\ \nonumber
&&m_{\chi_{3}^{0}}^{2}=\frac{1}{2}\left( M_{\Sigma}^{2}+M_{F}^{2}+2Y_{-}^{2}-\sqrt{(M_{F}^{2}-M_{\Sigma}^{2})^{2}+\Delta_{34}^{2}}\right)\,,\\ \nonumber
&&m_{\chi_{4}^{0}}^{2}=\frac{1}{2}\left( M_{\Sigma}^{2}+M_{F}^{2}+2Y_{-}^{2}+\sqrt{(M_{F}^{2}-M_{\Sigma}^{2})^{2}+\Delta_{34}^{2}}\right)\,,\\ \nonumber
&&m_{\chi_{5}^{0}}^{2}=\frac{1}{2}\left( M_{\Sigma}^{2}+M_{F}^{2}+2Y_{+}^{2}-\sqrt{(M_{F}^{2}-M_{\Sigma}^{2})^{2}+\Delta_{56}^{2}}\right)\,,\\
&&m_{\chi_{6}^{0}}^{2}=\frac{1}{2}\left( M_{\Sigma}^{2}+M_{F}^{2}+2Y_{+}^{2}+\sqrt{(M_{F}^{2}-M_{\Sigma}^{2})^{2}+\Delta_{56}^{2}}\right)\label{mchi2}\,,
\end{eqnarray}
The Majorana fermion mass eigenstates $\chi^{0}_{1,2,3,4,5,6}$ are related with $\Sigma^{c}_{0}$ and $F$ by the unitary transformation $V$
via
\begin{equation}
\label{eq:fermion-basis-scotogenic}
\begin{pmatrix} \Sigma_{1}^{0c} \\ F_{1} \\ \Sigma_{2}^{0c} \\ F_{3} \\ \Sigma_{3}^{0c} \\ F_{2} \end{pmatrix} =V \begin{pmatrix} \chi^{0}_{1} \\ \chi^{0}_{2} \\ \chi^{0}_{3} \\ \chi^{0}_{4}\\ \chi^{0}_{5} \\ \chi^{0}_{6}\end{pmatrix}\,.
\end{equation}

\begin{figure}[t!]
\centering
\includegraphics[width=0.98\textwidth]{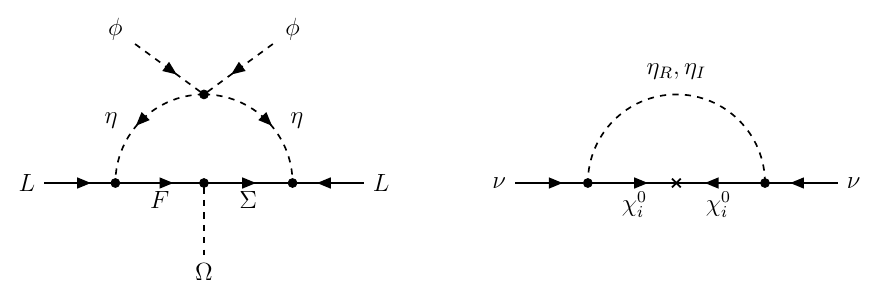}
\caption{\label{fig:nu-loop} Feynman diagrams for scotogenic neutrino mass generation in the weak- (left panel) and mass-eigenstate basis (right panel).}
\end{figure}

\subsubsection{Scotogenic neutrino masses}
\label{sec:scot-neutr-mass}

We now turn to neutrino masses. These arise radiatively,  at the one-loop level, as shown in figure~\ref{fig:nu-loop}, mediated by a ``dark sector'', within a scotogenic setup. In contrast to the ``flavour-blind'' singlet-triplet scotogenic model, the dark fermions now transform as $A_4$ triplets and all of the six dark fermions mediate the one-loop diagrams.
The interactions contributing to neutrino mass generation arise from the Yukawa terms involving $Y_{F}$ and $Y_{\Sigma}$.

In the mass-eigenstate basis of dark Majorana fermions, the relevant Lagrangian is of the following form
\begin{equation}
\label{eq:vertex}\mathcal{L}_{\nu}= -\frac{1}{\sqrt{2}}h_{\alpha i}\overline{\nu_{\alpha}}\eta_{R}\chi_{i}^{0}  + \frac{i}{\sqrt{2}}h_{\alpha i}\overline{\nu_{\alpha}}\eta_{I}\chi_{i}^{0}
-\frac{1}{\sqrt{2}}h_{\alpha i}^{*}\overline{\chi_{i}^{0}}\eta_{R}\nu_{\alpha} -\frac{i}{\sqrt{2}}h_{\alpha i}^{*} \overline{\chi_{i}^{0}}\eta_{I}\nu_{\alpha} \,,
\end{equation}
where $\alpha$ is the family index and the rectangular matrix $h$ is given by
\begin{equation}
h=\begin{pmatrix} \frac{Y_{\Sigma}}{\sqrt{2}} ~& Y_{F} ~& 0 ~& 0 ~& 0 ~& 0 \\
0 ~& 0 ~& \frac{Y_{\Sigma}}{\sqrt{2}}  ~& 0 ~& 0 ~& Y_{F} \\
0 ~& 0 ~& 0 ~& Y_{F} ~& \frac{Y_{\Sigma}}{\sqrt{2}} ~& 0
\end{pmatrix}V\,.
\end{equation}
Calculating the one-loop diagram in figure~\ref{fig:nu-loop}, we find that the radiatively generated neutrino mass matrix is given by
\begin{equation}
\left(m_{\nu}\right)_{\alpha\beta}=i\sum_{i}\frac{h_{\alpha i} h_{\beta i}}{32\pi^{2}}m_{\chi^{0}_{i}}\left( -\frac{m^{2}_{\eta_{R}}\text{ln}(\frac{m^{2}_{\eta_{R}}}{m^{2}_{\chi^{0}_{i}}})}{m^{2}_{\eta_{R}}-m^{2}_{\chi^{0}_{i}}} + \frac{m^{2}_{\eta_{I}}\text{ln}(\frac{m^{2}_{\eta_{I}}}{m^{2}_{\chi^{0}_{i}}})}{m^{2}_{\eta_{I}}-m^{2}_{\chi^{0}_{i}}}\right)\,,
\end{equation}
where $m_{\eta_R}$ and $m_{\eta_I}$ denote the masses of $\eta_R$ and $\eta_I$ respectively.
Note that $\eta_R$ and $\eta_I$ are the real and imaginary parts of the neutral field $\eta^{0}=(\eta_{R}+i\eta_I)/\sqrt{2}$. It is notable that the light neutrino mass matrix is predicted to be block-diagonal,
\begin{equation}
\label{eq:mnu-EXD1}m_{\nu}=\begin{pmatrix}
\times ~& 0 ~& 0 \\
0 ~& \times ~& \times \\
0 ~& \times ~& \times
\end{pmatrix} \,,
\end{equation}
where the symbol ``$\times$'' indicates a non-vanishing element. The reason is that only the flavon $\Omega$ is involved in the neutrino sector and its VEV preserves the $Z^{s}_2$ subgroup, so that the light neutrino mass matrix is invariant under the action of the $A_4$ generator $s$, i,e., $\rho^{\dagger}_{\mathbf{3}}(s)m_{\nu}\rho^{*}_{\mathbf{3}}(s)=m_{\nu}$. This implies the form in Eq.~\eqref{eq:mnu-EXD1}. The corresponding neutrino diagonalization matrix is of the form
\begin{equation}
U_{\nu}=\left(\begin{array}{ccc} 1 ~& 0 ~& 0 \\
0 ~& \cos\theta_{\nu} ~& \sin\theta_{\nu} e^{i\delta_{\nu}} \\
0 ~& -\sin\theta_{\nu} e^{-i\delta_{\nu}} ~& \cos\theta_{\nu}
\end{array} \right)\,,
\label{eq:U-nu}
\end{equation}
which satisfies $U_{\nu}^{\dagger}\mathcal{M}_{\nu}U_{\nu}^{*}=\text{diag}(m_{1},m_{2},m_{3})$.

One sees that at this level there is no solar mixing. However, including the contribution from the charged lepton sector one obtains a realistic lepton mixing matrix given as
\begin{equation}
\label{eq:UPMNS-EXD1}
U=\frac{1}{\sqrt{3}}\left( \begin{array}{ccc} \cos\theta_{\nu}-\sin\theta_{\nu}e^{-i\delta_{\nu}} ~& 1 ~& \cos\theta_{\nu}+\sin\theta_{\nu}e^{i\delta_{\nu}}\\
\omega^{2}\cos\theta_{\nu}-\omega \sin\theta_{\nu}e^{-i\delta_{\nu}} ~& 1 ~& \omega \cos\theta_{\nu}+\omega^{2}\sin\theta_{\nu}e^{i\delta_{\nu}}\\
\omega \cos\theta_{\nu}-\omega^{2}\sin\theta_{\nu}e^{-i\delta_{\nu}} ~& 1 ~& \omega^{2}\cos\theta_{\nu}+\omega \sin\theta_{\nu}e^{i\delta_{\nu}} \\ \end{array} \right)\,.
\end{equation}
with non-vanishing solar mixing angle. One sees the full lepton mixing matrix is predicted in terms of two free parameters $\theta_{\nu}$ and $\delta_{\nu}$.
Without loss of generality, these can be taken in the regions $0\leq\theta_{\nu}\leq\pi$ and $0\leq\delta_{\nu}\leq\pi$.
\begin{figure}[b!]
\centering
\begin{tabular}{cc}
\includegraphics[width=0.49\linewidth]{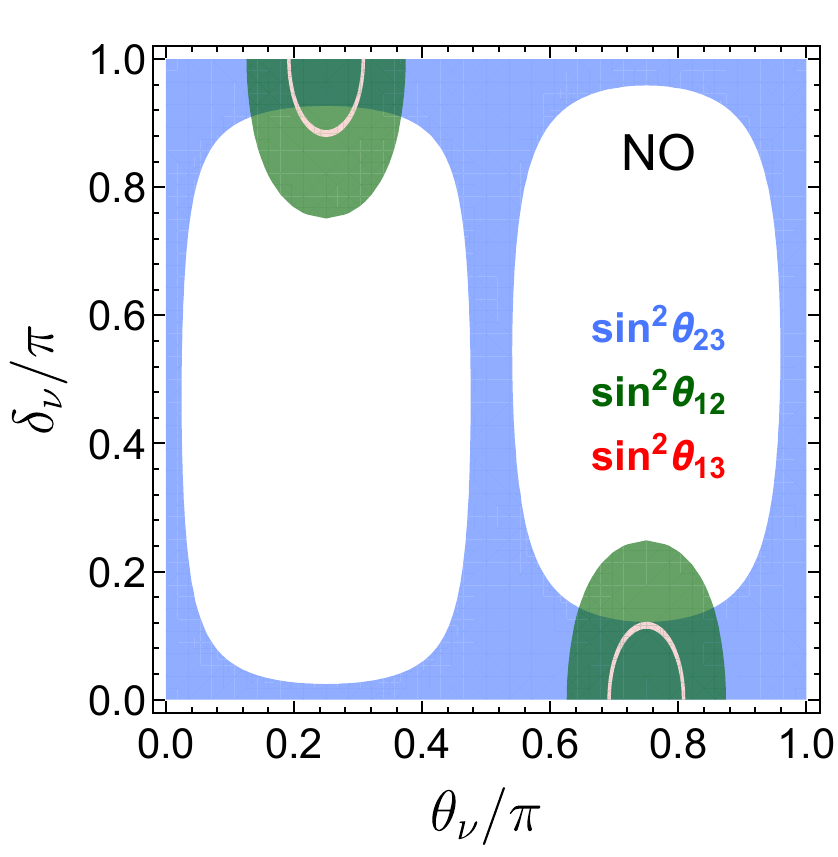}~
\includegraphics[width=0.49\linewidth]{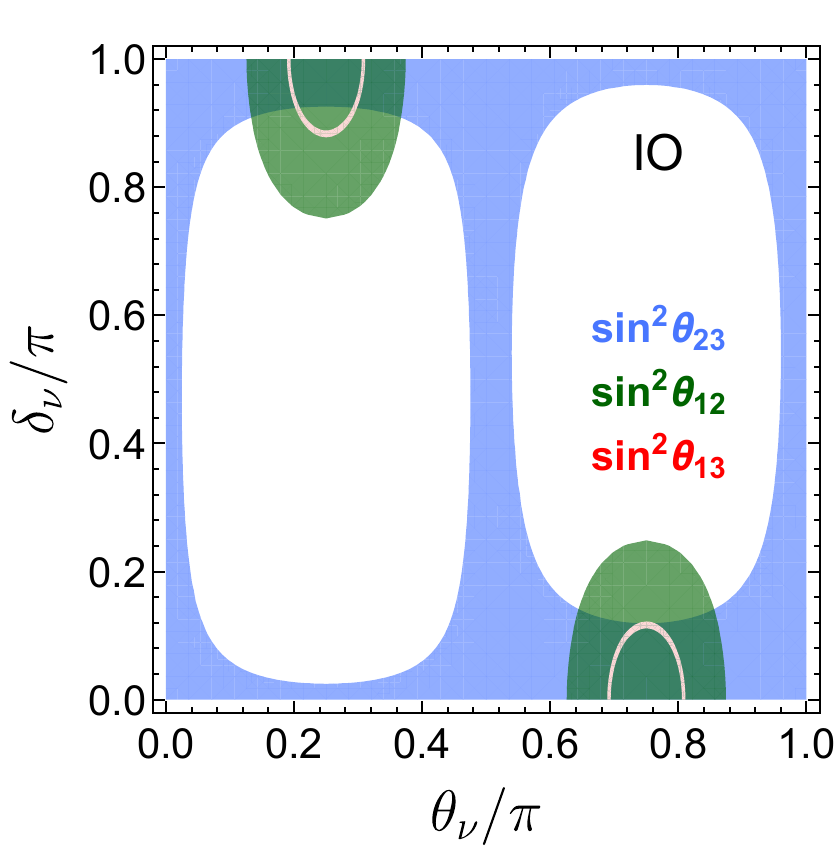}
\end{tabular}
\caption{\label{fig:osc-predictions-I}Contour plots of $\sin^2\theta_{12}$, $\sin^2\theta_{13}$, and $\sin^{2}\theta_{23}$ in the $\theta_{\nu}-\delta_{\nu}$ plane. The red, green and blue areas denote the allowed $3\sigma$ regions of $\sin^{2}\theta_{13}$, $\sin^{2}\theta_{12}$ and $\sin^{2}\theta_{23}$ respectively.}
\end{figure}

One also notes that the lepton mixing matrix $U$ has the so-called trimaximal {\bf TM2} form~\cite{Ding:2020vud}, since the second column is fixed to be $\frac{1}{\sqrt{3}}(1,1,1)^{T}$~\cite{Albright:2008rp,Albright:2010ap,He:2011gb}.
From the mixing matrix~\eqref{eq:UPMNS-EXD1} one can express the lepton mixing angles and the leptonic Jarlskog invariant in terms of just two free parameters
\begin{eqnarray}
\nonumber&&\sin^{2}\theta_{13}=\frac{1+\sin2\theta_{\nu}\cos\delta_{\nu}}{3}\,,~~~\sin^{2}\theta_{12}=\frac{1}{2-\sin2\theta_{\nu}\cos\delta_{\nu}}\,,\\
\nonumber&&\sin^{2}\theta_{23}=\frac{1}{2}-\frac{\sqrt{3}\sin2\theta_{\nu}\sin\delta_{\nu}}{4-2\sin2\theta_{\nu}\cos\delta_{\nu}}\,,~~~J_{CP}=\frac{\cos2\theta_{\nu}}{6\sqrt{3}}\,.
\end{eqnarray}
One predicts the following relations amongst the mixing angles and the CP phase,
\begin{equation}
\label{eq:osc-predictions}\hskip-0.02in
\cos^{2}\theta_{13}\sin^{2}\theta_{12}=\frac{1}{3}\,,~ \cos\delta_{CP}=\frac{2(3\cos ^{2}\theta_{12}\cos^{2}\theta_{23}+3\sin^{2}\theta_{12}\sin^{2}\theta_{13}\sin^{2}\theta_{23}-1)}{3\sin 2\theta_{23}\sin2\theta_{12}\sin\theta_{13}}\,.
\end{equation}
Using the $3\sigma$ range of the reactor angle $2.000\times10^{-2}\leq\sin^2\theta_{13}\leq2.405\times10^{-2}$ for NO and
$2.018\times10^{-2}\leq\sin^2\theta_{13}\leq2.424\times10^{-2}$ for IO~\cite{deSalas:2020pgw,10.5281/zenodo.4726908},
we predict narrow ranges for the solar mixing angle~\cite{Ding:2020vud},
\begin{equation}
\label{eq:sinSqtheta12-scotogenic}	\text{NO:}~~ 0.3401 \leq \sin^2\theta_{12}\leq 0.3415\,,~~~\text{IO:}~~  0.3402 \leq \sin^2\theta_{12}\leq 0.3416\,,
\end{equation}
These predictions are very close to the $1\sigma$ upper limits from the general neutrino oscillation global fit~\cite{deSalas:2020pgw,10.5281/zenodo.4726908} and
should be testable in forthcoming neutrino oscillation experiments.

\begin{figure}[hptb]
\centering
\begin{tabular}{cc}
\includegraphics[width=0.49\linewidth]{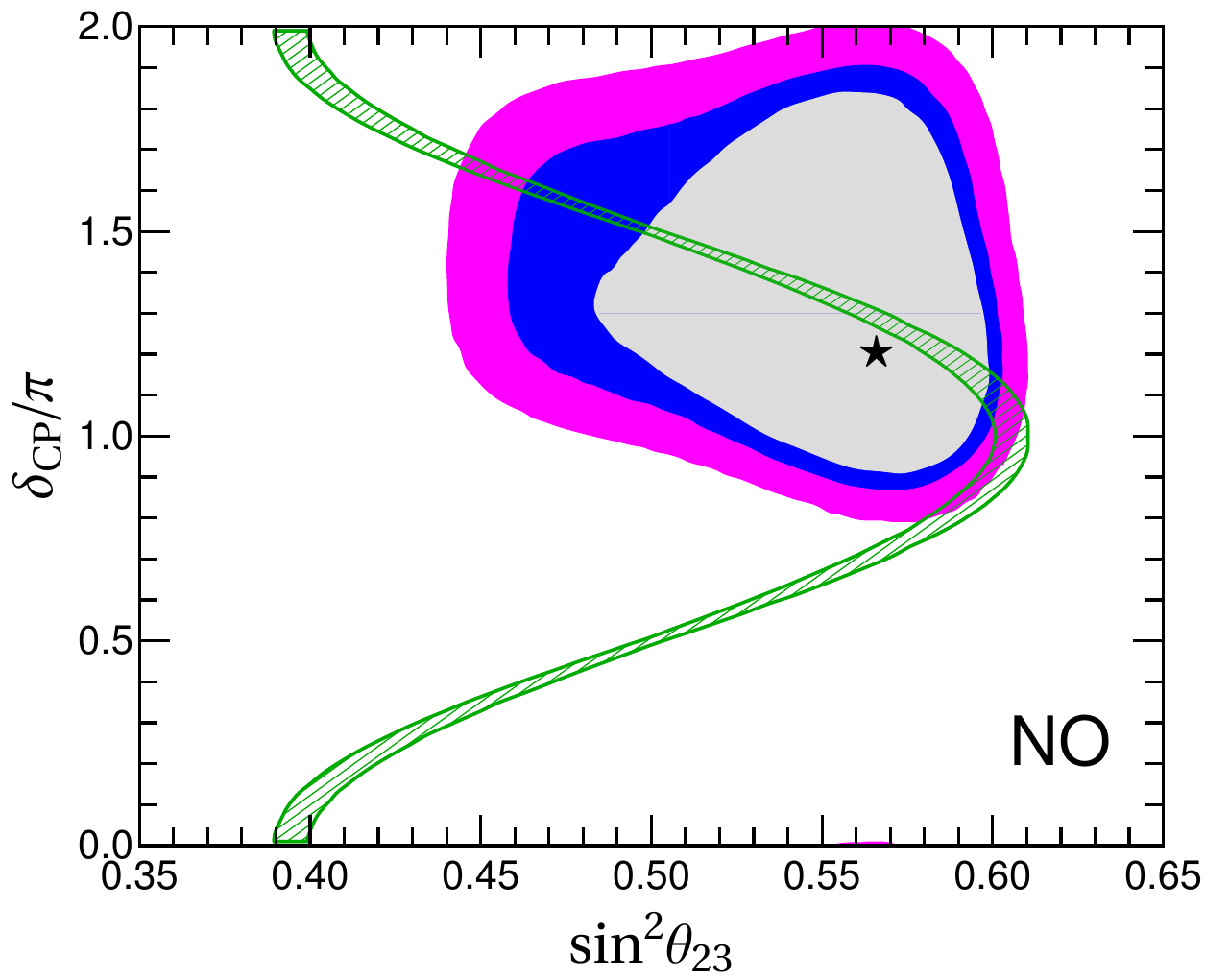}
\includegraphics[width=0.49\linewidth]{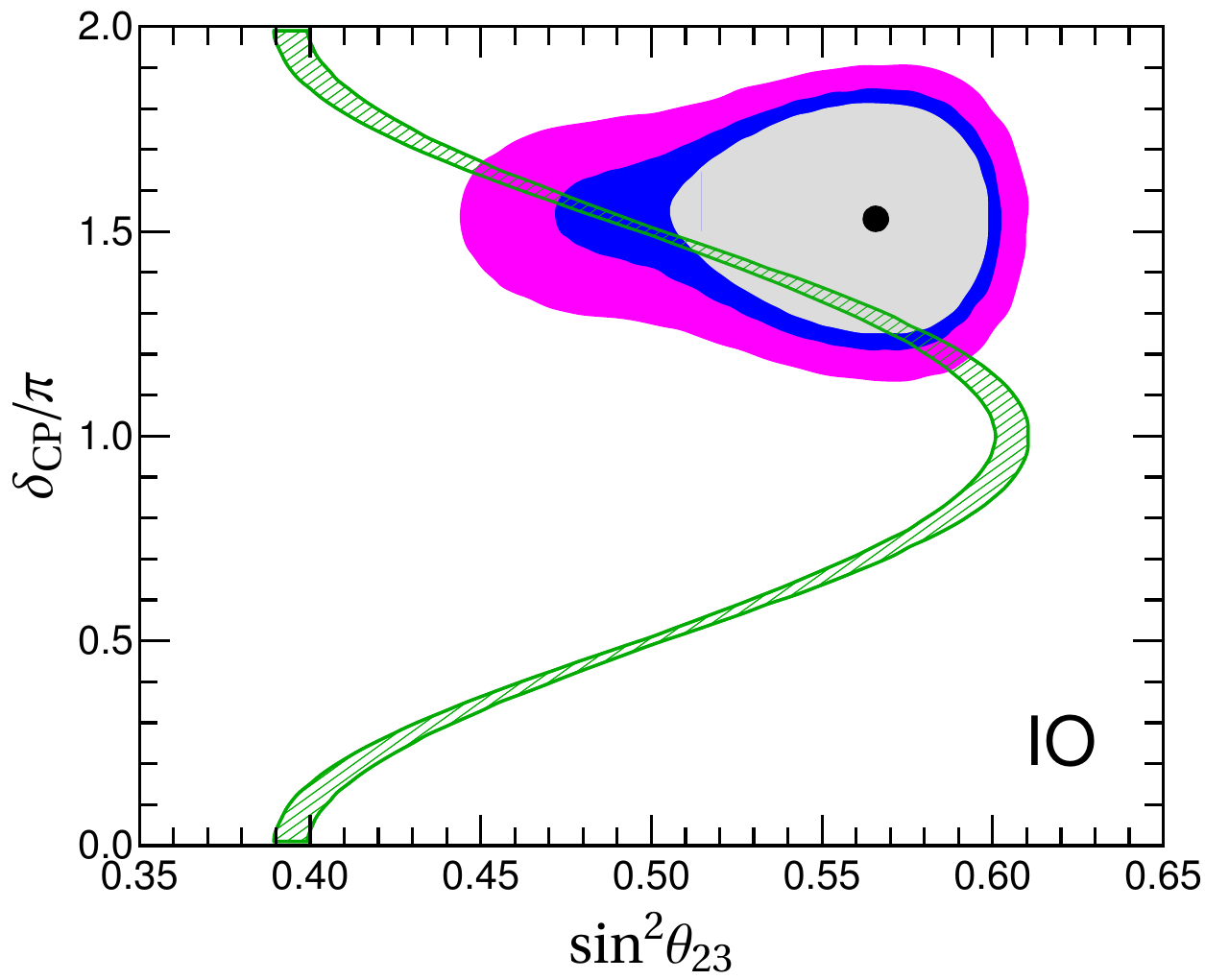}
\end{tabular}
\caption{\label{fig:osc-predictions-III} The hatched green bands indicate the predicted correlations between $\delta_{CP}$ and $\sin^{2}\theta_{23}$ for both neutrino mass orderings. The undisplayed parameters $\sin^{2}\theta_{13}$ and $\sin^{2}\theta_{12}$ are required to lie within their $3\sigma$ regions from the current oscillation global fit~\cite{deSalas:2020pgw,10.5281/zenodo.4726908}. The generic $1\sigma$, $2\sigma$ and  $3\sigma$ regions from the current neutrino oscillation global fit are indicated by the shaded areas~\cite{deSalas:2020pgw,10.5281/zenodo.4726908}. }
\end{figure}

To sum up we have obtained the three lepton mixing angles and the Dirac CP phase in terms of just two parameters $\theta_{\nu}$ and $\delta_{\nu}$ as given in figure~\ref{fig:osc-predictions-I}. The resulting predictions for the two most poorly determined oscillation parameters $\sin^2\theta_{23}$ and $\delta_{CP}$ are given in figure~\ref{fig:osc-predictions-III},
where the star and dot stand for the global best fit points for NO and IO respectively.

One sees also that the CP phase $\delta_{CP}$ is predicted to lie in a range narrower than that obtained in the generic global fits of neutrino oscillations. We also mention that in this model the smallness of the reactor angle $\theta_{13}$ and CP violation parameter $J_{CP}$ has a dynamical origin, given by the small ratio $v_{\Omega}/(M_{F}+M_{\Sigma})$ involving the triple Higgs VEV, see~\cite{Ding:2020vud} for details. Finally, as already mentioned, the incomplete fermion multiplet structure implies that one of the neutrinos is massless, leading to the \znbb decay predictions in figure~\ref{fig:dbd2}. All in all, this construction offers a serious benchmark theory for neutrino oscillations and dark matter.

\subsection{ A benchmark model with both flavour and CP symmetries }
\label{sec:benchmark-model-with}

In this section, we describe a model implementing both the $S_4$ flavour symmetry and the generalized CP symmetry~\cite{Ding:2013hpa}. See Refs.~\cite{Feruglio:2013hia,Li:2013jya} for alternative models. It realizes the breaking patterns of flavour and CP symmetry analyzed in subsection~\ref{sec:abelian-subgroup-g_l}. We adopt a supersymmetric (SUSY) formulation of the model in four dimensional space-time. The model is renormalizable at high energies, it gives rise to tri-bimaximal neutrino mixing at leading order, and the next-to-leading order contributions break the tri-bimaximal to a trimaximal mixing pattern. We introduce the auxiliary symmetry $Z_3\times Z_4$ so as to generate charged lepton mass hierarchies and forbid unwanted operators.

In order to construct the superpotential responsible for the alignment of the flavon vacuum expectation values, we use the standard supersymmetric driving field mechanism~\cite{Altarelli:2005yx}. We assume the existence of an $R$-symmetry $U(1)_R$ containing the usual $R$-parity, under which the Higgs and flavon fields are uncharged, while matter fields carry a $+1$ $R$-charge. In addition, the so-called driving fields are necessary and they carry two units of $R$-charge. Therefore all terms in the superpotential should be bilinear in the matter superfields or linear in the driving field.

\subsubsection{Flavon superpotential}
\label{subsubsec:S4CPmodel-vacuum-alignment}

The matter and flavon fields and their transformation properties under the flavour symmetry are summarized in table~\ref{tab:S4CP-model-matter-flavons}.
\begin{table} [h!]
\centering
\begin{tabular}{|c||c|c|c|c|c|c||c|c|c|c|c|c|}
\hline
Field &   ~~$L$~~    &   ~$N^c$~     &  ~$e^{c}$~     &   ~$\mu^c$~     &    ~$\tau^c$~  &  $H_{u,d}$ &  ~$\varphi_T$~   &  ~~$\eta$~~  &   ~$\varphi_S$~  & ~~$\phi$~~  &  ~~$\xi$~~  &   ~$\Delta$~   \\  \hline
$S_4$  &  $\mathbf{3}$ &  $\mathbf{3}$  &  $\mathbf{1}$    &    $\mathbf{1}'$   &    $\mathbf{1}$  &    $\mathbf{1}$  &  $\mathbf{3}$  &  $\mathbf{2}$   &   $\mathbf{3}'$  &   $\mathbf{2}$  &   $\mathbf{1}$  &  $\mathbf{1}'$ \\  \hline
$Z_3$  &  $\omega$   &     $\omega^2$   &   $\omega^2$   &  1  &  $\omega$  &  1   &  $\omega$  &   $\omega$  &   $\omega^2$  &   $\omega^2$  &  $\omega^2$    &   1 \\   \hline
  $Z_4$   &  1         &    1  &  $i$   &  $-1$   &   $-i$  & 1  &   $i$  &  $i$  &  1   &   1  &  1  &   $1$ \\
  \hline
$U(1)_R$ &  1  &   1  &   1  &  1   &  1  &  0  &  0  &   0   &   0     &  0 &  0   &   0    \\
\hline
\end{tabular}
\caption{Transformation properties of matter, Higgs and flavon fields.}
\label{tab:S4CP-model-matter-flavons}
\end{table}
On the other hand the driving fields in our model and their transformation rules under the flavour symmetry group are listed in table~\ref{tab:S4CPmodel-driving}.
\begin{table}[b!]
\centering
\begin{tabular}{|c||c|c|c|c|c|c||c|c|c|c|c|c|c|c|}
\hline
Field &  $\varphi^{0}_T$   &  $\zeta^0$   &   $\varphi^0_S$  &   $\widetilde{\varphi}^{\,0}_S$   &  $\xi^0$  &    $\Delta^0$  &   $\Omega_1$   &   $\Omega^{c}_1$   &   $\Omega_2$   &   $\Omega^c_2$  &  $\Omega_3$   &   $\Omega^{c}_3$  &   $\Sigma$   &   $\Sigma^c$   \\  \hline
$S_4$   &    $\mathbf{3}'$   &    $\mathbf{1}$   &   $\mathbf{3}'$   &   $\mathbf{3}$  &   $\mathbf{1}$  &   $\mathbf{1}$  &   $\mathbf{2}$  &   $\mathbf{2}$ &  $\mathbf{2}$   &  $\mathbf{2}$ &  $\mathbf{3}$  &   $\mathbf{3}$   &  $\mathbf{3}'$   &  $\mathbf{3}'$   \\ \hline
$Z_3$   &   $\omega$     &    $\omega$  &   $\omega^2$   &   $\omega^2$   &   $\omega^2$ &   1  &  1   &  1  &   $\omega$  &   $\omega^2$ & $\omega^2$   &   $\omega$  &   $\omega^2$   &  $\omega$  \\ \hline
$Z_4$   &   $-1$     &      $-1$   & 1  &  1  &  1  &  1 & $-1$   &  $-1$   &   $-i$  &   $i$    &  1   &   1  &  $1$  &  $1$      \\ \hline
$U(1)_R$  &  2  &  2  &   2  &  2  &  2  & 2  &   1   &  1   &  1   & 1  &  1  &  1  &  1  & 1   \\
\hline
\end{tabular}
\caption{Transformation rules for driving and messenger fields under the $S_4 \times Z_4\times Z_3$ and $U(1)_R$ symmetries.}
\label{tab:S4CPmodel-driving}
\end{table}
The most general driving superpotential $\mathcal{W}_d$ invariant under $S_4\times Z_3\times Z_4$ with $R=2$ is given by~\cite{Ding:2013hpa}
\begin{eqnarray}
\nonumber \mathcal{W}_d&=&
g_1\left(\varphi^{0}_T\left(\varphi_T\varphi_T\right)_{\mathbf{3}'}\right)_{\mathbf{1}}+
g_2\left(\varphi^{0}_T\left(\eta\varphi_T\right)_{\mathbf{3}'}\right)_{\mathbf{1}}
+g_3\zeta^{0}\left(\varphi_T\varphi_T\right)_{\mathbf{1}}
+g_4\zeta^{0}\left(\eta\eta\right)_{\mathbf{1}}\\
&&+f_1\left(\varphi^{0}_S\left(\varphi_S\varphi_S\right)_{\mathbf{3}'}\right)_{\mathbf{1}}\nonumber
+f_2\left(\varphi^{0}_S\left(\phi\varphi_S\right)_{\mathbf{3}'}\right)_{\mathbf{1}}
+f_3\left(\varphi^{0}_S\varphi_S\right)_{\mathbf{1}}\xi
+f_4\left(\widetilde{\varphi}^{\;0}_S\left(\phi\varphi_S\right)_{\mathbf{3}}\right)_{\mathbf{1}}\\
&&+f_5\xi^{0}\left(\varphi_S\varphi_S\right)_{\mathbf{1}}+f_6\xi^{0}\left(\phi\phi\right)_{\mathbf{1}}+f_7\xi^0\xi^2
\label{eq:alignment_renor}+M^2\Delta^0+f_8\Delta^0\Delta^2 \,.
\end{eqnarray}
Since we impose CP symmetry on the theory in the unbroken phase, all couplings in $w_d$ are real. In the limit of unbroken supersymmetry, the minimum of the scalar potential is determined by vanishing $F-$terms for the driving fields. As a result, the VEVs of the flavons are aligned as follows~\cite{Ding:2013hpa}
\begin{eqnarray}
\nonumber&&\langle\varphi_T\rangle=\left(\begin{array}{c}
0\\
1\\
0
\end{array}\right)v_T,~~~~ \langle\eta\rangle=\left(\begin{array}{c}
0\\
1
\end{array}\right)v_{\eta},~~~~\langle\Delta\rangle=v_{\Delta}\,,\\
\label{eq:VEV-S4CP}&&\langle\varphi_S\rangle=\left(\begin{array}{c}
1\\
1\\
1
\end{array}\right)v_S,~~~~ \langle\phi\rangle=\left(\begin{array}{c}
1\\
1
\end{array}\right)v_{\phi},~~~~  \langle\xi\rangle=v_{\xi}\,,
\end{eqnarray}
with
\begin{equation}
v_T=\frac{g_2}{2g_1}v_{\eta},~~~v^2_S=-\frac{1}{6f^2_2f_5}\left(f^2_3f_6+2f^2_2f_7\right)v^2_{\xi},~~~
v_{\phi}=-\frac{f_3}{2f_2}v_{\xi},~~~v^2_{\Delta}=-M^2/f_8\,,
\end{equation}
where $v_{\eta}$ and $v_{\xi}$ are undetermined, as they are related to a flat direction. The phase of $v_{\eta}$ can be absorbed into the lepton fields, thus we can take the VEVs $v_{\eta}$ and $v_T$ real without loss of generality. The common phase of $v_S$, $v_{\phi}$ and $v_{\xi}$ does not affect neutrino masses and mixing, because it can be factored out in the light neutrino mass matrix. As a consequence, $v_{\phi}$ and $v_{\xi}$ can be considered real while $v_S$ is real or purely imaginary depending on the coefficient $-\left(f^2_3f_6+2f^2_2f_7\right)/(f^2_2f_5)$
being positive or negative. Furthermore, the VEV $v_{\Delta}$ is real for $f_8<0$ and purely imaginary for $f_8>0$.

\subsubsection{The charged lepton sector}
\label{sec:charg-lept-sect}

\begin{figure}[t!]
\begin{center}
\includegraphics[width=1.01\linewidth]{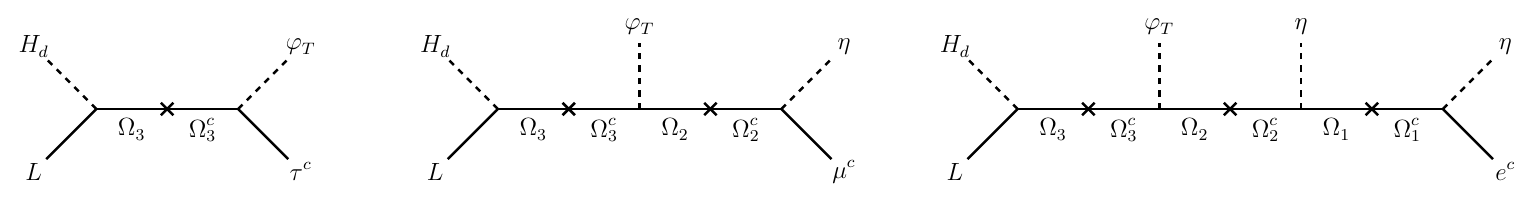}
\caption{\label{fig:charged_renor} Diagrams for charged-lepton-mass effective operators, crosses indicating fermion mass insertions.}
\end{center}
\end{figure}

The renormalizable Yukawa superpotential terms involving charged leptons is obtained by integrating out the three pairs of messengers $\Omega_i$ and $\Omega^{c}_i$ ($i=1,2,3$). These are chiral superfields with non-vanishing hypercharge $+2(-2)$ for $\Omega_i$ ($\Omega^{c}_i$). Given the field content and the symmetry assignments in tables~\ref{tab:S4CP-model-matter-flavons} and~\ref{tab:S4CPmodel-driving}, one can read off the superpotential for the charged leptons as follows,
\begin{eqnarray}
\nonumber \mathcal{W}_{\ell}&=&
z_1\left(L\Omega_3\right)_{\mathbf{1}}H_d
+z_2\left(\Omega^{c}_3\varphi_T\right)_{\mathbf{1}}\tau^c
+z_3\left(\left(\Omega^{c}_3\varphi_T\right)_{\mathbf{2}}\Omega_2\right)_{\mathbf{1}}
+z_4\left(\Omega^c_2\eta\right)_{\mathbf{1'}}\mu^c\\
&&+z_5\left(\left(\Omega^c_2\eta\right)_{\mathbf{2}}\Omega_1\right)_{\mathbf{1}}
\nonumber
+z_6\left(\Omega^c_1\eta\right)_{\mathbf{1}}e^c
+M_{\Omega_1}\left(\Omega_1\Omega^c_1\right)_{\mathbf{1}}
+z_7\Delta\left(\Omega_1\Omega^c_1\right)_{\mathbf{1}'}\\
&&+M_{\Omega_2}\left(\Omega_2\Omega^c_2\right)_{\mathbf{1}}
+z_8\Delta\left(\Omega_2\Omega^c_2\right)_{\mathbf{1}'}
+M_{\Omega_3}\left(\Omega_3\Omega^c_3\right)_{\mathbf{1}}\,,
\end{eqnarray}
where CP invariance requires all coupling constants $z_i$ and messenger masses $M_{\Omega_1}$, $M_{\Omega_2}$ and $M_{\Omega_3}$ to be real. The terms $\Delta\left(\Omega_1\Omega^c_1\right)_{\mathbf{1}'}$ and $\Delta\left(\Omega_2\Omega^c_2\right)_{\mathbf{1}'}$ lead to corrections to the $\Omega_1$ and $\Omega_2$ masses respectively. The mass scales of the messenger fields are much larger than the flavon VEVs, hence the contributions of these two operators can be safely neglected. Integrating out the messenger pairs $\Omega_i$ and $\Omega^c_i$ as shown in figure~\ref{fig:charged_renor}, we obtain the effective superpotential for the charged lepton masses as
\begin{equation}
\mathcal{W}^{eff}_{\ell}=-\frac{z_1z_2}{M_{\Omega_3}}\;\left(L\varphi_T\right)_{\mathbf{1}}H_d\tau^c+\frac{z_1z_3z_4}{M_{\Omega_2}M_{\Omega_3}}\;\left(\left(L\varphi_T\right)_{\mathbf{2}}\eta\right)_{\mathbf{1}'}\mu^c
-\frac{z_1z_3z_5z_6}{M_{\Omega_1}M_{\Omega_2}M_{\Omega_3}}\;\left(\left(L\varphi_T\right)_{\mathbf{2}}\left(\eta\eta\right)_{\mathbf{2}}\right)_{\mathbf{1}}H_de^c\,.
\end{equation}
Using the flavon VEVs in Eq.~\eqref{eq:VEV-S4CP}, we find a diagonal charged lepton mass matrix with
\begin{eqnarray}
\label{eq:charged_mass_renor}
\hskip-0.07in m_{\tau}=-z_1z_2\frac{v_T}{M_{\Omega_3}}v_d ,\quad
m_{\mu}=z_1z_3z_4\frac{v_Tv_{\eta}}{M_{\Omega_2}M_{\Omega_3}}v_d,\quad
m_e=-z_1z_3z_5z_6\frac{v_Tv^2_{\eta}}{M_{\Omega_1}M_{\Omega_2}M_{\Omega_3}}v_d\,,
\end{eqnarray}
where $v_d=\langle H_d\rangle$ is the VEV of the Higgs field $H_d$. We see that the electron, muon and tau masses are suppressed by different powers of $v_T$ and $v_{\eta}$, so that the mass hierarchies among the charged leptons are naturally reproduced.

\subsubsection{The neutrino sector}
\label{sec:neutrino-sector}

The renormalizable superpotential responsible for generating the light neutrino masses can be written as the sum of the leading-order terms and the relevant messenger terms:
\begin{equation}
\mathcal{W}_{\nu}=\mathcal{W}^{LO}_{\nu}+\mathcal{W}^{\Sigma}_{\nu}\ ,
\end{equation}
where
\begin{eqnarray}
\nonumber
\mathcal{W}^{LO}_{\nu}&=&y\left(LN^c\right)_{\mathbf{1}}H_u+y_1\left(\left(N^{c}N^c\right)_{\mathbf{3}'}\varphi_S\right)_{\mathbf{1}}+y_2\left(N^cN^c\right)_{\mathbf{1}}\xi+y_3\left(\left(N^cN^c\right)_{\mathbf{2}}\phi\right)_{\mathbf{1}}
\ ,\\
\mathcal{W}^{\Sigma}_{\nu}&=&x_1\left(\left(N^c\Sigma\right)_{\mathbf{3}'}\varphi_S\right)_{\mathbf{1}}+x_2\left(\left(N^c\Sigma\right)_{\mathbf{2}}\phi\right)_{\mathbf{1}}+x_3\left(N^c\Sigma^c\right)_{\mathbf{1}'}\Delta+M_{\Sigma}\left(\Sigma\Sigma^c\right)_{\mathbf{1}}\ .~~
\end{eqnarray}
where the couplings $x_i$ and $y_i$ are real due to CP invariance and the messenger field $\Sigma$ ($\Sigma^c$) is a chiral superfield with vanishing hypercharge. The first term of $\mathcal{W}^{LO}_{\nu}$ gives rise to a very simple form for the Dirac neutrino mass matrix,
\begin{equation}
m_D=y v_u\left(\begin{array}{ccc}
1~&0~&0  \\
0~&0~&1 \\
0~&1~&0
\end{array}\right)\,,
\end{equation}
with $v_u=\langle H_u\rangle$. On the other hand, the last three terms of $\mathcal{W}^{LO}_{\nu}$ lead to the mass matrix $m^{LO}_M$ of the right-handed neutrinos. Given the alignment of $\varphi_S$, $\phi$ and $\xi$ shown in Eq.~\eqref{eq:VEV-S4CP}, we have
\begin{equation}
m^{LO}_M=y_1v_{s}\left(\begin{array}{ccc}
2~& -1 ~& -1 \\
-1 ~&  2  ~&  -1  \\
-1 ~& -1  ~& 2
\end{array}\right)+y_2v_{\xi}\left(\begin{array}{ccc}
1 ~&0 ~&0 \\
0 ~&  0  ~& 1 \\
0 ~&  1 ~& 0
\end{array}\right)+y_3v_{\phi}\left(\begin{array}{ccc}
0  ~& 1  ~&  1  \\
1  ~&  1  ~&  0  \\
1  ~& 0  ~& 1
\end{array}\right).
\end{equation}
The effective light neutrino mass matrix is given by the simplest type-I seesaw formula~\cite{Schechter:1981cv} $$m^{LO}_{\nu}=-m_D(m^{LO}_M)^{-1}m^{T}_D$$
which is exactly diagonalized by the tri-bimaximal mixing matrix $U_{TBM}$,
\begin{equation}
U^{T}_{TBM}m^{LO}_{\nu}U_{TBM}=\text{diag}\left(m^{LO}_1,m^{LO}_2,m^{LO}_3\right)\,,
\end{equation}
so we obtain
\begin{equation}
\nonumber m^{LO}_1=-\frac{y^2v^2_u}{3y_1v_S+y_2v_{\xi}-y_3v_{\phi}},~
m^{LO}_2=-\frac{y^2v^2_u}{y_2v_{\xi}+2y_3v_{\phi}},~
m^{LO}_3=-\frac{y^2v^2_u}{3y_1v_S-y_2v_{\xi}+y_3v_{\phi}}\,.
\end{equation}
Note that the tri-bimaximal mixing pattern is produced at leading order. This follows from the fact that the VEVs of the flavons $\varphi_S$, $\phi$ and $\xi$ involved in the neutrino mass term are invariant under the action of the Klein subgroup generated by the tri-bimaximal $s$ and $u$ generators.

\begin{figure}[!h]
\begin{center}
\includegraphics[width=0.9\linewidth]{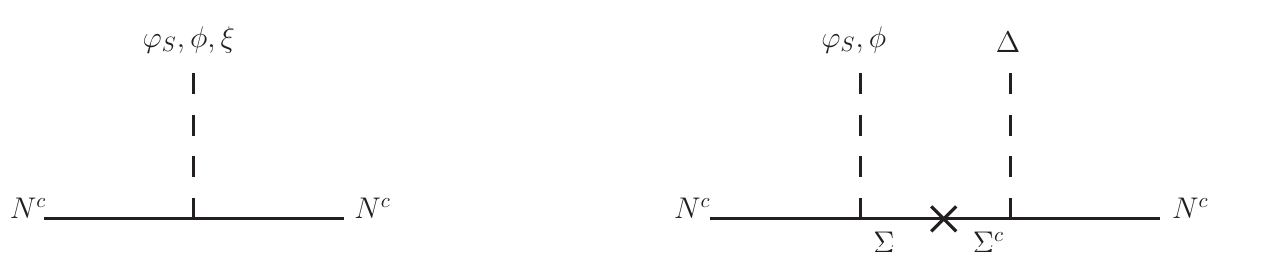}
\caption{The diagrams for the RH neutrino masses, where the cross indicates a fermionic mass insertion.}
\label{fig:neutrino_renor}
\end{center}
\end{figure}

The leading-order and next-to-leading order (NLO) contributions to the right-handed neutrino masses are depicted in figure~\ref{fig:neutrino_renor}.

Integrating out the messenger fields $\Sigma$ and $\Sigma^c$, we obtain the following NLO effective operator
\begin{equation}
\mathcal{W}^{NLO}_{\nu}=-\frac{x_2x_3}{M_{\Sigma}}\Delta\left(\left(N^cN^c\right)_{\mathbf{2}}\phi\right)_{\mathbf{1}'}\,.
\end{equation}
Notice that the VEV of the flavon $\Delta$ breaks the residual Klein symmetry down to a $Z_2$ subgroup generated by $s$ at NLO. Consequently, the corrected right-handed neutrino mass matrix is given as
\begin{eqnarray}
m_M &=&a\begin{pmatrix}
2~&-1~&-1\\-1~&2~&-1\\-1~&-1~&2\end{pmatrix}+b \begin{pmatrix}1~&0~&0\\0~&0~&1\\0~&1~&0 \end{pmatrix}+c\begin{pmatrix}0~&1~&1\\1~&1~&0\\1~&0~&1 \end{pmatrix}+d\begin{pmatrix}
0~&1~&-1\\1~&-1~&0\\-1~&0~&1\end{pmatrix}\,,
\end{eqnarray}
with
\begin{equation}
a=y_1v_S,~~~~ b=y_2v_{\xi},~~~~ c=y_3v_{\phi},~~~~
d=x_2x_3\frac{v_{\Delta}v_{\phi}}{M_{\Sigma}}\,.
\end{equation}
Therefore, up to an overall factor $y^2v^2_u$, the corrected light neutrino mass matrix obtained from the seesaw formula has the form
\begin{equation}
\label{eq:mnu-S4-CP}m_{\nu}=\alpha \begin{pmatrix} 2~&-1~&-1\\-1~&2~&-1\\-1~&-1~&2\end{pmatrix}
+\beta  \begin{pmatrix}1~&0~&0\\0~&0~&1\\0~&1~&0 \end{pmatrix}
+\gamma \begin{pmatrix}0~&1~&1\\1~&1~&0\\1~&0~&1 \end{pmatrix}
+\epsilon  \begin{pmatrix} 0~&1~&-1\\1~&-1~&0\\-1~&0~&1\end{pmatrix}\,,
\end{equation}
where the parameters $\alpha$, $\beta$, $\gamma$ and $\epsilon$ are given by,
\begin{eqnarray}
\nonumber&&\alpha=\frac{a}{-9 a^2+(b-c)^2+3 d^2}\,,~~~~ \beta=-\frac{1}{3(b+2 c)}+\frac{2 (b-c)}{3\left[9a^2-(b-c)^2-3d^2\right]}\,,\\
&&\gamma=-\frac{1}{3(b+2c)}-\frac{b-c}{3\left[9a^2-(b-c)^2-3d^2\right]}\,,~~~~\epsilon=\frac{d}{-9 a^2+(b-c)^2+3d^2}\,,
\end{eqnarray}
The first three terms in the light neutrino matrix of Eq.~\eqref{eq:mnu-S4-CP} preserve the tri-bimaximal mixing form. The last term, which is proportional to $\epsilon$, violates it.
The associated parameter $\epsilon$ is induced by the NLO contributions suppressed by $v_{\Delta}/M_{\Sigma}$ with respect to $\alpha$, $\beta$ and $\gamma$. This naturally accounts for the small reactor mixing angle $\theta_{13}$ and the small deviation from the maximal atmospheric mixing.

As shown in section~\ref{subsubsec:S4CPmodel-vacuum-alignment}, the VEVs $v_{\phi}$ and $v_{\xi}$ can be assumed real, while $v_{S}$ and $v_{\Delta}$ can be real or purely imaginary due to the generalized CP symmetry. If $v_{S}$ and $v_{\Delta}$ are real, the vacuum alignment of the flavons $\varphi_S$, $\phi$, $\xi$ and $\Delta$ is invariant under the CP transformation $X_{\mathbf{r}}=1$ and the residual flavour symmetry $Z^{s}_2$.
The parameters $\alpha$, $\beta$, $\gamma$ and $\epsilon$ are all real, the lepton mixing matrix has the form of Eq.~\eqref{UPMNS-S4CP-I} with the rotation angle $\theta$ given by
\begin{equation}
\tan2\theta=\frac{\sqrt{3}\epsilon}{\gamma-\beta}\,.
\end{equation}
The lepton mixing angles and CP violation phases are given in Eq.~\eqref{mixing-pars-S4CP-I}. The light neutrino masses are
\begin{eqnarray}
\nonumber m_1&=&\left|3\alpha-\text{sign}\left(\epsilon\sin2\theta\right)\sqrt{(\gamma-\beta)^2+3\epsilon^2}\right|\,,\\
\nonumber m_2&=&\left|\beta+2\gamma\right|\,, \\
\label{eq:nu-masses-S4CP-I}m_3&=&\left|3\alpha+\text{sign}\left(\epsilon\sin2\theta\right)\sqrt{(\gamma-\beta)^2+3\epsilon^2}\right|\,.
\end{eqnarray}
This model allows for both neutrino mass orderings, either NO or IO.

Moreover, if the VEV $v_{S}$ is real while $v_{\Delta}$ is pure imaginary, the neutrino sector would preserve the residual flavour symmetry $Z^s_2$ and CP symmetry $X_{\mathbf{r}}=u$ which corresponds to the $\mu-\tau$ refection symmetry. The parameters $\alpha$, $\beta$, $\gamma$ in Eq.~\eqref{eq:mnu-S4-CP} are real, and $\epsilon$ is pure imaginary.
We find that the lepton mixing matrix is given by Eq.~\eqref{UPMNS-S4CP-II} with the rotation angle $\theta$ given as
\begin{equation}
\tan2\theta=\frac{i\epsilon}{\sqrt{3}\alpha}\,.
\end{equation}
Both the atmospheric angle $\theta_{23}$ and the CP violation phase $\delta_{CP}$ are maximal, as shown in Eq.~\eqref{mixing-pars-S4CP-II}. The light neutrino masses are given by
\begin{eqnarray}
\nonumber m_1&=&\left|\beta-\gamma+\text{sign}\left(\alpha\cos2\theta\right)\sqrt{9\alpha^2-3\epsilon^2}\right|\,,\\
\nonumber m_2&=&\left|\beta+2\gamma\right|\,,\\
\label{eq:nu-masses-S4CP-II}m_3&=&\left|\beta-\gamma-\text{sign}\left(\alpha\cos2\theta\right)\sqrt{9\alpha^2-3\epsilon^2}\right|\,,
\end{eqnarray}
Again, this is consistent with both neutrino mass orderings, either NO or IO.\\[-.3cm]

All in all, this model implements the first two symmetry breaking patterns analyzed in detail in section~\ref{sec:abelian-subgroup-g_l}. The reader is addressed to that section for a discussion of the corresponding neutrino mixing, CP violation and \znbb decay predictions. See Ref.~\cite{Li:2013jya} for an $S_4$ model realizing the third breaking pattern and the lepton mixing matrix of Eq.~\eqref{UPMNS-S4CP-IV}.

\subsection{4-D family symmetries: other possible applications}
\label{sec:4-D-family}

Beyond the resolution of the flavor puzzle, there are other important drawbacks of the Standard Model, some of which have a solid experimental basis. These include the problem of accounting for a viable dark matter candidate and/or underpinning the mechanism responsible for neutrino mass generation. To close this chapter we wish to mention that the imposition of 4-D family symmetries can shed light on these drawbacks. Note also that the latter may be closely interconnected. An example is the strong CP problem (see below) and neutrino masses. While these are distinct issues, they both hint at new physics, and some theories beyond the Standard Model aim to address both issues simultaneously, irrespective of the flavor problem {\it per se}. For recent recent attempts see Refs.~\cite{Peinado:2019mrn,Dias:2020kbj,Batra:2023erw,Hati:2024ppg}. We now list some possible applications of flavor symmetries to tackle various SM drawbacks.

\subsubsection{Dirac nature of neutrinos from flavor symmetry}

Although neutrinos are generally expected to be Majorana fermions, pending the experimental discovery of lepton number violating processes, such as neutrinoless double beta decay, their true nature remains an open issue.
It could well happen that, as suggested in~\cite{Aranda:2013gga}, the Dirac nature of neutrinos could be the result of an underlying flavor symmetry.
Several recent improved scenarios incorporating also some sort of tree or radiative seesaw mechanism~\cite{CentellesChulia:2016rms,CentellesChulia:2016fxr,Borah:2017dmk,Bonilla:2017ekt} have been discussed.

\subsubsection{Stability of dark matter from flavor symmetry}

The idea here is that dark matter stability may be the result of an unbroken residual symmetry associated to some underlying family symmetry. For example, dark matter stability could be \textbf{accidental}, if the flavour-symmetry group is the double-cover group of the symmetry group of one of the regular geometric solids. In this case a stabilizing $\mathbb{Z}_2$ symmetry could emerge quite naturally~\cite{Lavoura:2012cv}. Dark matter may also be stabilized by a larger preserved $\mathbb{Z}_N$ symmetry, that could also be a subgroup of the family symmetry group, for example, a non-trivial center~\cite{Jurciukonis:2022oru}. In \cite{ChuliaCentelles:2022ogm} a model was introduced that extends the abobe ``accidental stability" mechanism beyond its initial proposal~\cite{Lavoura:2012cv}. Instead of having a $\mathbb{Z}_2$ center of the double cover of the flavor symmetry group (e.g. $\tilde{A}_4$, $\tilde{S}_4$, and $\tilde{A}_5$) as the dark matter stabilizer, the new model shows how larger centers, such as $\mathbb{Z}_3$ from groups like $\Sigma(81)$, can also serve as the dark matter stabilizing symmetry. The resulting aspects of \textit{accidentally stable} dark matter phenomenology are model-dependent. However, they may resemble what one expects from generic Higgs portal dark matter.

The \textbf{discrete dark matter mechanism} (DDM) was the proposal~\cite{Hirsch:2010ru} that the stabilizing symmetry for dark matter could emerge as a remnant from the spontaneous breakdown of a discrete non-Abelian flavor symmetry group accounting for the neutrino mixing pattern indicated by oscillation experiments~\cite{Boucenna:2011tj}. By construction, this breakdown gives rise to the light neutrino masses. This mechanism has been further explored in subsequent studies involving various discrete family symmetry groups, with several different implementations, for example,~\cite{Hirsch:2010ru,Boucenna:2011tj,Boucenna:2012qb,Lamprea:2016egz,DeLaVega:2018bkp,Meloni:2010sk,Adulpravitchai:2011ei,Meloni:2011cc,deAdelhartToorop:2011ad}.

\subsubsection{Scoto-seesaw mechanism and flavor symmetry}

The hierarchy observed between the solar and atmospheric neutrino mass-squared differences, namely $ \Delta m^2_{21} \ll \lvert  \Delta m^2_{31}\rvert$, may be the result of loop corrections. This is a characteristic feature of the proposal that the origin of neutrino masses is intrinsically supersymmetric~\cite{Hirsch:2000ef,Diaz:2003as,Hirsch:2004he}.

Recently this idea has been taken up within hybrid scoto-seesaw constructions~\cite{Rojas:2018wym,Aranda:2018lif,Mandal:2021yph} that put together the tree-level seesaw with the scotogenic neutrino mass generation approach. Here the idea is that the atmospheric squared mass splitting arises at tree level from seesaw mediators, while the solar splitting is calculable from radiative corrections associated to a dark scotogenic sector. Indeed, it has been shown how the scoto-seesaw mechanism can be closely interconnected with the discrete symmetry responsible for reproducing the observed pattern of neutrino oscillations~\cite{Barreiros:2020gxu}. Different realizations of the scoto-seesaw mechanism have been considered recently~\cite{Bonilla:2023pna,Kumar:2024zfb,Ganguly:2023jml}.

\subsubsection{Strong CP problem and flavor symmetry }

Understanding the puzzling absence of CP violation in strong interactions constitutes a challenge for the theory of quantum chromodynamics (QCD). The most well-known solution to such strong CP problem is to postulate a Peccei–Quinn (PQ) symmetry which is a QCD-anomalous global $U(1)$ symmetry~\cite{Peccei:1977hh}. Its spontaneous breaking results in a pseudo Nambu-Goldstone boson, called axion, providing a dynamical solution to the strong CP puzzle. The PQ symmetry may arise from flavor symmetries~\cite{Wilczek:1982rv}, in which case the axion scale is related to the flavor symmetry breaking scale. For example, the PQ symmetry could be identified with the Froggatt-Nielsen $U(1)$ flavor symmetry~\cite{Ema:2016ops,Calibbi:2016hwq}, or be a subgroup of a continuous flavor symmetry~\cite{Albrecht:2010xh,Arias-Aragon:2017eww,Linster:2018avp,Reig:2018ocz}. Moreover, the PQ symmetry could also arise accidentally from a discrete flavor symmetry~\cite{Dias:2002hz,Babu:2002ic,Bjorkeroth:2017tsz,Bjorkeroth:2018dzu}. In this case no new scalar fields are necessary to realize the PQ symmetry, and the same flavons already introduced to explain observed flavour patterns also give rise to the axion. As a consequence, the flavour structure seen in nature and the strong CP problem could potentially result from a single model, with the axion couplings fixed from the observed quark and lepton masses and mixing. Notice that one must ensure that the accidental PQ symmetry is protected to sufficiently high order, as the explicitly PQ-symmetry-violating high-dimensional operators involving flavons may shift the axion potential away from its CP-conserving minimum~\cite{Holman:1992us,Kamionkowski:1992mf,Barr:1992qq}.

Altogether, we have seen in this sub-section, that the imposition of four-dimensional family symmetries can have a variety of potential applications. However, in what follows we will refrain from delving into the arcane details of the associated model-building.

\clearpage

\section{Family symmetries in extra-dimensions }                
\label{sec:family-symmetry-from}                                

While the imposition of 4-D family symmetries provides a rather powerful technique for model-building in particle physics, with a variety of potential applications, it also suffers from important limitations. Indeed, an intrinsic drawback is the lack of a guiding principle to specify the nature of the family symmetry group itself. Another one is the lack of a clear general prescription to derive the observed pattern of fermion mass hierarchies, very odd by any standards. Why is the muon about 200 times heavier than the electron? Why is the top quark so heavy? Why are neutrinos so light? For all these reasons we find it not unlikely that the ultimate resolution of the flavor problem may require a more radical departure from our basic \sm set of assumptions.

It has been noted that the existence of extra space-time dimensions~\cite{Antoniadis:1998ig} provides an interesting way to address the so-called hierarchy problem~~\cite{Arkani-Hamed:1998jmv,Randall:1999ee,Arkani-Hamed:1998sfv}, by making the fundamental scale of gravity exponentially reduced from the Planck mass down to the TeV scale. This follows as a result of having the Standard Model fields localized near the boundary of the extra dimensions.

In this chapter we stress that also the fermion mass hierarchies can themselves be addressed through the localization of fermion profiles, which are fixed by the bulk mass parameters. Concerning the lepton and quark mixing angles, as we have shown in the previous section, they can be addressed by using a non-Abelian flavour symmetry within concrete 4-D models. Besides the intense activity using discrete family symmetries to build UV-complete 4-D gauge theories~\cite{Ishimori:2010au,Altarelli:2010gt,Morisi:2012fg,King:2013eh,King:2014nza,King:2017guk,Feruglio:2019ybq,Xing:2020ijf,Chauhan:2023faf},
the pattern of masses and mixing of leptons and quarks could also arise from the imposition of family symmetries in extra-dimensions~\cite{Altarelli:2005yp,Altarelli:2008bg,Csaki:2008qq,Chen:2009gy,Burrows:2009pi,Kadosh:2010rm,Kadosh:2011id,Kadosh:2013nra,Chen:2015jta,Chen:2020udk,Ding:2013eca,CarcamoHernandez:2015iao} such as holographic models~\cite{delAguila:2010vg,Hagedorn:2011un,Hagedorn:2011pw}. Concerning the structure of the four-dimensional family symmetry group, it has been noted that the existence of extra-dimensions may indeed shed light on its specific nature~\cite{deAnda:2018oik,deAnda:2018yfp}, providing a criterion for its choice.

In the first part of this chapter we examine the implementations of flavour symmetries in the context of extra-dimensional constructions, in such a way that the structure of both mass hierarchies as well as mixing angles could potentially be addressed in a comprehensive manner~\cite{Chen:2015jta,Chen:2020udk}. In the next sub-section we illustrate how the existence of extra dimensions may shed light on the nature of the family symmetry in four dimensions from the symmetries between the extra-dimensional branes. This idea was originally suggested within six-dimensional setups compactified on a torus~\cite{deAnda:2018oik,deAnda:2018yfp} and has now been implemented within realistic model setups~\cite{deAnda:2019jxw,deAnda:2020pti,deAnda:2020ssl,deAnda:2021jzc}.

\subsection{Warped flavordynamics with the $T^\prime$ family group }
\label{sec:warp-flav-dynam-1}

In this section we consider warped flavordynamics schemes, of which there have been two recent proposals in the literature~\cite{Chen:2015jta,Chen:2020udk}. The first warped flavordynamics model is based on the $\Delta(27)$ family symmetry, with neutrinos as Dirac fermions, and a predicted TM2 mixing pattern~\cite{Chen:2015jta}. An alternative proposal warped flavordynamics scheme is based on the $T^\prime$ family group~\cite{Chen:2020udk}. Here neutrinos are Majorana fermions with a predicted TM1 mixing pattern. In what follows we develop the key features of this second example, and refer the interested reader to the original work in~\cite{Chen:2015jta} for the other case.

Here we present our benchmark extra-dimensional model with $T'$ flavour symmetry~\cite{Chen:2020udk}, using a 5-D warped setup. The $T'$ group is the double covering of $A_4$. The relation between $T'$ and $A_4$ is quite similar to the familiar relation between SU(2) and SO(3). Although SU(2) and SO(3) possess the same Lie algebra, SO(3) has only odd-dimensional representations, while SU(2) possesses both even and odd-dimensional representations. Likewise $T'$ has three doublet representations $\mathbf{2}$, $\mathbf{2}'$ and $\mathbf{2}''$ besides the singlets $\mathbf{1}$, $\mathbf{1}'$, $\mathbf{1}''$ and triplet $\mathbf{3}$ of $A_4$.

We formulate our model in the framework of the Randall-Sundrum model~\cite{Randall:1999ee}. The bulk geometry is described by the following metric
\begin{equation}
ds^2=e^{-2ky}\eta_{\mu\nu}dx^{\mu}dx^{\nu}-dy^2\,.
\end{equation}
This extra dimension $y$ is compactified, and the two 3-branes with opposite tension are located at $y = 0$, the UV brane, and the infra-red (IR) brane at $y=L$.

In order to comply with electroweak precision measurement constraints, the electroweak symmetry of the model is promoted to
$\mathrm {G_{\text{bulk}}=SU(2)_L\otimes SU(2)_{R}\otimes U(1)_{B-L}}$~\cite{Agashe:2003zs,Agashe:2006at} and it is broken down to  the standard model gauge group $\mathrm {G_{\text{SM}}=SU(2)_{L}\otimes U(1)_{Y}}$ on the UV brane by the boundary conditions (BCs) of the gauge bosons. The Higgs field lives in the bulk, and it is in the $(\mathbf{2},\mathbf{2})$ bi-doublet representation of $\mathrm {SU(2)_L\otimes SU(2)_{R}}$. The Kaluza-Klein (KK) Higgs field decomposition is~\cite{Cacciapaglia:2006mz}
\begin{equation}
H(x^\mu,y)= H(x^\mu) \frac{f_H(y)}{\sqrt{L}} + \text{heavy KK Modes}\,,
\end{equation}
where $f_H(y)$ is the zero mode profile. For an adequate choice of BCs, we have
\begin{equation}
f_H(y)=\sqrt{\frac{2 k L (1-\beta )}{1-e^{-2(1-\beta )k L}}}e^{kL}e^{(2-\beta)k(y-L)}\,,
\end{equation}
with $\beta=\sqrt{4+m^2_H/k^2}$ and $m_H$ is the bulk mass of the Higgs field.

The three families of leptons and quarks transform under  $\mathrm {SU(2)_L\otimes SU(2)_{R}}$ in the following way,
\begin{eqnarray}
\label{eq:leptons-5D}\Psi_{\ell_i}&=&\left(\begin{array}{c}
\nu_{ i}^{[++]}\\
e_{i}^{[++]}
\end{array}\right)\sim (\mathbf{2},\mathbf{1})\,,~~
\Psi_{e_i}=\left(\begin{array}{c}
\tilde{\nu}_{i}^{\,[+-]}\\
e_{ i}^{[--]}
\end{array}\right)\sim(\mathbf{1},\mathbf{2})\,,~~
\Psi_{\nu_i}=\left(\begin{array}{c}
N_{i}^{[--]}\\
\tilde{e}_{i}^{\,[+-]} \end{array}\right)\sim(\mathbf{1},\mathbf{2})\,,~~\\
\label{eq:quarks-5D}\Psi_{Q_i}&=&\left(\begin{array}{c}
u_{i}^{[++]}\\
d_{i}^{[++]}
\end{array}\right)\sim (\mathbf{2},\mathbf{1})\,,~~
\Psi_{d_i}=\left(\begin{array}{c}
\tilde{u}_{i}^{\,[+-]}\\
d_{i}^{[--]}
\end{array}\right)\sim(\mathbf{1},\mathbf{2})\,,~~
\Psi_{u_i}=\left(\begin{array}{c}
u_{i}^{[--]}\\
\tilde{d}_{i}^{\,[+-]}
\end{array}\right)\sim(\mathbf{1},\mathbf{2})\,.~~
\end{eqnarray}
where the two signs in the bracket indicate Neumann ($+$) or Dirichlet ($-$) boundary conditions for the left-handed component of the corresponding field on UV and IR branes respectively. The Kaluza-Klein decomposition of a 5D fermion for the two different BCs are
\begin{eqnarray}
&&\psi^{[++]}(x^\mu,y)=\frac{e^{2ky}}{\sqrt{L}}\Big\{\psi_L(x^\mu)f_L(y,c_{L})+\text{heavy KK modes}\Big\}\,,\\
&&\psi^{[--]}(x^\mu,y)=\frac{e^{2ky}}{\sqrt{L}}\Big\{\psi_R(x^\mu)f_R(y,c_{R})+\text{heavy KK modes}\Big\}\,.
\end{eqnarray}
The 5D fields with $[++]$ BCs only have left-handed zero modes, while those with {$[--]$} BCs only have right-handed ones.
The functions $f_L(y,c_{L})$ and $f_R(y,c_{R})$ are the zero mode profiles~\cite{Grossman:1999ra,Gherghetta:2000qt}
\begin{equation}
f_L(y,c_{L})=\sqrt{\frac{(1-2 c_{L})kL}{e^{(1-2c_{L})kL}-1}} e^{-c_{L}ky}\,,
~~~f_R(y,c_R)=\sqrt{\frac{(1+2c_R)kL}{e^{(1+2c_R)kL}-1}} e^{ c_R ky}\,,
\end{equation}
where $c_{L}$ and $c_{R}$ denote the bulk mass of the 5D fermions in units of the AdS$_5$ curvature $k$. As usual, we adopt the zero mode approximation which identifies the standard model fields with the zero modes of corresponding 5D fields.

\subsubsection{Lepton masses and mixing }
\label{sec:lepton-sector}

The field content and the symmetry assignment are given in table~\ref{Tab:assignment_lepton}. The zero mode of $\Psi_{L}$ is the left-handed lepton doublet, and the zero modes of $\Psi_{e, \mu,\tau}$ and $\Psi_{\nu}$ are the right-handed charged leptons and neutrinos respectively.
The left-handed lepton fields are assumed to transform as a triplet under flavour symmetry group.

\begin{table}[hptb!]
\centering
\resizebox{1.0\textwidth}{!}{
\begin{tabular}{|c|c|c|c|c|c|c|c|c|c|c|}\hline
Field & $\Psi_{l}$ & $\Psi_{e}$ & $\Psi_{\mu}$ & $\Psi_{\tau}$ & $\Psi_{\nu}$ & $H$ & $\varphi_{l}(IR)$ & $\sigma_{l}(IR)$ & $\varphi_{\nu}(UV)$ & $\rho_{\nu}(UV)$  \\ \hline
$SU(2)_L\times SU(2)_R\times U(1)_{B-L}$ & $(2,1,-1)$ & $(1,2,-1)$ & $(1,2,-1)$ & $(1,2,-1)$ & $(1,2,-1)$ & $(2,2,0)$ & $(1,1,0)$ & $(1,1,0)$ & $(1,1,0)$ & $(1,1,0)$  \\ \hline
$T'$ & $\mathbf{3}$ & $\mathbf{1'}$ & $\mathbf{1''}$ & $\mathbf{1}$ & $\mathbf{3}$ & $\mathbf{1}$ &$\mathbf{2}$ & $\mathbf{1''}$ & $\mathbf{3}$ & $\mathbf{3}$   \\ \hline
$Z_{3}$ & $\omega^{2}$ & $1$ &  $1$ &  $1$ & $\omega^{2}$ & $1$ &  $\omega$ &  $\omega$ &  $\omega$ &  $\omega$ \\ \hline
$Z_{4}$ & $i$ & $i$ &  $i$ &  $i$ & $i$ & $1$ &  $-1$ & $-1$ &  $i$  &  $-i$  \\ \hline
\end{tabular}}
\caption{\label{Tab:assignment_lepton} Lepton and flavon fields under the $\mathrm {SU(2)_{L}\times SU(2)_{R}\times U(1)_{B-L}}$ gauge group and the $T'\times Z_3\times Z_4$ family symmetry, with $\omega = e^{2\pi i/3}$. Flavons $\varphi_l$, $\sigma_l$ and $\varphi_{\nu}$, $\rho_{\nu}$ lie on the IR and UV branes respectively. }
\end{table}

Note that in order to forbid dangerous terms,  besides the flavour group $T'$, we introduce the auxiliary symmetry $Z_3\times Z_4$. Four flavons $\varphi_l$, $\sigma_l$, $\varphi_{\nu}$ and $\rho_{\nu}$ are introduced to break the $T'$ family symmetry. The flavons $\varphi_l$ and $\sigma_l$ couple to the charged lepton sector, and are localized on the IR brane, while $\varphi_{\nu}$ and $\rho_{\nu}$ are localized on the UV brane. The vacuum expectation values of these flavons are aligned along the following directions
\begin{equation}
\label{eq:vacuum}
\vev{\varphi_{l}}=(1,0)v_{\varphi_{l}}\,,~~
\vev{\sigma_{l}}=v_{\sigma_{l}}\,,~~
\vev{\varphi_{\nu}}=(1, -2\omega^2, -2\omega)v_{\varphi_{\nu}}\,,~~
\vev{\rho_{\nu}}=(1, -2\omega, -2\omega^{2})v_{\rho_{\nu}}\,,
\end{equation}
with $\omega=e^{2i\pi/3}$. The above vacuum alignment can be the global minimum of the scalar potential in certain regions of parameters~\cite{Chen:2020udk}. At leading order, the lepton mass terms respecting both gauge symmetry as well as the flavour symmetry $T'\times Z_3\times Z_4$ take the following form
\begin{eqnarray}
\label{eq:charged_lepton-EXD}
\mathcal{L}^{l}_{Y}&=&\frac{\sqrt{G}}{\Lambda'^3}\Big[
 y _{e}(\varphi_{l}^{2}\overline{\Psi}_{l})_{\mathbf{1''}}H\Psi_{e}
+y_{\mu}(\varphi_{l}^{2}\overline{\Psi}_{l})_{\mathbf{1'}}H\Psi_{\mu}
+y_{\tau}(\varphi_{l}^2\overline{\Psi}_{l})_{\mathbf{1}}H\Psi_{\tau}\Big]
\delta(y-L)+\text{h.c.}\,,\\
\nonumber\mathcal{L}^{\nu}_{Y}&=&y_{\nu_{1}}\frac{\sqrt{G}}{\Lambda'}(\overline{\Psi}_{l}H\Psi_{\nu})_{\mathbf{1}}\delta(y-L) +\frac{1}{2}\frac{\sqrt{G}}{\Lambda^{2}}\Big[y_{\nu_{2}}(\overline{N^C}N)_{\mathbf{1}}(\varphi_{\nu}^{2})_{\mathbf{1}}+ y_{\nu_{3}}(\overline{N^C}N)_{\mathbf{1}} (\rho_{\nu}^{2})_{\mathbf{1}} \\
&~&\label{eq:neutrino-EXD}+y_{\nu_{4}}\left((\overline{N^C}N)_{\mathbf{3_{S}}}(\varphi_{\nu}^{2})_{\mathbf{3_{S}}}\right)_{\mathbf{1}} +y_{\nu_{5}}\left((\overline{N^C}N)_{\mathbf{3_{S}}}(\rho_{\nu}^{2})_{\mathbf{3_{S}}}\right)_{\mathbf{1}}\Big]\delta(y)+\text{h.c.}\,,
\end{eqnarray}
where $G=e^{-8ky}$ is the determinant of the 5D metric.

For the vacuum configuration in Eq.~\eqref{eq:vacuum}, the charged lepton mass matrix is diagonal and the three charged lepton masses are
\begin{equation}
m_{e}=\tilde{y}_{e}\frac{v_{\varphi_{l}}^2}{\Lambda'^2}\frac{v}{\sqrt{2}}\,,~~~
m_{\mu}=\tilde{y}_{\mu}\frac{v_{\varphi_{l}}^{2}}{\Lambda'^2}\frac{v}{\sqrt{2}}\,,~~~
m_{\tau}=\tilde{y}_{\tau}\frac{v_{\varphi_{l}}^2}{\Lambda'^2}\frac{v}{\sqrt{2}}\,,
\end{equation}
with
\begin{equation}
\tilde{y}_{e,\mu,\tau} = \frac{y_{e,\mu,\tau}}{L\Lambda'}f_L(L,c_\ell)f_R(L,c_{e, \mu, \tau})\,.
\end{equation}
Neutrino masses are generated by the type-I seesaw mechanism, and the large seesaw scale arises naturally, since the Majorana mass terms of the right-handed neutrinos are UV--localized. The first term in Eq.~\eqref{eq:neutrino-EXD} leads to a diagonal Dirac neutrino mass matrix $ m_{D}=\tilde{y}_{\nu_{1}}\frac{v}{\sqrt{2}}\mathbb{1}_3$ where $\tilde{y}_{\nu_{1}} = \frac{y_{\nu_1}}{L\Lambda'} f_L(L,c_\ell)f_R(L,c_{\nu})$ and $\mathbb{1}_3$ is the $3\times3$ unit matrix. The last four are the Majorana mass terms for the right-handed neutrinos, leading to the mass matrix
\begin{eqnarray}
\nonumber m_{N}&=&\left(\tilde{y}_{\nu_{2}}\frac{v_{\varphi_{\nu}}^{2}}{\Lambda}+\tilde{y}_{\nu_{3}}\frac{v_{\rho_{\nu}}^{2}}{\Lambda}\right)
\begin{pmatrix} 1 ~&~ 0 ~&~ 0 \\ 0 ~&~ 0 ~&~ 1 \\ 0 ~&~ 1 ~&~ 0 \end{pmatrix}+ \tilde{y}_{\nu_{4}}\frac{v_{\varphi_{\nu}}^{2}}{\Lambda} \begin{pmatrix} 2 ~&~ 2\omega ~&~ 2\omega^{2} \\
2\omega ~&~ -4\omega^{2} ~&~ -1 \\
2\omega^{2} ~&~ -1 ~&~ -4\omega  \end{pmatrix} \\
\label{eq:mD-mN} &&+\tilde{y}_{\nu_{5}}\frac{v_{\rho_{\nu}}^{2}}{\Lambda} \begin{pmatrix} 2 ~&~ 2\omega^{2} ~&~ 2\omega \\
2\omega^{2} ~&~ -4\omega ~&~ -1 \\
2\omega ~&~ -1 ~&~ -4\omega^{2} \end{pmatrix} \,,
\end{eqnarray}
where
\begin{equation}
\tilde{y}_{\nu_{2,3,4,5}} = \frac{y_{\nu_{2,3,4,5}}}{L\Lambda}f^2_R(0,c_{\nu})\,.
\end{equation}
The light neutrino mass matrix is given by the simple type-I seesaw formula
\begin{equation}
m_{\nu}=-m_{D}m_{N}^{-1}m_{D}^{T}\,.
\end{equation}
After performing a tri-bimaximal transformation on the neutrino fields, $m_{\nu}$ acquires block diagonal form,
\begin{equation}
\hskip-0.05in m'_{\nu}=U_{TBM}^{\dagger}m_{\nu}U_{TBM}^{*}=m_0
\left( \begin{array}{ccc}
\frac{-1}{1+3(y_{4}+y_{5})} ~& 0 ~& 0 \\
0 ~& \frac{1-3(y_{4} + y_{5})}{18(y_{4}-y_{5})^{2}+3(y_{4}+y_{5})-1} ~& \frac{3\sqrt{2}i(y_{4}-y_{5})}{18(y_{4}-y_{5})^{2}+3(y_{4}+y_{5})-1} \\
0 ~& \frac{3\sqrt{2}i(y_{4}-y_{5})}{18(y_{4}-y_{5})^{2}+3(y_{4}+y_{5})-1} ~& \frac{-1}{18(y_{4}-y_{5})^{2}+3(y_{4}+y_{5})-1}
\end{array} \right)\,,\label{eq:mnu-EXD-Tp}
\end{equation}
with $m_0=\frac{\tilde{y}_{\nu_{1}}^{2}\Lambda v^{2}}{2\left(\tilde{y}_{\nu_{2}}v_{\varphi_{\nu}}^{2}+\tilde{y}_{\nu_{3}}v_{\rho_{\nu}}^{2}\right)}$, $y_{4}=\frac{\tilde{y}_{\nu_{4}}v_{\varphi_{\nu}}^{2}}{\tilde{y}_{\nu_{2}}v_{\varphi_{\nu}}^{2}+\tilde{y}_{\nu_{3}}v_{\rho_{\nu}}^{2}}$ and $y_{5}=\frac{\tilde{y}_{\nu_{5}}v_{\rho_{\nu}}^{2}}{\tilde{y}_{\nu_{2}}v_{\varphi_{\nu}}^{2}+\tilde{y}_{\nu_{3}}v_{\rho_{\nu}}^{2}}$.\\[-.4cm]

One sees from Eq.~\eqref{eq:mnu-EXD-Tp} that the light neutrino mass matrix only depends on two complex parameters $y_4$ and $y_5$ and on the overall scale $m_0$. As a result one expects sharp predictions both for neutrino masses as well as mixing parameters. As shown in~\ref{app:diag}, the above block-diagonal matrix is exactly diagonalized as
\begin{equation}
U_{\nu}'^{\dagger}m_{\nu}'U_{\nu}'^{*}=\text{diag}(m_{1},m_{2},m_{3}),~~~~ U_{\nu}'=\begin{pmatrix} 1 & 0 & 0 \\ 0 & \cos\theta_{\nu} & \sin\theta_{\nu} e^{i\delta_{\nu}} \\ 0 & -\sin\theta_{\nu} e^{-i\delta_{\nu}} & \cos\theta_{\nu} \\ \end{pmatrix}\,,
\end{equation}
The charged lepton mass matrix $m_{l}$ is already in diagonal form, consequently the lepton mixing matrix is determined to be
\begin{equation}
\label{eq:UPMNS-EXD}U=U_{TBM}U_{\nu}'=\frac{1}{\sqrt{6}}\begin{pmatrix}
2 ~&\sqrt{2}\cos\theta_{\nu} ~& \sqrt{2}\sin\theta_{\nu} e^{i\delta_{\nu}} \\
-1 ~&\sqrt{2}\cos\theta_{\nu} -\sqrt{3}\sin\theta_{\nu}e^{-i\delta_{\nu}} ~& \sqrt{3}\cos\theta_{\nu}+\sqrt{2}\sin\theta_{\nu}e^{i\delta_{\nu}} \\
-1 ~& \sqrt{2}\cos\theta_{\nu} +\sqrt{3}\sin\theta_{\nu} e^{-i\delta_{\nu}}  ~&-\sqrt{3}\cos\theta_{\nu}+\sqrt{2}\sin\theta_{\nu} e^{i\delta_{\nu}} \end{pmatrix}\,.
\end{equation}
We notice that the first column of the lepton mixing matrix is fixed to be $(2, -1,-1)^{T}/\sqrt{6}$ which coincides with the first column of the TBM mixing pattern. In other words, the lepton mixing matrix has trimaximal TM1 form~\cite{Albright:2008rp,Albright:2010ap,He:2011gb}.

From Eq.~\eqref{eq:UPMNS-EXD} one can extract the mixing angles and the leptonic Jarlskog invariant in the usual way, to find
\begin{eqnarray}
\nonumber &&~~~~\sin^{2}\theta_{13}=\frac{1}{3}\sin^{2}\theta_{\nu}\,,~~~\sin^{2}\theta_{12}=\frac{1+\cos2\theta_{\nu}}{5+\cos2\theta_{\nu}}\,,\\
\label{eq:mix-param}&&\sin^{2}\theta_{23}=\frac{1}{2}+\frac{\sqrt{6}\sin2\theta_{\nu}\cos\delta_{\nu}}{5+\cos2\theta_{\nu}}\,,~~
J_{CP}=-\frac{\sin2\theta_{\nu}\sin\delta_{\nu}}{6\sqrt{6}}\,.
\end{eqnarray}
One sees from Eq.~\eqref{eq:mix-param} that all the three mixing angles and the Dirac CP phase $\delta_{CP}$ are given in terms of just two free parameters $\delta_{\nu}$ and $\theta_{\nu}$. The above relations imply two predicted correlations amongst the mixing angles and the Dirac CP violation phase,
\begin{equation}
\cos^{2}\theta_{12}\cos^{2}\theta_{13}=\frac{2}{3}\,,~~~~ \cos\delta_{CP}=\frac{(3\cos 2\theta_{12}-2)\cos 2\theta_{23}}{3\sin 2\theta_{23}\sin2\theta_{12}\sin\theta_{13}}\,,
\end{equation}
which characterizes the TM1 mixing pattern.

\begin{figure}[h!]
\centering
\begin{tabular}{cc}
\includegraphics[width=0.48\linewidth]{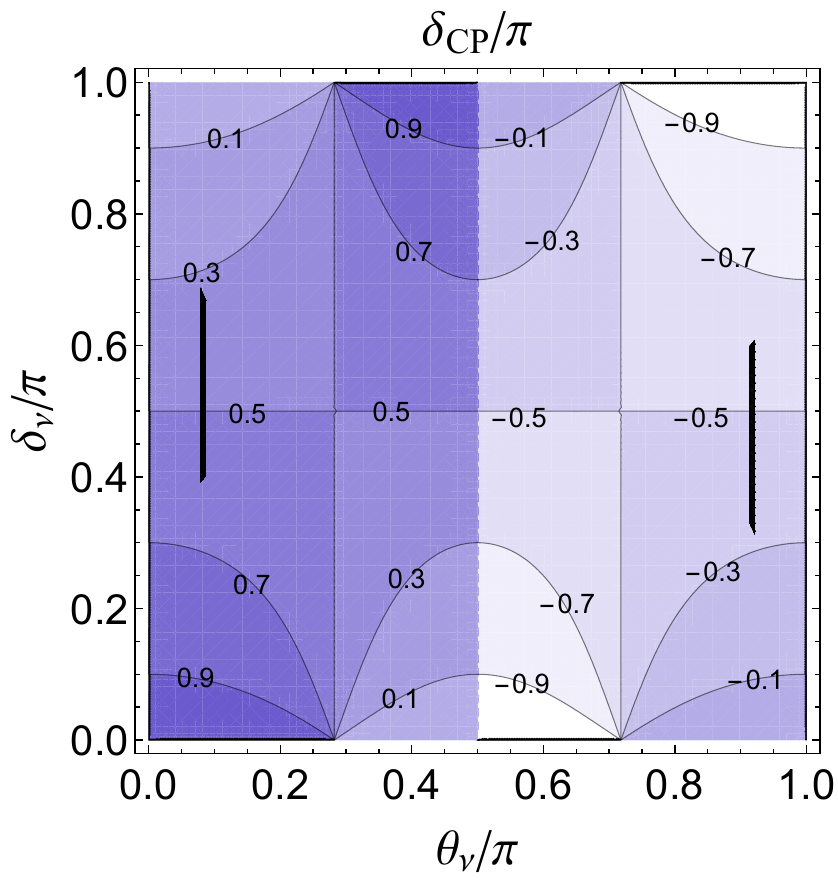}
\includegraphics[width=0.48\linewidth]{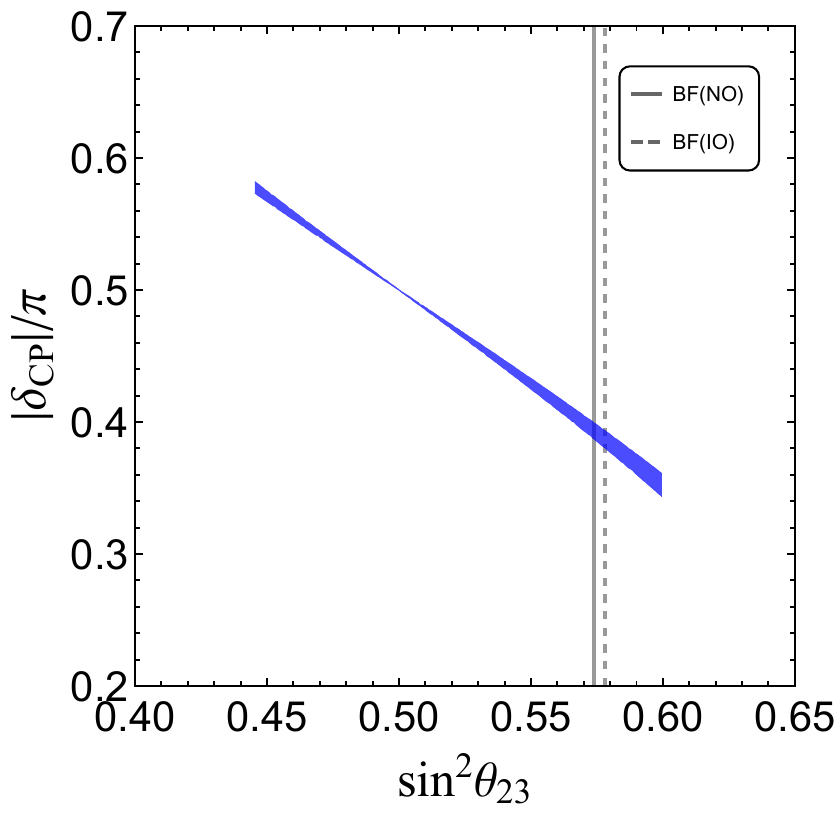}
\end{tabular}
\caption{\label{fig:deltaCP-contour-EXD} Contour plots of $\delta_{CP}$ in the $\theta_{\nu}-\delta_{\nu}$ plane (left) and predicted correlation between $|\delta_{CP}|$ and $\sin^{2}\theta_{23}$ (right).
The vertical solid/dashed lines in the right panel are the best-fit $\sin^2\theta_{23}$ values for NO/IO spectra, respectively~\cite{deSalas:2020pgw,10.5281/zenodo.4726908}. }
\end{figure}

In the left panel of figure~\ref{fig:deltaCP-contour-EXD} we display the contour plot of $\delta_{CP}$ in the plane $\delta_{\nu}$ versus $\theta_{\nu}$. The small black areas in the left panel indicate the regions in which all three lepton mixing angles lie within their experimentally allowed $3\sigma$ ranges. The right panel of figure~\ref{fig:deltaCP-contour-EXD} shows a very tight correlation between $|\delta_{CP}|$  and the magnitude of the atmospheric angle $\theta_{23}$. One sees that both octants are consistent, the sign of $\delta_{CP}$ can not be determined, and the absolute value the Dirac CP phase parameter $|\delta_{CP}|$ is predicted to lie in the restricted range $[0.325\pi, 0.592\pi]$. Upcoming long-baseline experiments will be able to test these predictions for $\theta_{23}$ and $\delta_{CP}$~\cite{DUNE:2015lol,Hyper-Kamiokande:2018ofw,Athar:2021xsd}.

\begin{figure}[h!]
\centering
\begin{tabular}{c}
\includegraphics[width=0.65\linewidth]{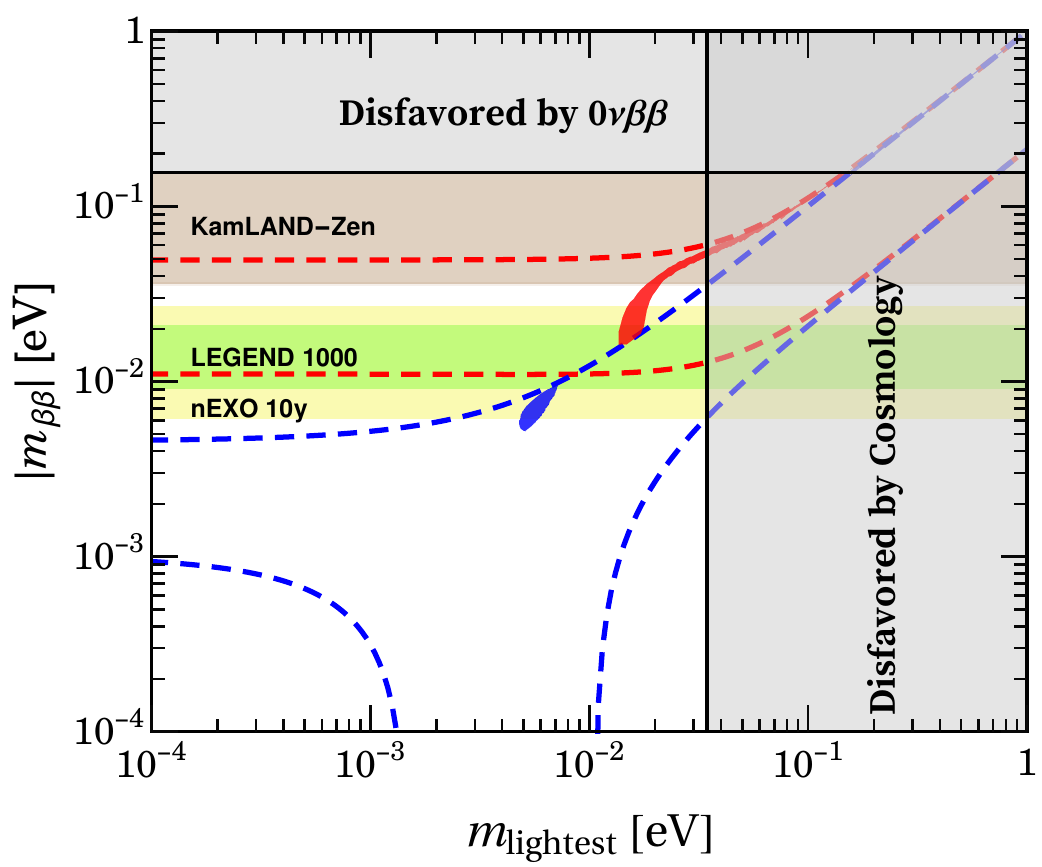}
\end{tabular}
\caption{Predicted mass parameter characterizing the \znbb decay amplitude; red and blue regions are for IO and NO neutrino mass spectra, respectively. Here we adopt the same convention as figure~\ref{fig:onbb-Z2xCP-S4} for different bands and boundaries.}
\label{fig:znbb-EXD-Tp}
\end{figure}

We now turn to neutrinoless double beta decay. In figure~\ref{fig:znbb-EXD-Tp} we display the expected values for the effective Majorana neutrino mass $|m_{\beta\beta}|$ characterizing the $0\nu\beta\beta$ decay amplitude. If the neutrino mass spectrum is inverted-ordered (IO), the effective Majorana mass has a lower limit $|m_{\beta\beta}|\geq 0.0162~\text{eV}$, while the lightest neutrino mass satisfies $m_{\text{lightest}}\geq 0.0133~\text{eV}$. For the case of normal-ordering (NO), the effective mass $|m_{\beta\beta}|$ lies in the narrow interval $[5.2\text{meV}, 9.6\text{meV}]$, and the allowed range of $m_{\text{lightest}}$ is $[4.8\text{meV}, 7.2\text{meV}]$. Clearly, as indicated in the figure, the predicted regions for the lightest neutrino mass and $|m_{\beta\beta}|$ are quite restricted. The existing experimental bounds as well as the estimated experimental sensitivities are also indicated by the horizontal bands~\cite{KamLAND-Zen:2022tow,LEGEND:2021bnm,nEXO:2021ujk} in figure~\ref{fig:znbb-EXD-Tp}. The predicted decay amplitudes do suggest a guaranteed \znbb discovery at the forthcoming round of experiments~\cite{Cirigliano:2022oqy}.

\begin{table}[!b]
\centering
\resizebox{1.0\textwidth}{!}{
\begin{tabular}{|c|c|c|c|c|c|c|c|c|c|c|}
\hline
Field & $\Psi_{UC}$ & $\Psi_{T}$ & $\Psi_{u}$ & $\Psi_{c}$ & $\Psi_{t}$ & $\Psi_{ds}$ & $\Psi_{b}$ & $H$ & $\varphi_{l}(IR)$ & $\sigma_{l}(IR)$\\\hline
$SU(2)_L\times SU(2)_R\times U(1)_{B-L}$ & $(2,1,1/3)$ & $(2,1,1/3)$ & $(1,2,1/3)$ & $(1,2,1/3)$ & $(1,2,1/3)$ & $(1,2,1/3)$ & $(1,2,1/3)$ & $(2,2,0)$ & $(1,1,0)$ & $(1,1,0)$ \\
\hline
$T'$ & $\mathbf{2}$ & $\mathbf{1}$ & $\mathbf{1'}$ & $\mathbf{1''}$ & $\mathbf{1'}$ & $\mathbf{2'}$ &$\mathbf{1''}$ & $\mathbf{1}$ & $\mathbf{2}$ & $\mathbf{1''}$ \\ \hline
$Z_{3}$ & $\omega^{2}$ & $\omega$ &  $1$ &  $\omega^{2}$ & $1$ & $\omega$ &  $\omega^{2}$ & $1$ & $\omega$ & $\omega$ \\ \hline
$Z_{4}$ & $1$ & $-1$ &  $1$ &  $-1$ & $-1$ & $1$ &  $-1$ & $1$ & $-1$ & $-1$  \\
\hline
\end{tabular}}
\caption{Classification of the quark fields under the bulk gauge group $\mathrm {SU(2)_{L}\times SU(2)_{R}\times U(1)_{B-L}}$ and the flavour symmetry $T'\times Z_3\times Z_4$. }
\label{tab:assignment-quark-EXD}
\end{table}

\subsubsection{Quark masses and CKM matrix }
\label{sec:quark-sector}

This model can be extended to include quarks, the transformation properties of the quark fields under the family symmetry $T'\times Z_3\times Z_4$ are listed in table~\ref{tab:assignment-quark-EXD}. Note that no new scalars are needed, beyond the flavons $\varphi_l$ and $\sigma_{l}$ characterizing the lepton sector.

The quark Yukawa interactions are localized on the IR brane and constrained by the $T'$ flavour symmetry to be of the following form,
\begin{eqnarray}
\nonumber \mathcal{L}^{d}_{Y}&=&\frac{\sqrt{G}}{\Lambda'^3}\Big[
y_{ds_1} (\overline{\Psi}_{UC}{H}\Psi_{ds})_{\mathbf{3}} \varphi_{l}^{\ast\,2}
 + y_{ds_2} (\overline{\Psi}_{UC}{H}\Psi_{ds})_{\mathbf{1'}} \sigma_l^{\ast\,2}
 + y_{b}' (\overline{\Psi}_{T}{H}\Psi_{b})_{\mathbf{1''}}  \sigma_l^{2}
\Big]\delta(y-L) + \text{h.c.} + \cdots \\
\nonumber \mathcal{L}^{u}_{Y}&=&\frac{\sqrt{G}}{\Lambda'^3}\Big[
 y_{u}' \Lambda' (\overline{\Psi}_{T}{H}\Psi_{u})_{\mathbf{1'}} \sigma_l
 + y_t \Lambda' (\overline{\Psi}_{UC}{H}\Psi_{t})_{\mathbf{2}} \varphi_{l}^{*}
 + y_u (\overline{\Psi}_{UC}{H}\Psi_{u})_{\mathbf{2}} \varphi_{l}\sigma_l\\
\label{eq:LYu}&&\qquad\qquad
 + y_{c}' (\overline{\Psi}_{T}{H}\Psi_{c})_{\mathbf{1''}} \sigma_{l}^{2}
 + y_t' (\overline{\Psi}_{T}{H}\Psi_{t})_{\mathbf{1'}} \sigma_{l}^{\ast 2}\Big]\delta(y-L)+ \text{h.c.}+\cdots\,,
\end{eqnarray}
for the down-type and up-type quark masses respectively. One can read out the mass matrices for the zero modes of the quark fields as
\begin{eqnarray}
\label{eq:mass-quark-down}
m^{d}&=&\frac{v}{\sqrt{2}}\begin{pmatrix}
\tilde{y}_{ds_2}v_{\sigma_l}^{\ast2}/\Lambda'^2 & 0 & 0 \\
\tilde{y}_{ds_1}v_{\varphi_l}^{\ast2}/\Lambda'^2 & \tilde{y}_{ds_2}v_{\sigma_l}^{\ast2}/\Lambda'^2 & 0 \\
0&0 & \tilde{y}_{b}'v_{\sigma_l}^{2}/\Lambda'^2
\end{pmatrix}\,,\\
\label{eq:mass-quark-up}m^{u}&=&\frac{v}{\sqrt{2}}\begin{pmatrix}
\tilde{y}_u v_{\varphi_l}v_{\sigma_l}/\Lambda'^2 & 0 & 0 \\
0 & 0 & \tilde{y}_t v_{\varphi_l}^\ast/\Lambda' \\
\tilde{y}_u' v_{\sigma_l}/\Lambda' &
\tilde{y}_c' v_{\sigma_{l}}^{2}/\Lambda'^{2} &
\tilde{y}_t' v_{\sigma_l}^{\ast2}/\Lambda'^2 \\
\end{pmatrix}\,.
\end{eqnarray}
where the parameters with tilde are given by
\begin{eqnarray}
\tilde{y}_{u,t,ds_{1,2}}=\frac{y_{u,t,ds_{1,2}}}{L\Lambda'}f_L(L,c_{UC})f_R(L,c_{u,t,ds})\,,~~~
\tilde{y}_{u,c,t,b}^\prime = \frac{y_{u,c,t,b}^\prime}{L\Lambda'}f_L(L,c_{T})f_R(L,c_{u,c,t,b})\,.~
\end{eqnarray}

Notice that the down-quark mass matrix is block diagonal and the (11) and (22) entries are exactly equal. Note also that the up-quark sector gives a negligible contribution to the Cabibbo angle $\theta_c$ but is responsible for generating $V_{ub}$ and $V_{cb}$. As a result this model gives rise to the well-known Gatto-Sartori relation $m_d/m_s\simeq \tan^2\theta_c$~\cite{Gatto:1968ss}.

\subsubsection{Global flavour fit results }

Following the procedure given in Appnedix~\ref{app:global-fitting}, we now perform a global fit of the masses and flavour mixing parameters of both quarks and leptons within this model. The fundamental 5D scales on the UV and IR branes are taken to be $\Lambda\simeq k\simeq2.44\times 10^{18}$ GeV and $\Lambda^\prime=ke^{-kL}\simeq 1.5$ TeV respectively. The vacuum expectation value of the Higgs field is fixed to its SM value $v\simeq246$ GeV, and we choose the flavon VEVs as $v_{\varphi_{l}}/\Lambda'=v_{\sigma_{l}}/\Lambda'=v_{\varphi_{\nu}}/\Lambda=v_{\rho_{\nu}}/\Lambda=0.2$.
In what follows we give a typical set of values for the free parameters.
The bulk mass parameters and the Yukawa coupling constants of the charged lepton and quarks are given by
\begin{eqnarray}
\nonumber &&c_l=0.460,~~
c_e=-0.725,~~
c_\mu=-0.553,~~
c_\tau=-0.117,\\
\nonumber&&c_{UC}=0.587, ~~
c_T=-0.980, ~~
c_u=-0.516, ~~
c_c=-0.555,\\
\nonumber&& c_t = 0.966,~~ c_{ds}=-0.503,~~c_b=-0.532, \\
\nonumber &&y_e=y_{\mu}=y_{\tau}=1.0,~~y_u = 6.321, ~~
y_t=6.20,~~y_u' = 4.00,\\
&&y_c'=1.00,~~y_t'=8.30, ~~
y_{ds1}=4.00,~~y_{ds2}=0.892,~~y_b'=4.00\,.
\end{eqnarray}
The values of the parameters in the neutrino sector depend on the neutrino mass ordering,
\begin{small}
\begin{eqnarray}
\nonumber&&\hskip-0.3in \text{NO}: c_\nu=-0.404,~~
y_{\nu1}=y_{\nu2}=y_{\nu3}=1,~~
y_{\nu4}=0.235+0.0770i,~~
y_{\nu5}=0.340+0.0710i,\\
&&\hskip-0.3in  \text{IO}: c_\nu=-0.383,~~
y_{\nu1}=y_{\nu2}=y_{\nu3}=1,~~
y_{\nu4}=-0.354+0.275i,~~
y_{\nu5}=-0.562+0.270i\,.
\end{eqnarray}
\end{small}
The resulting predictions for flavour observables such as lepton and quark mass and mixing parameters are all listed in table~\ref{Tab:fitted_fermion_parameters}. One sees that all \sm fermion masses and mixings can be very well reproduced.

\begin{table}[h!]
\centering
\begin{tabular}{|c|c|c|}
\hline
parameters
& best-fit $\pm$ $1\sigma$ & predictions\\
\hline
$\sin\theta_{12}^{q}$
& 0.22500$\pm0.00100$
& 0.22503  \\
\hline
$\sin\theta_{13}^{q}$
& 0.003675$\pm0.000095$
& 0.003668  \\
\hline
$\sin\theta_{23}^{q}$
& 0.04200$\pm0.00059$
& 0.04205  \\
\hline
$\delta^{q}_{CP}/^{\circ}$
& 66.9$\pm 2$
& 68.2  \\
\hline
$m_{u}~[\text{MeV}]$
& 2.16$^{+0.49}_{-0.26}$
& 2.16   \\
\hline
$m_{c}~[\text{GeV}]$
& 1.27$\pm 0.02$
& 1.27  \\
\hline
$m_{t}~[\text{GeV}]$
& 172.9$\pm 0.4$
& 172.90   \\
\hline
$m_{d}~[\text{MeV}]$
& 4.67$^{+0.48}_{-0.17}$
& 4.21   \\
\hline
$m_{s}~[\text{MeV}]$
& 93$^{+11}_{-5}$
& 93.00   \\
\hline
$m_{b}~[\text{GeV}]$
& 4.18$^{+0.03}_{-0.02}$
& 4.18   \\
\hline
\hline
$\sin^2\theta_{12}^{l} / 10^{-1}$ (NO)
& \multirow{2}{*}{3.20$^{+0.20}_{-0.16}$} & 3.19 \\
$\sin^2\theta_{12}^{l} / 10^{-1}$ (IO)
& & 3.18 \\
\hline
$\sin^2\theta_{23}^{l} / 10^{-1}$ (NO)
& 5.47$^{+0.20}_{-0.30}$ & 5.47  \\
$\sin^2\theta_{23}^{l} / 10^{-1}$ (IO)
& 5.51$^{+0.18}_{-0.30}$ & 5.51 \\
\hline
$\sin^2\theta_{13}^{l}/ 10^{-2}$ (NO)
& 2.160$^{+0.083}_{-0.069}$ & 2.160 \\
$\sin^2\theta_{13}^{l} / 10^{-2}$ (IO)
& 2.220$^{+0.074}_{-0.076}$ & 2.220 \\
\hline
$\delta_{CP}^{l}/\pi$ (NO)
& 1.32$^{+0.21}_{-0.15}$ & 1.567\\
$\delta_{CP}^{l}/\pi$ (IO)
& 1.56$^{+0.13}_{-0.15}$ & 1.571 \\
\hline
$m_{e}~[\text{MeV}]$ & $0.511 \pm 3.1\times 10^{-9}$ & $0.511$ \\
\hline
$m_{\mu}~[\text{MeV}]$ & $105.658 \pm 2.4\times 10^{-6}$ & $105.658$ \\
\hline
$m_{\tau}~[\text{MeV}]$ & $1776.86 \pm 0.12$ & $1776.86$ \\
\hline
$\Delta m_{21}^{2}~[10^{-5}\text{eV}^{2}]$ (NO)
& \multirow{2}{*}{ 7.55$^{+0.20}_{-0.16}$} & \multirow{2}{*}{ 7.55}  \\
$\Delta m_{21}^{2}~[10^{-5}\text{eV}^{2}]$ (IO)
&  & \\
\hline
$|\Delta m_{31}^{2}|~[10^{-3}\text{eV}^{2}]$ (NO)
&  2.50$\pm$0.03 & 2.50 \\
$|\Delta m_{31}^{2}|~[10^{-3}\text{eV}^{2}]$ (IO)
&  2.42$^{+0.03}_{-0.04}$ & 2.42\\
\hline
$\chi^{2}$ (NO) & \multirow{2}{*}{$-$} & 7.65 \\
$\chi^{2}$ (IO) &  & 7.66 \\
\hline
\end{tabular}
\caption{\label{Tab:fitted_fermion_parameters}
Global warped flavordynamics fit: neutrino oscillation parameters are taken from the global analysis in~\cite{deSalas:2020pgw,10.5281/zenodo.4726908}, while the quark parameters are taken from the Review of Particle Physics~\cite{Workman:2022ynf}.}
\end{table}

All in all the model provides a consistent scenario for the flavour problem, in which fermion mass hierarchies are accounted for by adequate choices of the bulk mass parameters, while quark and lepton mixing angles are restricted by the assumed $T^\prime$ flavour symmetry. Note that in this model neutrinos are Majorana particles, the tiny neutrino masses are generated by the type-I seesaw mechanism with ``right-handed'' neutrino masses in the range of $[10^{12}, 10^{13}]$ GeV and relatively sizeable rates for \znbb decay, accessible within the next round of experiments. For an alternative warped flavourdynamics construction along the similar lines, see Ref.~\cite{Chen:2015jta}. In that case the model uses the $\Delta(27)$ family symmetry, neutrinos are Dirac fermions, and the predicted neutrino mixing pattern is TM2.


\subsection{Family symmetry from 6-D orbifolds}                            
\label{sec:family-symmetry-cp}                                             

Underpinning the nature of the underlying family symmetry of particle physics amongst the huge plethora of possibilities constitutes a formidable task. As already seen in the previous section, a promising approach to the flavour problem is to imagine the existence of new dimensions in spacetime. Here we consider a six-dimensional setup compactified on a torus~\cite{deAnda:2018oik,deAnda:2018yfp} and implementing a realistic $A_4$ family symmetry, featuring the ``golden'' quark-lepton mass formula  \begin{equation}
\label{eq:golden} \frac{m_\tau}{\sqrt{m_\mu m_e}}\approx\frac{m_b}{\sqrt{m_s m_d}}\,.
\end{equation}
This formula was proposed in~\cite{Morisi:2011pt}, and emerges also in other 4-D flavour schemes such as those in Refs.~\cite{Bonilla:2014xla,Bonilla:2017ekt} and~\cite{Morisi:2013eca,King:2013hj}, as well as in implementations of the Peccei-Quinn symmetry~\cite{Reig:2018ocz}. It is remarkable that it also comes out neatly in scenarios where the family symmetry arises from the compactification of 6-dimensional orbifolds,
as proposed in Refs.~\cite{deAnda:2019jxw,deAnda:2020pti,deAnda:2020ssl,deAnda:2021jzc} and considered next. Fermions are nicely arranged in terms of the $A_4$ family symmetry. Different setups can be identified, with very interesting phenomenology. Indeed, they bring in the possibility of predicting neutrino mixing angles and CP phases, as well as providing a good global description of flavour observables.

\subsubsection{General preliminaries}
\label{sec:TheFrame}

Here we summarize the theory framework and survey its main features.
First of all we have a 6-dimensional version of the standard \SM gauge symmetry, supplemented with orbifold compactification, as outlined below.
In the full six-dimensional theory, the spacetime manifold $\mathcal{M}$ is identified as the direct product $\mathcal{M} = \mathbb{M}^4\times (T^2/\mathbb{Z}_2)$, where $\mathbb{M}^4$ is the four-dimensional Minkowski spacetime, and $T^2 / \mathbb{Z}_2$ is a one-parameter family (given by $\theta$) of 2-D toroidal orbifolds defined by the following relations satisfied by the extra-dimensional coordinates
\begin{align}
\left(x^5,x^6\right) &= \left( x^5 + 2 \pi R_1, x^6\right), \label{eq:id1} \\
\left(x^5,x^6\right) &= \left( x^5 + 2 \pi R_2 \cos \theta,  x^6 + 2 \pi R_2 \sin \theta\right). \label{eq:id2} \\
\left(x^5,x^6\right)&= \left(-x^5,-x^6\right) ,\label{eq:id3}
 \end{align}
The first two equations define a torus, with $\theta$ describing its twist angle, and the third equation defines the $\mathbb{Z}_2$ orbifolding. For simplicity we assume that the characteristic radii of the compact extra dimensions are similar, i.e.
\begin{equation}
 R_1 \sim R_2 \sim 1/M_c,
 \label{eq:cScale}
\end{equation}
in terms of the compactification scale $M_c$. Moreover, the twist angle is assumed to be $\theta = 2\pi /3$. To simplify the analysis we define the scaled complex coordinate $z= M_c(x_5+ i x_6)/(2\pi) $ and rewrite Eqs. (\ref{eq:id1})-(\ref{eq:id3}) as \\[-1cm]
\begin{align}
z &= z+1, \label{eq:id1c} \\
z &= z+\omega,\label{eq:id2c} \\
z&= -z, \label{eq:id3c}
 \end{align}
where $\omega$ is the cubic root of unity
\begin{equation}
\omega \equiv e^{i \theta}  = e^{i 2\pi / 3}.
\label{eq:omega}
\end{equation}

\begin{figure} \centering
\includegraphics[scale=0.8]{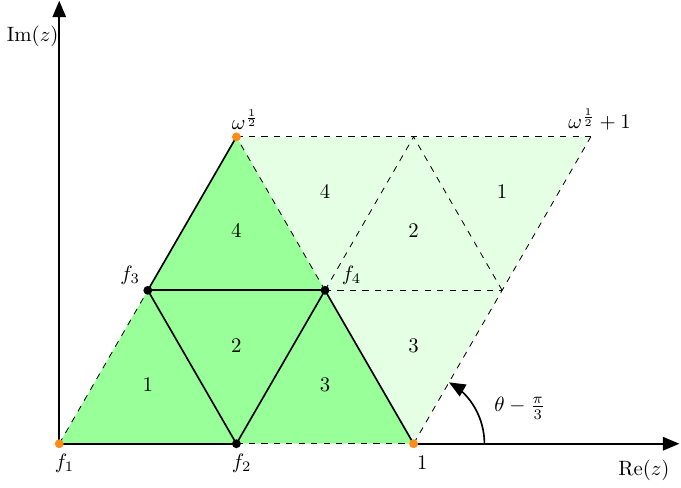}
\caption{The fundamental domain of the $T^2/\mathbb{Z}_2$ orbifold is the darkest region, obtained after the compactification of the corresponding domain of the twisted torus, which includes the lightest region. The resulting space is reminiscent of a tetrahedron, and can be visualized by identifying the three orange dots into a single vertex. The fixed points of the orbifold are located at the vertices of the tetrahedron. }
\label{fig:OrbifoldConstruction}
\end{figure}

A key feature of orbifolds is that they have singular points. In our case there are four of these, located at the points that remain fixed by the transformations in Eqs.~(\ref{eq:id1c})-(\ref{eq:id3c}), namely
\begin{equation}
f_1=0,\quad f_2=\frac{1}{2},\quad f_3=\frac{\omega}{2},\quad f_4=\frac{1+\omega}{2}.
 \label{eq:fixed}
\end{equation}

These fixed points define the location of 4-dimensional branes embedded in the 6-dimensional space $\mathcal{M}$. In figure~\ref{fig:OrbifoldConstruction} we display both the fundamental domain
of the twisted torus $T^2$ (light shaded green), as well as the fundamental domain of the $T^2/\mathbb{Z}_2$ orbifold (dark shaded green). After compactification, the continuous Poincar\'e symmetry of the two extra dimensions is broken, leaving a residual $A_4$ symmetry of the branes~\cite{Altarelli:2006kg}. The appearance of the discrete $A_4$ symmetry can be understood as the invariance under permutations displayed by the four fixed points of the orbifold. Any of these can be written in terms of two independent transformations
\begin{equation}
S: z\longrightarrow z+1/2, \quad T:z \longrightarrow \omega^2 z.
\label{eq:permut1}
\end{equation}
These can be also expressed as elements of the permutation group $S_4$.
\begin{equation}
 S=(12)(34), \quad T=(1)(243),
  \label{eq:permut2}
\end{equation}
This way $S$ and $T$ are related to the generators of the $A_4$ group, satisfying
\begin{equation}
S^2 =T^3=(ST)^3=1\,,
\label{eq:presentation}
\end{equation}
which are exactly the multiplication rules of the $A_4$ group in Eq.~\eqref{eq:A4-rule}.

The model is based on this remnant $A_4$ as a family symmetry. Same-charge fields located on the four different branes would transform into each other by the remnant $A_4$ transformations. These four branes transform as the reducible representation $\textbf{4}$, which decomposes as a sum of irreducible representations $\textbf{4}\to \textbf{3}+\textbf{1}$.
Thus, the brane-localized fields must transform under the flavour group as $A_4$ triplets or singlets, so the family symmetry is spontaneously broken.
Below we show how this can provide a realistic pattern for the three families of leptons and quarks in a rather predictive and economical way.

Notice that the assumption of extra dimensions implies the existence of infinitely many 4-D fields associated with every bulk field, called a Kaluza-Klein (KK) tower. Their masses $(n^2+m^2)M_c$ are determined by positive integers $n,m$. In our case, the fields in the bulk are the SM gauge fields $g_\mu, W_\mu, B_\mu$, the right-handed quarks $u_i^c$ and the gauge singlet scalar $\sigma$.

The tower of massive KK modes from the vector $\mathrm{SU(2)_L}$ triplets can affect the Peskin-Takeuchi oblique parameters $S$, $T$ and $U$ in an important way. The current experimental bound for our setup (2 non-universal extra dimensions) is~\cite{Deutschmann:2017bth,Ganguly:2018pzs}:
\begin{equation}
M_c>2.1\ {\rm TeV}\,.
\end{equation}
For a compactification scale sufficiently close to $2\ {\rm TeV}$, the electroweak precision tests could in principle probe the extra dimensions at the High Luminosity LHC run.

\subsubsection{Scotogenic orbifold}
\label{sec:simplest-model}

Our basic setup is a 6-dimensional extension of the Standard \SM Model, featuring the orbifold compactification described in the previous section,
and inheriting the $A_4$ discrete family symmetry in a natural manner. Its simplest model-realization includes three right-handed neutrinos, mediating neutrino mass generation through the type-I seesaw mechanism~\cite{deAnda:2019jxw,deAnda:2020pti}. Instead of pursuing such an approach, however, here we focus on a more complete, yet equally economical, scotogenic variant that also provides a
WIMP dark matter candidate~\cite{deAnda:2020ssl,deAnda:2021jzc}.

The field content and transformation properties of our benchmark scotogenic variant under the various symmetry groups are shown in table~\ref{tab:FieldContent}. Note that all fermionic fields, except for the right-handed quarks, transform as flavour triplets and are localized in the orbifold branes.
\begin{table}[H]
\centering
\begin{tabular}{@{}|c|ccc|c|c|c|c@{}}
 \multicolumn{7}{c}{\textbf{}}                 \vspace{2pt} \\ \hline
Field &                 $\mathrm{SU(3)_{C}}$       & $\mathrm{SU(2)_{L}}$ & $\mathrm{U(1)_{Y}}$ & $\mathbb{Z}_{4}$ & $A_4$ & Location  \\ \hline
   $L$                   &        $\bm{1}$     &   $\bm{2}$   &   $-1$               &       $1$    &   $\bm{3}$          &   Brane          \\
   $d^{c}$             &        $\bm{3}$     &   $\bm{1}$   &  $\sfrac{2}{3}$  &      $1$      &   $\bm{3}$           &   Brane         \\
   $e^c$                &       $\bm{1}$     &    $\bm{1}$   &     $  2   $     &           $1$       &   $\bm{3}$   &   Brane           \\
   $Q$                  &        $\bm{3}$     &    $\bm{2}$  &  $\sfrac{1}{3}$  &       $1$        &   $\bm{3}$     &   Brane           \\
   $u^c_{1,2,3}$   &        $\bm{3}$     &    $\bm{1}$  & $ -\sfrac{4}{3}$ &       $-1$         &   $\bm{1''}, \bm{1'}, \bm{1}$     &    Bulk        \\
   $F$                  &         $\bm{1}$    &     $\bm{1}$  & $ 0       $          &            $i$       &  $\bm{3}$   &    Brane          \\ \hline
   $H_u$              &        $\bm{1}$     &    $\bm{2}$   &    $1    $           &         $-1$        &  $\bm{3}$  & Brane       \\
   $H_d$             &         $\bm{1}$     &    $\bm{2}$   &   $-1   $           &            $1$        &  $\bm{3}$   &      Brane         \\
   $\eta$              &         $\bm{1}$     &    $\bm{2}$  &   $ 1  $             &              $-i$       &  $\bm{1}$  &    Brane          \\
   $\sigma$         &         $\bm{1}$      &    $\bm{1}$  &  $   0    $          &              $-1$      &    $\bm{3}$   &      Bulk      \\  \hline
\end{tabular}
\caption{Field representation content and symmetries of the scotogenic orbifold model~\cite{deAnda:2021jzc}.}
\label{tab:FieldContent}
\end{table}

\begin{figure}[H]
\centering
\includegraphics[height=6cm,width=0.5\textwidth]{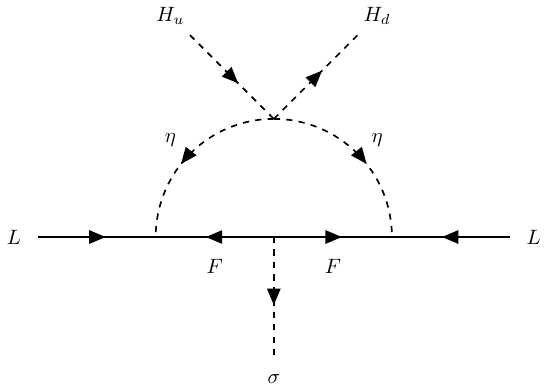}
\caption{One-loop diagram for Majorana neutrino masses, mediated by the ``dark sector'' particles~\cite{deAnda:2021jzc}.}
\label{fig:ScotoLoop}
\end{figure}

The model assumes an auxiliary lepton quarticity symmetry~\cite{CentellesChulia:2016rms,Srivastava:2017sno}. This $\mathbb{Z}_4$ is spontaneously broken to a residual $\mathbb{Z}_2$ symmetry, defining the ``dark sector'' which comprises the dark fermion $F$ and scalar $\eta$. Both transform non-trivially under the ``dark symmetry'' ensuring  stability of the lightest $\mathbb{Z}_2$-charged field. This makes it a potentially viable dark matter candidate, whose stability is directly related to the radiative origin of neutrino masses, Fig.~\ref{fig:ScotoLoop}.

The Higgs sector consists of two flavour-triplet weak iso-doublets, $H_u$ and $H_d$ and a SM singlet scalar $\sigma$ driving the spontaneous breaking of both lepton number~\cite{Chikashige:1980ui,Schechter:1981cv}, as well as family symmetry, a sort of ``flavoured'' Majoron scheme.

Given the defining symmetries of the model, one can write the most general effective Yukawa Lagrangian below the compactification scale. The Yukawa interaction terms of down-type quarks and charged leptons have the same structure, given by
\begin{equation}
\mathcal{L}^{\scriptscriptstyle{\text{Yukawa}}}_{\scriptstyle{H_d}}=y^{e}_1 \left( LH_de^c\right)_{\bm{1}_1} + y^e_2  \left(LH_de^c\right)_{\bm{1}_2} + y^{d}_1 \left(QH_dd^c\right)_{\bm{1}_1} + y^d_2  \left(QH_dd^c\right)_{\bm{1}_2} + \text{H.c.},
\label{eq:HdYukawa}
\end{equation}
while the transformation properties of the up-type quark fields under $A_4$ yield
\begin{equation}
\mathcal{L}^{\scriptscriptstyle{\text{Yukawa}}}_{\scriptstyle{H_u}}=y^{u}_1 \left( QH_u\right)_{\bm{1'}}u^c_1 + y^{u}_2 \left( QH_u\right)_{\bm{1''}}u^c_2 +y^{u}_3 \left( QH_u\right)_{\bm{1}}u^c_3 + \text{H.c.}
\label{eq:HuYukawa}
\end{equation}
Notice that the bold subscripts in each term indicate its transformation properties under the remnant $A_4$ family symmetry. The explicit expressions for the invariant multiplet products are given in~\ref{sec:group-theory-a_4}.

The dark fermion triplet $F$ couples to the scalar field $\sigma$. The latter acquires a vacuum expectation value which drives the spontaneous breaking of lepton number symmetry, $\mathbb{Z}_4$, and the $A_4$ family symmetry, giving rise to Majorana mass terms for the dark fermions,
\begin{equation}
\mathcal{L}^{\scriptscriptstyle{\text{Yukawa}}}_{\scriptstyle{\sigma}} =y^{\sigma} \left( F^{T}F \sigma\right)_{\bm{1}_{1}} + \text{H.c.}
\label{eq:SigmaYukawa}
\end{equation}
The dark scalar $\eta$ plays a crucial role in the model, as it couples with both dark fermions and neutrinos
\begin{equation}
\mathcal{L}^{\scriptscriptstyle{\text{Yukawa}}}_{\scriptsize{\eta}}= y^{\eta}_1 \left( L  \eta F \right)_{\bm{1}} + \text{H.c.}
 \label{eq:EtaYukawa}
\end{equation}
In the following, we will assume that all Yukawa couplings are real, and therefore that the model preserves a trivial CP symmetry.

\subsubsection{Symmetry breaking and fermion masses}
\label{sec:symmetry-breaking}

The scalar potential $V(H_d,H_u,\eta, \sigma)$ comprises all terms up to quartic interactions consistent with all the symmetries. Apart from the Higgs scalars $H_d,H_u,\sigma$ it contains also the dark scalar $\eta$ which does not develop a vacuum expectation value (VEV). The symmetry breakdown of the model proceeds in two steps. At high energies the electroweak singlet scalar field $\sigma$ develops a VEV compatible with the extra-dimensional boundary conditions. Subsequently, at lower energies, the electroweak Higgs doublets $H_u, H_d$ acquire VEVs according to the minimization of the scalar potential.

In order to describe the high-scale $A_4$ symmetry breaking produced by $\sigma$ we introduce a boundary condition $P$, consistent with the orbifold construction. It defines a non-trivial gauge/Poincar\'e twist of the orbifold, and must be a symmetry of the Lagrangian. We assume that the transformation $P$ acts trivially on the $A_4$ singlet bulk fields. Hence the only bulk field transforming non-trivially under $P$ is the flavour triplet scalar $\sigma$, obeying the boundary condition
\begin{equation}
\sigma(x,z)=P \sigma(x,-z).
\label{eq:bcOfSigma}
\end{equation}
so as to be consistent with Eq.~(\ref{eq:id3c}). The invariance of the kinetic term of the $\sigma$ field in the 6-D Lagrangian implies that $P \in SU(3)$, while the condition in Eq. (\ref{eq:bcOfSigma}) ensures that the matrix $P$ will leave invariant the interactions of fields in the brane. Thus the boundary condition matrix must satisfy
\begin{equation}
P \in SU(3), \qquad P^2= \bm{1}_{\scriptsize 3\times 3}, \quad P^{\dagger} = P\,.
\label{eq:Pcon}
\end{equation}

The boundary condition on the $\sigma$ field applies also to its VEV alignment. As a result, the masses of the dark fermions $F$ are a direct outcome of the boundary condition of the bulk field $ \sigma $ in the two extra dimensions
\begin{equation}
\braket{\sigma}=P \braket{\sigma}.
\label{eq:bcOfSigmaVEV}
\end{equation}
The boundary condition matrix $P$ is a property of the orbifold, its form is given explicitly in~\cite{deAnda:2021jzc}. Here we adopt the most general VEV alignment consistent with the spontaneous breaking of lepton number and the $A_4$ family symmetry, expressed as
\begin{equation}
\braket{\sigma}= v_{\sigma}\left( \begin{array}{c} \epsilon_1^{\sigma} \hspace{2pt} e^{i\varphi} \\ \epsilon_2^{\sigma} \\ 1 \end{array} \right), \qquad \text{with} \qquad  v_{\sigma}, \epsilon_1^{\sigma},  \epsilon_2^{\sigma} \in \mathbb{R} \quad \text{and} \quad 0\leq \varphi < \pi.
\label{eq:SigmaVEVBoundary}
\end{equation}

Here we will not present a dedicated analysis of the scalar potential, except to stress the importance of the $\lambda_5$ term for the neutrino mass generation mechanism, namely
\begin{equation}
V(H_{u},H_{d}, \eta) \supset \frac{1}{2} \lambda_{5} \left[ \left( {H_{d}}^{T} \left(i\sigma_{2}\right)\eta \right)_{\bm{3}} \left( H_{u}^{\dagger} \eta\right)_{\bm{3}} \right]_{\bm{1}} + \text{H.c.}
\label{eq:lambda5terms}
\end{equation}
Here $\lambda_5$ is a coupling constant and $\sigma_{2}$ is the second Pauli matrix. This term lifts the degeneracy of the mass eigenstates of the neutral components of $\eta$, denoted as $\sqrt{2}\,\text{Re}(\eta^0)$ and $\sqrt{2}\,\text{Im}(\eta^0)$, playing a key role in the scotogenic generation of neutrino masses at the one-loop level, as illustrated in Fig.~\ref{fig:ScotoLoop}. One can show that there is enough freedom in parameter space to drive the spontaneous breaking of the gauge symmetries down to $\mathrm{U(1)_{EM}}$.

\subsubsection{Quark and lepton masses}

Since $A_4$ breaks spontaneously at the $v_{\sigma}$ scale, in the second stage of spontaneous symmetry breaking, we assume that the weak iso-doublets $H_u$ and $H_d$ obtain the most general $A_4$ breaking VEVs consistent with trivial CP symmetry. For real $v_u,v_d,\epsilon^{u,d}_{1,2}$ it is given as
\begin{equation}\label{vevs}\begin{split}
\braket{H_u}=v_u\left(\begin{array}{c}\epsilon_1^u\\ \epsilon_2^u \\ 1\end{array}\right),\ \ \ \braket{H_d}=v_d \left(\begin{array}{c}\epsilon_1^d \\ \epsilon_2^d \\ 1\end{array}\right).
\end{split}\end{equation}
After spontaneous symmetry breaking, the up-quark mass matrix becomes
\begin{equation}\begin{split}
M_u&=v_u\left(\begin{array}{ccc} y_1^u\epsilon_1^u &y_2^u \epsilon_1^u & y_3^u\epsilon_1^u \\
y_1^u\epsilon_2^u \omega^2&  y_2^u\epsilon_2^u \omega &   y_3^u\epsilon^u_2 \\
y_1^u \omega& y_2^u \omega^2&y_3^u\end{array}\right),\\
\end{split}\label{eq:massmat-u}\end{equation}
while the down-quark and charged lepton mass matrices take the form
\begin{equation}\begin{split}
M_d&=v_d\left(\begin{array}{ccc} 0 & y_1^d\epsilon_1^d & y_2^d \epsilon_2^d \\
 y_2^d\epsilon_1^d & 0 &  y_1^d\\
 y_1^d\epsilon_2^d & y_2^d&0\end{array}\right),\\
M_e&=v_d\left(\begin{array}{ccc} 0 & y_1^e\epsilon_1^d & y_2^e \epsilon_2^d \\
 y_2^e\epsilon_1^d & 0 &  y_1^e\\
 y_1^e\epsilon_2^d & y_2^e&0\end{array}\right).
 \label{eq:massmat}
\end{split}\end{equation}
All Yukawa couplings in the last two equations are assumed to be real due to our imposition of trivial CP symmetry.

\subsubsection{Scotogenic neutrino masses}

The $A_4$ flavour symmetry structure of the Yukawa term in Eq.~\eqref{eq:SigmaYukawa} implies that the Majorana mass matrix of the dark fermions $M_{F}$ must have the following structure
\begin{equation}
M_F=y_{\sigma} v_{\sigma}\left(\begin{array}{ccc} 0 &1 &  \epsilon_2^{\sigma}\\
1 & 0 &  \epsilon_1^{\sigma} e^{i\varphi} \\
\epsilon_2^{\sigma} &   \epsilon_1^{\sigma} e^{i\varphi}& 0 \end{array}\right)\,.
\label{eq:DarkFermionsMassMatrix}
\end{equation}
In order to describe our one-loop scotogenic mechanism for neutrino masses we write the dark fermion $F$ fields in the mass eigenstate basis ($\tilde{F}$)
by performing the singular value decomposition of the dark fermion mass matrix $M_F$. Since the latter is symmetric, only one unitary matrix $V$ is needed in the Takagi decomposition~\cite{Schechter:1980gr},
\begin{equation}
y^{\sigma} \left( F^{T}F \sigma\right)_{\bm{1}_{1}}= F^{T} M_{F} F=F^{T} V^{T} D V F=\left(VF \right)^{T} D \left( VF\right) \equiv {\tilde{F} }^{T} D \tilde{F},
\label{eq:ChangeBasisDarkFermions}
\end{equation}
where $D=\mathrm{diag}(m_{F_1},m_{F_2},m_{F_3})$ and $\tilde{F} \equiv VF$ denotes the dark fermion triplet expressed in the mass eigenstate basis.
We can then rewrite Eq. (\ref{eq:EtaYukawa}) as
\begin{equation}
\mathcal{L}^{\scriptscriptstyle{\text{Yukawa}}}_{\scriptsize{\eta}}= y^{\eta}_1\eta  \left( L  V^{\dagger}\tilde{F} \right) + \text{H.c.}
 \label{eq:EtaYuakwaMass}
\end{equation}
As already mentioned, neutrino masses are forbidden at tree-level due to the auxiliary $\mathbb{Z}_4$ symmetry. However, thanks to the mediation of the dark fields $\eta$ and $F$, neutrino masses emerge at one-loop through the diagram depicted in Fig.~\ref{fig:ScotoLoop}, which has the basic scotogenic structure~\cite{Ma:2006km}.

Defining $y^{\eta}_1V^{\dagger} \equiv h$ in Eq.~\eqref{eq:ChangeBasisDarkFermions} we can write the expression for the one-loop neutrino mass matrix $M_{\nu}$ as
\begin{equation}
\left(M_{\nu}\right)_{ij} = \sum^3_{k} \frac{h_{ik} (h^{T})_{kj}}{16\pi^{2}} S(m_{F_k})\,,
\label{eq:NuMassMatrix}
\end{equation}
where $S(m_{F_k})$ is for the loop factor
\begin{equation}
S(m_{F_k})= m_{F_k} \left( \frac{m^2_R }{m^2_R- m^2_{F_k}}  \ln \frac{m^2_R}{m^2_{F_k}}- \frac{m^2_I }{m^2_I- m^2_{F_k}}  \ln \frac{m^2_I}{m^2_{F_k}}\right),
\label{eq:ScotoFactor}
\end{equation}
with $m_{R}=m(\sqrt{2}\hspace{2pt} \text{Re}  \hspace{2pt} \eta^{0})$, $m_{I}=m(\sqrt{2} \hspace{2pt} \text{Im}   \hspace{2pt} \eta^{0})$ with
\begin{equation}
m^{2}_{R} - m^{2}_{I} \equiv  2 \lambda_5  \left( \braket{H_u}_{\bm{3}} \braket{H_d}_{\bm{3}} \right)_{\bm{1}}\,.
\label{eq:lambda5Prop}
\end{equation}
Neutrino masses are not only loop-suppressed, but also symmetry-protected, as they vanish in the limit $\lambda_5 \to 0$, see Eq.~(\ref{eq:lambda5terms}).

After spontaneous symmetry breaking, the auxiliary $\mathbb{Z}_4$ breaks down to a residual $\mathbb{Z}_2$ that stabilizes the lightest dark particle. There are two possible dark matter candidates: the lightest state in the complex neutral scalar $\eta$, or the lightest Majorana fermion in the flavour triplet $F$. In either case, the phenomenology of dark matter is qualitatively similar to that of other scotogenic scenarios, which has been extensively discussed, see for example Ref.~\cite{Avila:2019hhv} and references therein.

\subsection{Flavour predictions of the scotogenic orbifold model}
\label{sec:analysis}

We now discuss in more detail the flavour predictions of our scotogenic orbifold model, both in the quark and lepton sectors. These follow directly from the $A_4$  family symmetry that results from the orbifold compactification of the extra dimensions.

\subsubsection{Golden quark-lepton mass relation}
\label{sec:golden}

This is a key feature of the model that results from the down-type quarks and charged lepton assignments under the $A_4$ flavour symmetry. Indeed, after spontaneous symmetry breaking these obtain masses from the same common Higgs doublet $H_d$, leading to the mass matrices in Eq.~\eqref{eq:massmat}. After diagonalization to physical mass-eigenstates, one obtains the golden quark-lepton mass relation in Eq.~\eqref{eq:golden}. This relation emerges in models with SO(3) family symmetry implementing a Peccei-Quinn symmetry~\cite{Reig:2018ocz}, and in the theories proposed in Refs.~\cite{Morisi:2013eca,King:2013hj} and~\cite{Morisi:2011pt,Bonilla:2014xla,Bonilla:2017ekt}.

In our case the golden relation is a common feature of the models with $A_4$ family symmetry arising from the compactification of 6-Dimensional orbifolds,
proposed in Refs.~\cite{deAnda:2019jxw,deAnda:2020pti} and further discussed in~\cite{deAnda:2020ssl,deAnda:2021jzc}. One can show that, given the current experimental measurements of the relevant masses, the golden relation holds with good precision, see for example, Fig.1~in~\cite{deAnda:2019jxw}. Besides, it constitutes a very robust prediction under the renormalization group evolution, as it involves only fermion mass ratios.
\begin{table}[!h]
\centering
\footnotesize
\renewcommand{\arraystretch}{0.78}
\begin{tabular}[t]{|lc|r|}
\hline
Parameter &\qquad& Value \\ \hline
$y^e_1v_d/\mathrm{GeV}$ &\quad& $-1.745$ \\
$y^e_2v_d/(10^{-1}\mathrm{GeV})$ && $1.021$ \\ \rule{0pt}{3ex}
$y^d_1v_d/(10^{-2}\mathrm{GeV})$ &&$-5.039$ \\
$y^d_2v_d /\mathrm{GeV}$ && $2.852$ \\ \rule{0pt}{3ex}
$y^u_{1} v_u/(10^{-1}\mathrm{GeV})$ &&$6.074$ \\
$y^u_2v_u/(10^2\mathrm{GeV})$ && $1.712$ \\
$y^u_3v_u/\mathrm{GeV}$ && $7.157$ \\ \rule{0pt}{3ex}
$\epsilon^u_1/10^{-4}$ && $7.055 $ \\
$\epsilon^u_2/10^{-2}$ && $-5.044$ \\ \rule{0pt}{3ex}
$\epsilon^d_1/10^{-3}$ && $-2.814$ \\
$\epsilon^d_2/10^{-3}$ && $5.833$ \\ \rule{0pt}{3ex}
$\epsilon^{\sigma}_1$ && $1.501$\\
$\epsilon^{\sigma}_2$ && $-0.654$\\
$\varphi$ && $3.527$\\
$(y^{\eta}_{1})^2 y_{\sigma} v_{\sigma}/(\mathrm{KeV})$ && $1.813$\\
$2 \lambda_5  \braket{H_u}  \braket{H_d} /(\mathrm{KeV})^2$ && $0.012$\\
\hline	
\end{tabular}
\hspace*{0.5cm}
\begin{tabular}[t]{ |l |c|c c |c| } \hline
\multirow{2}{*}{Observable}& \multicolumn{2}{c}{Data} & & \multirow{2}{*}{Model best fit}
\\ \cline{2-4}
& Central value & 1$\sigma$ range  &   & \\ \hline
$\theta_{12}^\ell$ $/^\circ$ & 34.3 & 33.3 $\to$ 35.3 && $33.0$  \\
$\theta_{13}^\ell$ $/^\circ$ & 8.45 & 8.31 $\to$ 8.61  && $8.52$  \\
$\theta_{23}^\ell$ $/^\circ$ & 49.26 & 48.47 $\to$ 50.05  && $50.44$ \\
$\delta^\ell$ $/^\circ$ & 194 & 172 $\to$ 218 && $192$  \\
$m_e$ $/ \mathrm{MeV}$ & 0.486 &  0.486 $\to$ 0.486 && $0.486$ \\
$m_\mu$ $/  \mathrm{GeV}$ & 0.102 & 0.102  $\to$ 0.102  &&  $0.102$ \\
$m_\tau$ $/ \mathrm{GeV}$ &1.745 & 1.743 $\to$1.747 && $1.745$ \\
$\Delta m_{21}^2 / (10^{-5} \, \mathrm{eV}^2 ) $ & 7.50  & 7.30 $\to$ 7.72 && $7.50$  \\
$\Delta m_{31}^2 / (10^{-3} \, \mathrm{eV}^2) $ & 2.55  & 2.52 $\to$ 2.57 &&  $2.54$ \\
$m_1$ $/\mathrm{meV}$  & & & & $135.35 $ \\
$m_2$ $/\mathrm{meV}$  & && & $ 135.63$ \\
$m_3$ $/\mathrm{meV}$  & && & $144.43 $ \\
$\phi_{12} $ $/^\circ$ & & && $87.01$  \\
$\phi_{13}$  $/^\circ$& & && $190.30$  \\
$\phi_{23}$  $/^\circ$& & && $271.05$  \\\hline
$\theta_{12}^q$ $/^\circ$ &13.04 & 12.99 $\to$ 13.09 &&  $13.04$ \\	
$\theta_{13}^q$ $/^\circ$ &0.20 & 0.19 $\to$ 0.22 && $0.20$  \\
$\theta_{23}^q$ $/^\circ$ &2.38& 2.32 $\to$ 2.44 && $2.38$  \\	
$\delta^q$ $/^\circ$ & 68.75 & 64.25 $\to$ 73.25  & & $60.23$\\
$m_u$ $/ \mathrm{MeV}$ & 1.28 & 0.76$\to$ 1.55 && $1.28$  \\	
$m_c$ $/ \mathrm{GeV}$ & 0.626 & 0.607 $\to$ 0.645 &&  $0.626$ \\	
$m_t$ $/\mathrm{GeV}$  	  & 171.6& 170 $\to$ 173  && $171.6$ \\
$m_d$ $/ \mathrm{MeV}$ & 2.74 & 2.57 $\to$ 3.15 &&  $2.49$ \\	
$m_s$ $/ \mathrm{MeV}$ & 54 & 51 $\to$ 57 && $54$ \\
$m_b$ $/ \mathrm{GeV}$	  & 2.85 &  2.83 $\to$   2.88 &&  $2.85$\\ \hline
$\chi^2$ & & & & $1.96$ \\\hline		
\end{tabular}
\caption{Flavour parameters and observables: measured versus predicted values for the best fit point.}
\label{tab:fit}
\end{table}

\subsubsection {Neutrino oscillation predictions}
\label{sec:leptonic-cp-phase}

\begin{figure}[]
\centering
\includegraphics[height=6cm,width=0.45\textwidth]{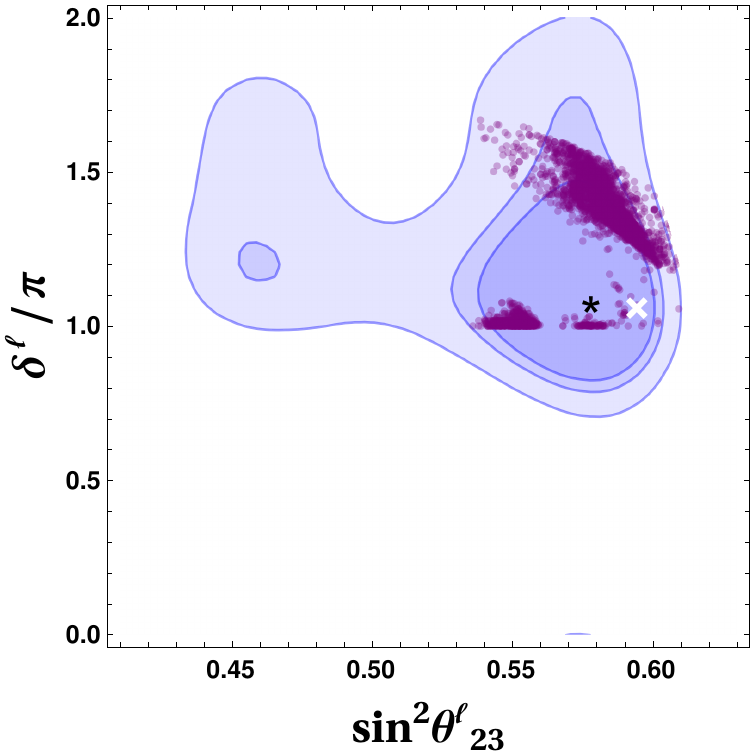}
\includegraphics[height=6cm,width=0.48\textwidth]{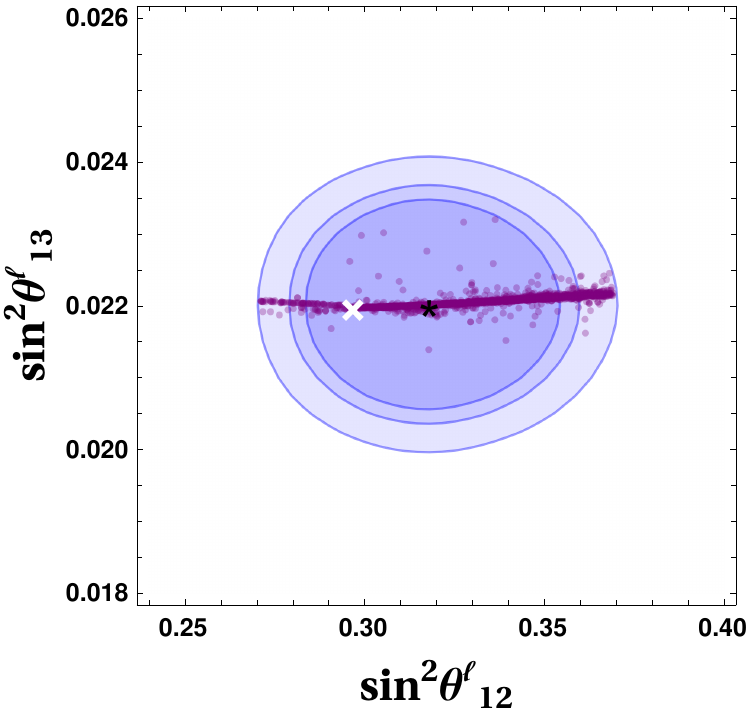}
\caption{Allowed values for the mixing angles and the leptonic Dirac CP phase. The purple points are compatible at $3\sigma$ with all flavour observables, while the blue shades are the generic 90, 95 and 99\% C.L. regions of the global oscillation fit in~\cite{deSalas:2020pgw,10.5281/zenodo.4726908}. The black star is the central point of the global oscillation fit, while the white cross stands for the best fit point in Table \ref{tab:fit}.}
\label{fig:Lepton-angles-phase}
\end{figure}

The above model is characterized by 16 independent parameters in the flavour sector, identified as follows: 8 real Yukawa couplings $y^{e,d}_{1,2}$, $y^u_{1,2,3}$, $y^{\eta}_1$, 6 real VEV ratios $\epsilon^{u,d}_{1,2}$, $\epsilon^\sigma_{1,2}$, one quartic coupling $\lambda_5$ and one CP violating phase $\varphi$ contained in $\vev{ \sigma}$. Due to the reduced number of parameters, the model makes strong flavour predictions. Following the procedure given in Appendix~\ref{app:global-fitting}, we perform a global flavour fit to the available experimental data. The results of our flavour fit are summarized in Table~\ref{tab:fit}. The minimum at $\chi^2\approx 2$ shows that the model reproduces the observed pattern of fermion masses and mixing rather well. From Table~\ref{tab:fit} one can read directly the predictions of the model concerning the mass of the lightest neutrino and the values of the CP phases characterizing the lepton sector. In order to identify the predictions concerning the oscillation parameters we have randomly varied the parameters
around the global best fit point in Table~\ref{tab:fit}, while requiring compatibility with all measured flavour observables at $3\sigma$. The results of the analysis are given in Fig.~\ref{fig:Lepton-angles-phase}, where the blue contours represent the 90, 95, and 99\% C.L. profiles from the Valencia global oscillation fit in~\cite{deSalas:2020pgw,10.5281/zenodo.4726908}, while the purple dots indicate regions compatible at $3\sigma$ with all experimental data. The best fit point of the global oscillation fit is marked with a black star, while that of the global flavour fit is indicated by a white cross. One sees from Fig. \ref{fig:Lepton-angles-phase} how the predicted values of the leptonic Dirac CP phase are restricted to the range $\delta^\ell \geq \pi$, while the atmospheric angle $\theta^\ell_{23}$ is required to lie in the higher octant. Besides, one sees from the right panel in Fig.~\ref{fig:Lepton-angles-phase} a sharp prediction for the reactor angle $\theta^\ell_{13}$, see also table~\ref{tab:fit}.

A scan of the parameter space of the model consistent at $3\sigma$ with all current experiments reveals that only the NO neutrino mass spectrum is possible. Indeed, the best fit point in table~\ref{tab:fit} has positive $\Delta m_{31}^2$, corresponding to NO, and a rather high absolute scale for the neutrino masses.

\subsubsection{Neutrinoless double beta decay predictions}
\label{sec:neutr-double-beta}

\begin{figure}[hptb]
\centering
\includegraphics[width=0.65\textwidth]{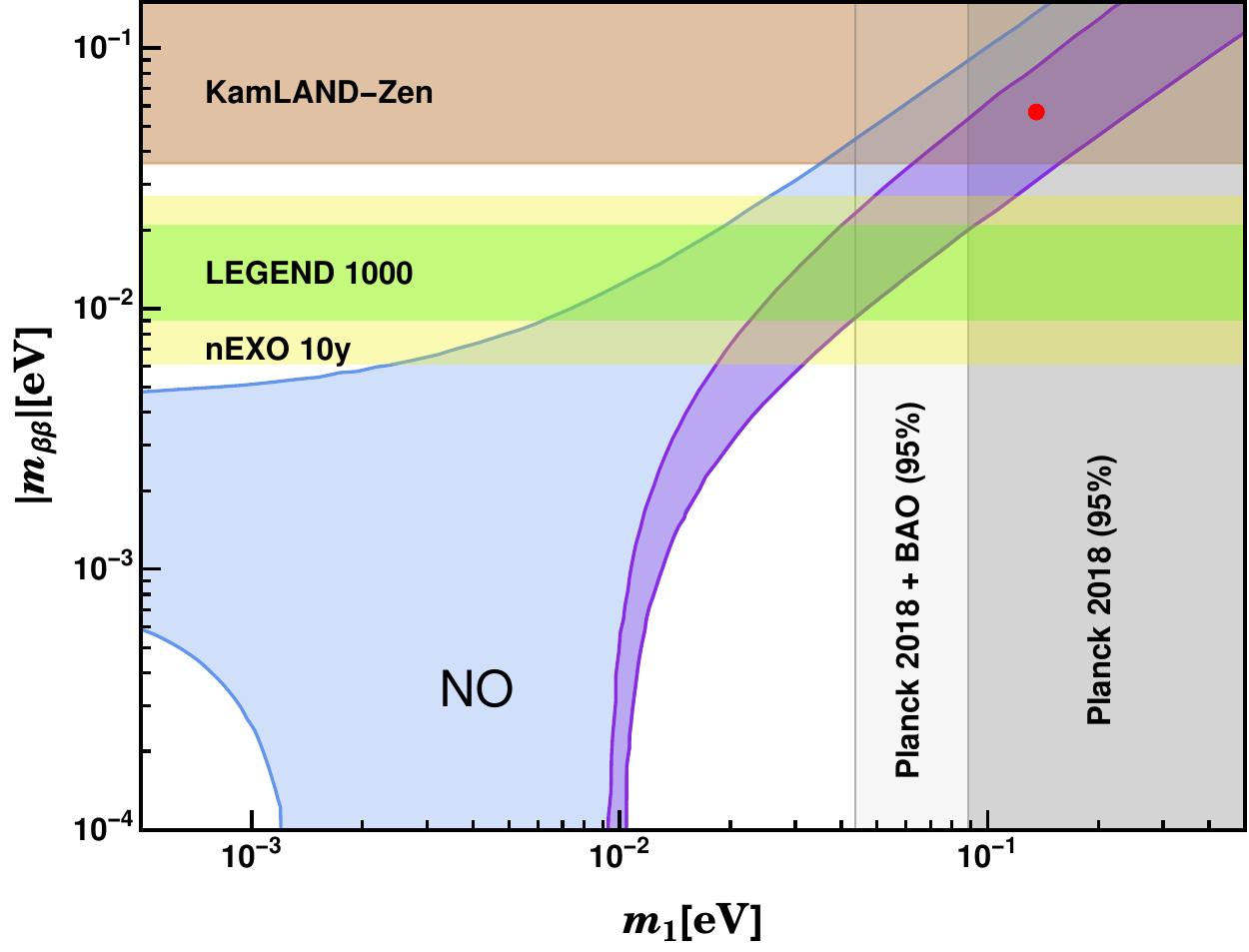}
\caption{Effective \znbb decay amplitude versus the lightest neutrino mass $m_1$. From the global fit one finds that only normal ordering is allowed. The blue region is the generic one consistent with oscillations at $2\sigma$. The purple region is the one allowed at $3\sigma$ around the global best fit point in table~\ref{tab:fit}, marked in red. The current KamLAND-Zen limit is shown in brown, and the projected sensitivities of future experiments LEGEND-1000 and nEXO are indicated with light yellow and light green horizontal bands respectively. The vertical gray bands represent the current sensitivity of cosmological data from the Planck collaboration (dark gray), and in combination with BAO data (light gray)~\cite{Planck:2018vyg,Gerbino:2022nvz}}.
\label{fig:neutrinoless1}
\end{figure}

Concerning neutrinoless double beta decay, using the Majorana phase and neutrino mass predictions of table~\ref{tab:fit} in Eq.~\eqref{eq:mbb},
one finds the preferred effective amplitude parameter:
\begin{equation}
|m_{\beta\beta}|= 58.08 \,\mathrm{meV}.
\end{equation}
A detailed analysis is presented in figure~\ref{fig:neutrinoless1}, where we show in purple the region of predicted $|m_{\beta\beta}|$ values as a function of the lightest neutrino mass $m_1$. To be conservative, we randomly varied the model parameters within the allowed $3\sigma$ range. The best fit point from table~\ref{tab:fit} is marked in red. One sees that the predicted central value of $|m_{\beta\beta}|$ lies inside the current exclusion band of Kamland-Zen $(36 - 156\;\mathrm{meV})$~\cite{KamLAND-Zen:2022tow}. This will also be probed by cosmological observations and possibly by future beta decay endpoint studies. We also display the projected sensitivities of the next round of \znbb experiments LEGEND~\cite{LEGEND:2017cdu} and nEXO~\cite{nEXO:2017nam} as the colored horizontal bands.

Notice that the central value of the lightest neutrino mass $m_1$ obtained from the global fit is disfavored by the latest results of the Planck collaboration on the sum of light neutrino masses~\cite{Planck:2018vyg}.
This tension is further enhanced by the addition of Baryon Acoustic Oscillations (BAO) data~\cite{Lattanzi:2020iik}, see vertical band in lighter gray in figure~\ref{fig:neutrinoless1}. Beyond the central prediction, however, there is a broad parameter region consistent both with measured flavour observables as well as with the cosmological bounds.

\clearpage

\section{Recent progress: modular symmetry}                                
\label{sec:modular-symmetry}                                               

In flavour symmetry models, complicated vacuum alignment assumptions are often required to break the flavor symmetry. The vacuum expectation values of the associated flavons should be conveniently oriented in family space. Moreover, higher-dimensional operators with flavon insertions and unknown coefficients are often present, affecting the resulting model predictions. In order to alleviate these shortcomings modular invariance as flavor symmetry has been recently proposed~\cite{Feruglio:2017spp}. In what follows we briefly sketch recent work, for dedicated reviews on modular symmetries see Refs.~\cite{Feruglio:2019ybq,Kobayashi:2023zzc,Ding:2023htn}.\par

\subsection{The modular group}
\begin{figure}[t!]
\centering
\includegraphics[width=0.6\textwidth]{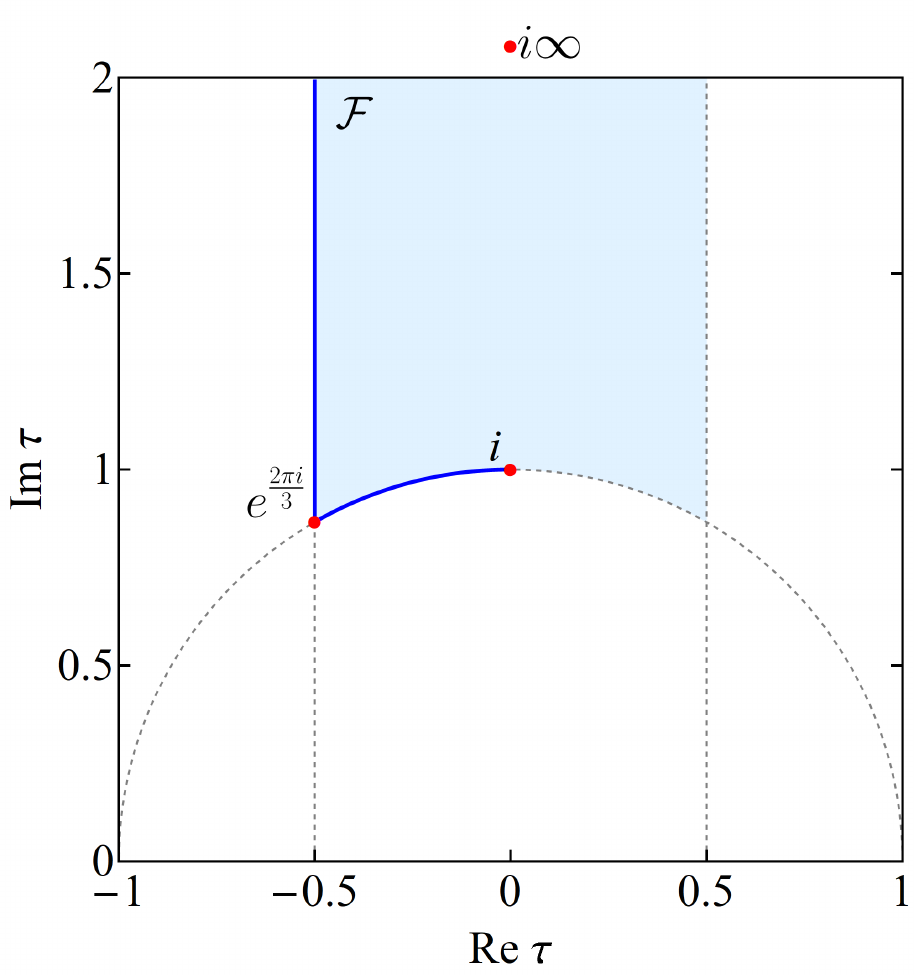}
\caption{\label{fig:fundamental-domain} The fundamental domain $\mathcal{F}$ of the modular group, where the red points denote the three fixed points $\tau_0=i, e^{2\pi i/3}, i\infty$ which preserve a residual modular symmetry.}
\end{figure}

The modular group $SL(2, \mathbb{Z})$ is the group of $2\times2$ integer matrices with determinant one,
\begin{equation}
SL(2, \mathbb{Z})=\left\{\begin{pmatrix}
a ~&~ b \\
c ~&~ d
\end{pmatrix}\Bigg|a, b,c,d\in\mathbb{Z}, ad-bc=1\right\}\,.
\end{equation}
The modular group $SL(2, \mathbb{Z})$ acts on the complex variable $\tau$ in the upper-half plane as the linear fractional transformation~\cite{cohen2017modular}
\begin{equation}
\label{eq:modular-trans}\gamma\tau=\frac{a\tau+b}{c\tau+d}~~\text{for}~~\gamma=\begin{pmatrix}
a ~&~ b \\
c ~&~ d
\end{pmatrix}\in SL(2, \mathbb{Z})~~\text{and}~~\text{Im}(\tau)>0\,.
\end{equation}
One sees that $\gamma$ and $-\gamma$ induce the same linear fraction transformation. The modular group can be generated by the two generators $S$ and $T$,
\begin{equation}
S=\begin{pmatrix}
0 ~&~ 1 \\
-1 ~&~ 0
\end{pmatrix}\,,~~~T=\begin{pmatrix}
1 ~&~ 1 \\
0 ~&~ 1
\end{pmatrix}\,,
\end{equation}
which lead to duality and shift symmetries of $\tau$ as follows,
\begin{equation}
\tau\xrightarrow{S} -\frac{1}{\tau},\qquad \tau\xrightarrow{T} \tau+1\,.
\end{equation}
Any value of $\tau$ in the upper half complex plane can be shifted in to the region of $-\frac{1}{2}\leq\text{Re}(\tau)<\frac{1}{2}$ by multiple $T$ transformations. Moreover, it can be mapped to the region $|\tau|\leq1$ by an $S$ transformation. As a consequence, the complex modulus $\tau$ could be restricted to the fundamental domain $\mathcal{F}$ of the modular group,
\begin{equation}
\mathcal{F}=\left\{\tau\Big|-\frac{1}{2}\leq\text{Re}(\tau)<\frac{1}{2}, \text{Im}(\tau)>0, |\tau|>1\right\}\cup\left\{\tau\Big|-\frac{1}{2}\leq\text{Re}(\tau)\leq0, \text{Im}(\tau)>0, |\tau|=1\right\}\,,
\end{equation}
which is displayed in figure~\ref{fig:fundamental-domain}. Every complex modulus $\tau$ can be mapped into the fundamental domain by a modular transformation of Eq.~\eqref{eq:modular-trans}, and no two points in $\mathcal{F}$ can be related by modular transformations.
Notice the right half boundary of $\mathcal{F}$ is not included into the fundamental domain, since it can related to the left half boundary by some modular transformation.
Moreover, no value of $\tau$ is left invariant by the whole modular group action of Eq.~\eqref{eq:modular-trans}. In the fundamental domain, there are only three fixed points $\tau_0=i, e^{2\pi i/3}, i\infty$ which break the modular group $SL(2, \mathbb{Z})$ partially. These are invariant under the modular transformations $S$, $ST$ and $T$ respectively~\cite{cohen2017modular,Novichkov:2018yse,Ding:2019gof}. Notice that each $\tau$ is trivially invariant under $S^2$ which is a negative identity matrix.

Modular symmetry provides a novel origin of discrete flavor symmetry through the quotient of $SL(2, \mathbb{Z})$ by the principal congruence subgroup of level $N$ which is defined as,
\begin{equation}
\Gamma(N)=\left\{\begin{pmatrix}
a ~&~ b \\
c ~&~ d
\end{pmatrix}\in SL(2, \mathbb{Z})\Bigg|a, d=1 \, (\text{mod}~N), b,c=0 \,(\text{mod}~N)
\right\}\,,
\end{equation}
which is a normal subgroup of finite index in $SL(2, \mathbb{Z})$. The finite modular groups are the quotient groups $\Gamma_N=SL(2, \mathbb{Z})/\pm \Gamma(N)$ which play the role of discrete flavor symmetry. Remarkably, for $N\leq5$ the finite modular groups $\Gamma_N$ are isomorphic to the permutation groups~\cite{deAdelhartToorop:2011re,Feruglio:2017spp}
\begin{equation}
\Gamma_2\cong S_3,~~~\Gamma_3\cong A_4,~~~\Gamma_4\cong S_4,~~~\Gamma_5\cong A_5\,,
\end{equation}
which have been widely used as traditional flavor symmetries.

\subsection{Modular invariance}
\label{sec:Modular-forms}

A key concept of the theory of modular symmetries~\cite{Feruglio:2017spp} is that of modular forms of level $N$ and weight $2k$, denoted $Y(\tau)$. These are holomorphic functions of the modulus $\tau$ with the following transformation property
\begin{equation}
Y(\gamma\tau)=(c\tau+d)^{2k}Y(\tau),\qquad \gamma=\begin{pmatrix}
a  &~ b\\
c  &~ d
\end{pmatrix}\in\Gamma(N)\,,
\end{equation}
where the integer $k\geq0$ determines the weight $2k$.  \par

There are only a finite number of linearly independent modular forms, denoted as $Y_i(\tau)$, and one can always choose a basis such that the transformation of the modular forms can be described by a unitary representation $\rho$ of $\Gamma_N$:
\begin{equation}
Y_{i}(\gamma\tau)=(c\tau+d)^{2k} \rho_{ij}(\gamma)Y_j(\tau)\,,
\end{equation}
where $\gamma=\begin{pmatrix}
a  ~& b \\
c  ~& d
\end{pmatrix}$ refers to  a representative element of $\Gamma_N$. Likewise, the modular transformation properties of matter fields $\varphi^{(I)}$ are completely specified by the weight $k_I$ and the unitary representation $\rho^{(I)}$ of the finite modular group $\Gamma_N$,
\begin{equation}
\varphi^{I}\to (c\tau+d)^{k_I}\rho^{(I)}\varphi^{(I)}\,.
\end{equation}
Note that the modular flavor symmetry requires supersymmetry to preserve the holomorphicity of the modular form. Then the superpotential can be expanded in a power series of the matter fields $\varphi^{(I)}$ as follows,
\begin{equation}
\mathcal{W}=\sum_{n}Y_{I_1I_2\ldots I_n}(\tau)\varphi^{(I_1)}\varphi^{(I_2)}\ldots \varphi^{(I_n)}\,.
\end{equation}
Modular invariance requires $Y_{I_1I_2\ldots I_n}(\tau)$ to be a modular form of weight $k_Y$ and level $N$ transforming in the representation $\rho$ of $\Gamma_N$,
\begin{equation}
Y_{I_1I_2\ldots I_n}(\gamma\tau)=(c\tau+d)^{k_Y}\rho(\gamma)Y_{I_1I_2\ldots I_n}(\tau)\,,
\end{equation}
with the conditions
\begin{eqnarray}
k_Y+k_{I_1}+k_{I_2}+\ldots++k_{I_n}=0,\qquad \rho\otimes\rho^{(I_1)}\otimes\rho^{(I_2)}\otimes\ldots \rho^{(I_n)}\supset\mathbf{1}\,,
\end{eqnarray}
where $\mathbf{1}$ denotes the invariant singlet representation of $\Gamma_N$. One sees that the modular flavor symmetry requires the Yukawa couplings to be modular forms $Y_{I_1I_2\ldots I_n}(\tau)$. In the simplest implementation, flavons are not necessary and the complex modulus $\tau$ is the unique symmetry breaking parameter, greatly simplifying the alignment problem. Moreover, modular invariance requires the Yukawa couplings and fermion mass matrices to be combinations of modular forms, holomorphic functions of $\tau$, all higher-dimensional operators are unambiguously determined in the limit of unbroken supersymmetry. As a result, flavour models with modular invariance depend on fewer parameters, enhancing their predictive power.

The above formalism of modular flavor symmetry has been extended to the double covers~\cite{Liu:2019khw,Novichkov:2020eep,Liu:2020akv,Wang:2020lxk,Yao:2020zml} or the metaplectic covers~\cite{Liu:2020msy,Yao:2020zml} of the finite group $\Gamma_N$, then the modular weights can take integer values or more general rational values.

\subsection{Generalized CP in modular symmetry}

The interplay of modular symmetry and generalized CP symmetry (gCP) has been studied~\cite{Novichkov:2019sqv,Kobayashi:2019uyt,Ding:2021iqp}. The consistency between the modular symmetry and gCP fixes the CP transformation of the modulus $\tau$ to be~\cite{Novichkov:2019sqv,Baur:2019kwi,Acharya:1995ag,Dent:2001cc,Giedt:2002ns}
\begin{equation}
\tau \xrightarrow{CP}-\tau^{*}
\end{equation}
up to modular transformations. For any modular transformation $$\gamma=\begin{pmatrix}
a ~& b \\
c ~& d\\
\end{pmatrix}\in SL(2, \mathbb{Z}),$$ one can check that the action of the transformation chain $CP \to \gamma \to CP^{-1}$ on $\tau$ is,
\begin{equation}
\tau \,\xrightarrow{CP}\, -\tau^{*}
\,\xrightarrow{\gamma}\, - \frac{a\tau^{*} + b}{c\tau^{*} + d}
\,\xrightarrow{CP^{-1}}\, \frac{a\tau - b}{-c\tau + d} \,.
\end{equation}
This implies that the CP transformation corresponds to an automorphism of the modular group, and it maps any modular transformation $\gamma$ to another modular transformation $u(\gamma)$,
\begin{equation}
\label{eq:CP-ganna-CP-chain} u(\gamma) \,\equiv\, CP\circ \gamma\circ CP^{-1} \,=\,
\begin{pmatrix}
  a ~&~ -b \\
  -c ~&~ d
 \end{pmatrix} \,.
\end{equation}
In particular, one has
\begin{equation}
 u(S)=CP\circ S\circ CP^{-1}=S^{-1},~~~ u(T)=CP\circ T\circ CP^{-1}=T^{-1}\,.
\end{equation}
We proceed to consider the gCP transformation of an arbitrary chiral superfield multiplet $\varphi(x)$ in the representation $\mathbf{r}$ of the finite modular group $\Gamma_N$,
\begin{equation}
\varphi(x) \,\xrightarrow{CP}\, X_\mathbf{r}\, \overline\varphi(x_P)\,,
\label{eq:Gcp-transf-modu}
\end{equation}
where a bar denotes the Hermitian conjugate superfield, and where $x=(t,\mathbf{x})$, $x_P=(t,-\mathbf{x})$ and $X_\mathbf{r}$ is a unitary matrix acting on flavour space.
Applying the consistency condition chain of Eq.~\eqref{eq:CP-ganna-CP-chain} to the matter multiplet $\varphi$, one find that the CP transformation matrix $X_\mathbf{r}$ has to satisfy the following constraint~\cite{Novichkov:2019sqv},
\begin{equation}
\label{eq:CP-modular-consis}
X_\mathbf{r}\, \rho_\mathbf{r}^{*}(\gamma) X_\mathbf{r}^{-1} = \rho_\mathbf{r}(u(\gamma))\,.
\end{equation}
It is sufficient to just consider the modular generators $\gamma=S, T$, namely
\begin{equation}
\label{eq:CP-modular-consis2}
X_\mathbf{r}\, \rho_\mathbf{r}^{*}(S) X_\mathbf{r}^{-1} = \rho_{\mathbf{r}}(S^{-1}),~~~~X_\mathbf{r}\, \rho_\mathbf{r}^{*}(T) X_\mathbf{r}^{-1} = \rho_{\mathbf{r}}(T^{-1})\,,
\end{equation}
which fixes $X_\mathbf{r}$ up to an overall phase by Schur's lemma, for each irreducible representation $\mathbf{r}$. In the basis where both $S$ and $T$ are represented by symmetric and unitary matrices,  $X_\mathbf{r}=  \mathbb{1}_\mathbf{r}$ solves the consistency condition. As a consequence, CP symmetry imposition enforces all coupling constants to be real in such basis. The VEV of $\tau$ is the unique source breaking both modular and gCP symmetries. The combination of modular and CP symmetries allows one to construct quite predictive flavor models. Indeed, it is remarkable that all the lepton masses, mixing angles and CP violation phases could be described in terms of only four real couplings plus the complex modulus $\tau$ in the minimal modular model of Refs.~\cite{Ding:2022nzn,Ding:2023ydy}. The gCP symmetry can also be consistently implemented in the context of multiple moduli~\cite{Ding:2021iqp}.

\subsection{Modular symmetry from top-down}

The idea of modular flavor symmetry is inspired by top-down considerations from string theory. The modular symmetry can naturally appear in orbifold compactifications of the heterotic string~\cite{Baur:2019kwi,Baur:2019iai} and magnetized toroidal compactification~\cite{Kobayashi:2016ovu,Kobayashi:2018rad,Kobayashi:2018bff,Kariyazono:2019ehj,Ohki:2020bpo,Kikuchi:2020frp,Almumin:2021fbk}. The double covers and metaplectic covers of the finite modular groups could be reproduced from top-down constructions.
Given the fact that string compactification generally yields several moduli, the modular invariance has been extended to involve multiple moduli based on the direct product of several $SL(2, \mathbb{Z})$~\cite{deMedeirosVarzielas:2019cyj} or the Symplectic modular group $Sp(2g, \mathbb{Z})$~\cite{Ding:2020zxw}, where $g$ is a generic positive integer called genus. Note that $Sp(2g, \mathbb{Z})$ can arise as the duality group in string Calabi-Yau compactifications~\cite{Ishiguro:2020nuf,Baur:2020yjl,Nilles:2021glx,Ishiguro:2021ccl}, and the group $Sp(2, \mathbb{Z})$ is isomorphic to the modular group $SL(2, \mathbb{Z})$.
Moreover, it has been found that the modular symmetry and traditional flavor symmetry appear together in top-down constructions. This leads to the concept of eclectic flavor group~\cite{Nilles:2020nnc,Nilles:2020kgo}, a maximal extension of the traditional flavor group by finite modular group. It is more predictive than the finite modular group and also the traditional flavor group by itself, combining the advantages of both approaches. In particular, the interplay of modular symmetry and traditional flavor symmetry can restrict the K\"ahler potential more severely than the modular symmetry by itself.
The possibility of eclectic flavor group can also be consistently combined with the gCP symmetry~\cite{Nilles:2020nnc}.  \par

\subsection{Quark-lepton mass relations from modular symmetry}

Quite generally, the problem of understanding the pattern of fermion masses and mixings presents a two-fold challenge. While predicting fermion mixings through the imposition of flavor symmetries is relativeley straightforward, formulating a convincing theory of fermion masses seems tougher. \par

It has recently been shown that realistic fermion mass relations can arise naturally in modular invariant models~\cite{Novichkov:2021evw,Feruglio:2021dte,Feruglio:2022koo,Feruglio:2023mii}, without relying on \textit{ad hoc} flavon alignments. As an example, Ref.~\cite{Chen:2023mwt} gave a set of viable fermion mass relations based on the $\Gamma_4\cong S_4$ symmetry. The new versions exhibit calculable deviations from the usual Golden Mass Relation in Eq.~\eqref{eq:golden}.
They were derived from modular flavor symmetry in a rather general manner,
relying only on the modular flavor group and its vector-valued modular forms, rather than \textit{ad hoc} flavon alignments~\footnote{They are determined by the Clebsch-Gordan coefficients of the given finite modular group as well as the expansion coefficients of its modular forms.}. The new relations were shown to be viable and experimentally testable, distinguishing modular models from the conventional flavon-based prediction in Eq.~\eqref{eq:golden}. \par
The method is largely model-independent, and can adapted also to obtain predictions in the up-quark sector and neutrinos. It may also prove useful for more comprehensive modular invariant models and for top-down constructions.

\subsection{Modular versus traditional flavor symmetry}

We now comment on the predictive power of traditional flavor symmetry and modular symmetry. As we have shown in section~\ref{sec:lepton_flavor_CP_symm},  traditional flavor symmetry in combination with generalized CP symmetry allows one to predict lepton mixing angles and CP violation phases in terms of just one or two real free parameters when certain residual symmetry is preserved. In general, however, lepton masses are unconstrained by the residual symmetry.  \par

In the case of modular symmetry, the fermion mass matrices exhibit certain symmetry-determined flavor structure through the modular forms which are functions of the modulus $\tau$. One finds that both fermion masses as well as flavor mixing parameters can be determined within specific models. However, one must resort to numerical analysis to reveal the possible correlations amongst the flavor parameters, since a residual modular symmetry with a single modulus has not led to phenomenologically viable models. Concerning vacuum alignment, it is notoriously difficult to dynamically realize the alignment required by the residual symmetry within traditional flavor symmetry constructions. On the other hand, determining the VEV of the modulus $\tau$ from a dynamical principle is also an open question, some possible schemes have been proposed to stabilize the modular VEV~\cite{Cvetic:1991qm,Gonzalo:2018guu,Kobayashi:2019xvz,Ishiguro:2020tmo,Novichkov:2022wvg,Leedom:2022zdm,Knapp-Perez:2023nty,Kikuchi:2023uqo,King:2023snq}.
In modular flavor models, the VEV of $\tau$ is usually treated as free parameter, and its value is determined by confronting the model predictions with experimental data.  \par

In short, modular flavor symmetries may  shed light on the ultimate symmetry underlying the flavor problem. They have been studied from both the bottom-up and top-down approaches.
There are still some unsolved problems such as modulus stabilization and the K\"ahler potential problem. However, invoking modular symmetry seems a useful tool towards formulating theories of flavor. A comprehensive description of these recent developments lies outside the scope of the present work, see Refs.~\cite{Feruglio:2019ybq,Kobayashi:2023zzc,Ding:2023htn} dedicated modular symmetry reviews.

\clearpage

\section{Summary and outlook}                                              
\label{sec:Outlook}                                                        

The flavour problem constitutes perhaps one of the major open issues and deepest challenges of particle physics. Even more so after the revolutionary discovery and confirmation of neutrino oscillations. The main legacy of these experiments has been to show that leptons mix rather differently from quarks, providing a key input on the flavour issue. Indeed, it seems unlikely that the peculiar pattern of neutrino mixing angles extracted from experiment is just a random coincidence. Rather, it should have its roots in some, perhaps subtle, underlying symmetry of nature.

By itself, the \sm lacks an organizing principle in terms of which to understand the ``flavours'' of the fundamental building blocks of matter. First of all, the flavour challenge comprises an understanding of family replication, i.e. why nature repeats itself three times. Moreover, it also requires an explanation for the pattern of fermion masses and mixings. This review examined the possibility that the flavour puzzle has a symmetry explanation.

To set up the stage we started with an introductory section in which we gave basic preliminaries on the description of lepton and quark mixing, as well as a recap on the status of neutrino oscillations, Fig.~\ref{fig:osc} and Fig.~\ref{fig:osc2}, and neutrinoless double beta decay, e.g. Fig.~\ref{fig:dbd1}. We discussed ways to have detectable \znbb decay rates, in particular theories where the lightest neutrino is massless or nearly so, see Fig.~\ref{fig:dbd2}, as occurs, for example, in the missing partner seesaw mechanism, where there are less ``right'' than ``left''-handed neutrinos.
We also commented on the detectability of \znbb decay in schemes with family symmetries, e.g. Fig.~\ref{fig:dbd3}, and also on the significance of a possible \znbb discovery for particle physics, Fig.~\ref{fig:dbd-bb}.

In Section~\ref{sec:orig-neutr-mass} we briefly discussed the origin of neutrino masses, both from point of view of effective theories, e.g. the Weinberg operator in Fig.~\ref{fig:neutrino-mass-Majorana}, as well as, for instance, its realization in terms of the seesaw paradigm, Fig.~\ref{fig:t1seesaw}, or through a ``radiative'' paradigm. For example,  dark matter could mediate neutrino mass generation via the ``scotogenic'' scenario. We also briefly discussed the idea that dark matter \textit{seeds} neutrino mass generation that proceeds \textit{a la seesaw}, as in the recent recently proposed low-scale seesaw variants dubbed ``dark inverse'' and ``dark linear seesaw'' Fig.~\ref{fig:darkseesaw}. Finally, we mentioned the \textit{flavour problem}, a major SM drawback and the main thread of this review, and how it may be approached by symmetry, as illustrated in Fig.~\ref{fig:flasy6}.

In Sect.~\ref{sec:flavor_symm} we described various phenomenological lepton mixing patterns, showing how most of them are at odds with current experimental data, especially from Daya Bay. In Sect.~\ref{sec:flavor_CP_bottom-up} we gave a bottom-up description of residual flavour and CP symmetries, both for the case of leptons as well as quarks. Indeed, the residual CP and flavor symmetries of quark and lepton mass terms are determined in terms of the experimentally measured mixing matrix, as given in Eqs.~(\ref{eq:Gl_remnant}, \ref{eq:Gnu_remnant_Maj}, \ref{eq:Gnu_remnant_Dirac}, \ref{eq:Xl_remnant}, \ref{eq:Xnu_remnant_Maj}, \ref{eq:Xnu_remnant_Dirac}) and Eqs.~(\ref{eq:Gu_Gd}, \ref{eq:Xu-Xd}).
On the other hand, they may arise from the breaking of flavour and CP symmetries at high energies. It is remarkable that one can fix the quark and lepton mixing matrices from the structure of the flavour symmetry group and the residual symmetries, irrespective of the flavour symmetry breaking dynamics. Residual symmetries can restrict the mixing matrix as summarized in table~\ref{tab:residual-sym-Nump}. Moreover, remnant symmetries are quite useful as a guide to construct concrete flavour models. They can also provide adequate revampings of various neutrino mixing patterns at odds with experiment, so as to yield generalizations that are not only viable, but also predictive, as discussed in Sect.~\ref{sec:revamp-lept-mixing}. For example, viable and predictive generalizations of the TBM mixing pattern are given in Figs.~\ref{fig:g1L} and \ref{fig:g1R}. On the other hand, similar generalizations of the Golden-ratio mixing pattern are given in Fig.~\ref{fig:revamp-GR}. Soon after the first Daya-Bay results indicating a nonzero value of $\theta_{13}$ the bi-large mixing pattern was suggested,
in which solar and atmospheric mixing angles as well as the Dirac CP phase are determined in terms of $\theta_{13}$, as seen in Figs.~\ref{fig:cor_Cons-BL1} and \ref{fig:cor_Cons-BL2}.

Within the family symmetry paradigm, the symmetry group is usually broken down to different subgroups in the neutrino and charged lepton sectors, the lepton mixing matrix arises from the mismatch of the symmetry breaking patterns, as shown in Fig.~\ref{fig:Gl-Gnu-flavor} in Sect.~\ref{sec:lepton_flavor_CP_symm}. If a Klein subgroup is preserved by the (Majorana) neutrino mass term and the residual abelian subgroup of the charged lepton sector distinguishes the three generations, the lepton mixing matrix would be completely fixed from group theory up to row and column permutations. On the other hand, the leptonic Dirac CP phase $\delta_{CP}$ is predicted to take on CP conserving values, while the Majorana phases cannot be constrained. Instead, if a $Z_2$ subgroup is preserved in the neutrino sector, then only one column of the lepton mixing matrix is determined and $\delta_{CP}$ can lie in a relative large region, e.g. see Fig.~\ref{fig:deltaCP_contour_Z2}. Generalized CP symmetries are very powerful to constrain the CP phase. For instance, the $\mu-\tau$ reflection symmetry on the neutrino fields enforces $\delta_{CP}$ to be maximal, while the predicted $\delta_{CP}$ values for the generalized $\mu-\tau$ reflection are displayed in Fig.~\ref{fig:sinDeltaCP_Theta}.

A flavour symmetry transformation can be generated by performing two CP transformations, the interplay of flavour symmetry and generalized CP symmetry is highly nontrivial, requiring that the consistency condition in Fig.~\ref{fig:consistency-flavor-CP} must hold. The inclusion of generalized CP symmetry in flavour symmetry provides richer symmetry breaking patterns. For example, one usually assumes that the flavour and CP symmetry are broken to an abelian subgroup in the charged lepton sector and $Z_2\times CP$ in the neutrino sector. In this case all the lepton mixing parameters depend only on a single real rotation angle $\theta$ in Fig.~\ref{fig:Zn-z2xCP-charge-leptons-nu}. The value of $\theta$ is fixed by the precisely measured reactor angle $\theta_{13}$, leading to a determination of the other lepton mixing angles and CP violation phases, see, e.g. Eq.~(\ref{mixing-pars-S4CP-I}). This also implies definite predictions for the \znbb decay amplitude, as shown in Fig.~\ref{fig:onbb-Z2xCP-S4}. Another possibility is that the remnant subgroup preserved by both neutrino and charged lepton mass terms has the structure $Z_2\times CP$.
In this case the lepton mixing matrix depends on two free rotation angles $\theta_{l}$ and $\theta_{\nu}$, see Fig.~\ref{fig:z2xCP-nu-charge-leptons}. All the mixing angles and CP violation phases as well as the effective \znbb mass parameter are predicted to lie in very restricted regions, as shown in Figs.~\ref{fig:Z2xCP-nu-char-S4-contour}, \ref{fig:contour_CP_phases}, \ref{fig:mee-Z2xCP-nu-charg-S4}. It is remarkable that this scheme can be extended to the quark sector, both quark and lepton mixings can be accommodated using a single flavour group, in terms of total four free parameters. Generalized flavour and CP symmetries may also be employed for the case of Dirac neutrinos. Similar results hold true, except that the remnant  flavour symmetry of neutrino sector can be an arbitrary abelian subgroup. In particular, the master formulas in Figs.~\ref{fig:Zn-z2xCP-charge-leptons-nu} and~\ref{fig:z2xCP-nu-charge-leptons} also hold for Dirac neutrinos.

In section~\ref{sec:test-symmetry} we gave a panoramic view illustrating possible tests of flavor symmetry models. The existence of mixing predictions is a characteristic feature of flavor symmetry models. This follows from the fact that typically they have less free parameters than physical observables, therefore correlations between light neutrino observables are predicted. This is explicitly shown in Eqs.~(\ref{eq:cor-RTBM-A}, \ref{eq:cor-RTBM-C}, \ref{eq:corr-revamp-cGR}, \ref{eq:correlation_trimaximal}, \ref{eq:corre_theta23_deltaCP}, \ref{mixing-pars-S4CP-I}, \ref{mixing-pars-S4CP-II}, \ref{eq:caseIV-Z2xCP-S4}, \ref{eq:quark-corr-Dn-CP}). For example the ranges of the CP violation $\delta_{\mathrm{CP}}$ obtained from Eq.~\eqref{eq:solar-sum-rule} are summarized in table~\ref{tab:deltaCP-ranges-solar-sum-rule}. The predictions of some typical flavor models are displayed in figure~\ref{fig:bf-predictions-models-Snowmass}. One can see that a precise measurement of the lepton mixing parameters at forthcoming neutrino facilities JUNO, DUNE and T2HK should provide a test of flavor models, see also figure~\ref{fig:chi2-vs-deltaCP}. On the other hand, neutrino mass sum rules are generally parameterized as Eq.~\eqref{eq:parametrization-numass-SR}, providing another type of correlation. It relates the three complex light neutrino mass eigenvalues amongst each other and thus also relate the Majorana CP phases to the neutrino masses. The mass sum rules lead to strong restrictions on the lightest neutrino mass and to distinct predictions for the effective mass $|m_{\beta\beta}|$ probed in $0\nu\beta\beta$ decay, as shown in figure~\ref{fig:mbb-sum-rules}. Finally, we commented on a recently developed toolkit to contrast flavor models with upcoming oscillation experiments.

So far we have confined ourselves to model-independent approaches to the flavour puzzle, in which the underlying theory is unspecified and only the predictive power of symmetry is explored. As a next step we turned to various UV-complete approaches to the flavour puzzle. In Sect.~\ref{sec:benchmark-models} we gave two benchmarks for UV-complete constructions in 4 dimensions. The first incorporates dark matter in a ``scotogenic'' manner, Fig.~\ref{fig:nu-loop} together with a successful family symmetry leading to the so-called trimaximal pattern TM2. The neutrino oscillation predictions are illustrated in Figs.~\ref{fig:osc-predictions-I} and \ref{fig:osc-predictions-III}. An alternative example given puts together family and CP symmetry within the same construction.

More ambitious approaches to the flavor problem have been proposed within extra spacetime dimensions. For example, in Sect.~\ref{sec:family-symmetry-from} we described a five-dimensional warped flavordynamics scenario in which mass hierarchies are accounted for by adequate choices of the bulk mass parameters, while quark and lepton mixing angles are restricted by the imposition of a family symmetry. We presented a $T^\prime$ model leading to Majorana neutrinos, with the TM1 mixing pattern and tight neutrino oscillation correlations, given in Fig.~\ref{fig:deltaCP-contour-EXD}. The \znbb decay rates lie within the sensitivities of the next round of experiments, as indicated in Fig.~\ref{fig:znbb-EXD-Tp}. Finally, one has a good global fit of all flavour observables, including quarks, see table~\ref{Tab:fitted_fermion_parameters}.

Furthermore, in Sect.~\ref{sec:family-symmetry-cp} we described orbifold compactification as a promising way to determine the structure of the family symmetry in four dimensions. The construction is illustrated in Fig.~\ref{fig:OrbifoldConstruction}. We illustrated the idea with a benchmark 6-dimensional scotogenic (see Fig.~\ref{fig:ScotoLoop}) orbifold scenario, in which a 4-dimensional $A_4$ flavour group emerges from the symmetries between the branes in extra dimensions. Predictions include the ``golden'' quark-lepton mass relation, Eq.~\eqref{eq:golden}, and a very good global description of all flavour observables, including quarks, as summarized in Table~\ref{tab:fit}. Concerning neutrino oscillations, the mass ordering and atmospheric octant are predicted, together with the reactor angle, see Fig.~\ref{fig:Lepton-angles-phase}. The lightest neutrino mass can be probed in neutrinoless double beta decay searches as well as cosmology, Fig.~\ref{fig:neutrinoless1}.

We have also stressed that adequate vacuum alignment is required in flavour symmetry models. Generally one must introduce additional large shaping symmetries and many new fields so as to cleverly design the flavon potential needed to obtain the correct vacuum alignment. In Sec.~\ref{sec:modular-symmetry} we discussed how modular symmetry provides an interesting way to overcome this drawback~\cite{Feruglio:2017spp}. In this case the role of flavor symmetry is played by the modular invariance, and the complex modulus $\tau$ could be the unique source of modular symmetry breaking. Modular invariance requires the Yukawa couplings to be modular forms, thus a small number of free parameters are involved in modular models. Such modular symmetry approach has been comprehensively reviewed elsewhere~\cite{Feruglio:2019ybq,Kobayashi:2023zzc,Ding:2023htn} within the bottom-up as well as top-down approaches.

Finally recent progress in the formulation of flavor models involving the use of modular symmetry is discussed in Sec.~\ref{sec:modular-symmetry}. This idea opens the door to a simpler description and deeper understanding of the flavor symmetry breaking required for obtaining viable flavor predictions without the need to invoke {\it flavons} in an {\it ad hoc} manner.

All in all, the legacy of the oscillation program over the past two decades has been a tremendous progress in the field, bringing neutrinos to the center of the particle physics stage. Indeed, addressing the dynamical origin of small neutrino masses touches the heart of the electroweak theory, such as the consistency of symmetry breaking. Moreover, the precise measurement of the neutrino mixing parameters could shed light into the flavor problem. One might have expected that this would bring a decisive boost towards the formulation of a comprehensive theory of fermion masses and mixings. It is somewhat frustrating, however, that so far no decisive flavor road map has emerged emerged. One can reproduce the observations in many different ways, within a wide range of models that go all the way from anarchy to discrete family symmetries. While the latter seems intellectually more appealing, we have not yet been able to underpin a convincing final theory of flavor. Despite many interesting ideas and the formulation of a plethora elegant models, the structure of the three families of fermions remains mysterious.

From the experimental viewpoint in the coming decade we expect a vibrant period for oscillation studies, within and beyond the minimum paradigm. Current neutrino facilities as well as future ones, such as JUNO, DUNE and T2HK should be capable of measuring the solar angle $\theta_{12}$, the atmospheric angle $\theta_{23}$ and the Dirac CP phase $\delta_{CP}$ with high sensitivity. The next generation of ton-scale $0\nu\beta\beta$ decay experiments will probe the Majorana nature of neutrinos, exploring the whole region associated with the inverted ordering spectrum. These experiments should provide important insights into the mysteries behind flavor mixing, fermion mass hierarchies and CP violation.

\appendix
\counterwithout{table}{section}

\clearpage

\section{The $A_4$ group }                                                 
\label{sec:group-theory-a_4}                                               

$A_4$ is the even permutation group of four objects, and it is isomorphic to the symmetry group of a regular tetrahedron. The $A_4$ group can be generated by two generators $s$ and $t$ which satisfy the following multiplication rules~\footnote{We use small $s$, $t$, $u$ for the generators of $A_4$ and $S_4$ flavour groups in order to avoid confusion with oblique parameters $S$, $T$, $U$.},
\begin{equation}
\label{eq:A4-rule}s^2=t^3=(st)^3=1\,.
\end{equation}
The 12 elements of $A_4$ belong to four conjugacy classes:
\begin{eqnarray}
\nonumber &&1C_1=\{1\}\,, \qquad \qquad 3C_2=\{s, tst^2, t^2st\}\,,\\
&&4C_3= \{t, st, ts, sts\}\,, \qquad  4C^{\prime}_3=\{ t^2, st^2, t^2s, st^2s\}\,,
\end{eqnarray}
where the conjugacy class is denoted by $kC_n$, $k$ is the number of elements belonging to it, and the subscript $n$ is the order of the elements contained in it. $A_4$ has a unique Klein group $K^{(s, tst^2)}_4$ and three $Z_3$ subgroups generated by $t$, $st$, $ts$ and $sts$ respectively.
The $A_4$ group has four inequivalent irreducible representations: three singlets $\mathbf{1}$, $\mathbf{1}^{\prime}$, $\mathbf{1}^{\prime\prime}$ and a triplet $\mathbf{3}$. Two different bases are used in the literature, the Ma-Rajasekaran (MR) basis~\cite{Ma:2001dn} and the Altarelli-Feruglio (AF) basis~\cite{Altarelli:2005yx}. They differ in the three-dimensional irreducible representation $\mathbf{3}$, the representation matrix of the generator $t$ is real in the MR basis~\cite{Ma:2001dn} while it is complex and diagonal in AF basis~\cite{Altarelli:2005yx}. The explicit form of the representation matrices are summarized in table~\ref{tab:A4-irr-decomp}. The two bases are related through a unitary transformation,
\begin{equation}
s_{AF}=V^{\dagger} s_{MR} V,~~~~t_{AF}=V^{\dagger} t_{MR} V\,,
\end{equation}
where
\begin{equation}
V=\frac{1}{\sqrt{3}}\begin{pmatrix}
1  ~&  1  ~&  1  \\
1  ~&  \omega  ~&  \omega^2  \\
1  ~&  \omega^2  ~&  \omega
\end{pmatrix}\,.
\end{equation}
If we have two triplets $\alpha\sim (\alpha_1, \alpha_2, \alpha_3)$ and $\beta\sim(b_1,b_2,b_3)$, their product decomposes as the sum
\begin{equation}
\mathbf{3}\otimes\mathbf{3}=\mathbf{1}\oplus\mathbf{1}'\oplus\mathbf{1}''\oplus\mathbf{3}_{\text{s}}\oplus\mathbf{3}_{\text{a}}\,,
\end{equation}
where $\mathbf{3}_{\text{s}}$ and $\mathbf{3}_{\text{a}}$ denote the symmetric and the antisymmetric triplet combinations respectively. The results for the contractions in the above two bases are summarized in table~\ref{tab:A4-irr-decomp}.

\begin{table}[t!]
\centering
\begin{tabular}{|c|c|c|c|c|} \hline \hline
\multicolumn{5}{|c|}{\texttt{Representation matrices of $A_4$ generators } } \\ \hline
  &   \multicolumn{2}{c|}{Ma-Rajasekaran basis}  &   \multicolumn{2}{c|}{Altarelli-Feruglio basis} \\ \cline{2-5}

  &  $s$   &  $t$  &  $s$  &  $t$  \\ \hline

$\mathbf{1}$  &  1  &  1  &  1  &  1\\ \hline

$\mathbf{1}'$  &  1  &  $\omega$ &  1  &  $\omega$  \\ \hline

$\mathbf{1}''$  &  1  &  $\omega^2$ &  1  &  $\omega^2$  \\ \hline

$\mathbf{3}$  &  $\begin{pmatrix}
1~&0~&0\\
0~&-1~&0\\
0~&0~&-1
\end{pmatrix} $   &  $\begin{pmatrix}
0~&1~&0\\
0~&0~&1\\
1~&0~&0
\end{pmatrix} $   &   $\frac{1}{3}
\begin{pmatrix}
-1~&2~&2\\
2~&-1~&2\\
2~&2~&-1
\end{pmatrix} $  &  $\begin{pmatrix}
1~&0~&0\\
0~&\omega~&0\\
0~&0~&\omega^2
\end{pmatrix} $  \rule[-4ex]{0pt}{10ex} \\ \hline \hline
\multicolumn{5}{|c|}{\texttt{Tensor products of two $A_4$ triplets} } \\ \hline
& \multicolumn{2}{c|}{Ma-Rajasekaran basis} & \multicolumn{2}{c|}{Altarelli-Feruglio basis}\\
\hline
$(\alpha\otimes\beta)_{\mathbf{1}}$ & \multicolumn{2}{c|}{$\alpha_1\beta_1+\alpha_2\beta_2+\alpha_3\beta_3$} & \multicolumn{2}{c|}{$\alpha_1\beta_1+\alpha_2\beta_3+\alpha_3\beta_2$}\rule[-2ex]{0pt}{5ex}\\
\hline
$(\alpha\otimes\beta)_{\mathbf{1}'}$ & \multicolumn{2}{c|}{$\alpha_1\beta_1+\omega^2\alpha_2\beta_2+\omega\alpha_3\beta_3$} & \multicolumn{2}{c|}{$\alpha_3\beta_3+\alpha_1\beta_2+\alpha_2\beta_1$} \rule[-2ex]{0pt}{5ex}\\
\hline
$(\alpha\otimes\beta)_{\mathbf{1}''}$ & \multicolumn{2}{c|}{$\alpha_1\beta_1+\omega\alpha_2\beta_2+\omega^2\alpha_3\beta_3$} & \multicolumn{2}{c|}{$\alpha_2\beta_2+\alpha_3\beta_1+\alpha_1\beta_3$} \rule[-2ex]{0pt}{5ex}\\
\hline
$(\alpha\otimes\beta)_{\mathbf{3}_{\text{s}}}$ & \multicolumn{2}{c|}{$\left(
\begin{array}{c}\alpha_2\beta_3+\alpha_3\beta_2\\ \alpha_3\beta_1+\alpha_1\beta_3\\ \alpha_1\beta_2+\alpha_2\beta_1\end{array}\right)$} & \multicolumn{2}{c|}{$\left(
\begin{array}{c}2\alpha_1\beta_1-\alpha_2\beta_3-\alpha_3\beta_2\\2\alpha_3\beta_3-\alpha_1\beta_2-\alpha_2\beta_1 \\ 2\alpha_2\beta_2-\alpha_3\beta_1-\alpha_1\beta_3 \end{array}\right)$} \rule[-4ex]{0pt}{10ex}\\
\hline
$(\alpha\otimes\beta)_{\mathbf{3}_{\text{a}}}$ & \multicolumn{2}{c|}{$\left(
\begin{array}{c}\alpha_2\beta_3-\alpha_3\beta_2\\ \alpha_3\beta_1-\alpha_1\beta_3\\ \alpha_1\beta_2-\alpha_2\beta_1\end{array}\right)$} & \multicolumn{2}{c|}{$\left(
\begin{array}{c}\alpha_2\beta_3-\alpha_3\beta_2\\ \alpha_1\beta_2-\alpha_2\beta_1 \\ \alpha_3\beta_1-\alpha_1\beta_3 \end{array}\right)$} \rule[-4ex]{0pt}{10ex}\\
\hline \hline
\end{tabular}
\caption{\label{tab:A4-irr-decomp}The representation matrices of the $A_4$ generators $s$ and $t$ in the different irreducible representations. We also give the tensor product rule of two $A_4$ triplets $\alpha=(\alpha_1,\alpha_2,\alpha_3)\sim\mathbf{3}$ and $\beta=(\beta_1,\beta_2,\beta_3)\sim\mathbf{3}$, where $\omega=e^{i2\pi/3}=-1/2+i \sqrt{3}/2$ is a cubic root of unity.}
\end{table}

\clearpage

\section{The $S_4$ group }                                                  %
\label{sec:ap_S4Group}                                                      %

$S_4$ is the permutation group of four distinct objects; geometrically it is the symmetry group of a regular octahedron.
Its generators $s$, $t$ and $u$ obey the following multiplication rules~\cite{Hagedorn:2010th,Ding:2013hpa,Lu:2016jit},
\begin{equation}
s^2=t^3=u^2=(st)^3=(su)^2=(tu)^2=(stu)^4=1\,.
\end{equation}
Note that the generators $s$ and $t$ alone generate the group $A_4$, while the generators $t$ and $u$ alone generate the group $S_3$.
The $S_4$ group elements can be divided into 5 conjugacy classes
\begin{eqnarray}
\nonumber 1C_1 &=& \left\{1 \right\}, \\
\nonumber 3C_2 &=& \left\{s, tst^2, t^2st \right\},  \\
\nonumber 6C_2^\prime &=& \left\{u, tu, su, ut, stsu, st^2su\right\},  \\
\nonumber 8C_3 &=& \left\{t, st, ts, sts, t^2, st^2, t^2s, st^2s\right\},  \\
6C_4 &=& \left\{stu,tsu,t^2su,st^2u,tst^2u,t^2stu\right\}\,,
\end{eqnarray}
The group structure of $S_4$ has been studied in detail in~\cite{Ding:2009iy}, it has thirty proper subgroups of orders 1, 2, 3, 4, 6, 8, 12 or 24. Since the number of inequivalent irreducible representations is equal to the number of conjugacy classes and the sum of the squares of the dimensions of the irreducible representations must be equal to the order of the group, it is easy to see that $S_4$ has two singlet irreducible representations $\mathbf{1}$ and $\mathbf{1^{\prime}}$, one doublet representation $\mathbf{2}$ and two triplet representations $\mathbf{3}$ and $\mathbf{3^{\prime}}$. In the singlet representations $\mathbf{1}$ and $\mathbf{1}'$, we have
\begin{eqnarray}
\nonumber&&\mathbf{1}~:~ s=t=u=1\,,\\
&&\mathbf{1}'~:~ s=t=1, \quad u=-1\,.
\label{eq:S4-rep-singlet}
\end{eqnarray}
For the doublet representation $\mathbf{2}$, the generators are represented by
\begin{equation}
\label{eq:S4-rep-doublet}s=\left(\begin{array}{cc}
1&~0 \\
0&~1
\end{array} \right),\quad
t=\left( \begin{array}{cc}
\omega&~0 \\
0&~\omega^2
\end{array} \right),\quad
u=\left(\begin{array}{cc}
0&~1 \\
1&~0
\end{array} \right)\,,
\end{equation}
with $\omega=e^{2\pi i/3}$. In the triplet representations $\mathbf{3}$ and $\mathbf{3}'$, the generators are
\begin{eqnarray}
\nonumber&&\hskip-0.25in\mathbf{3}: s=\frac{1}{3}\left(\begin{array}{ccc}
-1&~ 2  ~& 2  \\
2  &~ -1  ~& 2 \\
2 &~ 2 ~& -1
\end{array}\right),~~ t=\left( \begin{array}{ccc}
1 &~ 0 ~& 0 \\
0 &~ \omega^{2} ~& 0 \\
0 &~ 0 ~& \omega
\end{array}\right),~~ u=-\left( \begin{array}{ccc}
1 &~ 0 ~& 0 \\
0 &~ 0 ~& 1 \\
0 &~ 1 ~& 0
\end{array}\right)\,,\\
\label{eq:S4-rep-triplet}&&\hskip-0.25in\mathbf{3}': s=\frac{1}{3}\left(\begin{array}{ccc}
-1&~ 2  ~& 2  \\
2  &~ -1  ~& 2 \\
2 &~ 2 ~& -1
\end{array}\right),~~ t=\left( \begin{array}{ccc}
1 &~ 0 ~& 0 \\
0 &~ \omega^{2} ~& 0 \\
0 &~ 0 ~& \omega
\end{array}\right),~~ u=\left( \begin{array}{ccc}
1 &~ 0 ~& 0 \\
0 &~ 0 ~& 1 \\
0 &~ 1 ~& 0
\end{array}\right)\,.
\end{eqnarray}
Notice that the representations $\mathbf{3}$ and $\mathbf{3}'$ differ in the overall sign of the generator $u$. The character of an element is the trace of its representation matrix, then we can straightforwardly obtain the character table of $S_4$, as shown in table~\ref{tab:S4_characher}.
Moreover,  the decompositions of the tensor product of the $S_4$ irreducible representations are as follows,
\begin{table}[t!]
\begin{center}
\begin{tabular}{|c|c|c|c|c|c|} \hline
\texttt{Classes} & $1C_1$  & $3C_2$  & $6C_2^\prime$  & $8C_3$  &  $6C_4$ \\ \hline
$\mathbf{1}$  &  1  &  1  &  1  &  1  &  1  \\
$\mathbf{1}'$  &  1  &  1  &  $-1$  &  1  &  $-1$  \\
$\mathbf{2}$  &  2  &  2  &  0  &  $-1$  &  0  \\
$\mathbf{3}$  &  3  &  $-1$  &  $-1$  &  0  &  1  \\
$\mathbf{3}'$  &  3  &  $-1$  &  1  &  0  &  $-1$  \\ \hline
\end{tabular}
\caption{\label{tab:S4_characher}Character table of the $S_4$ group. }
\end{center}
\end{table}

\begin{eqnarray}
\nonumber && \bf{1}\otimes \mathbf{r}=\mathbf{r},\qquad \bf{1^\prime}\otimes \bf{1^\prime}=\bf{1},\qquad \bf{1^\prime}\otimes\bf{2}=\bf{2},\qquad \bf{1^\prime}\otimes\bf{3}=\bf{3^\prime},\qquad \bf{1^\prime}\otimes\bf{3^\prime}=\bf{3},  \\
\nonumber && \bf{2}\otimes\bf{2}=\bf{1}\oplus\bf{1^\prime}\oplus\bf{2},\qquad \bf{2}\otimes\bf{3}=\bf{2}\otimes\bf{3^\prime}=\bf{3}\otimes\bf{3^\prime},\\
&& \bf{3}\otimes\bf{3}=\bf{3^\prime}\otimes\bf{3^\prime}=\bf{1}\oplus\bf{2}\oplus\bf{3}\oplus\bf{3^\prime},\qquad \bf{3}\otimes\bf{3^\prime}=\bf{1^\prime}\oplus\bf{2}\oplus\bf{3}\oplus\bf{3^\prime}\,,
\end{eqnarray}
where $\mathbf{r}$ stands for any irreducible representation of $S_4$. We proceed to present the Clebsch-Gordan (CG) coefficients in the above basis. The entries of the two multiplets in the tensor product are denoted by $\alpha_i$ and $\beta_i$ respectively. For the product of the singlet $\mathbf{1^{\prime}}$ with a doublet or a triplet, we have~\cite{Hagedorn:2010th,Ding:2013hpa,Lu:2016jit}
\begin{equation}
\bf{1^\prime}\otimes\bf{2}=\bf{2}=\left(\begin{array}{c}
\alpha\beta_1 \\
-\alpha\beta_2
\end{array}\right),~~
\bf{1^\prime}\otimes\bf{3}=\bf{3^\prime} =\left(\begin{array}{c}
\alpha\beta_1  \\
\alpha\beta_2  \\
\alpha\beta_3
\end{array}\right),~~
\bf{1^\prime}\otimes\bf{3^\prime}=\bf{3}=\left(\begin{array}{c}
\alpha\beta_1  \\
\alpha\beta_2  \\
\alpha\beta_3
\end{array}\right)\,.
\end{equation}
The CG coefficients for the products involving the doublet   representation $\mathbf{2}$ are found to be
\begin{eqnarray}
\bf{2}\otimes\bf{2}=\bf{1}\oplus\bf{1^\prime}\oplus\bf{2}, \quad\quad &
\text{with}&\quad \left\{\begin{array}{l}
{\bf1}\;=\alpha_1\beta_2+\alpha_2\beta_1  \\
{\bf 1^\prime}=\alpha_1\beta_2-\alpha_2\beta_1  \\
{\bf2}\;=\left(\begin{array}{c}
\alpha_2\beta_2  \\
\alpha_1\beta_1
\end{array}\right)
\end{array}\right. \\
\bf{2}\otimes\bf{3}=\bf{3}\oplus\bf{3^\prime},\quad\quad &
\text{with}& \quad \left\{\begin{array}{l}
{\bf3}\;=\left(\begin{array}{c}
\alpha_1\beta_2+\alpha_2\beta_3 \\
\alpha_1\beta_3+\alpha_2\beta_1  \\
\alpha_1\beta_1+\alpha_2\beta_2
\end{array}\right) \\[0.35in]
{\bf3^\prime}=\left(\begin{array}{c}
\alpha_1\beta_2-\alpha_2\beta_3 \\
\alpha_1\beta_3-\alpha_2\beta_1  \\
\alpha_1\beta_1-\alpha_2\beta_2
\end{array}\right) \\
\end{array}\right.\\
\bf{2}\otimes\bf{3^\prime}=\bf{3}\oplus\bf{3^\prime}, \quad\quad &
\text{with}&\quad\left\{\begin{array}{l}
\mathbf{3}\;=\left(\begin{array}{c}
\alpha_1\beta_2-\alpha_2\beta_3 \\
\alpha_1\beta_3-\alpha_2\beta_1  \\
\alpha_1\beta_1-\alpha_2\beta_2
\end{array}\right) \\[0.35in]
\mathbf{3^\prime}=\left(\begin{array}{c}
\alpha_1\beta_2+\alpha_2\beta_3 \\
\alpha_1\beta_3+\alpha_2\beta_1  \\
\alpha_1\beta_1+\alpha_2\beta_2
\end{array}\right) \\
\end{array}\right.
\end{eqnarray}
Similarly, for the tensor products among the triplet representations $\mathbf{3}$ and $\mathbf{3^{\prime}}$, we have

\begin{eqnarray}
\bf{3}\otimes\bf{3}=\bf{3^\prime}\otimes\bf{3^\prime}=\bf{1}\oplus\bf{2}\oplus\bf{3}\oplus\bf{3^\prime},
     & \text{with}&
\left\{\begin{array}{l}
{\bf1}\;=\alpha_1\beta_1+\alpha_2\beta_3+\alpha_3\beta_2  \\[0.1in]
{\bf2}\;=\left(\begin{array}{c}
\alpha_2\beta_2+\alpha_1\beta_3+\alpha_3\beta_1  \\
\alpha_3\beta_3+\alpha_1\beta_2+\alpha_2\beta_1
\end{array}\right)\\[0.18in]
{\bf3}\;=\left(\begin{array}{c}
\alpha_2\beta_3-\alpha_3\beta_2  \\
\alpha_1\beta_2-\alpha_2\beta_1  \\
\alpha_3\beta_1-\alpha_1\beta_3
\end{array}\right) \\[0.35in]
{\bf3^\prime}=\left(\begin{array}{c}
2\alpha_1\beta_1-\alpha_2\beta_3-\alpha_3\beta_2  \\
2\alpha_3\beta_3-\alpha_1\beta_2-\alpha_2\beta_1  \\
2\alpha_2\beta_2-\alpha_1\beta_3-\alpha_3\beta_1
\end{array}\right) \\
\end{array}\right.\\
\bf{3}\otimes\bf{3^\prime}=\bf{1^\prime}\oplus\bf{2}\oplus\bf{3}\oplus\bf{3^\prime},
&\text{with}& \left\{\begin{array}{l}
{\bf1^\prime}\;=\alpha_1\beta_1+\alpha_2\beta_3+\alpha_3\beta_2  \\[0.1in]
{\bf2}\;=\left(\begin{array}{c}
\alpha_2\beta_2+\alpha_1\beta_3+\alpha_3\beta_1  \\
-(\alpha_3\beta_3+\alpha_1\beta_2+\alpha_2\beta_1)
\end{array}\right)\\[0.18in]
{\bf3}\;=\left(\begin{array}{c}
2\alpha_1\beta_1- \alpha_2\beta_3-\alpha_3\beta_2  \\
2\alpha_3\beta_3- \alpha_1\beta_2-\alpha_2\beta_1  \\
2\alpha_2\beta_2-\alpha_1\beta_3-\alpha_3\beta_1
\end{array}\right) \\[0.35in]
{\bf3^\prime}=\left(\begin{array}{c}
\alpha_2\beta_3-\alpha_3\beta_2  \\
\alpha_1\beta_2-\alpha_2\beta_1  \\
\alpha_3\beta_1-\alpha_1\beta_3
\end{array}\right) \\
\end{array}\right.
\end{eqnarray}

\clearpage

\section{The dihedral group }                                              
\label{sec:dihedral-Group}                                                 

The dihedral group $D_{n}$ is the symmetry group of an $n$-sided regular polygon, which includes rotations and reflections. For instance, $D_3$ is the symmetry group of the regular triangle, and consequently it is isomorphic to $S_3$. A regular polygon with $n$ sides is invariant under a rotation of multiples of $2\pi/n$ about the origin. If $n$ is odd, each reflection axis connects the midpoint of one side to the opposite vertex. If $n$ is even, there are $n/2$ reflection axes connecting the midpoints of opposite sides and $n/2$ axes of symmetry connecting opposite vertices. Consequently the group $D_{n}$ has $2n$ elements. The group $D_n$ can be generated by a rotation $R$ of order $n$ and a reflection $S$ of order 2, and they satisfy the following multiplication rules
\begin{equation}
R^{n}=S^{2}=(RS)^2=1\,,
\end{equation}
where $R$ denotes a rotation of $2\pi/n$ about the origin, and $S$ is a reflection across $n$ lines through the origin. Notice that $D_{1}$ is $Z_{2}$ group generated by $S$ and $D_{2}$ is isomorphic to $Z_{2}\times Z_{2}$. All the group elements of $D_{n}$ can be expressed as
\begin{equation}
 g=S^{\alpha}R^{\beta}\,
\end{equation}
where $\alpha=0, 1$ and $\beta=0, 1, \dots, n-1$. Then it is straightforward to determine the conjugacy classes of the dihedral group. Depending on whether the group index $n$ is even or odd, the $2n$ group elements of $D_n$ can be classified into three or five types of conjugacy classes.
\begin{itemize}

\item{odd $n$}
\begin{equation}
\begin{aligned}
1C_{1}&=\{1\}\,,\\
2C_{m}^{(\rho)}&=\{R^{\rho},R^{-\rho}\}\,,~~\text{with}~~\rho=1,\dots,\frac{n-1}{2}\,,\\
nC_{2}&=\{S,SR,SR^{2},\dots,SR^{n-1}\}\,,
\end{aligned}
\end{equation}
where $m$ refers to the order of the element $R^{\rho}$.

\item{even $n$}
\begin{equation}
\begin{aligned}
1C_{1}&=\{1\}\,,\\
1C_{2}&=\{R^{n/2}\}\,,\\
2C_{m}^{(\rho)}&=\{R^{\rho},R^{-\rho}\}\,,~~\text{with}~~\rho=1,\dots,\frac{n-2}{2}\,,\\
\frac{n}{2}C_{2}&=\{S,SR^{2},SR^{4},\dots,SR^{n-4},SR^{n-2}\}\,,\\
\frac{n}{2}C_{2}&=\{SR,SR^{3},\dots,SR^{n-3},SR^{n-1}\}\,.
\end{aligned}
\end{equation}
\end{itemize}
 The subgroups of $D_{n}$ are either dihedral groups or cyclic groups, and they are given by
\begin{eqnarray}
\nonumber Z_{j}&=&<R^{\frac{n}{j}}>\,\quad \text{with}~~j|n\,,\\
\nonumber Z_{2}^{(m)}&=&<SR^{m}>\,\quad \text{with}~~m=0,1,\dots,n-1\,,\\
D_{j}^{(m)}&=&<R^{\frac{n}{j}},SR^{m}>\,\quad\text{with}~~j|n,~ m=0,1,\dots,\frac{n}{j}-1\,,
\end{eqnarray}
where the elements inside the angle brackets denote the generators of the subgroups. We see that the total number of dihedral subgroups is the sum of positive divisors of $n$.

The group $D_{n}$ is a subgroup of $SO(2)$, and it has only one-dimensional and two-dimensional irreducible representations.
The representations of $D_{n}$ crucially depend on the value of the group index $n$.
\begin{itemize}
\item{odd $n$}

If $n$ is an odd integer, the group $D_{n}$ has two singlet representations $\mathbf{1}_{i}$ and $\frac{n-1}{2}$ doublet representations $\mathbf{2}_{j}$, where the subscripts $i$ and $j$ take values $i=1, 2$ and $j=1,\dots,\frac{n-1}{2}$. The sum of the squares of the dimensions of the irreducible representations is
\begin{equation}
1^2+1^2+2^{2}\times \frac{n-1}{2}=2n\,,
\end{equation}
which is exactly the number of elements of the $D_n$ group. In the singlet representations, the generators $R$ and $S$ are represented by
\begin{equation}
\mathbf{1}_{1}:~R=S=1\,, \quad \mathbf{1}_{2}:~R=1,~S=-1\,.
\end{equation}
For the doublet representations, we have
\begin{equation}
\mathbf{2}_{j}:~R=\begin{pmatrix} e^{2\pi i\frac{j}{n}} ~&~ 0 \\ 0 ~&~ e^{-2\pi i\frac{j}{n}} \end{pmatrix}\,, \quad
S=\begin{pmatrix} 0 ~&~ 1 \\ 1 ~&~ 0 \end{pmatrix}\,,
\end{equation}
with $j=1,\dots,\frac{n-1}{2}$.

\item{even $n$}

For the case that the index $n$ an even integer, the group $D_{n}$ has four singlet representations $\mathbf{1}_{i}$ with $i=1,2,3,4$ and $\frac{n}{2}-1$ doublet representations $\mathbf{2}_{j}$ with $j=1,\dots,\frac{n}{2}-1$. It is straightforward to obtain the generators $R$ and $S$ in the singlet representations
\begin{equation}
\begin{aligned}
&\mathbf{1}_{1}:~R=S=1\,, \qquad~~ ~~~~\mathbf{1}_{2}:~R=1,~S=-1\,,\\
&\mathbf{1}_{3}:~R=-1,~S=1\,, \qquad \mathbf{1}_{4}:~R=S=-1\,.
\end{aligned}
\end{equation}
The explicit forms of these generators in the irreducible two-dimensional representations are
\begin{equation}
\mathbf{2}_{j}:~R=\left(\begin{array}{cc} e^{2\pi i\frac{j}{n}} & 0 \\ 0 & e^{-2\pi i\frac{j}{n}} \end{array} \right)\,, \quad S=\left( \begin{array}{cc} 0 & 1 \\ 1 & 0 \end{array} \right)\,,
\end{equation}
with $j=1,\dots,\frac{n}{2}-1$. Notice that the doublet representation $\mathbf{2}_{j}$ and the complex conjugate $\bar{\mathbf{2}}_{j}$ are equivalent, and they are related through a similarity transformation, i.e., $R^{*}=VRV^{-1}$ and $S^{*}=VSV^{-1}$ with $V=\begin{pmatrix} 0 & 1 \\ 1 & 0 \end{pmatrix}$. Hence all the doublet representations of $D_n$ are real, although the representation matrix of $R$ is complex in our basis. Moreover, if $a=\left(a_1, a_2\right)^{T}$ is a doublet transforming as $\mathbf{2}_j$, then $\left(a^{*}_2, a^{*}_1\right)^{T}$ transform as $\mathbf{2}_j$ under $D_n$ as well.

\end{itemize}

The $D_n$ group has a class-inverting outer automorphism $\mathfrak{u}$, and its action on the generators $R$ and $S$ is
\begin{equation}
\label{eq:class_inverting_aut_Dn} R\stackrel{\mathfrak{u}}{\longmapsto}R^{-1}\,,\quad S\stackrel{\mathfrak{u}}{\longmapsto}S\,.
\end{equation}
The CP transformation corresponding to $\mathfrak{u}$ is denoted by
$X^0_{\mathbf{r}}$, its concrete form is determined by the following consistency conditions,
\begin{eqnarray}
\nonumber&&X^0_{\mathbf{r}}\rho^{*}_{\mathbf{r}}(R)X^{0\dagger}_{\mathbf{r}}=\rho_{\mathbf{r}}\left(\mathfrak{u}\left(R\right)\right)=\rho_{\mathbf{r}}\left(R^{-1}\right)\,,\\
\label{eq:cons_eq}&&X^0_{\mathbf{r}}\rho^{*}_{\mathbf{r}}(S)X^{0\dagger}_{\mathbf{r}}=\rho_{\mathbf{r}}\left(\mathfrak{u}\left(S\right)\right)=\rho_{\mathbf{r}}\left(S\right)\,.
\end{eqnarray}
Since both $\rho_{\mathbf{r}}(R)$ and $\rho_{\mathbf{r}}(S)$ are symmetric and unitary, $X^0_{\mathbf{r}}$ coincides the canonical CP transformation up to an overall irrelevant phase, i.e.,
\begin{equation}
\label{eq:gcp_trans}X^0_{\mathbf{r}}=\mathbb{1}\,.
\end{equation}
From $X^0_{\mathbf{r}}$ and the flavor symmetry transformations of $D_n$, we can obtain other generalized CP transformations $X_{\mathbf{r}}=\rho_{\mathbf{r}}(g)X^0_{\mathbf{r}}=\rho_{\mathbf{r}}(g),~ g\in D_{n}$, yet they don't impose any new restrictions.

\clearpage

\section{Diagonalization of a $2\times2$ complex symmetric matrix}         
\label{app:diag}                                                           

A generic complex symmetric $2\times2$ matrix can be written as
\begin{eqnarray}
  M&=&\left( \begin{array}{cc}
a       ~ &~ c  \\
c        ~&~ b
\end{array} \right)\,,
\end{eqnarray}
where $a$, $b$ and $~c$ are complex. It can be diagonalized by a  two dimensional unitary matrix $U$ via
\begin{equation}
 U^{T}MU={\rm diag}(\lambda_{1},\lambda_{2}),~~~U=\left( \begin{array}{cc}
{\rm cos}\theta & e^{i\phi}{\rm sin}\theta \\
-e^{-i\phi}{\rm sin}\theta &{\rm cos}\theta
\end{array} \right)
\left( \begin{array}{cc}
e^{-i\alpha} & 0 \\
0 &e^{-i\beta}
\end{array} \right)\,.
\end{equation}
The eigenvalues $\lambda_{1,2}$ are non-negative with
\begin{eqnarray}
\nonumber \lambda_{1}^{2}&=&\frac{1}{2}\left[|a|^{2}+|b|^{2}+2|c|^{2}
-\sqrt{(|b|^{2}-|a|^{2})^{2}+4|a^{\ast}c+bc^{\ast}|^{2}}\right] \\
\lambda_{2}^{2}&=&\frac{1}{2}\left[|a|^{2}+|b|^{2}+2|c|^{2}
+\sqrt{(|b|^{2}-|a|^{2})^{2}+4|a^{\ast}c+bc^{\ast}|^{2}}\right]
\end{eqnarray}
Without loss of generality, the rotation angle $\theta$ can be limited in the region $0\leq\theta\leq\pi/2$ and it satisfies
\begin{equation}
\tan2\theta=\frac{2|a^{\ast}c+bc^{\ast}|}{|b|^{2}-|a|^{2}}\,.
\end{equation}
The expressions of the phases $\phi$, $\alpha$ and $\beta$ are
\begin{eqnarray}
\nonumber\phi&=&\text{arg}(a^{\ast}c+bc^{\ast})\,,\\
\nonumber \alpha&=&\frac{1}{2}\text{arg}\left[a(|b|^{2}-|\lambda_{1}|^{2})-b^{\ast}c^{2}\right]\,,\\
\beta&=&\frac{1}{2}\text{arg}\left[b(|\lambda_{2}|^{2}-|a|^{2})+a^{\ast}c^{2}\right]\,.
\end{eqnarray}
The general case of an arbitrary dimension is given in Ref.~\cite{Schechter:1980gr}.

\clearpage

\section{Global fit of flavour observables of the extra-dimensional   models \label{app:global-fitting} }

In this Appendix, we shall present the details of global fitting to the extra dimensional models in section~\ref{sec:family-symmetry-from}. We adopt the symmetrical parametrization of the quark and lepton mixing matrices~\cite{Schechter:1980gr}, described in Sect.~\ref{sec:introduction}. For the case of quarks, choosing the PDG ordering prescription, the symmetrical form leads to the standard Cabibbo-Kobayashi-Maskawa (CKM) matrix in Eq.~\eqref{eq:CKM}. On the other hand, for the mixing of leptons, it is given by Eq.~\eqref{eq:lepton-mixing}~\footnote{In both cases we assume unitarity, neglecting therefore possible mixing with exotic fermions that can be relevant, say, within a low-scale seesaw scheme~\cite{Escrihuela:2015wra}.}. In both cases the mixing matrix description is supplemented by the PDG factor ordering convention. However, for the leptonic case the symmetrical form provides a much neater description of CP violation than the PDG form.
In this case the Dirac CP phase that enters in neutrino oscillations, i.e. the leptonic analogue of the quark Jarlskog invariant, is identified with the ``rephasing-invariant'' combination $\delta^{\ell}=\phi_{13}-\phi_{12}-\phi_{23}$ given in Eq.~\eqref{eq:dell}.
Notice that this phase must not be present in the effective mass parameter $|m_{\beta\beta}|$ characterizing the amplitude for neutrinoless double beta decay, which involves only the two Majorana phases, as shown in Eq.~\eqref{eq:mbb}. Hence the symmetrical presentation is more transparent for describing the \znbb decay amplitude~\cite{Rodejohann:2011vc}.

The number of free parameters in both extra dimensional models of section~\ref{sec:family-symmetry-from} is less than the number of the flavour observables, consequently these models make strong flavour predictions. In order to explore the predictivity of the models, one performs a global flavour fit to the available experimental data, by minimizing the chi-square function defined as
\begin{equation}
\chi^2=\sum (\mu_{\text{exp}}-\mu_{\text{model}})^2/\sigma^{2}_{\text{exp}},
\end{equation}
where the sum runs through the 19 measured physical parameters, i.e. 6 quark masses, 3 CKM mixing angles, 1 CKM CP phase, 3 charged lepton masses, 3 lepton mixing angles, the Dirac lepton CP violating phase, and 2 neutrino squared mass splittings. Note that we have only limits on the lightest neutrino mass from experiment and no direct information on the Majorana \znbb phases. The fit is performed by scanning the values of all independent model parameters that provide a description of the above 22 flavour observables.

In our chi-square minimization with respect to the independent model parameters, all quark and charged-lepton masses were evaluated at the same energy scale, chosen as $M_Z$~\cite{Antusch:2013jca}. This assumption was shown to be consistent with the so-called golden mass relation discussed in Sec.~\ref{sec:family-symmetry-cp}. Coming to the neutrino oscillation parameters, these were extracted from the global fit in~\cite{deSalas:2020pgw,10.5281/zenodo.4726908}, neglecting the effect of running to $M_Z$~\cite{Antusch:2013jca,Xing:2007fb}. The remaining observables were taken from the PDG~\cite{Workman:2022ynf}. In order to extract the flavour observables from the mass matrices predicted by the models, we use the Mathematica Mixing Parameter Tools package~\cite{Antusch:2005gp}.

\clearpage

\black
\section*{Acknowledgements}

\noindent

This review is based on the fruitful collaboration with many friends and colleagues worldwide. We have also benefited from useful discussions with many others. We express our sincere gratitude to all of them. This work was supported by the Spanish grants PID2020-113775GB-I00 and CEX2023-001292-S (AEI/10.13039/501100011033), as well as CIPROM/2021/054 (Generalitat Valenciana). It also received support from the National Natural Science Foundation of China, under grants number 12375104, 11975224 and 11835013.



\small


\end{document}